\documentclass[a4paper,11pt]{article}
\usepackage{jheppub}
\usepackage{mathrsfs}
\usepackage{amsfonts}
\usepackage{setspace}
\usepackage{cellspace}
\usepackage{amsmath,amssymb,bm}
\usepackage[colorlinks=true,linkcolor=blue]{hyperref}
\usepackage{xcolor}
\usepackage{epsfig}
\usepackage{slashed}
\usepackage{caption}
\usepackage{hhline,multirow,tabularx}  
\usepackage{dcolumn}    
\usepackage{url}        
\usepackage{braket} 
\definecolor{cyan}{rgb}{0.0, 1.0, 1.0}
\definecolor{applegreen}{rgb}{0.55, 0.71, 0.0}
\definecolor{arylideyellow}{rgb}{0.91, 0.84, 0.42}
\definecolor{bananayellow}{rgb}{1.0, 0.88, 0.21}
\definecolor{burlywood}{rgb}{0.87, 0.72, 0.53}
\definecolor{buff}{rgb}{0.94, 0.86, 0.51}
\definecolor{blond}{rgb}{0.98, 0.94, 0.75}
\definecolor{bisque}{rgb}{1.0, 0.89, 0.77}
\definecolor{bananamania}{rgb}{0.98, 0.91, 0.71}
\definecolor{apricot}{rgb}{0.98, 0.81, 0.69}
\definecolor{almond}{rgb}{0.94, 0.87, 0.8}
\usepackage{rotating}
\usepackage{float}
\usepackage{aliascnt}
\newaliascnt{eqfloat}{equation}
\newfloat{eqfloat}{h}{eqflts}
\floatname{eqfloat}{Equation}

\newcommand*{\ORGeqfloat}{}
\let\ORGeqfloat\eqfloat
\def\eqfloat{%
  \let\ORIGINALcaption\caption
  \def\caption{%
    \addtocounter{equation}{-1}%
    \ORIGINALcaption
  }%
  \ORGeqfloat
}

\title{Effects of threshold resummation for large-$x$ PDF in large momentum effective theory}
\author[a]{Xiangdong Ji}
\author[b]{Yizhuang Liu}
\author[a]{Yushan Su}
\author[c]{Rui Zhang}

\affiliation[a]{Department of Physics, University of Maryland, College Park, MD 20742, USA}
\affiliation[b]{Institute of Theoretical Physics,
Jagiellonian University, 30-348 Kraków, Poland}
\affiliation[c]{Physics Division, Argonne National Laboratory, Lemont, IL 60439, USA}

\emailAdd{xji@umd.edu}
\emailAdd{yizhuang.liu@uj.edu.pl}
\emailAdd{ysu12345@umd.edu}
\emailAdd{ruizhang@anl.gov}

\abstract{Parton distribution functions (PDFs) at large $x$ are challenging to extract from experimental data, yet they are essential for understanding hadron structure and searching for new physics beyond the Standard Model. Within the framework of the large momentum $P^z$ expansion of lattice quasi-PDFs, we investigate large $x$ PDFs, where the matching coefficient is factorized into the hard kernel, related to the active quark momentum $x P^z$, and the threshold soft function, associated with the spectator momentum $(1-x) P^z$. The renormalization group equation of the soft function enables the resummation of the threshold double logarithms $\alpha^{k} \ln^{2k}(1-x)$, which is crucial for a reliable and controllable calculation of large $x$ PDFs. Our analysis with pion valence PDFs indicates that perturbative matching breaks down when the spectator momentum $(1-x)P^z$ approaches $\Lambda_{\rm QCD}$, but remains valid when both $x P^z$ and $(1-x)P^z$ are much larger than $\Lambda_{\rm QCD}$. Additionally, we incorporate leading renormalon resummation within the threshold framework, demonstrating good perturbative convergence in the region where both spectator and active quark momenta are perturbative scales. }

\date{\today}

\begin{document}
\maketitle
\flushbottom
\section{Introduction}
Parton distribution functions (PDFs) are universal objects for understanding the dynamics of high-energy collisions and the internal structures of hadron. They can be extracted through the global fit with experimental data, such as ATLASpdf21~\cite{ATLAS:2021vod}, NNPDF4.0~\cite{NNPDF:2021njg}, MSHT20~\cite{Bailey:2020ooq}, CT18~\cite{Hou:2019efy}, ABMP16~\cite{Alekhin:2017kpj}, HERAPDF2.0~\cite{H1:2015ubc} and JR14~\cite{Jimenez-Delgado:2014twa}. 

PDFs at large $x$ are important in understanding the properties of strong interaction, including isospin dependence~\cite{Farrar:1975yb,Brodsky:1994kg,Isgur:1998yb,Afnan:2003vh,Tropiano:2018quk,JeffersonLabHallATritium:2021usd}, EMC effect~\cite{EuropeanMuon:1983wih,Ke:2023xeo}, color confinement~\cite{Holt:2010vj} and spin structure~\cite{JeffersonLabHallA:2016neg,STAR:2019yqm,Friscic:2021oti,Lagerquist:2022tml}, as well as searching for new physics beyond the Standard Model~\cite{Kuhlmann:1999sf,CMS:2012ftr,ATLAS:2011juz,Brady:2011hb}. For example, the ratio of neutron to proton structure functions is expected to be 2/3 according to SU(6) symmetry~\cite{Afnan:2003vh} while the data have been known to deviate from 2/3 at large $x$, which indicates nontrivial dynamical effects at large $x$ that break SU(6) symmetry. Another example is a jet event excess detected in~\cite{CDF:1996yow}, which was later explained by a larger than expected gluon distribution at large $x$~\cite{Huston:1995tw}. 

Despite their importance, the extraction of large $x$ PDFs from experimental data suffers from large uncertainties (e.g. $5\sim10\%$ relative uncertainties for $x \gtrsim 0.7$ for quark PDFs). One reason is due to a lack of experimental data at large $x$, and the data that do exist often come with large errors. This issue stems from the fact that PDFs tend to zero as $x \rightarrow 1$, which leads to lower luminosity at high $x$, reducing the precision of measurements in this region. Another source of uncertainty is the presence of threshold logarithms, which take the form $\alpha^{k} \ln^{2k}(1-x)$, in the perturbative hard kernels used to match PDFs to experimental structure functions, such as $C_2$ in the deeply inelastic scattering (DIS) factorization: $F_2 = C_2 \otimes f$. Near the endpoint region where $x \rightarrow 1$, the threshold logarithms become large due to incomplete infrared (IR) cancellation between the real and virtual diagrams, and thus, they have significant influence in this region. One needs to resum these large threshold logarithms to make controlled predictions. Many works have been done in proposing a theoretical framework to resum the threshold logarithms~\cite{Sterman:1986aj,Catani:1989ne,Bauer:2002nz,Manohar:2003vb,Pecjak:2005uh,Chay:2005rz,Idilbi:2006dg,Chen:2006vd,Becher:2006mr,Bonvini:2012az,Bauer:2000yr,Bauer:2001yt,Becher:2014oda,Becher:2007ty,Das:2019btv,Das:2020adl,AH:2020cok,Das:2023bfi,Banerjee:2017cfc,Banerjee:2018vvb,Moch:2005ba} in Mellin or momentum space for various experimental processes. The threshold resummation has been regularly applied to global fittings of PDFs to experimental data~\cite{Corcella:2005us,Aicher:2010cb,Bonvini:2015ira,Westmark:2017uig,Barry:2021osv}. 

The large $x$ PDFs can also be calculated using Large Momentum Effective Theory (LaMET)~\cite{Ji:2013dva,Ji:2014gla,Ji:2024oka}, which is a systematic approach to access the parton physics through large momentum expansion of the Euclidean observables, calculable from the first principles of quantum chromodynamics (QCD), such as lattice QCD. Since its proposal, LaMET has a wide range of applications, including calculations of quark distribution functions~\cite{Xiong:2013bka,Lin:2014zya,Alexandrou:2015rja,Chen:2016utp,Alexandrou:2016jqi,Alexandrou:2018pbm,Chen:2018xof,Lin:2018pvv,LatticeParton:2018gjr,Alexandrou:2018eet,Liu:2018hxv,Chen:2018fwa,Izubuchi:2018srq,Izubuchi:2019lyk,Shugert:2020tgq,Chai:2020nxw,Lin:2020ssv,Fan:2020nzz,Gao:2021hxl,Gao:2021dbh,Gao:2022iex,Su:2022fiu,LatticeParton:2022xsd,Gao:2022uhg,Chou:2022drv,Gao:2023lny,Gao:2023ktu,Chen:2024rgi,Holligan:2024umc,Holligan:2024wpv}, gluon distribution functions~\cite{Fan:2018dxu,Good:2024iur}, generalized parton distributions~\cite{Chen:2019lcm,Alexandrou:2019dax,Lin:2020rxa,Alexandrou:2020zbe,Lin:2021brq,Scapellato:2022mai,Bhattacharya:2022aob,Bhattacharya:2023nmv,Bhattacharya:2023jsc,Lin:2023gxz,Holligan:2023jqh,Ding:2024hkz}, distribution amplitudes~\cite{Zhang:2017bzy,Chen:2017gck,Zhang:2020gaj,Hua:2020gnw, LatticeParton:2022zqc,Gao:2022vyh,Xu:2022guw,Holligan:2023rex,Deng:2023csv,Han:2023xbl,Han:2023hgy,Han:2024min,Han:2024ucv,Baker:2024zcd,Cloet:2024vbv,Han:2024cht,Deng:2024dkd}, transverse-momentum-dependent distributions~\cite{Ji:2014hxa,Shanahan:2019zcq,Shanahan:2020zxr,Zhang:2020dbb,Ji:2021znw,LatticePartonLPC:2022eev,Liu:2022nnk,Zhang:2022xuw,Deng:2022gzi,Zhu:2022bja,LatticePartonCollaborationLPC:2022myp,Rodini:2022wic,Shu:2023cot,LatticeParton:2023xdl,delRio:2023pse,LatticePartonLPC:2023pdv,Alexandrou:2023ucc,Avkhadiev:2023poz,Zhao:2023ptv,Avkhadiev:2024mgd,Bollweg:2024zet,Spanoudes:2024kpb}, and double parton distribution functions~\cite{Zhang:2023wea,Jaarsma:2023woo}. Recent reviews on LaMET can be found in Refs.~\cite{Cichy:2018mum,Ji:2020ect}. Compared to global fits, LaMET offers several advantages for extracting large $x$ PDFs. First, LaMET is a model-independent approach while global fits depend on specific parametrizations of PDF, such as $\sim x^{\alpha} (1-x)^{\beta}$. Second, LaMET utilizes the full range of data from lattice QCD, addressing the lack of experimental data at large $x$. 

However, the quasi-PDF matching kernels calculated in LaMET also contain threshold logarithms. To achieve controlled predictions of perturbative matching at large $x$, it is required to resum them. Initial studies focused on resumming the next-to-leading logarithms in Mellin space~\cite{Gao:2021hxl}, which demonstrated that threshold logarithms arise from the leading divergences in the 3D momentum distribution as $x \rightarrow 1$. Their work avoids the Landau pole, and indicates that the effects of threshold logarithms are marginal for the accessible moments when applied to lattice data. 
More recently, the factorization and resummation valid at the leading $(1-x)$ expansion in both the momentum space and the coordinate space were worked out to NNLO in Ref.~\cite{Ji:2023pba}. Underlying this factorization scheme is the scale separation $x P^z \gg (1-x) P^z \gg \Lambda_{\rm QCD}$, where $x P^z$ is the active quark momentum and $(1-x) P^z$ is the spectator momentum. The matching kernel is factorized into the heavy-light Sudakov hard kernel\footnote{It is the same hard kernel for quasi-TMD factorizations~\cite{Ji:2019ewn,Ji:2021znw}.}, concerning the hard scale $x P^z$, and the threshold soft function\footnote{This is called ``jet function" in Ref.~\cite{Ji:2023pba}.}, related to the semi-hard scale $(1-x) P^z$. Using the renormalization group (RG) equation for the soft function, one can resum the threshold logarithms of the form $\alpha^{k} \ln^{2k}(1-x)$. This momentum-space threshold resummation (TR) formalism has been applied to extract distribution amplitudes from lattice data~\cite{Baker:2024zcd, Cloet:2024vbv}. The factorization scheme in Ref.~\cite{Ji:2023pba} can also be converted to the Mellin space and obtain fully factorized asymptotic expansions~\cite{Liu:2023onm}. 

This work extends and complements Ref.~\cite{Ji:2023pba} in several key aspects. 
First, we check the threshold asymptotics for quark-nonsinglet light-cone PDF and quasi-PDF for an on-shell external quark state with finite quark mass $m$ acting as a regulator for IR collinear divergences. Based on this, we propose the IR-free definitions for the matching coefficients by including light-cone matrix elements that are scaleless and identical to $1$ when calculated purely in dimensional regularization~\cite{Ji:2023pba,Liu:2023onm}. This clarifies the physics of collinear and soft parts
in the matrix elements.
Second, we explore the effects of threshold resummation using the pion valence PDF, which reveals that perturbative matching breaks down when $(1-x)P^z \sim \Lambda_{\rm QCD}$, but remains valid when both $ (1-x) P^z $ and $ x P^z $ are much larger than $\Lambda_{\rm QCD}$.
Third, we implement leading renormalon resummation (LRR) within the threshold resummation framework, demonstrating good perturbative convergence. Our formalism is expected to achieve the leading (linear) power accuracy~\cite{Zhang:2023bxs}. 

The rest of the paper is organized as follows. In Sec.~\ref{sec:theory}, we present the theoretical framework for the threshold resummation. In Sec.~\ref{sec:pionPDF}, we perform the numerical tests on the pion valence PDF case. In Sec.~\ref{sec:renormalon}, we implement the threshold resummation together with the leading renormalon resummation. We conclude in Sec.~\ref{sec:conclu}. Some technical details are collected in the appendices.

\section{Formalism for resumming spectator large logarithms}\label{sec:theory}

In this section, we provide the theoretical formalism to perform the threshold resummation for the quasi-PDF matching kernel under LaMET. We then discuss the physical scale choices for the resummed form, with particular emphasis on the choice of the semi-hard scale. Finally, we address the combination of the threshold-resummed kernel with the low-order fixed-order result to improve perturbative accuracy.

\subsection{LaMET expansion}\label{sec:LaMexp}
Our goal is to perform the first-principle calculation of the light-cone PDF $f(x,\mu)$, defined with a quark-bilinear light-like correlator sandwiched by the hadron states,
\begin{align}\label{eq:LCpdf}
    f(x,\mu) \equiv \int_{-\infty}^{+\infty} \frac{d \xi^{-}}{4\pi} e^{- i x P^{+} \xi^{-}} 
    \langle H(P) | \bar{\psi}(\xi^{-}) \gamma^{+} W_{\bar{n}}(\xi^{-},0) \psi(0) | H(P) \rangle \ ,
\end{align}
where $P^{\mu} = (P^{t},0,0,P^{z})$ denotes the hadron momentum, and $\xi^{\mu}=(t,0,0,z)$ is the spacetime coordinate. Our conventions for the light-cone vectors are $n = (1,0,0,1)/\sqrt{2}$ and $\bar{n} = (1,0,0,-1)/\sqrt{2}$ in Cartesian coordinates. $n^{\perp}$ is used to denote the transverse direction. We denote $P^{+} = P\cdot \bar{n} = (P^{t}+P^{z})/\sqrt{2}$ and $\xi^{-} = \xi \cdot n = (t-z)/\sqrt{2}$. The lightlike gauge link is defined as
\begin{align}
    W_{\bar{n}}(\xi^{-},0) \equiv \mathcal{P} \exp\left[-i g \int_{0}^{\xi^{-}} d \eta^{-} A^{+}(\eta^{-})\right] \ ,
\end{align}
where the variable ``0" is the starting point, the subscript $\bar{n}$ denotes the direction of the gauge link, and ``$\xi^{-} \bar{n}$" is the end point. $\mathcal{P}$ is a path-ordering operator. The light-cone PDF is independent of the external momentum $P^{\mu}$ but depends on renormalization scale $\mu$. The Dokshitzer-Gribov-Lipatov-Altarelli-Parisi (DGLAP) $\mu$ evolution has been calculated up to three-loop~\cite{Moch:2004pa,Vogt:2004mw,Blumlein:2021enk} and known approximately at four-loop accuracy~\cite{Moch:2017uml,Falcioni:2023vqq,Falcioni:2023luc,Moch:2023tdj,Moch:2021qrk,Falcioni:2023tzp,Gehrmann:2023cqm,Gehrmann:2023iah,Falcioni:2024qpd}. 

The light-cone correlator cannot be directly simulated on a Euclidean lattice. However, it has been argued that it can be approximated by an equal-time correlator with a large Lorentz boost, and the latter
can be calculated in lattice QCD~\cite{Ji:2013dva,Ji:2014gla,Ji:2020ect,Ji:2024oka}. The boost
can be accomplished through a large momentum $P^z$ hadron
state. The result is a physical longitudinal momentum distribution, referred to as the quasi-PDF in the literature,
\begin{align}\label{eq:qpdf}
    \tilde f(y,P^z,\mu) \equiv \int_{-\infty}^{+\infty} \frac{d z}{4\pi} e^{i y P^z z} 
    \langle H(P) | \bar{\psi}(z) \gamma^{t} W_{n_{z}}(z,0) \psi(0) | H(P) \rangle \ ,
\end{align}
where $P^z \gg \Lambda_{\rm QCD}$, and the gauge link is along a spacelike direction $n_z = (0,0,0,1)$,
\begin{align}
    W_{n_{z}}(z, 0) \equiv \mathcal{P} \exp\left[i g \int_{0}^{z} d z' A^{z}(z')\right] \ .
\end{align}
An all-order reduced diagram analysis showing that Eqs.~(\ref{eq:LCpdf}) and~(\ref{eq:qpdf}) share the same IR/hadron structure physics can be found in Ref.~\cite{Ma:2014jla, Ji:2020ect}. However, they are different in perturbative ultra-violet (UV) physics. 

The exact relation between light-cone PDF and quasi-PDF can be obtained through the large momentum expansion~\cite{Ji:2024oka},
\begin{align}\label{eq:LaMETmatching}
f(x,\mu) = \int_{-\infty}^{+\infty} \frac{d y}{|y|} {\cal C}\left(\frac{x}{y},\frac{|y| P^z}{\mu}\right) \tilde f(y,P^z,\mu) + ... \ ,
\end{align}
where the matching kernel ${\cal C}\bigg(\frac{x}{y},\frac{|y| P^z}{\mu}\bigg)$ is free of IR physics and perturbatively calculable. The matching relation is valid at the operator level, and the matching kernels remain independent of the external states, provided these states are large-momentum on-shell states in the desired polarization. Consequently, the matching kernel can be determined using quark matrix elements. The one-loop result can be found in Refs.~\cite{Xiong:2013bka,Izubuchi:2018srq,Ji:2020ect}, and the two-loop result is available in Refs.~\cite{Li:2020xml,Chen:2020ody}. The three-loop result in Mellin space is presented in Ref.~\cite{Cheng:2024wyu}. The power corrections denoted by $...$ depend on whether the linear renormalons are consistently regularized, which will be discussed in Sec.~\ref{sec:renormalon}.

The above relation has a factorization interpretation. The PDF $ f(x,\mu) $ contains the renormalization scale $ \mu $ for the light-cone UV divergence and the non-perturbative scale $ \Lambda_{\rm QCD} $ for the collinear mode. The physics associated with these two scales is factorized into the expansion kernel $ {\cal C}\left(\frac{x}{y},\frac{|y| P^z}{\mu}\right) $ and the quasi-PDF $ \tilde{f}(y,P^z,\mu) $. The hadron momentum $ P^z $ acts like a ``factorization scale" that separates the UV and IR physics, and its evolution effects should cancel out in the factorization formula. The $\mu $-dependence in $ \tilde{f}(y,P^z,\mu) $, which is related to the renormalization of heavy-light quark currents~\cite{Ji:1991pr, Chetyrkin:2003vi, Braun:2020ymy, Grozin:2023dlk} conspires with the $ \mu $-dependence in $ {\cal C}\left(\frac{x}{y},\frac{|y| P^z}{\mu}\right) $, giving rise to DGLAP evolution of the light-cone PDF.

Eq.~(\ref{eq:LaMETmatching}) can also be understood as a scheme conversion for the light-cone UV divergence. In the PDF \( f(x,\mu) \), the light-cone divergence is regulated using dimensional regularization and renormalized in the \(\overline{\rm MS}\) scheme. However, in the quasi-PDF \( \tilde{f}(y,P^z,\mu) \), the finite large momentum \( P^z \) acts as a natural regulator for the light-cone divergence. Thus, the matching kernel \( {\cal C}\left(\frac{x}{y},\frac{|y| P^z}{\mu}\right) \) serves as a conversion factor, translating the finite-momentum scheme to the light-cone scheme.


\subsection{Threshold factorization}
In the matching kernel in Eq.~(\ref{eq:LaMETmatching}), threshold logarithms at fixed orders behave as $\sim \alpha^k \ln^{2k-1}\left(1-x/y\right)$, which significantly impact the large-$x$ PDF. To calculate the latter in a reliable and controlled manner, a systematic approach is needed to resum the threshold logarithms. However, the scale $\mu$ dependence does not appear in these logarithms, preventing their resummation using standard RG equations. To resolve this issue, the threshold factorization is proposed, which effectively splits the threshold logarithm as $\ln\left(1-x/y\right)$ into $\ln\left(\mu/y P^z\right)$ in the hard kernel plus $\ln\left((y-x) P^z/\mu\right)$ in the soft function. This separation introduces new physical quantities and their $\mu$-dependences, enabling the resummation of the threshold logs using the RG equations for these quantities. The matching coefficients for the threshold factorization have been calculated purely under dimensional regularization up to two loops in Ref.~\cite{Ji:2023pba}, where scaleless integrals are set to 0. Here, we provide the IR-free definitions for the matching coefficient by taking those integrals into account in the quark mass IR regulator.

\begin{figure}
    \centering
    \includegraphics[width=0.927\linewidth]{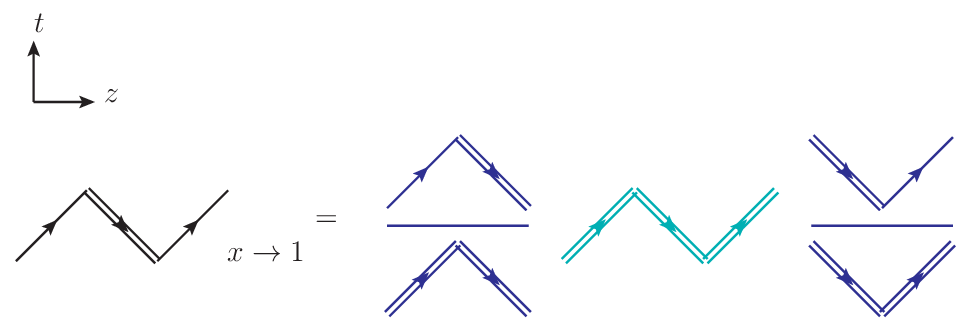}
    \caption{The factorization of the threshold light-cone PDF $q(x, m/\mu)|_{x \rightarrow 1}$ (left hand side) into the light-cone collinear function $J$ (blue) and soft-collinear function $W_{C}$ (green), shown as Eq.~(\ref{eq:TFLCPDF}). The single lines represent the incoming or outgoing quarks moving with momentum $p$. The double lines indicate the gauge links along the quark's velocity $v$ or light-cone $\bar{n}$ directions. For simplicity, perturbative corrections are omitted in this diagram, although they are taken into account in the factorization theorem. }
    \label{fig:LCPDF_TR}
\end{figure}

For readers' convenience, we review the factorization for threshold light-cone PDF itself discussed in Ref.~\cite{Becher:2006mr}. This factorization is derived from the soft-collinear effective theory (SCET), which we will interpret with QCD fields. Consider the light-cone PDF in a free massive on-shell unpolarized quark state,
\begin{align}\label{eq:LCpdfq}
    q\left( x, \frac{m}{\mu} \right) = \int_{-\infty}^{+\infty} \frac{d \xi^{-}}{4\pi} e^{- i x p^{+} \xi^{-}} 
    \langle q(p) | \bar{\psi}(\xi^{-}) \gamma^{+} W_{\bar{n}}(\xi^{-},0) \psi(0) | q(p) \rangle \ ,    
\end{align}
where the quark momentum $p^\mu=(\sqrt{p_z^2+m^2},0,0,p^z)$ in Cartesian coordinate, with $p^z \gg m$.  The threshold light-cone PDF $q\left( x, m/\mu\right) |_{x \rightarrow 1}$ contains two distinct IR modes~\cite{Becher:2006mr} in perturbation theory, in the light-cone coordinate $(+,-,\perp)$ scaling as
\begin{align}\label{eq:trpc}
\textit{\rm collinear mode:} \, &\sim p^+ \left(1, \delta^2, \delta \right) \ , \nonumber\\ 
\textit{\rm soft-collinear mode:} \, &\sim p^+ \epsilon \left(1, \delta^2, \delta \right) \ ,
\end{align}
where $\epsilon \equiv 1-x$, and $\delta p^+ \sim m$ with $\delta \ll 1$\footnote{While $\lambda$ is commonly used as the power counting parameter, we opt for $\delta$ to prevent any potential overlap with the notation for light-cone distance, ensuring clarity in our formulation.}. If $\epsilon \sim 1$, the two modes are indistinguishable, and one obtains the PDF in the standard collinear factorization. In the threshold limit $\epsilon \ll 1$, they are factorized into the collinear function $J$ and soft-collinear function $W_C$,
\begin{align}\label{eq:TFLCPDF}
q\left( x, \frac{m}{\mu} \right) \bigg|_{x \rightarrow 1} 
= \int_{-\infty}^{+\infty} \frac{d \lambda}{2\pi} e^{i (x-1) \lambda} 
J\left(\frac{m}{\mu}\right)
W_C\left(\frac{\lambda\mu}{m}\right) \ ,
\end{align}
where $e^{-i \lambda}$ is the phase factor arising from the location difference between the quark fields, and the dimensionless light-cone distance is $\lambda = - p^{+} \xi^{-}$. An illustration diagram is shown in Fig.~\ref{fig:LCPDF_TR}. 

\begin{figure}
    \centering
    \includegraphics[width=0.927\linewidth]{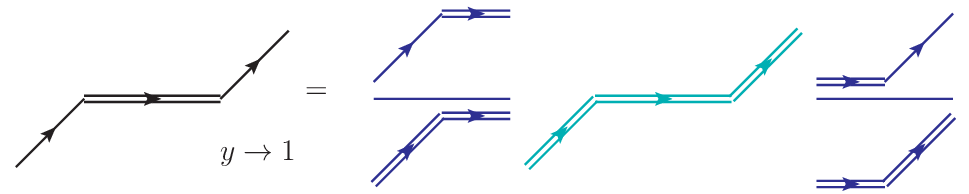}
    \caption{The factorization of the threshold quasi PDF $\tilde{q}(y,p^z/\mu,m/\mu)|_{y \rightarrow 1}$ (left hand side) into the quasi collinear function $\tilde{J}$ (blue) and soft-collinear function $\tilde{W}_{C}$ (green), shown as Eq.~(\ref{eq:TFqPDF}). The notations are similar to Fig.~\ref{fig:LCPDF_TR} except that the intermediate gauge links are along the spacelike $n_z$ direction instead of light-cone $\bar{n}$. }
    \label{fig:QPDF_TR}
\end{figure}

The light-cone collinear function $J$ is defined as
\begin{align}\label{eq:LCjet}
J\left(\frac{m}{\mu}\right) (2p^+)
\equiv 
&\frac{\langle q(p)| \bar{\psi}(0) W_{\bar{n}}\left(0, \infty\right)| \Omega \rangle }{\langle \Omega| W_{v}\left(\infty, 0\right) W_{\bar{n}}\left(0, \infty \right) | \Omega \rangle}  \gamma^{+} \frac{\langle \Omega| W_{\bar{n}}\left(-\infty, 0\right) \psi(0)| q(p) \rangle}{\langle \Omega| W_{\bar{n}}\left(-\infty , 0\right) W_{v}\left(0,-\infty\right) | \Omega \rangle} \ ,
\end{align}
where $| \Omega \rangle$ denotes the vacuum. In the numerator, the incoming/outgoing quark, moving as a fast color charge, is accompanied by collinear gluons with momenta approximately $\sim p^+\left(1, \delta^2, \delta \right)$, nearly parallel to the quark, which defines the collinear modes. However, the numerator also includes contributions from the soft mode with momenta $\sim p^+\left( \delta, \delta, \delta \right)$. These soft contributions can be analyzed using the eikonal approximation, where the incoming/outgoing quark is effectively replaced by a gauge link along its velocity $v^{\mu}=p^{\mu}/m$. This substitution is captured in the denominator, subtracting off the soft mode in the numerator, and ensuring that the ratio only reflects the collinear mode. This is related to the ``zero-bin subtraction"~\cite{Becher:2006mr, Manohar:2006nz} in SCET language. $J$ depends on the kinematic variable $p^2=m^2$, which represents the scale of the collinear mode.

The light-cone soft-collinear function is defined as
\begin{align}\label{eq:LCsoft}
&\int_{-\infty}^{+\infty} \frac{d \lambda}{2 \pi} e^{i (x-1) \lambda} W_C\left(\frac{\lambda \mu}{m}\right) \nonumber\\
&\equiv \int_{-\infty}^{+\infty} \frac{d \lambda}{2 \pi} e^{i (x-1) \lambda} \langle \Omega |  W_{v}(+\infty,\xi^{-}) W_{\bar{n}}(\xi^{-},0) W_{v}(0,-\infty) | \Omega \rangle \ ,
\end{align}
where $\xi^{-} = -\lambda/p^{+}$. The interactions between the soft-collinear gluons $\sim p^+ \epsilon \left(1, \delta^2, \delta \right)$ and incoming/outgoing quarks are equivalently described by the gauge links along quarks' trajectories in the $v$ direction. $W_C$ contains the coordinate-space variable $\lambda/m$, whose Fourier conjugate is $(1-x)m$, representing the scale of the soft-collinear mode. The soft-collinear function has a spectral representation:
\begin{align}
&p^{+} \sum_{n}\int d\Pi_{n_{\vec{k}}} \delta\left[k^{+}-(1-x)p^{+}\right] \, \left| \langle n_{\vec{k}}| W_{\bar{n}}(-\infty,0) W_{v}(0,-\infty) | \Omega \rangle \right|^2 \ ,
\end{align}
where $\Pi_{n_{\vec{k}}}$ denotes the phase space of the final state particles $n_{\vec{k}}$, with total momentum $(1-x)p^{+}$. This indicates that the phase space for real emissions is suppressed, leading to the suppression of the soft-collinear scale $(1-x) m$ by a factor of $(1-x)$ relative to the collinear scale $m$.  

Similar factorizations as Eq.~(\ref{eq:TFLCPDF}) have also been studied using the reduced diagram analysis in Refs.~\cite{Korchemsky:1988si,Berger:2002sv,Ji:2004hz}. For readers' convenience, we provide a one-loop verification of Eq.~(\ref{eq:TFLCPDF}) in Appendix~\ref{sec:1loopfmLC}. 

We now extend the above factorization to the quasi-PDF. Consider the quasi-PDF in a free massive on-shell unpolarized quark state,
\begin{align}\label{eq:qpdfq}
    \tilde q\left( y, \frac{p^z}{\mu}, \frac{m}{\mu}\right) \equiv \int_{-\infty}^{+\infty} \frac{d z}{4\pi} e^{i y p^z z} 
    \langle q(p) | \bar{\psi}(z) \gamma^{t} W_{n_{z}}(z,0) \psi(0) | q(p) \rangle \ . 
\end{align}
Similar logic applies in the threshold limit, as illustrated in Fig.~\ref{fig:QPDF_TR},
\begin{align}\label{eq:TFqPDF}
\tilde q\left( y, \frac{p^z}{\mu}, \frac{m}{\mu}\right) \bigg|_{y \rightarrow 1} 
= \int_{-\infty}^{+\infty} \frac{d \lambda}{2\pi} e^{i (y-1) \lambda}
\tilde{J}\left(\frac{s_z p^z}{\mu}, \frac{m}{\mu}\right) \tilde W_C\left(\frac{\lambda^2 \mu^2}{p_z^2}, \frac{\lambda \mu}{m} \right) + ... \ ,
\end{align}
where $...$ denotes the power corrections such as ${\cal O}\left( m^2/p_z^2 \right)$, which will be discussed further in Sec.~\ref{sec:renormalon}. The definitions of the quasi collinear and soft-collinear functions are
\begin{align}\label{eq:qjet}
\tilde{J}\left(\frac{s_z p^z}{\mu}, \frac{m}{\mu}\right) (2p^z)
\equiv 
&\frac{\langle q(p)| \bar{\psi}(0) W_{s_z n_z}\left(0, -\infty \right) | \Omega \rangle}{\langle \Omega| W_{v}\left(\infty, 0\right) W_{s_z n_z}\left(0, -\infty\right) | \Omega \rangle} \gamma^{z} \nonumber\\
&\times \frac{\langle \Omega| W_{s_z n_z}\left(\infty, 0\right) \psi(0)| q(p) \rangle}{\langle \Omega| W_{s_z n_z}\left(\infty, 0\right) W_{v}\left(0,-\infty\right) | \Omega \rangle} \ ,
\end{align}
\begin{align}\label{eq:qsoft}
& \int_{-\infty}^{+\infty} \frac{d \lambda}{2 \pi} e^{i (y-1) \lambda}  \tilde W_C\left(\frac{\lambda^2 \mu^2}{p_z^2}, \frac{\lambda \mu}{m} \right) \nonumber\\
&\equiv \int_{-\infty}^{+\infty} \frac{d \lambda}{2 \pi} e^{i (y-1) \lambda}  \langle \Omega | W_{v}(+\infty, z) W_{n_z}(z, 0) W_{v}(0, -\infty) | \Omega \rangle \ ,
\end{align}
where $s_z \equiv {\rm sign}(z)$ is related to the direction of the gauge link, and $z=\lambda/p^z$. They are analogous to the light-cone PDF case, except that the intermediate gauge links follow the $n_z$ direction instead of the light-cone direction $\bar{n}$. The $\gamma^{z}$ in Eq.~(\ref{eq:qjet}) can also be chosen as $\gamma^t$, which does not influence the leading twist result. 

The quasi collinear function $\tilde J$ captures the collinear mode $\sim p^+ \left(1, \delta^2, \delta \right)$ of the threshold quasi PDF $\tilde q\left(y,p^z/\mu,m/\mu\right) |_{y \rightarrow 1}$, which is reflected in the variable $p^2=m^2$. In addition, it contains a hard mode $\sim p^{+} \left(1,1,1\right)$, with the scale $p^z$.   

On the other hand, the quasi soft collinear function $\tilde{W}_{C}$ contains the soft-collinear mode $\sim p^+ \epsilon \left(1, \delta^2, \delta \right)$, whose scale $(1-y)m$ is the Fourier conjugate of the coordinate-space variable $\lambda/m$. Additionally, the semi-hard mode $\sim p^+\epsilon(1,1,1)$ also exists, reflected as the semi-hard scale $(1-y)^2 p_z^2$, which is the Fourier conjugate of the coordinate-space variable $\lambda^2/p_z^2$.

The factorization Eq.~(\ref{eq:TFqPDF}) is checked at one-loop in Appendix~\ref{sec:1loopfmq}. 

\begin{figure}
    \centering
    \includegraphics[width=1\linewidth]{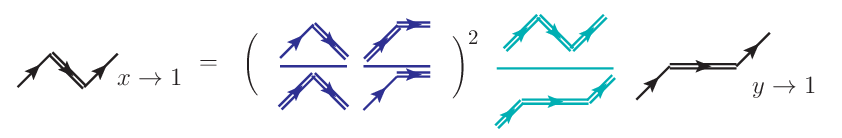}
    \caption{The matching relation between threshold light-cone PDF $q\left(x, m/\mu\right) |_{x \rightarrow 1}$ and quasi PDF $\tilde{q}\left(y, p^z/\mu, m/\mu\right) |_{y \rightarrow 1}$ shown as Eq.~(\ref{eq:thrmatchingq}), which is obtained by comparing Figs.~\ref{fig:LCPDF_TR} and~\ref{fig:QPDF_TR}. The collinear (blue) and soft-collinear (green) functions are matched, respectively. The RG equation of matching coefficient for the soft-collinear functions is crucial for resumming the threshold logarithms. }
    \label{fig:LCQPDF_TR_Match}
\end{figure}

Based on Eqs.~(\ref{eq:TFLCPDF}) and~(\ref{eq:TFqPDF}), the matching relation between $q\left(x, m/\mu\right) |_{x \rightarrow 1}$ and $\tilde q\left(y, p^z/\mu, m/\mu\right) |_{y \rightarrow 1}$ can be derived (Fig.~\ref{fig:LCQPDF_TR_Match}),
\begin{align}\label{eq:thrmatchingq}
q\left(x, \frac{m}{\mu}\right) \bigg|_{x \rightarrow 1} =  \int_{-\infty}^{+\infty} \frac{d y}{|y|} \mathcal{C}\left(\frac{x}{y},\frac{p^z}{\mu}\right)_{\rm sg} \tilde q\left(y, \frac{p^z}{\mu}, \frac{m}{\mu}\right) \bigg|_{y \rightarrow 1} + ... \ ,
\end{align}
where the threshold matching kernel is defined using the ratios of light-cone to quasi functions,
\begin{align}\label{eq:MKthr}
\mathcal{C}\left(\xi,\frac{p^z}{\mu}\right)_{\rm sg}  
\equiv \int_{-\infty}^{+\infty} \frac{d \lambda}{2\pi} e^{i \lambda (\xi - 1)} 
\Delta J\left(\tilde{L}_{p^z},\alpha(\mu)\right) S\left(l_z,\alpha(\mu)\right) \ ,
\end{align}
where the collinear and soft-collinear functions are matched in $\Delta J$ and $S$\footnote{For the convenience of discussing the RG evolutions in the next subsection, the variable $\alpha(\mu)$ is explicitly included as an argument for $\Delta J$ and $S$. However, it is omitted in $J$, $\tilde J$, $W_C$ and $\tilde{W}_C$ to streamline the notation.}, respectively. The subscript ``sg" refers to ``singular".

The ratio of the collinear functions in Eqs.~(\ref{eq:LCjet}) and~(\ref{eq:qjet}) is expressed as
\begin{align}\label{eq:DJdef}
\Delta J\left(\tilde{L}_{p^z},\alpha(\mu)\right)
\equiv \frac{ J\left(m/\mu\right) }{ \tilde{J}\left(s_{\lambda} | p^z |/\mu, \, m/\mu \right) } \ ,
\end{align}
where $\tilde{L}_{p^z}=\ln(-2 i s_{\lambda} |p^z|/\mu)$, and $s_{\lambda} \equiv {\rm sign}(\lambda)$ is related to the direction of the gauge link. In this ratio, the collinear mode $\sim p^{+}(1,\delta^2,\delta)$ is canceled because the intermediate lightlike and spacelike eikonal lines are equivalent under the collinear approximation $k^t \sim k^z$~\cite{Ji:2020ect,Ji:2024oka}. The perturbative hard mode $\sim p^{+} (1,1,1)$ is left, which gives the hard scale logarithm $\ln\left(p_z^2/\mu^2\right)$. For the convenience of performing RG resummation in the next subsection, we further write $\Delta J$ in terms of its absolute value $H$ and phase angle $\Phi$,
\begin{align}\label{eq:Hdef}
H\bigg(L_{p_z},\alpha(\mu)\bigg) \exp\left[i s_{\lambda} \Phi(L_{p_z},\alpha(\mu)) \right] \equiv \Delta J\left(\tilde{L}_{p^z},\alpha(\mu)\right)  \ ,
\end{align}
where $L_{p_z} = \ln\left(4 p_z^2/\mu^2\right)$. The phase depends on the coordinate space parameter $s_\lambda$, which is related to $\sim 1/(1-\xi)$ in the momentum space matching kernel $\mathcal{C}\left(\xi,\frac{p^z}{\mu}\right)_{\rm sg}$, through the Fourier transformation identities Eq.~(\ref{eq:FTodd}). 

The ratio of the soft-collinear functions in Eqs.~(\ref{eq:LCsoft}) and~(\ref{eq:qsoft}) is given by
\begin{align}\label{eq:Sdef}
S\left(l_z,\alpha(\mu)\right) \equiv \frac{ W_C\left(\lambda \mu/m \right) }{\tilde W_C\left(\lambda^2 \mu^2/p_z^2, \, \lambda \mu/m \right) }  \ ,
\end{align}
where $l_z=\ln\left(\bar{\lambda}^2\mu^2/4 p_z^2\right)$, and $\bar{\lambda} \equiv \lambda e^{\gamma_E}$.
For the same reason, the ratio $W_C/\tilde{W}_C$ cancels the soft-collinear mode $\sim p^{+}\epsilon(1,\delta^2,\delta)$. The semi-hard gluons with momentum $\sim p^{+} (\epsilon,\epsilon,\epsilon)$ are left,  leading to the threshold logarithm $\ln\left(\epsilon^2 p_z^2/\mu^2\right)$. To calculate the semi-hard mode in perturbation theory, we must have $p^z \gg |1-x| p^z \gg \Lambda_{\rm QCD}$, or equivalently $1 \gg \epsilon \gg \delta$. This ratio $S$ is called the threshold soft factor and will play a crucial role in resumming the threshold logs. The perturbative results up to two loops are provided in Appendix~\ref{sec:fix}.

The matching relation Eq.~(\ref{eq:thrmatchingq}), though derived from the quark matrix elements in Eqs.~(\ref{eq:LCpdfq}) and~(\ref{eq:qpdfq}), can be generalized to the hadron matrix elements defined in Eqs.~(\ref{eq:LCpdf}) and~(\ref{eq:qpdf}),
\begin{align}\label{eq:thrmatching}
f(x,\mu) |_{x \rightarrow 1} =  \int \frac{d y}{|y|} \mathcal{C}\left(\frac{x}{y},\frac{|y|P^z}{\mu}\right)_{\rm sg} \tilde f(y,P^z,\mu) |_{y \rightarrow 1} + ... \ ,
\end{align}   
because the IR modes are canceled in the ratios in Eqs.~(\ref{eq:DJdef}) and~(\ref{eq:Sdef}), and the remaining UV physics captured by $\mathcal{C}\left(\frac{x}{y},\frac{|y|P^z}{\mu}\right)_{\rm sg}$ is perturbatively calculable, which is obtained by replacing $p^z$ in Eq.~(\ref{eq:MKthr}) with $|y|P^z$~\cite{Izubuchi:2018srq}. The term $...$ represents the power corrections, which will be discussed in Sec.~\ref{sec:renormalon}.

The threshold matching relation Eq.~(\ref{eq:thrmatching}), with the definitions of the matching kernel in Eq.~(\ref{eq:MKthr}), collinear functions in Eqs.~(\ref{eq:LCjet})(\ref{eq:qjet}) and soft-collinear factors in Eqs.~(\ref{eq:LCsoft})(\ref{eq:qsoft}), is verified in the following ways:
\begin{itemize}
    \item The cancellation of collinear modes regularized by $m$ in the ratio $J/\tilde{J}$ is confirmed up to one-loop in Appendix~\ref{sec:1loopfmlcq}. Here, $J$ and $\tilde{J}$ contain the collinear divergences associated with the threshold light-cone and quasi PDFs, respectively. The observed cancellation suggests that threshold light cone and quasi PDFs share common collinear modes.
    
    \item The cancellation of soft-collinear modes in the ratio $W_C/\tilde{W}_C$ is checked up to one-loop in Appendix~\ref{sec:1loopfmlcq}. Again, this cancellation suggests that the threshold light-cone and quasi PDFs share the same soft-collinear modes.

    \item A consistency check for the UV part (the coefficient function) has been performed up to two-loop in Ref.~\cite{Ji:2023pba}. The threshold matching kernel $\mathcal{C}\left(\xi,\frac{|y| P^z}{\mu}\right)_{\rm sg}$ can be computed using the definition from Eq.~(\ref{eq:MKthr}), but with massless quark and pure dimensional regularization, as described in Ref.~\cite{Ji:2023pba}. On the other hand, it can also be obtained directly from the threshold expansion $\xi \rightarrow 1$ of the full matching kernel in Eq.~(\ref{eq:LaMETmatching}),
    \begin{align}\label{eq:threxp}
    {\cal C}\left(\xi,\frac{|y| P^z}{\mu}\right) &= {\cal C}\left(\xi,\frac{|y| P^z}{\mu}\right)_{\rm sg} \nonumber\\
    &\quad + \sum_{m=0} C_{m}\left(\ln(|\xi-1|),{\rm sign}(\xi-1),\frac{|y| P^z}{\mu}\right) (\xi-1)^{m} \ ,
    \end{align}
    where ${\cal C}\left(\xi,\frac{|y| P^z}{\mu}\right)_{\rm sg}$ contains the singular term $\sim 1/(1-\xi)$, which is the leading power of $1-\xi$.
    The consistency between the two methods has been verified up to two-loop in Ref.~\cite{Ji:2023pba}. In particular, the scale $\mu$ dependence of Eq.~(\ref{eq:thrmatching}) has been checked up to two-loop.
    
\end{itemize}

The threshold factorization provides a systematic method to separate and resum logarithmic terms. One may regard the factorization in the following way: In the perturbative quasi-PDF expression in Eq.~(\ref{eq:qpdfq1loopfmtr}), the threshold logarithms $\ln(-i \bar{\lambda})$ (or $\ln(1-y)$ in momentum space) need to be resummed. However, since these logarithms lack explicit dependence on the scale $\mu$, resummation using RG equations is challenging.
The threshold factorization for the quasi-PDF in Eq.~(\ref{eq:TFqPDF}) resolves this issue by splitting the logarithmic term $\ln(-i \bar{\lambda})$ into two parts: $ \ln\left(-2 i s_z p^z m/\mu^2\right) $ in the quasi-collinear function in Eq.~(\ref{eq:qZ1loopfm}) and $\ln\left(2 i s_z p^z m/\mu^2 \bar{\lambda}^2\right)$ in the quasi soft-collinear function in Eq.~(\ref{eq:qWC1loopfm}). Furthermore, in the matching relation Eq.~(\ref{eq:thrmatchingq}), the logarithmic terms in the quasi soft-collinear function are further separated into $\ln(m/\mu \bar{\lambda})$ in the light-cone soft-collinear function Eq.~(\ref{eq:WC1loopfm}) and $\ln(p^z/\mu \bar{\lambda})$ (or $\ln(p^z(1-\xi)/\mu)$ in momentum space) in the threshold soft function Eq.~(\ref{eq:WCoverWCt}). By introducing scale $\mu$-dependence in the threshold soft function, resummation of the threshold logarithms via the RG equations becomes possible.

\subsection{Threshold resummation}\label{sec:TRMS}
Based on the threshold factorization discussed in the previous subsection, the renormalization group (RG) equations are established and used to resum the threshold logarithms.

The renormalization group (RG) equations for the factorized ingredients defined in Eqs.~(\ref{eq:Hdef}) and~(\ref{eq:Sdef}) are given by~\cite{Ji:2023pba}
\begin{align}\label{eq:RGEHab}
\frac{d}{d\ln \mu} \ln   H\bigg(L_{p_z},\alpha(\mu)\bigg) = -\Gamma_{\rm cusp}(\alpha)L_{p_z} - \tilde {\gamma}_{H}(\alpha) \ ,
\end{align}   
\begin{align}\label{eq:RGEA}
\frac{d \, \Phi\bigg(L_{p_z},\alpha(\mu)\bigg)}{d\ln \mu} = \pi \, \Gamma_{\rm cusp}(\alpha) \ ,
\end{align}
\begin{align}\label{eq:RGEtiledJ}
\frac{d \ln S\bigg(l_z,\alpha(\mu)\bigg)}{d\ln \mu}= - \Gamma_{\rm cusp}(\alpha) l_z +  \tilde \gamma_{S}(\alpha) \ ,
\end{align}
where the cusp anomalous dimension up to four loops and single log anomalous dimensions up to three loops are collected in Appendix~\ref{sec:ano}. The anomalous dimensions $\tilde \gamma_S$ and $\tilde \gamma_H$ are related to $\gamma_C $, $\gamma_{\tilde C}$, $\gamma_J$, and $\gamma_{\tilde J}$ in Appendix~\ref{sec:1loopfm}. 

In the threshold region $x \rightarrow 1$, there are two perturbative scales $p^z$ and $(1-x)p^z$, and their log ratio $\ln(1-x)$ becomes large. The large logs
do not disappear by a simple choice of $\mu$. Therefore, we need to 
solve the above equations by some starting scales $\mu_h\sim p^z$ for $H$ and $\Phi$ and $\mu_i\sim (1-x)p^z$ for $S$, evolving them to some common scale $\mu$. Large double logs are resummed in the solution of the RG equation, and a reliable prediction for the large $x$ PDF can be achieved. 

By solving the above RG equations following the same method as Ref.~\cite{Becher:2006mr} and performing the Fourier transformation, one obtains the resumed form for the matching kernel in Eqs.~(\ref{eq:MKthr}) and~(\ref{eq:thrmatching}) where {\it hard and semi-hard scales} are fully factorized,
\begin{align}\label{eq:resummedform}
&  {\cal C}\left(\frac{x}{y},\frac{|y| P^z}{\mu},\mu_h,\mu_i\right)_{\rm sg}= \int \frac{d y'}{|y'|}   {\cal S}\left( \frac{x}{y'},\frac{|y'|P^z}{\mu},\mu_i\right)    {\cal H}\left( \frac{y'}{y},\frac{|y|P^z}{\mu},\mu_h\right) \ ,
\end{align}
where $H$ and $\Phi$ defined in Eq.~(\ref{eq:Hdef}) are included in ${\cal H}$, and $S$ defined in Eq.~(\ref{eq:Sdef}) is absorbed in ${\cal S}$. The RG resummed soft function in momentum space introduces an intermediate or semi-hard scale $\mu_i$~\cite{Becher:2006mr}, 
\begin{align}\label{eq:Jmom}
{\cal S}\left( \frac{x}{y},\frac{|y|P^z}{\mu},\mu_i\right)
= \int \frac{d y'}{|y'|} P_{S}\left(\frac{x}{y'},\frac{|y'|P^z}{\mu},\mu_i\right) {\cal S}\left( \frac{y'}{y},\frac{|y|P^z}{\mu_i},\mu_i\right) \nonumber\\
\times \theta\left(\gamma-\left|1-\frac{x}{y'}\right|\right) \theta\left(\gamma-\left|1-\frac{y'}{y}\right|\right) \ ,
\end{align}
where the threshold evolution factor is
\begin{align}\label{eq:TRevo}
&P_{S}\left(\xi,\frac{p^z}{\mu},\mu_i\right)
= \exp \bigg[-2   S_{\rm sud.}(\mu_i,\mu)+   a_{S}(\mu_i,\mu)\bigg] \int_{-\infty}^{+\infty} \frac{d\lambda}{2\pi} e^{i \lambda (\xi - 1)} \left( \frac{|\bar{\lambda}|\mu_i}{2 p^z} \right)^{-\eta} \nonumber\\
&=\exp \bigg[-2   S_{\rm sud.}(\mu_i,\mu) +   a_{S}(\mu_i,\mu)\bigg] \left[\frac{\sin \left(\frac{  \eta \pi}{2}\right)}{|1-\xi|}\left(\frac{2|1-\xi||p^z|}{\mu_i}\right)^{\eta}\right]_{*} \frac{ \Gamma(1-\eta) \mathrm{e}^{-\eta \gamma_E}}{\pi} \ ,
\end{align}
and the fixed-order soft function is
\begin{align}\label{eq:sffixed}
&{\cal S}\left( \xi,\frac{p^z}{\mu_i},\mu_i\right) = \int_{-\infty}^{+\infty} \frac{d\lambda}{2\pi} e^{i \lambda (\xi - 1)}  S\bigg(l_z=\ln \frac{\bar{\lambda}^2 \mu_i^2 }{4 p_z^2},\alpha(\mu_i)\bigg) \nonumber\\
& =  S\bigg(l_z=-2\partial_{\eta_0},\alpha(\mu_i)\bigg) \left[\frac{\sin \left(\frac{  \eta_0 \pi}{2}\right)}{|1-\xi|}\left(\frac{2|1-\xi||p^z|}{\mu_i}\right)^{\eta_0}\right]_{*} \frac{ \Gamma(1-\eta_0) \mathrm{e}^{-\eta_0 \gamma_E}}{\pi} \Bigg|_{\eta_0 \rightarrow 0} \ ,
\end{align}
where the Fourier transformation can be performed based on Appendix~\ref{sec:FTofLogsasymp}.
The RG resummed hard kernel in momentum space starts from some
high-energy scale $\mu_h$,
\begin{align}\label{eq:Hmom} 
{\cal H}\left( \xi,\frac{p^z}{\mu},\mu_h\right) 
=&\exp \bigg[2   S_{\rm sud.}(\mu_h,\mu)-   a_{H}(\mu_h,\mu)\bigg] \left(\frac{\mu_h}{2 p^z}\right)^{2a_{\Gamma}(\mu_h,\mu)} \theta\left(\gamma-|1-\xi|\right)  \nonumber\\
&\times   H\left(\ln \frac{4p_z^2}{\mu_h^2},\alpha(\mu_h)\right) \int_{-\infty}^{+\infty} \frac{d\lambda}{2\pi} e^{i \lambda (\xi-1)} e^{i s_{\lambda} \hat \Phi(p^z,\mu_h,\mu)} \nonumber\\
=&\exp \bigg[2  S_{\rm sud.}(\mu_h,\mu)-  a_{H}(\mu_h,\mu)\bigg] \left(\frac{\mu_h}{2 p^z}\right)^{2   a_{\Gamma}(\mu_h,\mu)}  \theta\left(\gamma-|1-\xi|\right) \nonumber\\
&\times   H\left(\ln \frac{4p_z^2}{\mu_h^2},\alpha(\mu_h)\right) \left[ \cos(\hat \Phi(p^z,\mu_h,\mu)) \delta(1-\xi) + \frac{\sin(\hat \Phi(p^z,\mu_h,\mu))}{\pi(1-\xi)}\right] \ .
\end{align}
Various quantities are defined as follows
\begin{align}
&\hat \Phi(p^z,\mu_h,\mu) = \pi a_{\Gamma}(\mu_h,\mu)+  \Phi\bigg(L_{p_z}=\ln \frac{4p_z^2}{\mu_h^2},\alpha(\mu_h)\bigg) \nonumber\\
&  \eta=2   a_{\Gamma}(\mu_i,\mu) \ .
\end{align} 
The star function is 
\begin{align}\label{eq:starf}
&{\cal C}(\xi)_{*} 
= {\cal C}(\xi) - \sum_{k=0}^{n}\frac{(-1)^k\delta^{(k)}(1-\xi)}{k!}\int_{-\infty}^{+\infty} d \xi' (1-\xi')^k {\cal C}(\xi') \ ,
\end{align}
where $n$ is an integer with $n<-\eta<n+1$ for $\eta<0$. For $\eta>0$, the subtraction terms involving $\delta^{(k)}(1-\xi)$ are not required. For $-1<\eta<1$ (which is applicable in our calculation), the star function can be written as
\begin{align}
\left[\frac{\sin \left(\frac{\eta \pi}{2}\right)}{|1-\xi|}\left(\frac{2|1-\xi||p^z|}{\mu_i}\right)^\eta\right]_{*} = {\cal P}\frac{\sin \left(\frac{\eta \pi}{2}\right)}{|1-\xi|}\left(\frac{2|1-\xi||p^z|}{\mu_i}\right)^\eta + \delta(1-\xi) \frac{2\sin \left(\frac{\eta \pi}{2}\right)}{\eta}\left(\frac{2 |p^z|}{\mu_i}\right)^\eta \ ,
\end{align}
where ${\cal P}$ is the principal value prescription for the plus function {\it defined} as follows
\begin{align}\label{eq:plusPV}
{\cal P} \frac{C(\xi)}{|\xi-1|} = \frac{C(\xi)}{|\xi-1|} - \delta(1-\xi) \int_{0}^{2} d\xi' \frac{C(\xi')}{|\xi'-1|}   \ .
\end{align}
The evolution factors regarding double logs are
\begin{align}\label{eq:aGamma}
  S_{\rm sud.}(\nu,\mu)=\int_{\alpha(\nu)}^{\alpha(\mu)}\frac{\Gamma_{\rm cusp}(\alpha)d\alpha}{\beta(\alpha)}\int_{\alpha(\nu)}^{\alpha}\frac{d\alpha'}{\beta(\alpha')} \ , \   a_{\rm \Gamma}(\nu,\mu)=\int_{\alpha(\nu)}^{\alpha(\mu)} d\alpha\frac{\Gamma_{\rm cusp}(\alpha)}{\beta(\alpha)} \ .
\end{align}
The single log evolution factors~\cite{Becher:2006mr} read
\begin{align}\label{eq:othera}
&  a_{H}(\nu,\mu)=\int_{\alpha(\nu)}^{\alpha(\mu)} d\alpha\frac{  \gamma_H(\alpha)}{\beta(\alpha)} \ ,  \ 
  a_{S}(\nu,\mu)=\int_{\alpha(\nu)}^{\alpha(\mu)} d\alpha\frac{  \gamma_{S}(\alpha)}{\beta(\alpha)} \ ,
\end{align} 
where our convention for the beta function is 
\begin{align}\label{eq:beta}
    \frac{d\alpha}{d\ln \mu}= \beta(\alpha) =-\beta_0 \alpha^2 - \beta_1 \alpha^3 +... \ ,  \ 
    \beta_0=\frac{11C_A}{6\pi}-\frac{n_f}{3\pi} \ ,
\end{align}
where $C_A=N_C=3$, number of color, and $n_f$ is the number of active quark flavor which is taken to be 3 in our work.

Note when $\mu=\mu_h=\mu_i$, the resummation formula, Eq.~(\ref{eq:resummedform}), reduces to the singular part of the fixed-order result. 

In Eqs.~(\ref{eq:Jmom}) and~(\ref{eq:Hmom}), the threshold cuts $\theta\left(\gamma-\left|1-x/y'\right|\right)$, $\theta\left(\gamma-\left|1-y'/y\right|\right)$ and $\theta\left(\gamma-\left|1-\xi\right|\right)$, which are not derived from the RG equations, are manually introduced to limit the regions away from the threshold. These cuts do not affect the factorization and RG evolution in the leading threshold expansion $\sim 1/(1-x/y)$ in Eq.~(\ref{eq:resummedform}). Furthermore, the cuts for the soft function ensure that the total gluon momentum, in both the fixed-order soft function and higher-order resummed parts, remains much smaller than the hard scale, in line with the power counting of the factorization theorem. Further discussion on the implementation of the threshold cuts will be provided in Sec.~\ref{sec:Rfix}.

\subsection{Choice of resummation scales}\label{sec:scalcho}

As shown in Sec.~\ref{sec:TRMS}, the result of threshold resummation for Eq.~(\ref{eq:resummedform}) depends parametrically on the two scales, $\mu_h$ and $\mu_i$, defining the RG evolution boundary. Therefore, the effects of resummation depend on judicious choices of these two scales to account for large logs appearing in higher-order perturbation series. 

The choice of the hard scale $\mu_h$ has been discussed in Ref.~\cite{Su:2022fiu} where by analogy with DIS, the most natural hard scale in quasi-PDF is $2 |y| P^z$, where $y$ is the momentum fraction. To calculate the light-cone PDF, the scale effectively becomes the physical parton momentum $\mu_h=2 |x| P^z$, which can be seen in the one-loop inversion discussed in Appendix~C in Ref.~\cite{Su:2022fiu}. 

The choice of the semi-hard scale $\mu_i$ is less straightforward. In the intermediate steps of calculations, the large double logs come from partonic momentum $2|1-\xi||x| P^z = 2|x-y| P^z$, which appears as a natural resummation scale, as shown in Eq.~(\ref{eq:Jmom}). However, the integration over $x$ will go through a region $\mu_i \sim 0$, which causes artificial factorial growth in perturbation series. Similar observations have been made early in~\cite{Catani:1996yz,Becher:2006mr}. In Ref.~\cite{Becher:2006mr}, they have shown explicitly that this choice is the source of the spurious non-perturbative power contributions.

To avoid the above pathology, one may choose the intermediate scale $\mu_i$ in the matching kernel ${\cal C}\left(\frac{x}{y},\frac{|y| P^z}{\mu},\mu_h,\mu_i\right)_{\rm sg}$ as the external physical scale in the light-cone PDF, which is, in this case, $\mu_i = 2|1-x|P^z$ by analogy to the DIS threshold factorization: 
In the $x_B \rightarrow 1$ limit of DIS structure function~\cite{Becher:2006mr}, the hadronic final states are predominantly collinear in momentum, and the final phase space integration is limited. The invariant mass of hadronic system $M_{X} = Q \sqrt{(1-x_B)/x_B}$ is much smaller than the photon virtuality $Q$,  and the corresponding resummation of the threshold logarithms $\sim \ln\left(M_{X}^2/Q^2\right)$ involves the physical semi-hard scale $M_{X}$. Similarly, in the $x \rightarrow 1$ limit of light-cone PDF, the phase space of the hadronic ``final state" is severely constrained, and the corresponding ``final-state'' momentum $2|1-x|P^z$ becomes much smaller than $2|x|P^z$. Therefore, the threshold resummation for light-cone PDF shall involve $\mu_i = 2|1-x|P^z$. 

As a demonstration, the momentum scale $2|1-x|P^z$ in large logs can be seen by convoluting the fixed order soft function Eq.~(\ref{eq:sffixed}) with quasi-PDF for $0<x<y<1$,
\begin{align}
&\int_{x}^{1} \frac{d y}{|y|} {\cal S}\left( \frac{x}{y},\frac{|y|P^z}{\mu_i},\mu_i\right) \tilde f(y,P^z)   \nonumber\\
&=\int_{x}^{1} \frac{d y}{|y|} {\cal S}\left( \frac{x}{y},\frac{|y|P^z}{\mu_i},\mu_i\right) \sum_{n=0} \frac{\tilde f^{(n)}(x,P^z)}{n!}(y-x)^n   \nonumber\\
&= S\bigg(l_z=-2 \partial_{\eta_{0}},\alpha(\mu_i)\bigg) \frac{ \sin \left(\frac{\eta_{0} \pi}{2}\right) \Gamma(1-\eta_{0}) e^{-\eta_{0} \gamma_E}}{\pi} \nonumber\\
&\quad \times \sum_{n=0} \frac{\tilde f^{(n)}(x,P^z)}{n!} \left(\frac{2|1-x|P^z}{\mu_i}\right)^{\eta_{0}} \frac{(1-x)^n}{n+\eta_{0}} \Bigg|_{\eta_0 \rightarrow 0} \nonumber\\ 
&=\left(\frac{2|1-x|P^z}{\mu_i}\right)^{\eta_{0}} S\bigg(l_z=-2 \partial_{\eta_{0}} + { \color{red} \ln\frac{\mu_i^2}{4 (1-x)^2 P_z^2} },\alpha(\mu_i)\bigg) \nonumber\\ 
&\quad \times \frac{ \sin \left(\frac{\eta_{0} \pi}{2}\right) \Gamma(1-\eta_{0}) e^{-\eta_{0} \gamma_E}}{\pi} \sum_{n=0} \frac{\tilde f^{(n)}(x,P^z)}{n!}  \frac{(1-x)^n}{n+\eta_{0}} \Bigg|_{\eta_0 \rightarrow 0} \ ,
\end{align}
where the log term $\sim \ln\left(\frac{\mu_i^2}{4(1-x)^2 P_z^2}\right)$ appears in the first argument of the soft function $S$ after the convolution. If $\mu_i$ differs from $2|1-x|P^z$ significantly, the log term $\sim \ln\left(\frac{\mu_i^2}{4(1-x)^2 P_z^2}\right)$ becomes too large. Thus, choosing $\mu_i = 2|1-x|P^z$ eliminates these large logs for better perturbative accuracy. The convolution for $y>1$ is not considered because the quasi-PDF there is suppressed by $O(\alpha(P^z))$ compared to $0<y<1$. The integral range for $y<x$ is also not discussed because complete IR cancellation occurs and threshold logarithms are not significant there. Similar calculations in the context of DIS threshold factorization have been discussed in Ref.~\cite{Becher:2006mr}.

\subsection{Conversion to the hybrid renormalization scheme}
In this subsection, we convert the threshold factorization and resummation to the hybrid renormalization scheme defined in~\cite{Ji:2020brr}, which is more convenient for lattice QCD calculations. This scheme introduces some controllable UV effects to the lattice matrix elements, which are then transferred to the quasi-PDF in Eq.~(\ref{eq:qpdf}) after Fourier transformation. Since the light-cone PDF is independent of these UV effects, they are absorbed into the perturbative kernels.

In the hybrid scheme, renormalization is performed at short and long distances in different ways: 
\begin{align}\label{eq:hybridrenorm}
h^{h}(z,P^z) = \frac{h^{\rm lat}(z,a,P^z)}{h^{\rm lat}(z,a,P^z=0)}\theta(z_s-|z|)+\frac{h^{\rm lat}(z,a,P^z)}{Z^{R}(z,a,\mu) h^{\rm \overline{MS}}(z_s,\mu,P^z=0)}\theta(|z|-z_s) \ ,
\end{align}
where $h^{\rm lat}(z,a,P^z) \equiv \langle H(P)| \bar{\psi}(z) \gamma^{t} W_{n_z}(z,0) \psi (0) |H(P) \rangle / \langle H(P)| \bar{\psi}(0) \gamma^{t} \psi (0) |H(P) \rangle$ is the lattice matrix element, which depends on the lattice spacing $a$. $z_{s}$ is the hybrid cutoff that satisfies $a \ll z_s \ll 1/\Lambda_{\rm QCD}$ to ensure the validity of continuum perturbation theory. $Z^{R}(z,a,\mu)$ is the renormalization factor used to eliminate both linear and logarithmic divergences, which can be extracted through self-renormalization~\cite{LatticePartonCollaborationLPC:2021xdx} or other methods. $Z^{R}(z,a,\mu)$ also includes the mass renormalization parameter $m_0$, which is obtained by setting the renormalized lattice zero momentum matrix element equal to the perturbative zero momentum matrix element at short distances:
\begin{align}\label{eq:Hyrenormfit}
    \frac{h^{\rm lat}(z,a,P^z=0)}{Z^{R}(z,a,\mu)} = h^{\rm \overline{MS}}(z,\mu,P^z=0) \ ,
\end{align}
where the perturbative zero momentum matrix elements $h^{\rm \overline{MS}}(z,\mu,P^z=0)$ are provided in Eq.~(\ref{eq:p0m}) up to two loops. The extra UV effects introduced in the hybrid renormalized matrix elements, compared to $\overline{\rm MS}$ ones, are
\begin{align}\label{eq:UVhybrid}
    \frac{1}{h^{\rm \overline{MS}}(z,\mu,P^z=0)}\theta(z_s-|z|) + \frac{1}{h^{\rm \overline{MS}}(z_s,\mu,P^z=0)}\theta(|z|-z_s) \ .
\end{align}
Next, the Fourier transformation is applied to obtain the quasi-PDF in the hybrid scheme,
\begin{align}\label{eq:FT}
    \tilde f^{h}(y,P^z) = P^z \int_{-\infty}^{+\infty} \frac{d z}{2\pi} e^{i z P^z y} h^{h}(z,P^z) \ ,
\end{align}
where the UV effects in Eq.~(\ref{eq:UVhybrid}) are transferred to the quasi-PDF. Finally, LaMET expansion is used to obtain the light-cone PDF, 
\begin{align}\label{eq:matchhybrid}
f(x,\mu)=\int_{-\infty}^{+\infty} \frac{d y}{|y|}{\cal C}^{h}\bigg(\frac{x}{y},\frac{|y| P^z}{\mu}\bigg)\tilde f^{h}\left(y,P^z\right) + ... \ ,
\end{align}
where ${\cal C}^{h}\bigg(\frac{x}{y},\frac{|y| P^z}{\mu}\bigg)$ is the matching kernel in the hybrid scheme. Compared to the kernel in Eq.~(\ref{eq:LaMETmatching}), this kernel includes hybrid counterterms to cancel the extra UV effects in the quasi-PDF, which can be found in Ref.~\cite{Su:2022fiu} up to NNLO for the isovector combination.

To convert the threshold matching kernel in Eqs.~(\ref{eq:MKthr}),~(\ref{eq:thrmatching}) and~(\ref{eq:resummedform}) to the hybrid scheme, we only consider the long tail correction $h^{\rm \overline{MS}}(z_s,\mu,P^z=0)$ in Eq.~(\ref{eq:UVhybrid}). This is because the threshold limit $\xi \rightarrow 1$ in momentum space corresponds to $z \rightarrow \infty$ in coordinate space. This correction can be absorbed into the hard kernel, as $z_s \sim \frac{1}{2 p^z}$ during the moderate and large $x$ range in practical calculations, 
\begin{align}\label{eq:Hhybrid}
H^{h} \bigg(L_{p_z},\alpha(\mu)\bigg) = H\bigg(L_{p_z},\alpha(\mu)\bigg) h^{\rm \overline{MS}}(z_s,\mu,P^z=0) \ .
\end{align}
With the hybrid scheme hard kernel $H^{h} \bigg(L_{p_z},\alpha(\mu)\bigg)$, the singular term in the matching kernel~\cite{Ji:2020brr} can be reproduced up to NLO,
\begin{align}\label{eq:ChybridTL}
&{\cal C}^h\left(\xi,\frac{p^z}{\mu}\right)_{\rm sg}=H^h\bigg(L_{p_z},\alpha(\mu)\bigg) \int_{-\infty}^{+\infty} \frac{d \lambda}{2 \pi} e^{i \lambda (\xi-1)} e^{i s_{\lambda} \Phi(L_{p_z},\alpha(\mu))} S\left(\ln\frac{\bar{\lambda}^2\mu^2}{4 p_z^2},\alpha(\mu)\right) \nonumber \\ 
&=\delta(1-\xi)\bigg[1-\frac{\alpha(\mu) C_F}{2\pi}\left(\frac{\pi^2}{3}-\frac{5}{2}-\frac{3}{2}\ln\frac{z_s^2\mu^2 e^{2\gamma_E}}{4}\right)\bigg]\nonumber \\ 
&\quad -\frac{\alpha(\mu)C_F}{2\pi}\bigg[{\cal P}\left(\frac{\ln[ (1-\xi)^2]}{|1-\xi|}\right)+(L_{p_z}-1)\bigg(\frac{1}{1-\xi} +{\cal P}\left(\frac{1}{|1-\xi|} \right)\bigg)\bigg] \ .
\end{align}
Here, the Fourier transformation with respect to $\lambda$ is performed based on the identities in Appendix~\ref{sec:FTofLogsasymp}. The terms ${\cal P}\left(\frac{\ln[ (1-\xi)^2]}{|1-\xi|}\right)$ and ${\cal P}\left(\frac{1}{|1-\xi|} \right)$ correspond to the Fourier transformation of even functions with respect to $\lambda$, such as $\ln^2 \lambda^2$ and $\ln \lambda^2$. For these terms, the singularity at $\xi=1$ is not canceled between $\xi<1$ and $\xi>1$ during the convolution, so the plus function definition Eq.~(\ref{eq:plusPV}) is necessary to ensure a finite result. The $\frac{1}{1-\xi}$ term corresponds to the Fourier transformation of odd functions with respect to $\lambda$, such as ${\rm sign}(\lambda)$. The singularity at $\xi=1$ is canceled between $\xi<1$ and $\xi>1$ during the convolution, so the plus function is unnecessary. We have also checked that Eqs.~(\ref{eq:Hhybrid}) and~(\ref{eq:ChybridTL}) can reproduce the real emission part of the threshold logarithms in the hybrid corrections at NNLO. 

The renormalization group equation for the hybrid scheme hard kernel is
\begin{align}
\frac{d}{d\ln \mu} \ln H^h\bigg(L_{p_z},\alpha(\mu)\bigg)=-\Gamma_{\rm cusp}(\alpha)L_{p_z}-(\tilde {\gamma}_{H}(\alpha)-2\gamma_F(\alpha)) \ ,
\end{align}
and the evolution factor $a_H^h$ in the hybrid scheme is
\begin{align}\label{eq:ahevo}
a_H^h(\nu,\mu)=\int_{\alpha(\nu)}^{\alpha(\mu)} d\alpha\frac{\tilde {\gamma}_{H}(\alpha)-2\gamma_F(\alpha)}{\beta(\alpha)} \ ,
\end{align}
where the results of $\gamma_F$ are collected in Appendix~\ref{sec:ano}. 

To obtain the hybrid scheme version of the threshold resummed matching kernel, Eq.~(\ref{eq:resummedform}) is modified to
\begin{align}\label{eq:resummedformhybrid}
&{\cal C}^{h}\left(\frac{x}{y},\frac{|y|P^z}{\mu},\mu_h,\mu_i\right)_{\rm sg}= \int \frac{d y'}{|y'|} {\cal S}\left( \frac{x}{y'},\frac{|y'|P^z}{\mu},\mu_i\right) {\cal H}^{h}\left( \frac{y'}{y},\frac{|y|P^z}{\mu},\mu_h\right) \ ,
\end{align}
where the soft function ${\cal S}\left( \frac{x}{y'},\frac{|y'|P^z}{\mu},\mu_i\right)$ is defined in Eq.~(\ref{eq:Jmom}). The hybrid scheme hard kernel ${\cal H}^{h}\left( \frac{y'}{y},\frac{|y|P^z}{\mu},\mu_h\right)$ is
\begin{align}\label{eq:Hmomhybrid}
{\cal H}^{h}\left( \xi,\frac{p^z}{\mu},\mu_h\right) 
=& \exp \bigg[2S_{\rm sud.}(\mu_h,\mu)-a^{h}_{H}(\mu_h,\mu)\bigg]H^{h}\left(\ln \frac{4p_z^2}{\mu_h^2},\alpha(\mu_h)\right) \left(\frac{\mu_h}{2 p^z}\right)^{2 a_{\Gamma}(\mu_h,\mu)} \nonumber\\
&\times \left[ \cos\left(\hat \Phi(p^z,\mu_h,\mu)\right) \delta(1-\xi) + \frac{\sin\left(\hat \Phi(p^z,\mu_h,\mu)\right)}{\pi(1-\xi)}\right] \theta\left(\gamma-|1-\xi|\right)\ ,
\end{align}
which is obtained by replacing $H\left(L_{p_z},\alpha(\mu)\right) \rightarrow H^h\left(L_{p_z},\alpha(\mu)\right)$ and $a_H(\nu,\mu) \rightarrow a_H^h(\nu,\mu)$ in Eq.~(\ref{eq:Hmom}).

\subsection{Resummation with accuracy at fixed orders}\label{sec:Rfix}

The resummation formalism concerns the 
singular threshold log contributions at $\xi=1$ at large orders in perturbation theory, where we do not have exact results. 
Very often, however, we do have exact results, including regular terms at lower fixed orders. 
In this subsection, the known fixed-order result without the threshold expansion is utilized to improve the accuracy of the resummation formalism. 

Eq.~(\ref{eq:matchhybrid}) with threshold resummation and fixed order accuracy is given by  
\begin{align}\label{eq:matchhybridtr}
f(x,\mu)  =  & \int_{x}^{1} d x' \hat{\cal P} \exp{\left[\int_{\mu_h}^{\mu} \frac{d\mu'}{\mu'} P \right]\left(\frac{x}{x'}\right)} \, \\
&\times \int \frac{dy}{|y|} \left[ {\cal C}^{h}\left(\frac{x'}{y},\frac{|y| P^z}{\mu_h},\mu_h,\mu_i=2|1-x|P^z\right)_{\rm sg} + \Delta {\cal C}\bigg(\frac{x'}{y},\frac{y P^z}{\mu_h}\bigg) \right] \tilde f^{h}\left(y,P^z\right) \ ,  \nonumber
\end{align}
where the hard scale $\mu_h=2|x|P^z$. In the second line, the light-cone PDF at the scale $\mu_h$ is calculated, with the resummation of the threshold logarithms through evolution from $\mu_i$ to $\mu_h$ in ${\cal C}^{h}\left(\frac{x'}{y},\frac{y P^z}{\mu_h},\mu_h,\mu_i\right)_{\rm sg}$, as defined in Eq.~(\ref{eq:resummedformhybrid}). The regular term correction up to a certain fixed order can be obtained through 
\begin{align}\label{eq:matchhybridcor}
\Delta {\cal C}\bigg(\frac{x'}{y},\frac{y P^z}{\mu_h}\bigg) = {\cal C}^{h}\bigg(\frac{x'}{y},\frac{y P^z}{\mu_h}\bigg) - {\cal C}^{h}\left(\frac{x'}{y},\frac{y P^z}{\mu_h},\mu_h,\mu_h\right)_{\rm sg} \ ,
\end{align}
where ${\cal C}^{h}\bigg(\frac{x'}{y},\frac{y P^z}{\mu_h}\bigg)$ is the fixed order kernel without threshold expansion in Eq.~(\ref{eq:matchhybrid}), and the subtraction term ${\cal C}^{h}\left(\frac{x'}{y},\frac{y P^z}{\mu_h},\mu_h,\mu_h\right)_{\rm sg}$ avoids double counting. The singular terms up to a certain fixed order cancel exactly in the above equation, although there are singular terms that are not of the threshold log type $\sim \ln\left(\mu_i/\mu_h\right)$, present beyond the order of accuracy, which are not important. 

In the first line of Eq.~(\ref{eq:matchhybridtr}), the DGLAP evolution from $\mu_h$ to $\mu$ is performed. $\int_{x}^{1} d x' \hat{\cal P} \exp{\left[\int_{\mu_h}^{\mu} \frac{d\mu'}{\mu'} P \right]\left(\frac{x}{x'}\right)}$ denotes the full DGLAP evolution~\cite{Moch:2004pa,Vogt:2004mw} from scale $\mu_h$ to $\mu$, where $P$ is the DGLAP kernel and $\hat{\cal P}$ is the path ordering operator for the integral with $\mu'$.

The singular terms in Eqs.~(\ref{eq:matchhybridtr}) and~(\ref{eq:matchhybridcor}) are applied under the constraint $|1-\xi| < \gamma$, as indicated by the unit step functions in Eqs.~(\ref{eq:Jmom}) and~(\ref{eq:Hmomhybrid}). These terms are the leading power under the threshold expansion $\xi \rightarrow 1$, which may not hold for $|1-\xi| > 1$, implying that the threshold cut $\gamma$ should be smaller than $1$. On the other hand, $\gamma$ should be much larger than $\Lambda_{\rm QCD}/P^z$ to ensure proper IR cancellation between the real and virtual contributions. In this work, we mostly constrain the singular terms within $|\xi-1| < 0.4$. Other threshold cut choices for $\gamma$, slightly bigger or smaller, do not impact the result appreciably, as will be explored in the next section. 

Besides, we resum the threshold logs only for $\mu_h > \mu_i$ since the threshold factorization is established for large $x$ light-cone PDFs, which may not be applicable in the small $x$ region. In the small $x$, the integral range for the real part is not restricted, and the effects of threshold logs are expected to be negligible. For $\mu_h < \mu_i$, the threshold resummation is turned off and we only resum the DGLAP logs following the method described in Ref.~\cite{Su:2022fiu}. At $\mu_h=\mu_i$, the results with and without threshold resummation are continuously connected in the formalism. The effect of our method is similar to the profile function method in Refs.~\cite{Abbate:2010xh,Hoang:2014wka}, as both approaches turn off threshold resummation in non-threshold regions (such as the $x < 0.5$ region in our case and the far tail regions in their cases). 
The two key differences are as follows: 
We have only two scales (hard and soft), whereas they discuss three scales (hard, jet, and soft). Consequently, the concept of merging the hard and soft scales to define a jet scale does not arise in our method; 
We require only the continuity of the light-cone PDF, whereas they impose continuity on both the profile function and its first derivative. The discontinuity in the first derivative at $x = 0.5$ in our method is a higher-order effect, and our numerical tests will demonstrate that it is negligible in the next section. 

\begin{table}[htbp]
\begin{equation}
\begin{array}{|c|c|c|c|c|c|}
\hline \text { RG-impr. PT } & \text { Log. Approx. } & \text { Accuracy } \sim \alpha^n L^k & \Gamma_{\text {cusp }} & \tilde \gamma_H, \gamma_F, \tilde \gamma_S & {H}^{h}, \Phi, {S} \\
\hline- & \mathrm{LL} & k = 2 n \, (\alpha^{-1}) & \text {1-loop } & \text {tree-level } & \text {tree-leve } \\
\hline \mathrm{LO} & \text { NLL } & 2n-1 \leq k \leq 2 n \, (\alpha^{0}) \quad & \text {2-loop } & \text {1-loop } & \text {tree-leve } \\
\hline \mathrm{NLO} & \text { NNLL } & 2n-3 \leq k \leq 2 n \, (\alpha^{1}) \quad &  \text {3-loop } & \text {2-loop } & \text {1-loop } \\
\hline \text { NNLO } & \text { NNNLL } & 2n-5 \leq k \leq 2 n \, (\alpha^{2}) \quad & \text {4-loop } &  \text {3-loop } & \text {2-loop } \\
\hline
\end{array} \nonumber
\end{equation}
\caption{Different approximation schemes for the matching kernel ${\cal C}^{h}_{\rm sg}$ in the threshold limit.}
\label{appschS}
\end{table}

\begin{table}[htbp]
\begin{equation}
\begin{array}{|c|c|c|c|c|c|}
\hline \text { RG-impr. PT } & \text { Log. Approx. } & \text { Accuracy } \sim \alpha^n L^k & \text{DGLAP kernel} & \beta & {\cal C}^{h}\\
\hline \mathrm{LO} & \text { LL } & n=k  & \text {1-loop } & \text {1-loop } & \text {tree-leve } \\
\hline \mathrm{NLO} & \text { NLL } & n-1 \leq k \leq n  &  \text {2-loop } & \text {2-loop } & \text {1-loop } \\
\hline \text { NNLO } & \text { NNLL } & n-2 \leq k \leq n  & \text {3-loop } &  \text {3-loop } & \text {2-loop } \\
\hline
\end{array} \nonumber
\end{equation}
\caption{Different approximation schemes for the full matching kernel ${\cal C}^{h}$ and DGLAP evolution kernel $P$.}
\label{appschM}
\end{table}
The log accuracy of the RG-improved perturbation theory depends on the order of anomalous dimensions and fixed order perturbation series. Tab.~\ref{appschS} shows different approximation schemes for the double log resummation~\cite{Becher:2006mr,Becher:2007ty}, which are related to ${\cal C}^{h}_{\rm sg}$ in Eq.~(\ref{eq:matchhybridtr}). In this context, the log accuracy is counted in coordinate space for convenience, though this can be done in momentum space based on the Fourier transformation formulas in Appendix~\ref{sec:FTofLogsasymp}. $L$ denotes $\ln \left(\mu_h^2/\mu^2\right)$ or $\ln\left(4 p_z^2/\mu^2\right)$ for the hard kernel, and $\ln\left(\mu_i^2/\mu^2\right)$ or $\ln\left(z^2 \mu^2 e^{2\gamma_E}/4\right)$ for the soft function in coordinate space. In the round brackets, we count the accuracy on the exponential factor, while outside the round brackets, we do it by expanding the exponential factor with $\alpha(\mu)$. Tab.~\ref{appschM} discusses single-log resummation, relevant to ${\cal C}^{h}$ and $P$ in Eq.~(\ref{eq:matchhybridtr}). In this case, $L$ means the DGLAP logs of type $\ln \left(\mu_h^2/\mu^2\right)$, which are resumed during the evolution from $\mu_h$ to $\mu$.

\section{Threshold resummation effects in pion PDF}\label{sec:pionPDF}

We conduct numerical studies of the threshold resummation on lattice QCD matrix elements for the valence quark distribution of a pion at lattice spacing $a$= 0.04\,fm and momenta $P^{z}$ = 1.9 and 2.4 GeV. These matrix elements have been calculated by the BNL/ANL collaboration and analyzed in their recent papers~\cite{Izubuchi:2019lyk,Gao:2020ito,Gao:2021hxl,Gao:2021dbh,Gao:2022iex,Gao:2022ytj}. A detailed explanation of their data is available in Ref.~\cite{Gao:2020ito}.

\begin{figure}[htbp]
    \centering
    \includegraphics[height=6.18cm]{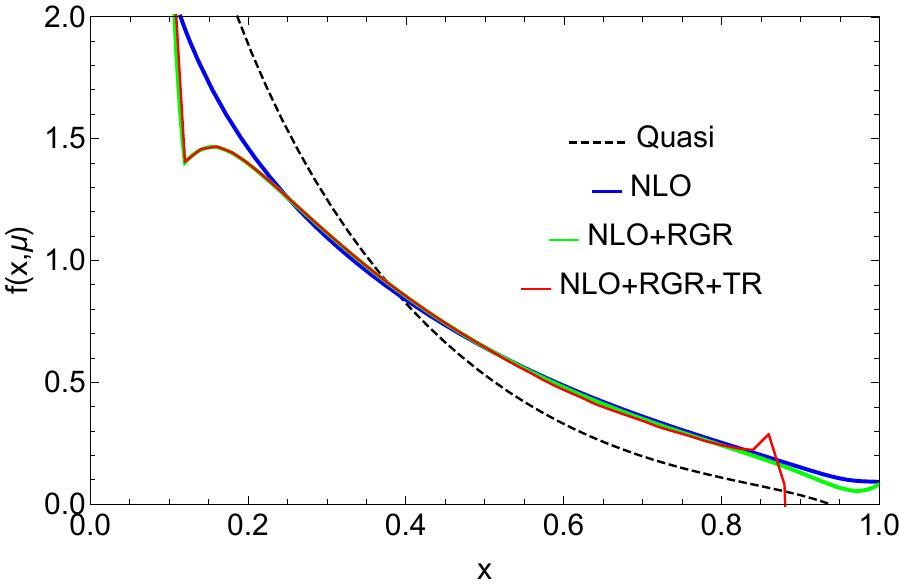}
    \caption{The light cone PDFs calculated from different types of matching kernels convoluted with the quasi-PDF. The dashed curve represents the quasi-PDF (or the light-cone PDF calculated with the LO matching kernel). The blue curve shows the NLO fixed-order result without any resummation. The green curve corresponds to the DGLAP-resummed result with $\mu_h = 2 |x| P^z$. The red curve incorporates both DGLAP and threshold logs resummation with $\mu_i=2|1-x|P^z$, $\mu_h = 2 |x| P^z$ and threshold cut $\gamma=0.4$. $P^z=1.94 \, {\rm GeV}$, $\mu=2 \, {\rm GeV}$ and $z_s = 0.12 \, {\rm fm}$.}
    \label{fig:MCfull1}
\end{figure}

The renormalization of the lattice data is performed using the hybrid scheme discussed in Sec.~\ref{sec:theory}, as shown in Eq.~(\ref{eq:hybridrenorm}). We extrapolate the result to the large $\lambda$ region~\cite{Ji:2020brr} assuming that the exponential decay of the correlation in lattice data has already set in. The Fourier transformation Eq.~(\ref{eq:FT}) is performed to obtain the quasi-PDF, shown as the dashed curve in Fig.~\ref{fig:MCfull1}. The matching is performed on the quasi-PDF to obtain the light cone PDF, ignoring power corrections in $1/P^z$.

To study the effects of threshold resummation, we compare the light-cone PDFs obtained using various matching kernels, as shown in Fig.~\ref{fig:MCfull1}. The NLO curve represents the light-cone PDF obtained with the one-loop fixed order matching without any resummation, based on Eq.~(\ref{eq:matchhybrid}). The NLO+RGR curve shows the light-cone PDF calculated with the NLO DGLAP resummation~\cite{Su:2022fiu}, which mostly influences small $x$ regions (e.g. $2 x P^z < 0.8$ GeV) compared to the NLO fixed order one. The NLO+RGR+TR curve is computed using both DGLAP and threshold logarithms resummation, based on Eq.~(\ref{eq:matchhybridtr}). As illustrated, the NLO+RGR+TR curve is almost consistent with the NLO+RGR curve in the moderate $x<0.8$ region where $2(1-x)P^z > 0.8$ GeV. In the large $x$ region, threshold resummation effects start to become important, but very quickly, the perturbation series breaks down.  

\begin{figure}[htbp]
    \includegraphics[height=5.0cm]{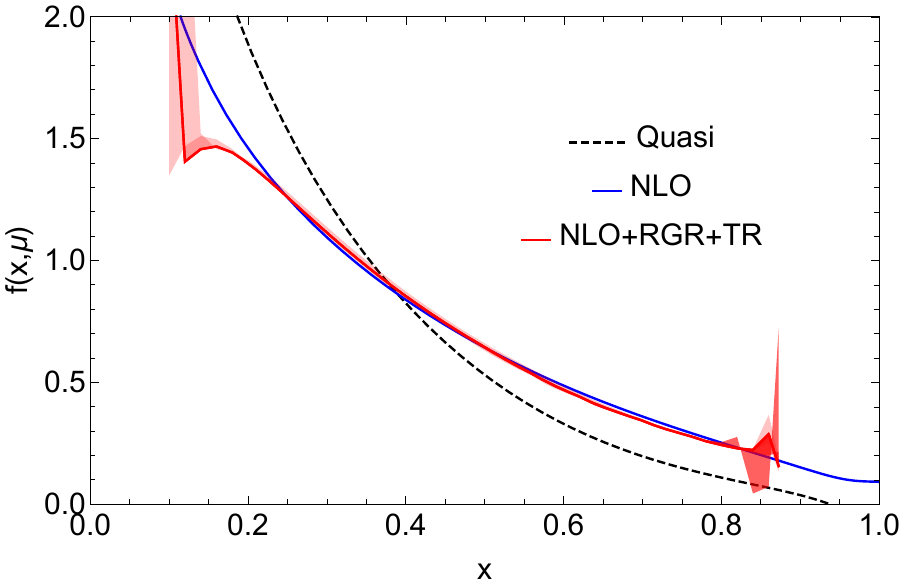}
    \includegraphics[height=5.0cm]{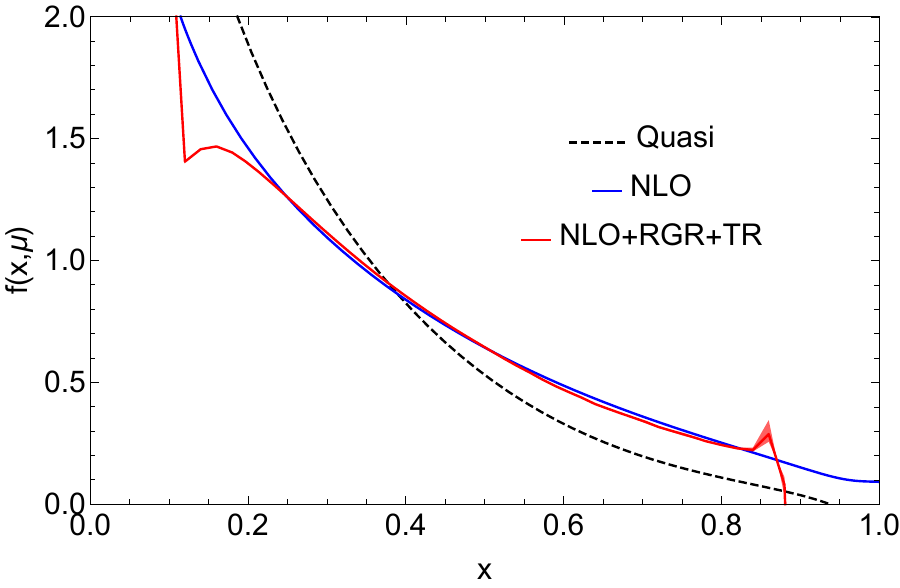}
    \caption{Similar to NLO+RGR+TR case in Fig.~\ref{fig:MCfull1}, but including uncertainties from scale variations and threshold cuts. 
    In the left panel, the dark red band represents the variation of the semi-hard scale $\mu_i=2|1-x|P^z r_1$ with $r_1=0.8 \sim 1.2$. The light red band shows the variation of the hard scale $\mu_h = 2 |x| P^z r_2$ with $r_2=0.8 \sim 1.2$. 
    In the right panel, the red band shows the variation of the threshold cuts, ranging from $\gamma=0.3\sim0.5$. }
    \label{fig:MCfullerr}
\end{figure}
Since the physical scale choice is an order of magnitude estimate rather than an exact result, we vary the hard and semi-hard scales to account for the uncertainties in the perturbation series, as shown in the left panel of Fig.~\ref{fig:MCfullerr}. In the moderate $x$-region (e.g. $2 x P^z > 0.8$ GeV and $2 (1-x) P^z > 0.8$ GeV), the effective couplings $\alpha(2 x P^z)$ and $\alpha(2 (1-x) P^z)$ for the hard and semi-hard scales are reasonably small, keeping the perturbative uncertainties under control. However, in the large $x$-region (e.g. $2 (1-x) P^z < 0.8$ GeV), the effective coupling $\alpha(2 (1-x) P^z)$ for the semi-hard scale becomes too large to be perturbative, leading to large uncertainties and indicating that the perturbative matching breaks down. In the right panel of Fig.~\ref{fig:MCfullerr}, we explore the impact of varying the threshold cuts, which are found to be negligible across the entire $x$-range.

\begin{figure}[htbp]
    \centering
    \includegraphics[height=6.18cm]{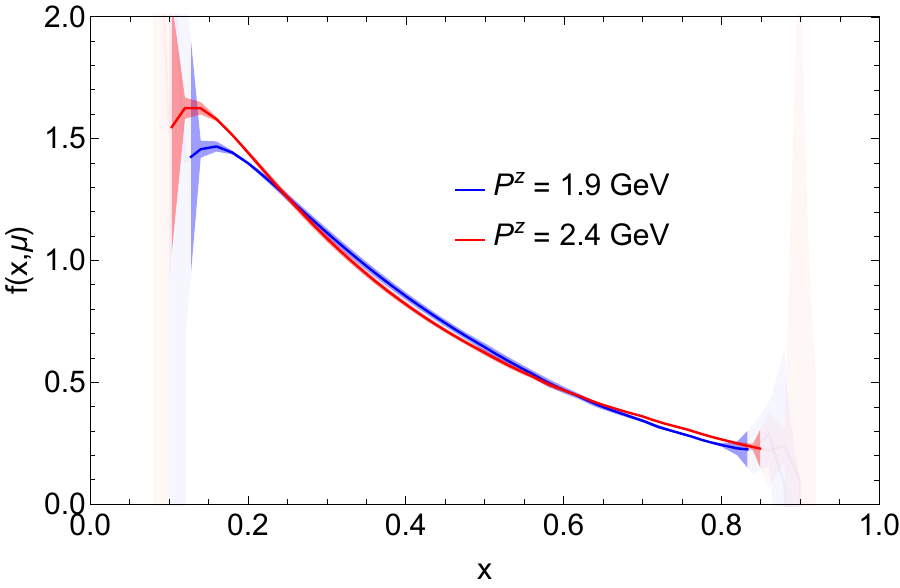}
    \caption{A comparison of light-cone PDFs calculated from different momenta under NLO+RGR+TR. The bands represent the scale and threshold cut variation effects, which are faded when the relative errors reach 33$\%$. $\mu=2 \, {\rm GeV}$ and $z_s = 0.12 \, {\rm fm}$.}
    \label{fig:ComparePz}
\end{figure}

\begin{figure}[htbp]
    \centering
    \includegraphics[height=6.18cm]{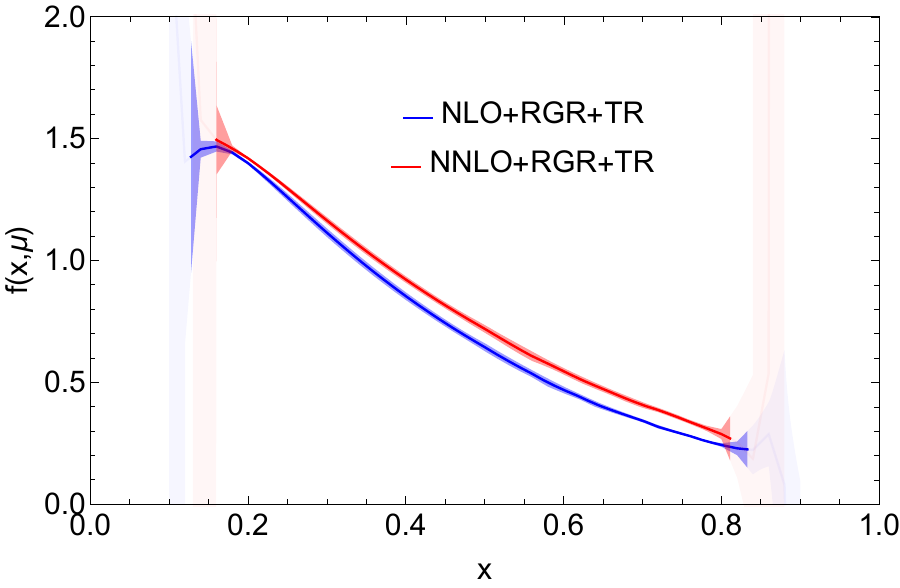}
    \caption{A comparison of light-cone PDFs calculated with NLO+RGR+TR and NNLO+RGR+TR matching kernels, without the control of renormalon. The calculations are performed with $P^z=1.94 \, {\rm GeV}$, $\mu=2 \, {\rm GeV}$, and $z_s = 0.12 \, {\rm fm}$.}
    \label{fig:NLO_NNLO_wo_renormalon}
\end{figure}
In Fig.~\ref{fig:ComparePz}, we compare the light-cone PDFs calculated from two different momenta $P^z=1.9$ GeV and $2.4$ GeV, using NLO+RGR+TR matching. The perturbative uncertainties from the hard scale variation $\sigma_h$, semi-hard scale variation $\sigma_i$, and threshold cut variation $\sigma_{\gamma}$ are combined as $\sqrt{\sigma_h^2+\sigma_i^2+\sigma_{\gamma}^2}$, which are represented by the bands. As the momentum increases, the reliable $x$-region extends.

The NLO+RGR+TR and NNLO+RGR+TR results are compared in Fig.~\ref{fig:NLO_NNLO_wo_renormalon}. They deviate in the moderate $x$ range (e.g. $0.3<x<0.7$), which may suggest poor perturbative convergence. This poor convergence could potentially be attributed to the leading renormalon effect~\cite{Zhang:2023bxs}, which will be further discussed in the next section.

\section{Threshold resummation with leading power accuracy}\label{sec:renormalon}

Using quasi-PDF as the starting point to make LaMET expansions, it is observed that the perturbative matching coefficients are numerically large in the first few orders. Moreover, for the threshold expansion, which is the main interest of this paper, large $\alpha$-expansion coefficients are artificially generated in the leading term in $1-\xi$.  Therefore, to have the expansions under control, one must study their convergences in $\alpha$. According to previous studies, the origin of these large perturbative coefficients is intimately connected to the linear power corrections of type $\mathcal{O}\left(\frac{\Lambda_{\rm QCD}}{|x| P^z}\right)$ and $\mathcal{O}\left(\frac{\Lambda_{\rm QCD}}{|1-x| P^z}\right)$, which are also large for $P^z\sim 2$ GeV in current lattice simulations. 

This section focuses on improving the $\alpha$-expansion convergence and at the same time, achieving the leading (linear) power accuracy in the threshold region. 
We show that the perturbative convergence is improved after resuming the leading renormalon contribution $C_n\sim n!\beta_0^n \alpha^{n+1}$ in the higher-order terms. Starting from the quasi-PDF after proper renormalization of linear divergences, the lightcone PDF $f(x,\mu)$ calculated through matching is expected to achieve the linear power accuracy $\mathcal{O}\left(\frac{\Lambda_{\rm QCD}}{|x| P^z}\right)$ and $\mathcal{O}\left(\frac{\Lambda_{\rm QCD}}{|1-x| P^z}\right)$, which correspond to the renormalon series in the hard-kernel phase angle and threshold soft function, respectively. 

\subsection{Factorial divergence and linear power accuracy}

In Eq.~(\ref{eq:matchhybrid}), the light-cone PDF $f(x,\mu)$ can be calculated from the quasi-PDF $\tilde f^{h}(y,P^z)$ through a triple expansion including large momentum expansion in $1/P^z$, threshold expansion in $1-\xi$ and perturbative expansion in $\alpha$,
\begin{align}\label{eq:OPEtri}
    f(x,\mu) = &\left[ \sum_{m=-1} \sum_{n=0} \left(1-\frac{x}{y}\right)^{m} \alpha^n(\mu) C_{m,n}\left( \frac{x}{y}, \frac{|y| P^z}{\mu} \right) \right] \otimes \tilde f^{h} \left(y, P^z\right) \nonumber\\
    & + \sum_{d=1} \mathcal{O}\left[ \frac{\Lambda_{\rm QCD}^d}{\left(|x| P^z\right)^d}, \frac{\Lambda_{\rm QCD}^d}{\left(|1-x| P^z\right)^d} \right] \ ,
\end{align}
where the first line represents the leading twist in the large momentum expansion, with the perturbative matching kernel expanded in $1-x/y$ and $\alpha$. The convolution with respect to $y$ is denoted by ``$\otimes$". In the leading threshold expansion ($m=-1$), the plus function definition and the corresponding $\delta(1-\xi)$ are omitted for simplicity. The second line corresponds to the power corrections in the large momentum expansion. Our goal is to achieve the {\it leading (linear) power accuracy} $\mathcal{O}\left[ \frac{\Lambda_{\rm QCD}}{|x| P^z}, \frac{\Lambda_{\rm QCD}}{|1-x| P^z} \right]$ for the calculation of light-cone PDF.

The leading threshold expansion ($m=-1$ in Eq.~(\ref{eq:OPEtri})) in the leading twist perturbation series exhibits factorial divergences~\cite{Beneke:1994sw, Beneke:1998ui, Beneke:2000kc,Shifman:2013uka},
\begin{align}
\sum_{n=0} \left(1-\frac{x}{y}\right)^{-1} \alpha^n(\mu) C_{-1,n}\left( \frac{x}{y}, \frac{|y| P^z}{\mu} \right) \ , 
\end{align}
where $C_{-1,n} \sim n!$ at large orders. This implies that the perturbation series diverges even for small values of $\alpha$. The factorial growth arises from contributions of ``IR momenta $\ll P^z$" or ``UV momenta $\gg P^z$" in the loop integral, such as $\int d^4 k \, \ln^n\left(-k^2/\mu^2\right)$, which are not the physics we seek at the $P^z$ scale. 

To address this, one can introduce a regularization prescription $\tau$ to tame the factorial divergence, which effectively regulates the unwanted physics. Different choices of the $\tau$ prescription lead to results that differ only by power corrections. This can be understood intuitively as ambiguities in defining the boundaries between ``the IR scale $\ll P^z$" and ``the $P^z$ scale", or ``the UV scale $\gg P^z$" and ``the $P^z$ scale". 

As a universal non-perturbative scaling function, the light-cone PDF $f(x,\mu)$ should be independent of the prescription $\tau$ used to regulate the factorial divergence in the leading twist leading threshold perturbation series. Thus, there must be $\tau$-dependences in the quasi-PDF, higher threshold, or higher twist expansions that cancel out the $\tau$ dependence in the leading twist leading threshold perturbation series,
\begin{align}\label{eq:OPEtritau}
    f(x,\mu)  = & \left[\sum_{n=0} \left(1-\frac{x}{y}\right)^{-1} \alpha^n(\mu) C_{-1,n}\left( \frac{x}{y}, \frac{|y| P^z}{\mu} \right) \right]_{\tau} \otimes \tilde f^{h} \left(y, P^z\right)_{\tau} \nonumber\\
    &+ \left[ \sum_{m=0} \sum_{n=0} \left(1-\frac{x}{y}\right)^{m} \alpha^n(\mu) C_{m,n}\left( \frac{x}{y}, \frac{|y| P^z}{\mu} \right) \right]_{\tau} \otimes \tilde f^{h} \left(y, P^z\right)_{\tau} \nonumber\\
    &+ \sum_{d=2} \mathcal{O}\left[ \frac{\Lambda_{\rm QCD}^d}{\left(|x| P^z\right)^d} , \frac{\Lambda_{\rm QCD}^d}{\left(|1-x| P^z\right)^d} \right]_{\tau} .
\end{align}
There is no sharp boundary between different terms in the triple expansion. Instead, they conspire with each other, and distinguishing between them is always ambiguous.

Note that the linear power corrections of type $\mathcal{O}\left[ \frac{\Lambda_{\rm QCD}}{|x| P^z} \right]$ and $\mathcal{O}\left[ \frac{\Lambda_{\rm QCD}}{|1-x| P^z} \right]$ do not appear in the above equation. The reason is that these are not genuine non-perturbative power corrections that can be constructed using higher-twist operators. The threshold renormalon ambiguity $\mathcal{O}\left[ \frac{\Lambda_{\rm QCD}}{|x| P^z} \right]$ in the hard kernel phase angle is an artifact of threshold expansion, which is expected to be canceled between the leading and higher threshold expansions regularized under the same prescription $\tau$~\cite{Liu:2023onm}. The mass renormalon ambiguity $\mathcal{O}\left[ \frac{\Lambda_{\rm QCD}}{|1-x| P^z} \right]$ in the soft function is canceled by the mass renormalization ambiguity of quasi-PDF under the same $\tau$~\cite{Braun:2018brg,Zhang:2023bxs}. It is important to note that the entire mass renormalon series for the quasi-PDF, discussed in Ref.~\cite{Zhang:2023bxs}, is absorbed into the soft function in the threshold limit. 

Thus, in an idealized situation, to achieve the leading (linear) power accuracy in calculating the light-cone PDF, the following conditions need to be satisfied:
\begin{itemize}
\item Obtain all orders in the leading-twist perturbation theory to achieve logarithmic accuracy;
\item Regulate the leading-twist perturbation theory under a prescription $\tau$, including both leading and higher threshold expansions;
\item Regulate the quasi-PDF using the same prescription $\tau$.
\end{itemize}
For this purpose, power corrections for $d \geq 2$ are not needed. 

In practice, it is difficult to obtain all-order leading-twist perturbation series. However, the phenomenon of ``leading renormalon dominance"\footnote{The perturbation series can be decomposed into the leading renormalon series and the remaining part. The leading renormalon series can be determined up to all orders of $\alpha$, while the remaining part cannot. However, as shown in Tab.~1 of Ref.~\cite{Pineda:2001zq}, the remaining part appears to be negligible starting from $n=1$ or $2$, a phenomenon we refer to as ``leading renormalon dominance". } has been observed in the perturbation series for transferring the $\overline{\rm MS}$ mass to the pole mass, starting from two or three loops~\cite{Pineda:2001zq}. LaMET matching kernels share some common features with the pole mass series, such as the presence of the same mass renormalon and both being defined in the $\overline{\rm MS}$ scheme. Therefore, we assume ``leading renormalon dominance" for the LaMET matching kernel starting from a low order of $\alpha$. 

To achieve linear power accuracy, we resum the leading (linear) renormalon series under a certain regularization prescription $\tau$\footnote{There are several approaches to resum a renormalon series, such as minimum term truncation and Borel summation. We choose the latter one in this paper.}, called \textit{leading (linear) renormalon resummation} (LRR)~\cite{Zhang:2023bxs}, and the prescription $\tau$ is chosen to be the same as that for quasi-PDF. The remaining part is added up to NLO or NNLO, based on known fixed-order results, with the assumption that higher-order contributions in the rest part should be negligible.

In the next few subsections, we discuss in detail on the leading (linear) renormalon series, the renormalon cancellation pattern and the strategy to achieve the leading (linear) power accuracy, for threshold and mass renormalons, respectively. We will explore how these renormalon effects manifest in the perturbative expansions and how their proper resummation and cancellation can lead to more accurate predictions.

\subsection{Threshold renormalon in hard-kernel phase angle}\label{sec:TRH}
This subsection provides detailed explanations for the following questions: What is the leading renormalon series for the hard-kernel phase angle? What is the renormalon cancellation pattern?
How do we deal with this renormalon to achieve the linear power accuracy of type $\mathcal{O}\left(\frac{\Lambda_{\rm QCD}}{|x| P^z}\right)$?
The first two questions have already been explored in Ref.~\cite{Liu:2023onm}, which we will explain in a pedagogical manner. Based on those insights, we address the third question. 

\subsubsection{The leading renormalon series}
The perturbative expansion of the phase angle $\Phi(L_{p_z}, \alpha(\mu))$ defined in Eq.~(\ref{eq:Hdef}) reads
\begin{align}\label{eq:Phaseanglepert}
    \Phi(L_{p_z}, \alpha(\mu)) = \sum_{n=0}^{\infty} a_{n}\left(L_{p_z}\right) \, \alpha^{n+1}(\mu) \ ,
\end{align}
where the coefficients $a_{n}\left(L_{p_z}\right)$, which include the logarithmic terms $L_{p_z} =\ln(4p_z^2/\mu^2)$, have been determined up to two-loop order, as shown in Appendix~\ref{sec:fix}. The phase angle $\Phi(L_{p_z}, \alpha(\mu))$ also satisfies the RG equation given by Eq.~(\ref{eq:RGEA}), with the four-loop cusp anomalous dimension provided in Appendix~\ref{sec:ano}. 

At large $n$, the coefficients $a_{n}\left(L_{p_z}\right)$ are dominated by contributions from IR momenta in the loop integral, such as $\int d^4 k \, \ln^n\left(-k^2/\mu^2\right)$. Due to these IR momenta, the $a_{n}(L_{p_z})$ coefficients exhibit a factorial growth factor $n!$, leading to a divergent perturbative series. This factorial growth has been confirmed through calculations in the large $\beta_0$ limit~\cite{Liu:2023onm}. 

The above factorial divergence is also called renormalon divergence~\cite{Beneke:1994sw, Beneke:1998ui, Beneke:2000kc,Shifman:2013uka}. To study the renormalon divergence, a Borel transform $B[\Phi]\left(t,L_{p_z}\right)$ and Borel sum $\tilde \Phi(L_{p_z}, \alpha(\mu))$ are typically defined as
\begin{align}\label{eq:BorelTran}
    &B[\Phi]\left(t,L_{p_z}\right) = \sum_{n=0}^{\infty} a_{n}\left(L_{p_z}\right) \, \frac{t^n}{n!} ,\
    &\tilde \Phi(L_{p_z}, \alpha(\mu)) = \int_{0}^{\infty} d t \, e^{-t/\alpha(\mu)} B[\Phi]\left(t,L_{p_z}\right) .\
\end{align}
The Borel transform maps the renormalon divergences of $\Phi(L_{p_z}, \alpha(\mu))$ into singularity poles or branch cuts of $B[\Phi]\left(t, L_{p_z}\right)$ on the Borel plane. The Borel sum $\tilde \Phi(L_{p_z}, \alpha(\mu))$ allows us to regulate the renormalon divergences by choosing a specific integral contour in the complex plane of $t$, called a summation scheme $\tau$. 

Since there are singularity poles or branch cuts on the Borel plane, different contour choices lead to different results~\cite{Beneke:1994sw, Beneke:1998ui, Beneke:2000kc,Shifman:2013uka}. The resulting difference is called renormalon ambiguity, which manifests as a power correction. In the hard kernel, the ``on-shell" spacelike heavy quark generates a background potential that interacts with the light quark. Since the perturbation series of the static potential contains a linear renormalon~\cite{Pineda:2002se}, it is expected that the hard kernel also exhibits a linear renormalon. The renormalon ambiguity should be of the order $\mathcal{O}\left(\frac{\Lambda_{\rm QCD}}{|x| P^z}\right)$, where $x P^z$ is the only scale in the hard kernel. The linear renormalon series should be purely real if the heavy quark is timelike. After the analytical continuation, the spacelike hard kernel in our threshold resummation formalism contains the linear renormalon series in its imaginary part, which can be absorbed into the phase angle, as checked in the large $\beta_0$ limit~\cite{Liu:2023onm}.

To perform the leading renormalon resummation (LRR), we need to know the leading renormalon series up to all orders. It can be extracted from the closest singularity near the origin on the Borel plane. In the large $\beta_0$ limit, its Borel transform is given by~\cite{Liu:2023onm} 
\begin{align}
B[\Phi](t(u),L_{p_z}) = - \frac{C_F}{2 u} \left( \frac{e^{ \left( \frac{5}{3}-L_{p_z} \right) u}}{u-1}\frac{1}{\cos(\pi u)}+1\right) \ ,
\end{align}
where $u=\beta_0 t/2$. Our convention for the beta function is shown in Eq.~(\ref{eq:beta}). For the closest singularity near the origin on the positive real axis, the Borel transform behaves as
\begin{align}\label{eq:ABhalf}
B[\Phi]\left(t(u), L_{p_z}\right) \stackrel{u \sim 1/2}{=}  -\frac{C_F e^{6/5}}{\pi} \frac{1}{u-1/2} \frac{\mu}{p^z} + ... \ ,
\end{align}
where we omit the terms that are analytic near $u=1/2$. Following the same logic in Ref.~\cite{Pineda:2001zq}, we obtain the dominant contribution to the perturbative expansion at large orders from the small $u$ expansion of Eq.~(\ref{eq:ABhalf}) compared with Eq.~(\ref{eq:BorelTran}),
\begin{align}\label{eq:asymA}
a_n(L_{p_z}) \stackrel{n \rightarrow \infty}{=} 2\frac{\mu}{p^z} r_{n} \ ,
\end{align}
where $r_n$ in the large $\beta_0$ limit is $r_n = \beta_0^n n! C_F e^{5/6}/\pi $~\cite{Liu:2023onm}. Note that the overall amplitude $C_F e^{5/6}/\pi$ here is the same as that for the mass renormalon series in the large $\beta_0$ limit~\cite{Beneke:1994bc,Bigi:1994em,Beneke:1994sw}.

We provide an interpretation for the power dependence $\mu/p^z$ in Eq.~(\ref{eq:asymA}). At large orders $n$ of the perturbation series, numerous logs arise whose coefficients are constrained by both renormalon effects and the RG equation. As $n$ approaches infinity, the number of these logarithmic terms also grows indefinitely. When summed, they produce the power dependence $\mu/p^z$. To demonstrate this, we begin with the renormalon series expressed in terms of logarithms,
\begin{align}\label{eq:ALog}
a_n(L_{p_z}) \stackrel{n \rightarrow \infty}{=} \sum_{m=0}^{n} c^{\rm asym}_{n,m} (-L_{p_z})^m \ .
\end{align}
The constant term is extracted from Eq.~(\ref{eq:asymA}) by evaluating at the physical scale $\mu=2 p^z$,
\begin{align}\label{eq:Acn0}
    c^{\rm asym}_{n,0} = 4 r_{n} \ .
\end{align}
The coefficients for the log terms in the large $n$ limit can be derived from the RG equation Eq.~(\ref{eq:RGEA}),
\begin{align}\label{eq:Acnm}
    c^{\rm asym}_{n,m} = \frac{1}{2^m m!} c^{\rm asym}_{n,0} \ .
\end{align}
Note that the anomalous dimension in Eq.~(\ref{eq:RGEA}) does not enter the large $n$ asymptotic form. 
The logarithmic renormalon series in Eq.~(\ref{eq:ALog}) can be simplified based on Eqs.~(\ref{eq:Acn0}) and~(\ref{eq:Acnm}), 
\begin{align}
a_n(L_{p_z}) \stackrel{n \rightarrow \infty}{=} \sum_{m=0}^{n} \frac{1}{2^m m!} (4 r_{n}) (-L_{p_z})^m =  (4 r_{n}) \frac{\mu}{2 p^z}\ ,
\end{align}
where an infinite number of log terms are summed up in the $n \rightarrow \infty$ limit, ending up with the power dependence $\mu/p^z$. 

To improve the resummation accuracy, we study the leading renormalon series beyond the large $\beta_0$ limit. Since the threshold renormalon involves the interaction between a light quark and an ``on-shell" heavy quark, a conjecture was proposed in Ref.~\cite{Liu:2023onm} that the coefficients $r_n$ beyond large $\beta_0$ are the same as those for the mass renormalon series or static potential~\cite{Beneke:1994rs,Pineda:2001zq,Pineda:2002se,Bali:2013pla,Ayala:2014yxa},
\begin{align}\label{eq:asymcoe}
r_{n} = N_m \beta_0^n \frac{\Gamma(n+1+b)}{\Gamma(1+b)} \left( 1+ \frac{b}{b+n} c_1 + \frac{b(b-1)}{(n+b)(n+b-1)} c_2 + ...\right),
\end{align}
where $b = \frac{\beta_1}{\beta_0^2}$, $c_1 = \frac{\beta_1^2 - \beta_0 \beta_2}{b \beta_0^4}$ and $c_2 = \frac{-\beta _3 \beta _0^4+2 \beta _1 \beta _2 \beta _0^3+\left(\beta _2^2-\beta _1^3\right)\beta _0^2-2 \beta _1^2 \beta _2 \beta _0+\beta _1^4}{2 (b-1) b \beta _0^8}$. Our convention for the beta function is shown in Eq.~(\ref{eq:beta}). Note that the coefficients $b$, $c_1$ and $c_2$ are determined by requiring that the renormalon ambiguity is scale independent~\cite{Beneke:1994rs}. The overall amplitude can be determined~\cite{Pineda:2001zq} from the perturbation series between the pole mass and $\overline{\rm MS}$ mass~\cite{Melnikov:2000qh},
\begin{align}
    N_m = 
    \begin{cases}
        0.622 & n_f = 0 \\
        0.609 & n_f = 1 \\
        0.593 & n_f = 2 \\
        0.575 & n_f = 3 \\
        0.552 & n_f = 4 \\
        0.524 & n_f = 5 
    \end{cases} \ ,
\end{align}
where the $N_m=0.622$ for $n_f=0$ is consistent with that obtained from the linear divergence series $N_m = 0.660(56)$~\cite{Bali:2013pla}.\footnote{The computation of the normalization factor $N_m$ has also been carried out using R-evolution techniques in Ref.~\cite{Hoang:2017suc}.} The pole mass series and linear divergence series approach the asymptotic form Eq.~(\ref{eq:asymcoe}) at different rates. As shown in Ref.~\cite{Pineda:2001zq}, the pole mass series is almost consistent with the asymptotic form near $n \sim 2$. Whereas in Ref.~\cite{Bali:2013pla}, the linear divergence series starts to be dominated by the asymptotic form near $n \sim 10$.

\subsubsection{Threshold renormalon cancellation pattern}
The threshold renormalon ambiguity $\mathcal{O}\left(\frac{\Lambda_{\rm QCD}}{|x| P^z}\right)$ does not exist in the full matching kernel, and it arises purely from the threshold expansion as $\xi \rightarrow 1$. This ambiguity is canceled between the leading and next-to-leading threshold expansions under the same prescription $\tau$. These arguments have been confirmed in the large $\beta_0$ limit~\cite{Liu:2023onm}, which we will briefly revisit for pedagogical purposes.

To study the threshold renormalon, only the sail diagram (Fig.~1(a) or (b) in~\cite{Braun:2018brg}) of quasi-PDF in a massless free quark state is needed. The box diagram (Fig.~1(c) in~\cite{Braun:2018brg}) is irrelevant to the threshold limit. The self-energy diagram (Fig.~1(d) in~\cite{Braun:2018brg}) is related to mass renormalon, which will be discussed in the next subsection. As calculated in~\cite{Braun:2018brg,Liu:2023onm}, the Borel transform of the factorial growing part of the sail diagram in coordinate space, under dimensional regularization with the UV or IR divergences subtracted in the $\overline{\rm MS}$ scheme, is
\begin{align}\label{eq:SailL}
&B[\tilde f_{\rm sail}]\left(t(u),\lambda, z^2\mu^2\right) \nonumber\\
=& \frac{C_F e^{-i \lambda }}{4 \pi u} \left\{ e^{5u/3} \left(\frac{z^2 \mu ^2}{4}\right)^u \Gamma(1-u) \Gamma(1+2 u) 
\left[ \, _2\tilde{F}_2(1,2 u;2+u,1+2 u;i \lambda ) \right. \right. \nonumber\\
&\left. \left. -2 i \lambda  \left(\, _2\tilde{F}_2(1,1+2 u;2+u,2+2u;i \lambda )-\, _2\tilde{F}_2(2,1+2 u;3+u,2+2 u;i \lambda )\right) \right]  \right. \nonumber\\
&\left.-\left[ 1 + \frac{2 (i-\lambda -i e^{i \lambda})}{\lambda} +   2 \ln\left(-i \lambda e^{\gamma_E} \right) + 2 \Gamma(0,-i \lambda) \right] \right\} \ ,
\end{align}
where $z=\lambda/p^z$. $_2\tilde{F}_2$ is the regularized generalized hypergeometric function, and $\Gamma(0,-i \lambda)$ is the incomplete gamma function. Eq.~(\ref{eq:SailL}) is the left hand side of Eq.~(3.26) in Ref.~\cite{Liu:2023onm}. There is no singularity pole at $u = 1/2$, which means the equation does not contain a linear renormalon ambiguity. 

As discussed in~\cite{Liu:2023onm}, the threshold expansion $\xi \rightarrow 1$ corresponds to $\lambda \rightarrow \infty$ in coordinate space,
\begin{align}
B[\tilde f_{\rm sail}]\left(t(u),\lambda,z^2\mu^2\right) 
= B_0(u,\lambda,z^2\mu^2)+\frac{1}{\lambda} B_1(u,\lambda,z^2\mu^2)+\mathcal{O}\left(\frac{1}{\lambda^2}\right) \ ,
\end{align}
where the large $\lambda$ power counting is performed at $u=0$. The leading threshold expansion is given by
\begin{align}\label{eq:SailLThr}
B_0(u,\lambda,z^2\mu^2) & = 
\frac{e^{-i \lambda} C_F}{4 \pi u^2}\left(e^{\frac{5 u}{3}}\left(\frac{i|z| \mu}{2 \lambda}\right)^{2 u} \frac{2 \pi u}{(u-1) \sin 2 \pi u}+\left(\frac{8}{3}+2 \ln \frac{i|z| \mu}{2 \lambda}\right) u+1\right) \nonumber\\
& \quad + \frac{e^{-i \lambda} C_F}{4 \pi u^2}\left(e^{\frac{5 u}{3}}\left(\frac{z^2 \mu^2}{4}\right)^u \frac{\Gamma(1-u)}{\Gamma(1+u)}-\left(\frac{5}{3}+\ln \frac{e^{2 \gamma_E} z^2 \mu^2}{4}\right) u-1\right) \nonumber\\
&\stackrel{u \sim 1/2}{=} i \, s_{\lambda} \, e^{-i \lambda} \frac{C_F e^{5/6}}{\pi} \frac{1}{2 u -1} \frac{\mu}{p^z} \ ,
\end{align}
where $s_{\lambda}$ is defined below Eq.~(\ref{eq:DJdef}). A singularity pole at $u=1/2$ appears in the equation, indicating that a linear renormalon ambiguity arises after the threshold expansion. This is the threshold renormalon in the hard kernel phase angle. 

The next leading threshold expansion is
\begin{align}
\frac{1}{\lambda} B_1(u,\lambda,z^2\mu^2) & 
= i \frac{1}{\lambda} \frac{e^{-i \lambda} C_F}{2 \pi u}\left(e^{\frac{5 u}{3}}\left(\frac{z^2 \mu^2}{4}\right)^u \frac{\Gamma(1-u)}{(1-2 u) \Gamma(u+1)}-1\right) \nonumber\\
&\stackrel{u \sim 1/2}{=} - i \, s_{\lambda} \, e^{-i \lambda} \frac{C_F e^{5/6}}{\pi} \frac{1}{2 u -1} \frac{\mu}{p^z} \ ,
\end{align}
which contains a singularity pole at $u=1/2$, and its residue is exactly the opposite of that in Eq.~(\ref{eq:SailLThr}). This shows that the threshold renormalon is canceled between the leading and next-to-leading threshold expansion when they are regularized under the same prescription.

\subsubsection{Resuming the threshold renormalon series}
To improve the perturbative convergence and eliminate the linear power correction of type $\mathcal{O}\left(\frac{\Lambda_{\rm QCD}}{|x| P^z}\right)$, we perform LRR for both leading and next-to-leading threshold expansions under the same prescription $\tau$, in the following two steps.

First, the LRR is implemented for the hard kernel phase angle. The results up to NLO and NNLO are
\begin{align}\label{eq:ANLOLRR}
    &\Phi^{\rm NLO+LRR}\bigg(L_{p_z}=\ln \frac{4p_z^2}{\mu^2},\alpha(\mu)\bigg) = a_{0}\left(L_{p_z}\right) \, \alpha(\mu) \nonumber\\
    &- 2\frac{\mu}{p^z} r_{0} \alpha(\mu) + {\rm p.v.}\left(2\frac{\mu}{p^z} \sum_{n=0}^\infty r_{n} \alpha^{n+1}(\mu) \right),
\end{align}
\begin{align}\label{eq:ANNLOLRR}
    &\Phi^{\rm NNLO+LRR}\bigg(L_{p_z}=\ln \frac{4p_z^2}{\mu^2},\alpha(\mu)\bigg) = a_{0}\left(L_{p_z}\right) \, \alpha(\mu) + a_{1}\left(L_{p_z}\right) \, \alpha^2(\mu) \nonumber\\
    &- 2\frac{\mu}{p^z} r_{0} \alpha(\mu) - 2\frac{\mu}{p^z} r_{1} \alpha^2(\mu) + {\rm p.v.}\left(2\frac{\mu}{p^z} \sum_{n=0}^\infty r_{n} \alpha^{n+1}(\mu) \right),
\end{align}
where $a_{0}\left(L_{p_z}\right) \, \alpha(\mu)$ and $a_{1}\left(L_{p_z}\right) \, \alpha^2(\mu)$ are the fixed order results at NLO and NNLO, following the notation in Eq.~(\ref{eq:Phaseanglepert}). The coefficients $r_{n}$ is the leading renormalon series beyond large $\beta_0$ limit in Eq.~(\ref{eq:asymcoe}). The subtraction terms are introduced to avoid double counting. ``${\rm p.v.}$" means resuming the renormalon series in the Borel plane using the principle value prescription~\cite{Chyla:1990na,Cvetic:2002qf,Ayala:2019uaw}, 
\begin{align}\label{eq:pvrenormalon}
{\rm p.v.}\sum_{n=0}^\infty r_{n} \alpha^{n+1}(\mu) = {\rm Re}\left[ \lim_{\epsilon \rightarrow 0^{+}} \int_{0}^{\infty} d t \, e^{-t/\alpha(\mu)} \sum_{n=0}^{\infty} r_n \, \frac{(t+i \epsilon)^n}{n!} \right],
\end{align}
where the prescription $ i \epsilon$ is introduced to regulate the branch cut on the Borel plane, and the principle value prescription is defined as the real part of the integral.\footnote{Compared with the $R$-scheme papers~\cite{Hoang:2008yj,Hoang:2009yr,Benitez-Rathgeb:2022yqb}, the renormalon series or subtraction terms in Eqs.~(\ref{eq:ANLOLRR}) and~(\ref{eq:ANNLOLRR}) amount to taking $R=\mu$ for simplicity. However, this is not mandatory. In general, $R$ can be varied around the same order of $\mu$ as one of the sources of perturbative uncertainties. The $R$-variation effects have not been considered in this paper and can be investigated in a future work. } 

Then, the phase angle $\Phi\left(L_{p_z}=\ln \frac{4p_z^2}{\mu_h^2},\alpha(\mu_h)\right)$ in the hard kernel ${\cal H}^{h}\left( \xi,\frac{p^z}{\mu},\mu_h\right)$ from Eq.~(\ref{eq:Hmomhybrid}) is replaced with the LRR versions provided in Eqs.~(\ref{eq:ANLOLRR}) and (\ref{eq:ANNLOLRR}), setting $\mu=\mu_h$. The LRR-improved ${\cal H}^{h}\left( \xi,\frac{p^z}{\mu},\mu_h\right)$ is then used for both ${\cal C}^{h}\left(\frac{x'}{y},\frac{y P^z}{\mu_h},\mu_h,\mu_i\right)_{\rm sg}$ and
${\cal C}^{h}\left(\frac{x'}{y},\frac{y P^z}{\mu_h},\mu_h,\mu_h\right)_{\rm sg}$ in the matching formula from Eq.~(\ref{eq:matchhybridtr}). 

Since the leading and next-to-leading threshold expansions are regularized in the same summation scheme $\tau$, the leading power correction of type $\mathcal{O}\left(\frac{\Lambda_{\rm QCD}}{|x| P^z}\right)$ should cancel in Eq.~(\ref{eq:matchhybridtr}), which can be demonstrated in the large $\beta_0$ limit. The singular term defined in Eq.~(\ref{eq:resummedformhybrid}) can be expressed as follows
\begin{align}\label{eq:resummedformhybridLbeta0}
{\cal C}^{h}\left(\frac{x}{y},\frac{|y|P^z}{\mu},\mu_h,\mu_i\right)_{\rm sg} = & {\cal S}\left( \frac{x}{y},\frac{|y|P^z}{\mu},\mu_i\right) +  {\cal H}^{h}\left( \frac{x}{y},\frac{|y|P^z}{\mu},\mu_h\right)_{\tau} \nonumber\\
& - \delta\left(1-\frac{x}{y}\right) + \mathcal{O}\left(\frac{1}{\beta_{0}^2}\right) \ ,
\end{align}
where the soft function ${\cal S}$ and hard kernel ${\cal H}^{h}$ can be added rather than convoluted because their product contributes only subleading terms at $\mathcal{O}\left(1/\beta_0^2\right)$ in the large $\beta_0$ limit~\cite{Liu:2023onm,Liu:2024omb}. ${\cal H}^{h}$ is regularized under the prescription $\tau$, such as ``p.v." defined in Eq.~(\ref{eq:pvrenormalon}). The summation scheme of ${\cal S}$ is not specified here as it does not affect the current discussion and will be studied in Sec.~\ref{sec:MRJ}. The tree-level $\delta$-function is subtracted to avoid double counting, although it is irrelevant for this discussion. Substituting Eq.~(\ref{eq:resummedformhybridLbeta0}) into the second line of Eq.~(\ref{eq:matchhybridtr}) gives
\begin{align}\label{eq:LPALbeta0}
&f\left(x',\mu_h\right)=\int \frac{dy}{|y|} \left[ {\cal S}\left( \frac{x'}{y},\frac{|y|P^z}{\mu_h},\mu_i=2|1-x|P^z\right) + {\cal H}^{h}\left( \frac{x'}{y},\frac{|y|P^z}{\mu_h},\mu_h\right)_{\tau} \right.\nonumber\\
&\left.\quad + {\cal C}^{h}\bigg(\frac{x'}{y},\frac{|y| P^z}{\mu_h}\bigg) - {\cal S}\left( \frac{x'}{y},\frac{|y|P^z}{\mu_h},\mu_i=\mu_h\right) - {\cal H}^{h}\left( \frac{x'}{y},\frac{|y|P^z}{\mu_h},\mu_h\right)_{\tau} \right]\tilde f^{h}(y,P^z) \ ,
\end{align}
where the first line corresponds to the threshold resummed kernel ${\cal C}^{h}\left(\frac{x'}{y},\frac{y P^z}{\mu_h},\mu_h,\mu_i\right)_{\rm sg}$, and the second line comes from the regular term correction $\Delta {\cal C}\bigg(\frac{x'}{y},\frac{y P^z}{\mu_h}\bigg)$, which contains the next-to-leading threshold expansion. 
The hard kernels ${\cal H}^{h}$ cancel in Eq.~(\ref{eq:LPALbeta0}), indicating that leading power accuracy of $\mathcal{O}\left(\frac{\Lambda_{\rm QCD}}{|x| P^z}\right)$ can be achieved in the large $\beta_0$ limit, either with LRR under the same prescription $\tau$ or even without performing LRR. Note that the large $\beta_0$ limit is used solely for theoretical discussions but not for practical calculations.

Beyond the large $\beta_0$ limit, the effects of the hard kernel are not fully canceled. However, if the assumption of ``leading renormalon dominance" holds, we expect that performing LRR will improve the perturbative convergence.




\subsection{Mass renormalon in the soft function}\label{sec:MRJ}

In this subsection, we discuss the asymptotic form and cancellation pattern for the mass renormalon in the soft function. The strategy to perform the LRR and achieve the leading power accuracy of $\mathcal{O}\left(\frac{\Lambda_{\rm QCD}}{|1-x| P^z}\right)$ is similar to that in Ref.~\cite{Zhang:2023bxs}, but more accurate in the large $x$ region. 

The threshold soft function defined in Eq.~(\ref{eq:Sdef}) can be expanded in $\alpha$,
\begin{align}
S(l_z,\alpha(\mu)) = 1 + \sum_{n=0}^{\infty} s_{n}(l_z) \, \alpha^{n+1}(\mu) \ ,
\end{align}
where the coefficients $s_{n}\left(l_z\right)$ contain the log terms $l_z =\ln\left( \mu^2 z^2 e^{2 \gamma_E} / 4 \right)$ with $z = \lambda/p^z$ and have been determined up to two-loop order, as shown in Appendix~\ref{sec:fix}. The soft function $S(l_z,\alpha(\mu))$ satisfies the RG equation given in Eq.~(\ref{eq:RGEtiledJ}), with the relevant anomalous dimensions listed in Appendix~\ref{sec:ano}. 

The mass renormalon in the soft function is exactly the same as that in the quasi-PDF~\cite{Braun:2018brg, Liu:2020rqi, Zhang:2023bxs} because the Wilson link self-energy of the soft function is identical to that of the quasi-PDF\footnote{Also note that the Wilson link self-energy of quasi-PDF only contributes to the leading threshold expansion, which can be shown by doing the Fourier transformation $\sim \int d \lambda \, \ln^n\left(\lambda\right) \, e^{i \lambda (\xi-1)}$, detailed in Appendix~\ref{sec:FTofLogsasymp}. Here $\ln^n\left(\lambda\right)$ is the only possible $\lambda$-dependence in the Wilson link self-energy diagram in coordinate space. }. 
This means they share the same mass renormalon series,
\begin{align}\label{eq:Jasymp}
    s_{n}(l_z)  \stackrel{n \rightarrow \infty}{=} - |z|\mu \, r_{n} \ ,
\end{align}
where $r_{n}$ is the coefficient from Eq.~(\ref{eq:asymcoe}). And, this renormalon series conspires with the same finite mass counterterm during the renormalization of linear divergence. 

Thus, we follow a similar strategy as Ref.~\cite{Zhang:2023bxs} to achieve the leading power accuracy. Specifically, we perform the LRR for the soft function $S$ based on the leading renormalon series, regularized under a certain prescription, such as the principle value prescription. At the same time, the renormalized lattice matrix element $h^h$ is converted to the same prescription in Eq.~(\ref{eq:hybridrenorm}), where the renormalization factor $Z_R$ contains the mass renormalization parameter $m_0$, which converts lattice data to the desired renormalon regularization scheme, such as the principle value prescription. $m_0$ is calculated by fitting lattice data to the LRR version of $\overline{\rm MS}$ perturbation series at short distances in Eq.~(\ref{eq:Hyrenormfit}). See Ref.~\cite{Zhang:2023bxs} for more details on the determination of $m_0$.

However, there are two major differences between our strategy and Ref.~\cite{Zhang:2023bxs}. The first is that we exponentiate the mass renormalon series,
\begin{align}\label{eq:SLRR}
&S^{\rm NLO+LRR}(l_z,\alpha(\mu)) = S^{\rm NLOsub}(l_z,\alpha(\mu)) \exp\left[-|z|\mu \left({\rm p.v.}\sum_{n=0} r_{n} \alpha^{n+1}(\mu)\right) \right] \ , \nonumber\\
&S^{\rm NNLO+LRR}(l_z,\alpha(\mu)) = S^{\rm NNLOsub}(l_z,\alpha(\mu))\exp\left[-|z|\mu \left({\rm p.v.}\sum_{n=0} r_{n} \alpha^{n+1}(\mu)\right) \right] \ ,
\end{align}
where the ``p.v." is defined in Eq.~(\ref{eq:pvrenormalon}). The renormalon series expanded up to a certain fixed order is subtracted from the original fixed order result to avoid double counting,
\begin{align}\label{eq:Ssub}
&S^{\rm NLOsub}(l_z,\alpha(\mu)) = 1 + s_0(l_z) \alpha(\mu) + |z|\mu r_0 \alpha(\mu) \ , \nonumber\\
&S^{\rm NNLOsub}(l_z,\alpha(\mu)) = 1 + s_0(l_z) \alpha(\mu) + |z|\mu r_0 \alpha(\mu) \nonumber\\
& + s_1(l_z) \alpha^2(\mu) + |z|\mu r_1 \alpha^2(\mu)
+ s_0(l_z) |z|\mu r_0 \alpha^2(\mu) + \frac{1}{2} z^2 \mu^2 r_0^2 \alpha^2(\mu) \ ,
\end{align}
where $s_0(l_z) |z|\mu r_0 \alpha^2(\mu)$ is the mixing term between the renormalon and original series, and $z^2 \mu^2 r_0^2 \alpha^2(\mu)/2$ comes from the Taylor expansion of the exponential factor up to NNLO. In contrast, in Ref.~\cite{Zhang:2023bxs}, the mass renormalon series is combined additively with the fixed-order calculation. 

The second difference is that the mass renormalon series is evaluated at the semi-hard scale $\mu_i=2|1-x|P^z$ in the threshold resumed kernel ${\cal C}^{h}\left(\frac{x'}{y},\frac{y P^z}{\mu_h},\mu_h,\mu_i\right)_{\rm sg}$. To perform the LRR in the threshold resummation, we replace $S$ in Eq.~(\ref{eq:sffixed}) with the LRR versions in Eq.~(\ref{eq:SLRR}). The LRR soft function is applied to both ${\cal C}^{h}\left(\frac{x'}{y},\frac{y P^z}{\mu_h},\mu_h,\mu_i\right)_{\rm sg}$ and
${\cal C}^{h}\left(\frac{x'}{y},\frac{y P^z}{\mu_h},\mu_h,\mu_h\right)_{\rm sg}$ in the matching formula Eq.~(\ref{eq:matchhybridtr}).


In Ref.~\cite{Zhang:2023bxs}, the authors study the mass renormalon under collinear factorization where $ 2 |1-x| P^z \sim 2 |x| P^z \gg \Lambda_{\rm QCD} $. In the perturbative matching kernel, the mass renormalon series is resummed at the physical scale $ 2 |x| P^z $, and the corresponding power correction is interpreted as $\sim \frac{\Lambda_{\rm QCD}}{|x|P^z}$. This approach seems reasonable in the moderate $x$ range where the hard scale $ 2 |x| P^z $ and the semi-hard scale $ 2 |1-x| P^z $ are similar to each other. However, in the large $x$ region, it is not accurate because the physical scale for the mass renormalon series can no longer be approximated as $ 2 |x| P^z $. Under threshold factorization, where $ 2 |x| P^z \gg 2 |1-x| P^z \gg \Lambda_{\rm QCD}$, the mass renormalon series exists only in the soft function. Therefore, its physical scale should be the semi-hard scale $\mu_i=2|1-x|P^z$, and the corresponding power correction is $\sim \frac{\Lambda_{\rm QCD}}{|1-x|P^z}$. 

\subsection{Numerical test on the leading renormalon resummation}\label{sec:LPL}

\begin{figure}[htbp]
    \centering
    \includegraphics[height=6.18cm]{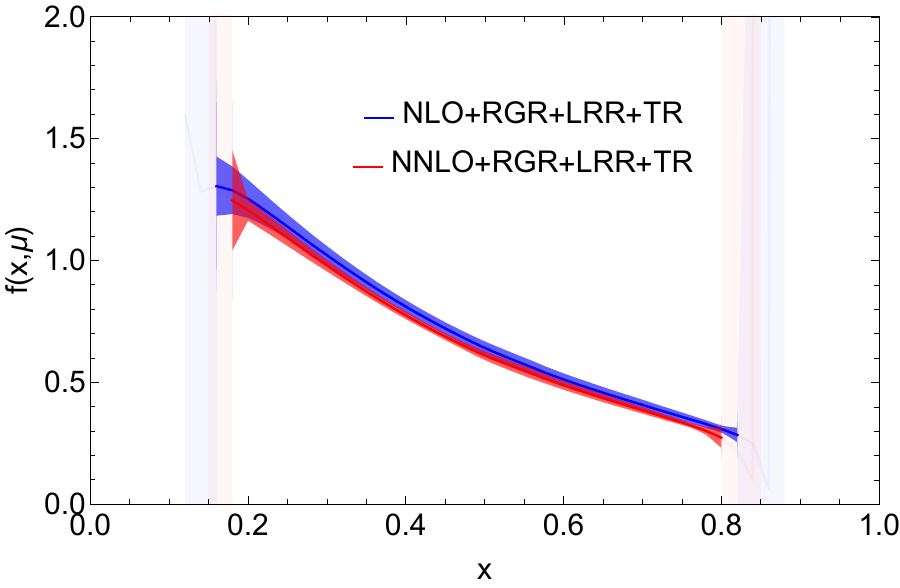}
    \caption{Similar to Fig.~\ref{fig:NLO_NNLO_wo_renormalon} except that the leading renormalon resummation (LRR) is performed here. The bands show the uncertainties from the scale and threshold cut variations. $P^z=1.94 \, {\rm GeV}$, $\mu=2 \, {\rm GeV}$, and $z_s = 0.12 \, {\rm fm}$.}
    \label{fig:MCfullLRRNLOandNNLO}
\end{figure}


In this subsection, we test the LRR version of Eq.~(\ref{eq:matchhybridtr}). The LRR is performed for the singular terms ${\cal C}^{h}\bigg(\frac{x'}{y},\frac{|y| P^z}{\mu_h},\mu_h,\mu_i\bigg)_{\rm sg}$ and ${\cal C}^{h}\bigg(\frac{x'}{y},\frac{|y| P^z}{\mu_h},\mu_h,\mu_h\bigg)_{\rm sg}$ using the methods discussed in the previous subsections. For the fixed order matching kernel ${\cal C}^{h}\bigg(\frac{x'}{y},\frac{|y| P^z}{\mu_h}\bigg)$ without threshold expansion, we apply the LRR mostly following the method in~\cite{Zhang:2023bxs}, except that the mass renormalon series is exponentiated like Eq.~(\ref{eq:SLRR}),
\begin{align}
{\cal C}^{h}_{\rm LRR}\bigg(\frac{x'}{y},\frac{|y| P^z}{\mu_h}\bigg)
= \int_{-\infty}^{+\infty} \frac{d y'}{|y'|} {\cal C}^{h}_{\rm sub}\bigg(\frac{x'}{y'},\frac{|y'| P^z}{\mu_h}\bigg) {\cal M}^{h}\bigg(\frac{y'}{y},\frac{|y| P^z}{\mu_h}\bigg) \ ,
\end{align}
where the mass renormalon part of the hybrid scheme kernel is
\begin{align}
{\cal M}^{h}\bigg(\xi,\frac{p^z}{\mu}\bigg)
= \int_{-\infty}^{+\infty} \frac{d\lambda}{2\pi} e^{i \lambda (\xi-1)} e^{-\theta\left(\left|\frac{\lambda}{p^z}\right|-z_s\right)
\, \left(\left|\frac{\lambda}{p^z}\right|-z_s\right) 
\, \mu \, \left({\rm p.v.}\sum_{n=0}^{\infty} r_{n} \alpha^{n+1}(\mu)\right) } \ ,
\end{align}
where $r_{n}$ is the coefficient in Eq.~(\ref{eq:asymcoe}) and ``p.v." is defined in Eq.~(\ref{eq:pvrenormalon}). ${\cal C}^{h}_{\rm sub}$ is the fixed order matching kernel subtracted by the expanded renormalon series up to the same order to avoid double counting, similar to Eq.~(\ref{eq:Ssub}) but with the hybrid cutoff $z_s$. 

Again, one can show that the renormalon ambiguity $\sim \frac{\Lambda_{\rm QCD}}{|1-x|P^z}$ is canceled between the matching kernel and quasi-PDF regularized in the same prescription $\tau$ in the large $\beta_0$ limit, similar to the discussions around Eq.~(\ref{eq:LPALbeta0}). First, consider the $\overline{\rm MS}$ scheme version of Eq.~(\ref{eq:matchhybridtr}) for simplicity. In this case, the physical scale for the soft function in the singular part ${\cal C}\left(\frac{x'}{y},\frac{y P^z}{\mu_h},\mu_h,\mu_i\right)_{\rm sg}$ and for the threshold quasi-PDF is the same semi-hard scale. This ensures the cancellation of the mass renormalon between these two terms. The regular part, ${\cal C}\bigg(\frac{x'}{y},\frac{|y| P^z}{\mu_h}\bigg) - {\cal C}\left(\frac{x'}{y},\frac{y P^z}{\mu_h},\mu_h,\mu_h\right)_{\rm sg}$, does not contain the mass renormalon and is therefore irrelevant to this discussion. After being converted to the hybrid scheme, the ${\cal C}\bigg(\frac{x'}{y},\frac{|y| P^z}{\mu_h}\bigg)$ in the regular part is adjusted consistently with the quasi-PDF. Therefore, the renormalon cancellation remains ensured.

The comparison between NLO+RGR+LRR+TR and NNLO+RGR+LRR+TR is shown in Fig.~\ref{fig:MCfullLRRNLOandNNLO}. The perturbative convergence in the moderate $x$ range (e.g. $2 x P^z > 0.8$ GeV and $2(1-x)P^z > 0.8$ GeV) is improved with LRR, compared to Fig.~\ref{fig:NLO_NNLO_wo_renormalon} without LRR. This suggests that we have a reliable prediction in the moderate $x$ range after considering threshold resummation with leading renormalon resummation. In the end point regions (e.g. $2 x P^z < 0.8$ GeV or $2(1-x)P^z < 0.8$ GeV), the NNLO+RGR+LRR+TR case has larger uncertainties than NLO+RGR+LRR+TR, indicating that perturbative matching breaks down there.


The comparison between Figs.~\ref{fig:NLO_NNLO_wo_renormalon} and~\ref{fig:MCfullLRRNLOandNNLO} confirms our assumption on the ``leading renormalon dominance" starting from NNLO, implying that the leading twist accuracy may be under control, though higher-order perturbative calculations are needed to provide further justification. It can be shown that the power corrections of type $\mathcal{O}\left[ \frac{\Lambda_{\rm QCD}}{|x| P^z} \right]$ and $\mathcal{O}\left[ \frac{\Lambda_{\rm QCD}}{|1-x| P^z} \right]$ are eliminated in the full formula, at least in the large $\beta_0$ limit. These serve as evidence that our formalism achieves linear power accuracy.

\section{Conclusions}\label{sec:conclu}


We introduce the IR-free defintions for the hard kernel and threshold soft factors, which clarifies the collinear and soft-collinear physics in the matrix elements discussed in Ref.~\cite{Ji:2023pba}. These improvements provide a more precise understanding of the contributions from different momentum modes in the threshold region. 
 
We implement threshold resummation with leading renormalon resummation for the pion valence PDF case under LaMET. The results indicate the breakdown of perturbative matching near the end-point region (e.g. $ 2|1-x|P^z < 0.8 \, {\rm GeV}$) and demonstrate good perturbative convergence in the moderate $x$ range (e.g. $2|x|P^z > 0.8 \, {\rm GeV}$ and $2|1-x|P^z > 0.8 \, {\rm GeV}$). 

Importantly, because the threshold resummation for the matching kernel of the quark PDF is independent of the specific gamma structures or external states, the formalism and methods developed here are broadly applicable. They can be generalized to various quark polarizations in any large-momentum hadron states, expanding the utility of these techniques across a range of hadronic systems in lattice QCD.

\acknowledgments
We thank Yong Zhao, Xiang Gao, Andreas Schäfer, and Feng Yuan for useful discussions and comments. 
Y. L. is supported by the Priority Research Area SciMat and DigiWorlds under the program Excellence Initiative - Research University at the Jagiellonian University in Krak\'{o}w. Y.S. and R.Z. are partially supported by Quark-Gluon Tomography (QGT) collaboration, which is supported by the U.S. Department of Energy (DOE) topical collaboration program (DE-SC0023646). R.Z. is also supported by The U.S. Department of Energy, Office of Science, Office of Nuclear Physics through Contract No.~DE-SC0012704, Contract No.~DE-AC02-06CH11357, and within the frameworks of Scientific Discovery through Advanced Computing (SciDAC) award Fundamental Nuclear Physics at the Exascale and Beyond.

\appendix

\section{Threshold factorizations at one-loop}\label{sec:1loopfm}
This section checks Eqs.~(\ref{eq:TFLCPDF}),~(\ref{eq:TFqPDF}), and~(\ref{eq:thrmatching}) with one-loop calculation using quark pole mass as IR regulator. 

\subsection{Threshold light-cone PDF}\label{sec:1loopfmLC}

The one-loop correction to Eq.~(\ref{eq:LCpdfq}) is given by
\begin{align}\label{eq:LCpdfq1loopfmtot}
q^{(1)}\left(x, \frac{m}{\mu}\right)
= & \frac{\alpha(\mu) C_F}{2 \pi} \left[ \left(1+x^2\right) \left(\frac{-2\ln \left((1-x) m/\mu\right)-1}{1-x}\right)_{+}\theta (0\leq x\leq 1) \right. \nonumber\\
&\left. + \delta (1-x) \left(-3\ln\left(\frac{m}{\mu}\right)+2\right) \right] \ ,
\end{align}
where $m$ is the pole mass acting as regulator for IR (collinear) singularity, and other UV (including light-cone) divergences are subtracted in $\overline{\rm MS}$ scheme. $C_F=4/3$ is the Casimir constant in the fundamental representation. $\theta(0 \leq x \leq  1)$ is the unitstep function, which is equal to one within the range $0 \leq x \leq 1$ and vanishes elsewhere. The plus function is defined as
\begin{align}\label{eq:pluspdf}
\left(w(x)\right)_{+} = w(x) - \delta(1-x)\int_{0}^{1} d x' w(x') \ . 
\end{align}
The log terms $\ln\left((1-x) m/\mu\right)$ and $\ln\left(m/\mu\right)$ agree with the result in Eq.~(10) in Ref.~\cite{Xiong:2013bka}. However, the constant terms differ due to the use of different UV divergence regulators. 

In the threshold limit $x \rightarrow 1$, the above result becomes
\begin{align}\label{eq:LCpdfq1loopfmtr}
&q^{(1)}\left(x, \frac{m}{\mu}\right) \bigg|_{x \rightarrow 1}
= \int_{-\infty}^{+\infty} \frac{d \lambda}{2\pi} e^{i \lambda (x-1)} \frac{\alpha(\mu) C_F}{2 \pi} \nonumber\\
&\quad \times \left[ - 2 \ln^2\left(-i \bar{\lambda}\right) + 2 \ln\left(-i \bar{\lambda}\right) \left( 2 \ln\left(\frac{m}{\mu}\right) + 1 \right) - 3 \ln\left(\frac{m}{\mu}\right)+2-\frac{\pi ^2}{3} \right] \nonumber\\
&= \frac{\alpha(\mu) C_F}{2 \pi} \left[ 2 \left(\frac{- 2 \ln\left((1-x) m/\mu\right) - 1}{1-x} \right)_{+} \theta(1-x) + \delta(1-x) \left(-3\ln\left(\frac{m}{\mu}\right)+2\right) \right] \ ,
\end{align}
where $\bar{\lambda} \equiv \lambda e^{\gamma_E}$, and $\lambda \equiv - (\xi^{-} \bar{n}) \cdot p = - \xi^{-} p^{+}$ is the dimensionless light-cone distance. Using the identities provided in Appendix~\ref{sec:FTofLogsasymp}, the Fourier transformation of $\ln^2\left(-i \bar{\lambda}\right)$ and $\ln\left(-i \bar{\lambda}\right)$ only contributes to the region $x \leq 1$, as indicated by the unit step function $\theta(1-x)$, which is 1 for $x \leq 1$ and 0 for $x>1$.

The one-loop result for the collinear function defined in Eq.~(\ref{eq:LCjet}) is
\begin{align}\label{eq:Z1loopfm}
J^{(1)}\left(\frac{m}{\mu}\right) = \frac{\alpha(\mu) C_F}{2\pi} 
\left[ 2 \ln^2\left(\frac{m}{\mu}\right)
- \ln\left(\frac{m}{\mu}\right)
+ 2 + \frac{\pi ^2}{12} \right] \ ,
\end{align}
which contains double logs in pole mass and is consistent with Ref.~\cite{Beneke:2023nmj}. The one-loop result for the soft-collinear function defined in Eq.~(\ref{eq:LCsoft}) is
\begin{align}\label{eq:WC1loopfm}
W_C^{(1)}\left(\frac{\lambda\mu}{m}\right) 
= & \frac{\alpha(\mu) C_F}{2\pi} \left[-2 \ln^2\left(-i \bar{\lambda} \frac{\mu}{m}\right) + 2 \ln\left(-i \bar{\lambda} \frac{\mu}{m} \right) -\frac{5 \pi ^2}{12} \right] \nonumber\\
= & \frac{\alpha(\mu) C_F}{2\pi} \left[ - 2 \ln^2\left(-i \bar{\lambda}\right) + 2 \ln\left(-i \bar{\lambda}\right) \left( 2\ln\left(\frac{m}{\mu}\right) + 1 \right) \right. \nonumber\\
&\left. - 2 \ln^2\left(\frac{m}{\mu}\right) - 2 \ln\left(\frac{m}{\mu}\right) - \frac{5\pi^2}{12} \right] \ ,
\end{align}
where in the second line, we made the substitution $\ln\left(-i \bar{\lambda} \mu/m\right) = \ln(-i \bar{\lambda}) - \ln\left(m/\mu\right)$, for checking the factorization. 

By comparing Eqs.~(\ref{eq:LCpdfq1loopfmtr}),~(\ref{eq:Z1loopfm}) and~(\ref{eq:WC1loopfm}), we check Eq.~(\ref{eq:TFLCPDF}) up to one-loop. 

One-loop results support RG equations for $J$ and $W_C$ 
\begin{align}
&\frac{d \ln\left[J\left(m/\mu\right)\right]}{d \ln \mu} 
= K\left(\frac{\mu}{m}\right) + \gamma_{J}(\mu) \ , \\
&
\frac{d\ln \left[W_C\left( \lambda \mu/m \right)\right]}{d\ln \mu} = -2 \Gamma_{\rm cusp}(\mu) \ln\left(-i \bar{\lambda} \right) -K\left(\frac{\mu}{m}\right) + \gamma_{C}(\mu) 
\ , 
\end{align}
where $\gamma_{J}(\mu) = \frac{\alpha(\mu) C_F}{2\pi}$, $\gamma_C(\mu) = \frac{\alpha(\mu) C_F}{\pi}$ and $\Gamma_{\rm cusp}(\mu) = \frac{\alpha(\mu) C_F}{\pi}$ are obtained from one-loop results. The one-loop result for the IR-sensitive perturbative kernel~\cite{Collins:2011zzd} is given by
\begin{align}
K\left(\frac{\mu}{m}\right) =  2 \Gamma_{\rm cusp}(\mu) \ln\frac{\mu}{m} \ ,
\end{align}
which satisfies
the RG equation
\begin{equation}
    \frac{dK}{d\ln \mu} = 2\Gamma_{\rm cusp}\ . 
\end{equation}
Combining the RG equations for $J$ and $W_C$, we obtain the RG equation
for the light-cone PDF in coordinate space,
\begin{equation}
    \frac{d \ln q\left(\lambda,m/\mu\right)e^{i\lambda}}{d\ln \mu} = -2\Gamma_{\rm cusp}(\mu) \ln(-i\bar{\lambda}) + \gamma_J(\mu) + \gamma_C(\mu) \ .
\end{equation}
The momentum space version has been used to study the evolution of the light-cone PDF in the $x \rightarrow 1$ region~\cite{Becher:2006mr}.

\subsection{Threshold quasi-PDF}\label{sec:1loopfmq}

The one-loop correction to Eq.~(\ref{eq:qpdfq}) in the threshold limit $y \rightarrow 1$ is
\begin{align}\label{eq:qpdfq1loopfmtr}
&\tilde q^{(1)}\left(y,\frac{p^z}{\mu},\frac{m}{\mu}\right) \bigg|_{m^2/p_z^2 \rightarrow 0, \, y \rightarrow 1}
= \int_{-\infty}^{+\infty} \frac{d\lambda}{2\pi} e^{i \lambda (y-1)} \frac{\alpha(\mu) C_F}{2\pi} \nonumber\\
&\quad \times \left[  4 \ln \left(-i \bar{\lambda} \right) \left( \ln\left(\frac{m}{-2 i s_{z} p^z}\right) + 1 \right) - 3 \ln\left(\frac{m}{\mu}\right) + 2-\frac{2 \pi ^2}{3} \right] \nonumber\\
&= \frac{\alpha(\mu) C_F}{2\pi} \left[ 2 \left(\frac{-\ln \left(m^2/4 p_z^2\right)-2}{1-y}\right)_{+} \theta (1-y) - \mathcal{P} \frac{\ln \left((1-y)^2\right)}{1-y} \right.\nonumber\\
&\left. \quad + \left(-3 \ln \left(\frac{m}{\mu}\right)+2+\frac{\pi ^2}{3}\right) \delta (1-y)  \right] \ ,
\end{align}
where $s_z \equiv {\rm sign}(z)$ defined below Eq.~(\ref{eq:qjet}) is related to the direction of the intermediate gauge link. $\bar{\lambda} \equiv \lambda e^{\gamma_E}$, and $\lambda \equiv - (z n_z) \cdot p = z p^z$ is the dimensionless distance. The plus function is defined in Eq.~(\ref{eq:pluspdf}). $\mathcal{P}$ denotes the principal value prescription defined in Eq.~(\ref{eq:plusPV}). A similar result has been calculated in Ref.~\cite{Xiong:2013bka} but in a different UV regulator.   

The one-loop correction to the quasi collinear function $\tilde{J}$ defined in Eq.~(\ref{eq:qjet}) gives
\begin{align}\label{eq:qZ1loopfm}
&\tilde{J}^{(1)}\left(\frac{s_{z} p^z}{\mu}, \frac{m}{\mu}\right) \bigg|_{m^2/p_z^2 \rightarrow 0} =  \frac{\alpha(\mu) C_F}{2\pi} \left[ 2 \ln^2\left(\frac{m}{\mu}\right) - \ln\left(\frac{m}{\mu}\right) \right. \nonumber\\
&\left.\quad -2 \ln^2\left(\frac{- 2 i s_{z} p^z }{\mu}\right) + 2 \ln\left(\frac{-2 i s_{z} p^z }{\mu}\right) - \frac{\pi^2}{3} \right]  \nonumber\\
&= \frac{\alpha(\mu) C_F}{2\pi} \left[  2 \ln\left(\frac{-2 i s_z p^z m}{\mu^2}\right) \left(\ln\left(\frac{m}{-2 i s_z p^z}\right)+1\right) - 3 \ln\left(\frac{m}{\mu}\right) - \frac{\pi^2}{3}\right] \ .
\end{align}
The one-loop result of the quasi soft-collinear function defined in Eq.~(\ref{eq:qsoft}) is
\begin{align}\label{eq:qWC1loopfm}
&\tilde W_C^{(1)}\left(\frac{\lambda^2 \mu^2}{p_z^2}, \frac{\lambda \mu}{m}\right) \bigg|_{m^2/p_z^2 \rightarrow 0} = \frac{\alpha(\mu) C_F}{2\pi} \left[ - 2 \ln^2\left(-i \bar{\lambda}  \frac{\mu}{m}\right) + 2 \ln\left(-i \bar{\lambda} \frac{\mu}{m} \right) \right.\nonumber\\
&\left.\quad  + \frac{1}{2}\ln^2\left(\frac{\bar{\lambda}^2 \mu^2}{4 p_z^2}\right) + \ln\left(\frac{\bar{\lambda}^2 \mu^2}{4 p_z^2}\right) + 2 -\frac{\pi^2}{3}  \right] \nonumber\\
&= \frac{\alpha(\mu) C_F}{2\pi} \left[ -2 \ln\left(\frac{2 i s_z p^z m}{\mu^2 \bar{\lambda}^2}\right) \left( \ln\left(\frac{m}{-2 i s_z p^z}\right) + 1 \right) + 2 - \frac{\pi^2}{3}\right] \ ,
\end{align}
where the log terms are combined for the convenience of verifying the factorization.

By comparing Eqs.~(\ref{eq:qpdfq1loopfmtr}),~(\ref{eq:qZ1loopfm}) and~(\ref{eq:qWC1loopfm}), we check Eq.~(\ref{eq:TFqPDF}) up to one-loop. 

One-loop results support RG equations for $\tilde{J}$ and $\tilde{W}_C$ 
\begin{align}
&\frac{d \ln\left[\tilde{J}\left( s_z p^z/\mu, m/\mu \right)\right]}{d \ln \mu} 
= 2 \Gamma_{\rm cusp}(\mu) \ln\left(\frac{-2 i s_z p^z}{\mu}\right) +  K\left(\frac{\mu}{m}\right) + \gamma_{\tilde{J}}(\mu) \ , \\
& \frac{d\ln \left[\tilde W_C\left( \lambda^2 \mu^2/p_z^2, \, \lambda \mu/m \right)\right]}{d\ln \mu} 
=  \Gamma_{\rm cusp}(\mu) \ln\left(\frac{
\bar{\lambda}^2\mu^2}{4p_z^2}\right) - 2 \Gamma_{\rm cusp}(\mu) \ln\left(-i \bar{\lambda} \right) -  K\left(\frac{\mu}{m}\right) + 
 \gamma_{\tilde{C}}(\mu) \ , 
\end{align}
where $\gamma_{\tilde{J}}(\mu) = -\frac{\alpha(\mu) C_F}{2\pi}$ and $\gamma_{\tilde{C}}(\mu) = 2\frac{\alpha(\mu) C_F}{\pi}$ at one-loop. It is clear that the right-hand sides of both equations are independent of manifest $\ln \mu$. Combining the above, one gets a simple RG equation for quasi-PDF, only related to heavy-light quark anomalous dimension~\cite{Ji:1991pr, Chetyrkin:2003vi, Braun:2020ymy, Grozin:2023dlk}, in both momentum and 
coordinate spaces.

\subsection{Compare light-cone and quasi}\label{sec:1loopfmlcq}
We check Eq.~(\ref{eq:thrmatching}) at one-loop, including the consistency check of the collinear and soft collinear modes, respectively, between threshold light-cone and quasi-PDFs. The UV difference has also been cross-checked. 

Based on Eqs.~(\ref{eq:Z1loopfm}) and~(\ref{eq:qZ1loopfm}), the one-loop correction to $\Delta J=J/\tilde{J}$ is
\begin{align}\label{eq:ZoverZt}
& \Delta J^{(1)} = \frac{\alpha(\mu) C_F}{2\pi} \left[ 2 \ln^2\left(\frac{-2 i s_z p^z}{\mu }\right) - 2 \ln\left(\frac{-2 i s_z p^z}{\mu }\right) + 2 + \frac{5 \pi ^2}{12} \right] \ ,
\end{align}
where the collinear singularities in $m$ are canceled, which means threshold light-cone and quasi-PDFs share the same non-perturbative collinear physics. The logs in  $\ln\left(-2 i s_z p^z/\mu\right)$ comes from contributions of collinear modes associated with hard scale $p^z$ and $\mu$. The result is consistent with the calculation purely under dimensional regularization and massless quark state in Ref.~\cite{Ji:2023pba}. 

$\Delta J$ has a simple RG equation, 
\begin{equation}
   \frac{d\ln \Delta J}{d\ln \mu} = -  2 \Gamma_{\rm cusp}(\mu) \ln\left(\frac{-2 i s_z p^z}{\mu}\right) + \gamma_J(\mu)-\gamma_{\tilde J}(\mu)  \ . 
\end{equation}
The scale $\mu$ dependence of $\Delta J$ is independent of $K$, and pertains only to UV physics, as shown in Eqs.~(\ref{eq:RGEHab}) and~(\ref{eq:RGEA}).

Based on Eqs.~(\ref{eq:WC1loopfm}) and~(\ref{eq:qWC1loopfm}), the one-loop correction to $S=W_C/\tilde{W}_C$ is
\begin{align}\label{eq:WCoverWCt}
&S^{(1)} =\frac{\alpha(\mu) C_F}{2 \pi}\left[ -\frac{1}{2}\ln^2\left(\frac{\bar{\lambda}^2 \mu^2}{4 p_z^2}\right) - \ln\left(\frac{\bar{\lambda}^2 \mu^2 }{4 p_z^2}\right) - \frac{\pi^2}{12} -  2  \right] \ ,
\end{align}
where the soft-collinear logs $\ln\left(-i \bar{\lambda} \mu/m \right)$ are canceled, indicating that threshold light-cone and quasi-PDFs share the same soft-collinear mode.  
The left-over semi-hard logs $\ln\left(\bar{\lambda}^2 \mu^2/4 p_z^2\right)$ are related to the threshold log that we would like to resum. Again, the result agrees with the calculation purely under dimensional regularization and lightlike incoming/outgoing gauge links in Ref.~\cite{Ji:2023pba}.

The RG equations of $S=W_C/\tilde{W}_C$ is 
\begin{align}
&\frac{d\ln S}{d\ln \mu} = -\Gamma_{\rm cusp}(\mu) \ln\left(\frac{
\bar{\lambda}^2\mu^2}{4p_z^2}\right)  + \gamma_{C}(\mu)-\gamma_{\tilde C}(\mu) \ .
\end{align}
Similarly, the RG equation for the ratio $W_C / \tilde{W}_C$ is independent of the Collins-Soper kernel $K$, as indicated in Eq.~(\ref{eq:RGEtiledJ}). The cancellation of $K$ in the ratio is essential for resumming the threshold logarithms within perturbation theory.

The total one-loop result for the threshold matching kernel $\mathcal{C}\left(\xi,\frac{p^z}{\mu}\right)_{\rm sg}$ in Eq.~(\ref{eq:MKthr}) can be calculated based on the ratios of threshold collinear and soft-collinear functions, Eqs.~(\ref{eq:ZoverZt}) and~(\ref{eq:WCoverWCt}),
\begin{align}\label{eq:TRmatchUV}
&\int_{-\infty}^{+\infty} \frac{d\lambda}{2\pi} e^{i \lambda (\xi-1)} \left[\Delta J+S \right]^{(1)} \nonumber\\
&= - \frac{\alpha(\mu) C_F}{2 \pi} \left[ {\cal P}\frac{\ln\left( (1-\xi)^2 \right)}{|1-\xi|} + \left({\cal P} \frac{1}{|1-\xi|} + {\cal P}\frac{1}{1-\xi}\right)\left(\ln\left(\frac{4 p_z^2}{\mu^2}\right)-1\right) + \frac{\pi^2}{3} \delta(1-\xi)
\right] \ ,
\end{align}
where $\mathcal{P}$ denotes the principal value prescription defined in Eq.~(\ref{eq:plusPV}). On the other hand, the above result can be obtained through the threshold expansion $\xi \rightarrow 1$ (see Eq.~(\ref{eq:threxp})) of the one-loop matching kernel $\mathcal{C}\left(\xi,\frac{p^z}{\mu}\right)$, which can be found in Refs.~\cite{Xiong:2013bka,Izubuchi:2018srq,Ji:2020ect}. 
The two approaches are consistent, which confirms the UV part of Eq.~(\ref{eq:thrmatching}) up to one-loop.

\section{Perturbative expansion for matching kernels up to NNLO}\label{sec:fix}
In this Appendix, we collect the fixed-order perturbation series up to NNLO, defined in Eqs.~(\ref{eq:Hdef}) and~(\ref{eq:Sdef}). They are used as the initial conditions evaluated at the physical scales in the resumed perturbative objects, Eqs.~(\ref{eq:Jmom}), (\ref{eq:Hmom}) and (\ref{eq:Hmomhybrid}), for the RG-improved perturbation theory up to NNLO, see Table~\ref{appschS}.  

Our conventions for the fixed-order perturbation series in $\overline{\rm MS}$ scheme are
\begin{align}
    &H(L_{p_z},\alpha(\mu)) = 1 - \alpha(\mu) \, H^{(1)}(L_{p_z}) + \alpha^2(\mu) \, \left[-H^{(2)}(L_{p_z})+H^{(1)}(L_{p_z})^2\right]\ ,  \nonumber\\
    &\Phi(L_{p_z},\alpha(\mu)) = - \alpha(\mu) \, \Phi^{(1)}(L_{p_z}) - \alpha^2(\mu) \, \Phi^{(2)}(L_{p_z}) \ , \nonumber\\    
    &S(l_z,\alpha(\mu)) = 1 - \alpha(\mu) \, S^{(1)}(l_z) + \alpha^2(\mu) \, \left[-S^{(2)}(l_z)+S^{(1)}(l_z)^2\right] \ , 
\end{align}
where the hard kernel $H(L_{p_z},\alpha(\mu))$ and its phase $\Phi(L_{p_z},\alpha(\mu))$ are in momentum space, and $S(l_z,\alpha(\mu))$ is in coordinate space and real. 
The NLO results are~\cite{Ji:2021znw,Ji:2023pba}
\begin{align}
&H^{(1)}(L_{p_z}) = \frac{C_F}{2\pi}\left( -\frac{L_{p_z}^2}{2} + L_{p_z} + \frac{\pi^2}{12} -2 \right) \ , \nonumber\\
&\Phi^{(1)}(L_{p_z}) = \frac{C_F}{2} L_{p_z} -\frac{C_F}{2} \ , \nonumber\\
&S^{(1)}(l_z) = \frac{C_F}{2\pi} \left( \frac{l_z^2}{2} + l_z + \frac{\pi^2}{12} + 2 \right) \ , 
\end{align}
where $L_{p_z}=\ln\left(4 p_z^2/\mu^2\right)$ with $p^z=|y|P^z$, and the coordinate space log, $l_z=\ln\left(\bar{\lambda}^2\mu^2/4 p_z^2\right)$ with $\bar{\lambda} = \lambda e^{\gamma_E}$. $C_F=4/3$ is the Casimir constant in the fundamental representation.

The NNLO result for the magnitude (absolute value) of the hard kernel is determined in Refs.~\cite{Ji:2023pba,delRio:2023pse},
\begin{align}
&H^{(2)}(L_{p_z}) = \frac{C_F^2 L_{p_z}^4}{32 \pi ^2} + \left(\frac{11 C_A C_F}{144 \pi ^2}-\frac{C_F n_f}{72 \pi ^2}-\frac{C_F^2}{8 \pi^2}\right) L_{p_z}^3 \nonumber\\ 
&+ \left[ C_A C_F \left(-\frac{25}{36 \pi ^2}+\frac{1}{48} \right) +\frac{C_F n_f}{9 \pi ^2} + C_F^2 \left(\frac{3}{8 \pi ^2}-\frac{1}{96}\right) \right] L_{p_z}^2 \nonumber\\
&+ \left[C_A C_F \left(-\frac{11 \zeta(3) }{8 \pi ^2}+\frac{475}{216 \pi ^2}+\frac{11}{144}\right)
+C_F n_f \left(-\frac{19}{54 \pi ^2}-\frac{1}{72}\right)
+C_F^2\left(\frac{3 \zeta(3)}{2 \pi^2}-\frac{1}{4 \pi ^2}-\frac{13}{48}\right)\right] L_{p_z} \nonumber\\
& +\bigg(\frac{241 \zeta(3)}{144 \pi ^2}+\frac{11 \pi^2}{320} -\frac{971 }{324\pi ^2}-\frac{559 }{1728}\bigg)C_A C_F+\bigg(\frac{\zeta(3) }{72 \pi ^2}+\frac{41 }{81 \pi^2}+\frac{17 }{864}\bigg) C_F n_f\nonumber\\
&+\bigg(-\frac{15 \zeta(3)}{8 \pi ^2}-\frac{95 \pi^2 }{1152}-\frac{3 }{4 \pi ^2}+\frac{29 }{24}\bigg)C_F^2 \ , 
\end{align}
where $C_A = 3$, $n_f$ is the number of active quark flavors, and $\zeta(3)$ is the Riemann zeta function.
The same order result for the phase angle of the hard kernel is determined in Ref.~\cite{Ji:2023pba},
\begin{align}
&\Phi^{(2)}(L_{p_z}) = \left(\frac{C_F n_f}{24 \pi }-\frac{11 C_A C_F}{48 \pi }\right) L_{p_z}^2 
+ \left[C_A C_F \left(-\frac{\pi}{24}+\frac{25}{18 \pi }\right)-\frac{2 C_F n_f}{9 \pi }\right] L_{p_z} \nonumber\\
&+ C_A C_F \left(\frac{11 \zeta(3) }{8 \pi }-\frac{11 \pi}{48} -\frac{475}{216 \pi}\right)
+  C_F n_f \left(\frac{\pi}{24} + \frac{19}{54 \pi }\right)
+ C_F^2\left(-\frac{3 \zeta(3) }{2 \pi}+\frac{7 \pi}{24}-\frac{1}{4 \pi }\right) \ .
\end{align}
And finally, the NNLO result for the time-like soft function is calculated in~\cite{Jain:2008gb}. Our space-like soft function is real and is determined through the analytical continuation~\cite{Ji:2023pba} from the time-like soft function,
\begin{align}
&S^{(2)}(l_z) = \frac{C_F^2}{32 \pi ^2}l_z^4 + \left(\frac{11 C_A C_F}{144 \pi ^2}-\frac{C_F n_f}{72 \pi ^2}+\frac{C_F^2}{8 \pi^2}\right)l_z^3 \nonumber\\
&+ \left[ C_A C_F \left(\frac{25}{36 \pi ^2}-\frac{1}{48}\right)
-\frac{C_F n_f}{9 \pi ^2}
+ C_F^2 \left(\frac{3}{8 \pi ^2}+\frac{1}{96}\right) \right]l_z^2 \nonumber\\
&+ \left[C_A C_F \left(-\frac{5 \zeta(3)}{8 \pi ^2}+\frac{547 }{216 \pi ^2}-\frac{1}{24}\right)
-\frac{47 C_F n_f}{108 \pi ^2}
+C_F^2 \left(\frac{1}{2 \pi^2}+\frac{1}{48}\right) \right] l_z \nonumber\\
& + C_A C_F\left(-\frac{101 \zeta(3)}{144 \pi ^2}-\frac{17 \pi^2}{2880}+\frac{2959}{648\pi ^2}+\frac{139}{1728}\right)
+ C_F n_f \left(\frac{\zeta(3)}{72 \pi ^2}-\frac{281}{324\pi ^2}-\frac{5}{864}\right)\nonumber\\
&+C_F^2 \left( \frac{\pi ^2}{1152}+\frac{1}{2 \pi^2}+\frac{1}{24} \right) \ . 
\end{align}
The above results are in $\overline{\rm MS}$ scheme. In practical calculations, one uses the hybrid scheme, where one rescales the hard kernel using a coordinate space correlation at a perturbative $z=z_s$ as Eq.~(\ref{eq:Hhybrid}),
\begin{align}
H^{h} \bigg(L_{p_z},\alpha(\mu)\bigg) =  H\bigg(L_{p_z},\alpha(\mu)\bigg) h^{\rm \overline{MS}}(z_s,\mu,P^z=0) \ ,
\end{align}
where we use the upper index $^{h}$ to denote the objects in hybrid scheme. $h^{\rm \overline{MS}}(z,P^z=0)$ is the perturbative zero-momentum matrix element $\langle P_z=0 | \bar{\psi}(z) \gamma^{t} U(z,0) \psi (0) |P_z=0 \rangle$ in $\overline{\rm MS}$ scheme, and the NNLO result can be obtained from Ref.~\cite{Li:2020xml}
\begin{align}\label{eq:p0m}
&h^{\rm \overline{MS}}(z,\mu,P^z=0) 
=1+\frac{\alpha C_{F}}{2 \pi}\left(\frac{3}{2} l_z +\frac{5}{2}\right) \nonumber\\
&+\left(\frac{\alpha}{2\pi}\right)^2\left[\left(\frac{11 C_A C_F}{8}-\frac{C_F n_f T_F}{2}+\frac{9 C_F^2}{8}\right) l_z^2 \right.\nonumber\\
&\left.+\left(\left(\frac{53}{8}-\frac{\pi ^2}{6}\right) C_A C_F+\left(\frac{25}{8}+\frac{2\pi ^2}{3}\right) C_F^2 -\frac{5 C_F n_f T_F}{2}\right) l_z \right. \nonumber\\
&\left.+2 \left(-4 \zeta(3) +\frac{223}{192} +\frac{\pi^2}{9}\right) C_F^2 +2\left(\zeta(3) +\frac{4877}{576}-\frac{5\pi ^2}{24}\right) C_A C_F-\frac{469 C_F n_f T_F}{72}\right] \  , 
\end{align}
where $T_F=1/2$. The NNNLO result can be found in Ref.~\cite{Cheng:2024wyu}.

\section{Relevant anomalous dimensions up to three loops}\label{sec:ano}
In this Appendix, we collect the cusp anomalous dimension up to four-loop order and single-log anomalous dimensions up to three-loop order, which are used in the evolution factors, Eqs.~(\ref{eq:aGamma}), (\ref{eq:othera}) and (\ref{eq:ahevo}), for the RG-improved perturbation theory up to NNLO, see Table~\ref{appschS}. 

Our conventions for the cusp anomalous dimension and single log anomalous dimensions are
\begin{align}
&\Gamma_{\rm cusp} = \Gamma_1 \alpha + \Gamma_2 \alpha^2 + \Gamma_3 \alpha^3 + \Gamma_4 \alpha^4 + ...\nonumber\\
&\gamma = \gamma_1 \alpha + \gamma_2 \alpha^2 + \gamma_3 \alpha^3 + ...
\end{align}
The cusp anomalous dimension $\Gamma_{\rm cusp}$ appears in Eqs.~(\ref{eq:RGEHab}), (\ref{eq:RGEA}) and~(\ref{eq:RGEtiledJ}). It is used in the evolution factors in Eq.~(\ref{eq:aGamma}). The results up to four-loop order are collected here
\begin{align}
&\Gamma_1 = \frac{C_F}{\pi} \nonumber\\
&\Gamma_2 = \left(\frac{67}{36 \pi ^2}-\frac{1}{12}\right) C_A C_F-\frac{5 C_F n_f}{18 \pi ^2} \nonumber\\
&\Gamma_3 = C_A C_F n_f\left( -\frac{7 \zeta (3)}{12 \pi ^3}+\frac{5}{108 \pi }-\frac{209}{432 \pi ^3} \right) + C_A^2 C_F \left(\frac{11 \zeta (3)}{24 \pi ^3}+\frac{11 \pi}{720}  -\frac{67}{216 \pi }+\frac{245}{96 \pi ^3}\right) \nonumber\\
& \quad \quad +  C_F^2 n_f\left(\frac{\zeta (3)}{2 \pi ^3}-\frac{55}{96 \pi ^3}\right)-\frac{C_F n_f^2}{108 \pi ^3} \nonumber\\
&\Gamma_4 = 0.000131226 n_f^3 + 0.00784294 n_f^2 - 0.207401 n_f + 0.830195 \ ,
\end{align}
where the two-loop cusp anomalous dimension was obtained in Ref.~\cite{Korchemskaya:1992je}. The three-loop result was calculated in Ref.~\cite{Moch:2004pa}. The four-loop result was obtained in Ref.~\cite{Henn:2019swt} and confirmed in Ref.~\cite{vonManteuffel:2020vjv}. A review on the cusp anomalous dimension can be found in Ref.~\cite{Grozin:2022umo}.

The single log anomalous dimension $\tilde \gamma_H$ appears in Eq.~(\ref{eq:RGEHab}) for the absolute value of the hard kernel in $\overline{\rm MS}$ scheme. It is used for the evolution factor $a_H$ in Eq.~(\ref{eq:othera}). According to the universality of anomalous dimensions discussed in Ref.~\cite{Ji:2023pba}, $\tilde \gamma_H$ is related to $\gamma_F$ (for heavy-light current UV anomalous dimension~\cite{Ji:1991pr, Chetyrkin:2003vi}), $\gamma_V$ (for light-light Sudakov hard kernel~\cite{Moch:2005id,Becher:2006mr}), $\gamma_s$ (for light-light Wilson line cusp~\cite{Korchemskaya:1992je}) and $\gamma_{HL}$ (for heavy-light Wilson line cusp~\cite{Korchemsky:1992xv}) as follows
\begin{align}\label{eq:gammaHtNNNLO}
\tilde \gamma_H =  \gamma_V + 2\gamma_F + 2 \gamma_{HL} - \gamma_s \ ,
\end{align}
where $\gamma_{HL}$ (for heavy-light Wilson line cusp~\cite{Korchemsky:1992xv}) can be expressed with $\gamma_{HH}$ (for heavy-heavy Wilson line cusp~\cite{Ji:2019ewn,Ji:2020ect,Ji:2021znw}) and $\gamma_s$,
\begin{align}
\gamma_{HL} = \frac{\gamma_{HH}+\gamma_s}{2} \ .
\end{align}
Based on the three-loop results of $\gamma_V$\cite{Becher:2006mr}, $\gamma_F$\cite{Braun:2020ymy}, $\gamma_{HH}$\cite{Bruser:2019yjk} and $\gamma_s$\cite{Moult:2022xzt}, we obtain the results for $\tilde \gamma_H$ up to three-loop order,
\begin{align}
&\tilde \gamma_{H1} = -\frac{C_F}{\pi } \nonumber\\
&\tilde \gamma_{H2} =
C_A C_F\left(\frac{11 \zeta (3)}{4 \pi ^2}-\frac{277}{108 \pi ^2}-\frac{11}{48}\right)
+C_F n_f\left(\frac{10}{27 \pi ^2}+\frac{1}{24}\right)
+C_F^2\left(-\frac{3 \zeta (3)}{\pi^2}-\frac{1}{2 \pi ^2}+\frac{7}{12}\right) \nonumber\\
&\tilde \gamma_{H3}= \frac{\left(-108 \zeta (3)+844-135 \pi ^2\right) n_f^2}{8748 \pi ^3}+\frac{\left(-54360 \zeta(3)+91285+8775 \pi ^2+51 \pi ^4\right) n_f}{29160 \pi ^3} \nonumber\\
&\quad \quad \quad +\frac{1199160 \zeta (3)-46320 \pi^2 \zeta (3)-772200 \zeta (5)-781735+64695 \pi ^2-2863 \pi ^4}{19440 \pi ^3} \ .
\end{align}
The single log anomalous dimension $\tilde \gamma_S$ appears in Eq.~(\ref{eq:RGEtiledJ}) for the soft function. It is used for the evolution factor $a_S$ in Eq.~(\ref{eq:othera}). According to Ref.~\cite{Ji:2023pba}, $\tilde \gamma_S$ can be obtained based on the following relation,
\begin{align}
\tilde \gamma_S = 2 \gamma_{HL} - 2\gamma_s \ .
\end{align}
The results are collected here,
\begin{align}
&\tilde \gamma_{S1} = -\frac{C_F}{\pi } \nonumber\\
&\tilde \gamma_{S2} = 
C_A C_F\left(\frac{5 \zeta (3)}{4 \pi ^2}-\frac{349}{108 \pi ^2}+\frac{23}{144}\right)
+C_F n_f\left(\frac{29}{54 \pi ^2}-\frac{1}{72} \right)\nonumber\\
&\tilde \gamma_{S3}= \frac{\left(-756 \zeta (3)+628+45 \pi ^2\right) n_f^2}{8748 \pi ^3}+\frac{\left(-1980 \zeta(3)+134665-16545 \pi ^2+828 \pi ^4\right) n_f}{29160 \pi ^3} \nonumber\\
&\quad \quad \quad +\frac{999540 \zeta (3)-45360 \pi^2 \zeta (3)-437400 \zeta (5)-961735+176955 \pi ^2-10692 \pi ^4}{19440 \pi ^3} \ .
\end{align}
To calculate the evolution factor $a_H^h$ for the absolute value of the hard kernel in hybrid scheme, one needs both $\tilde \gamma_H$ and $\gamma_F$, see Eq.~(\ref{eq:ahevo}). For readers' convenience, we collect the results for $\gamma_F$ up to three loops~\cite{Ji:1991pr, Chetyrkin:2003vi,Braun:2020ymy} as well,
\begin{align}
&\gamma_{F1} = \frac{3 C_F}{4 \pi } \nonumber\\
&\gamma_{F2} = C_A C_F\left(\frac{49}{96 \pi ^2}-\frac{1}{24}\right)
-\frac{5 C_F n_f}{48 \pi ^2} 
+ C_F^2\left(-\frac{5}{32\pi ^2}+\frac{1}{6}\right) \nonumber\\
&\gamma_{F3} = -\frac{35 n_f^2}{1296\pi^3}+\frac{\left(-747 \zeta (3)-129-49 \pi ^2\right) n_f}{972 \pi ^3}+\frac{-19224 \zeta (3)-4941+8232 \pi ^2+1520 \pi ^4}{15552 \pi ^3} \ .
\end{align}
Note that it has been calculated up to four-loop order recently~\cite{Grozin:2023dlk}.

\section{Fourier transformation of coordinate-space logarithms and asymptotic analysis}\label{sec:FTofLogsasymp}

In this Appendix, we study the Fourier transformation of 
logarithmic correlations in coordinate space $\lambda$ of type $\ln^{n}\left(|\lambda|r e^{\gamma_E} \right)$ with $n=0,1,2,...$, \, $r$ as
a real positive parameter and $\gamma_E$ as the Euler–Mascheroni constant,
\begin{align}\label{eq:FTevenlog}
I(\xi, n,r) \equiv \int_{-\infty}^{\infty} \frac{d \lambda}{2 \pi} \ln^{n}\left(|\lambda| r e^{\gamma_E} \right) \exp \left(i \lambda (\xi-1)\right) \ ,
\end{align}
which appears in the perturbative calculations in the matching kernels in the threshold limit, such as Eqs.~(\ref{eq:sffixed}),~(\ref{eq:ChybridTL}),~(\ref{eq:LCpdfq1loopfmtr}),~(\ref{eq:qpdfq1loopfmtr}) and~(\ref{eq:TRmatchUV}). We also explore the large $n$ asymptotic forms of the result.

The result of the above Fourier transformation is a distribution, which is well-defined in the convolution with a Schwartz function $\phi(\xi)$, 
\begin{align}\label{eq:condist}
    \int_{-\infty}^{+\infty} d\xi \, I(\xi, n,r) \phi(\xi) 
    = \int_{-\infty}^{+\infty} \frac{d\lambda}{2\pi} \, \ln^{n}\left(|\lambda| r e^{\gamma_E} \right) h(\lambda) \ ,
\end{align}
where $\phi(\xi) = \int_{-\infty}^{+\infty} \frac{d \lambda}{2\pi} e^{-i \lambda (\xi-1)} h(\lambda)$ connects the Schwartz function in momentum space $\xi$ and coordinate space $\lambda$. $I(\xi, n,r)$ can be defined as the distribution whose momentum space convolution with any Schwartz function on the left-hand side can reproduce the coordinate space convolution on the right-hand side.

We obtain $I(\xi,n,r)$ through the following identity,
\begin{align}\label{eq:FTforalpha}
&I(\xi, n,r) = \frac{d^n}{d\alpha^n}\left[ \int \frac{d \lambda}{2 \pi} \left(|\lambda| r e^{\gamma_E} \right)^{\alpha} \exp \left(i \lambda (\xi-1)\right) \right] \bigg|_{\alpha \rightarrow 0} \nonumber\\
&= \frac{d^n}{d\alpha^n}\left[ \frac{-\sin\left(\frac{\pi \alpha}{2}\right)}{\pi} \Gamma (\alpha +1) \left({\cal P} \frac{1}{|\xi-1|^{\alpha+1}}-\frac{2 \delta (\xi -1)}{\alpha }\right) r^{\alpha } e^{\gamma_E  \alpha } \right] \Bigg|_{\alpha \rightarrow 0} \ , 
\end{align}
where ${\cal P}$ is the principal value prescription defined in Eq.~(\ref{eq:plusPV}) where the integration limits for the $\delta$-function term are correlated with 
the $\delta(\xi-1)$ term above. 

The results for $n=0,...,4$ are
\begingroup
\allowdisplaybreaks
\begin{align}
&I(\xi, 0,r) = \delta(\xi-1) \nonumber\\
&I(\xi, 1,r) = {\cal P}\frac{-1}{2|\xi-1|} + \ln r \delta(\xi-1) \nonumber\\
&I(\xi, 2,r) = {\cal P}\frac{-\ln\frac{r}{|\xi-1|}}{|\xi-1|} + \left(\ln^2 r + \frac{\pi^2}{12}\right)\delta(\xi-1) \nonumber\\
&I(\xi, 3,r) = {\cal P}\frac{-\frac{3}{2} \ln ^2\left(\frac{r}{\left| \xi -1\right| }\right)-\frac{\pi ^2}{8}}{\left| \xi -1\right|}+ \left(\ln^3 r+\frac{\pi ^2}{4} \ln r - 2 \zeta (3)\right) \delta (\xi -1)\nonumber\\
&I(\xi, 4,r) = {\cal P}\frac{-2 \ln ^3\left(\frac{r}{\left| \xi -1\right| }\right)-\frac{\pi^2}{2} \ln \left(\frac{r}{\left| \xi-1\right| }\right)+4 \zeta (3)}{\left| \xi -1\right| } \nonumber\\
&\quad \quad \quad \quad + \left(\ln^4 r +\frac{\pi^2}{2} \ln^2 r - 8 \zeta (3) \ln r+\frac{19 \pi ^4}{240}\right) \delta (\xi -1) \ ,
\end{align}
\endgroup
where $\zeta(3)$ is a Riemann zeta function. One can check numerically that the above results satisfy Eq.~(\ref{eq:condist}). In general, we can decompose $I(\xi, n,r)$ into the principal value and delta function parts, 
\begin{align}\label{eq:Inr}
I(\xi, n,r) = {\cal P} \frac{I_{p}(\xi, n,r)}{|\xi-1|} + \delta(\xi-1)I_{\delta}(n,r) \ , 
\end{align}  
where $I_p(\xi, n, r)$ and $I_\delta (n,r)$ can be derived from recursion relations as follows.

Based on the definition Eq.~(\ref{eq:FTevenlog}), one can easily show 
\begin{align}
\frac{d}{d\ln r} I(\xi, n,r) = n \, I(\xi, n-1,r) \ ,
\end{align} 
which leads to the recursive relations,
\begin{align}\label{eq:Irecursive}
&I_{p}(\xi, n,r) = n \int^{\ln r}_{\ln |\xi-1|} d(\ln r') \, I_p(\xi, n-1,r') + C_p(n) \nonumber\\
&I_{\delta}(n,r) = n \int^{\ln r}_{0} d(\ln r') \, I_{\delta}(n-1,r') + C_{\delta}(n) \ ,
\end{align}
where $C_p(n)$ and $C_\delta (n)$ constants are obtained by setting $r=|\xi-1|$ or $r=1$ respectively in Eq.~(\ref{eq:FTforalpha}),
\begin{align}\label{eq:Cp&Cdelta}
&C_p(n) = \frac{d^n}{d\alpha^n} \left(\frac{-\sin\left(\frac{\pi \alpha}{2}\right)}{\pi} \Gamma (\alpha +1) e^{\gamma_E  \alpha }  \right) \Bigg|_{\alpha \rightarrow 0} \nonumber\\
&C_{\delta}(n) = \frac{d^n}{d\alpha^n}\left( \frac{\sin\left(\frac{\pi \alpha}{2}\right)}{\pi} \Gamma (\alpha +1) \frac{2}{\alpha} e^{\gamma_E  \alpha } \right) \Bigg|_{\alpha \rightarrow 0} \  .
\end{align}
The solutions to Eq.~(\ref{eq:Irecursive}) are
\begin{align}\label{eq:Ip&Idelta}
&I_{p}(\xi, n,r) = \sum_{m=0}^{n} \binom{n}{m} C_{p}(n-m) \ln^{m}\left(\frac{r}{|\xi-1|}\right) \nonumber\\
&I_{\delta}(n,r) = \sum_{m=0}^{n} \binom{n}{m} C_{\delta}(n-m) \ln^{m}\left(r\right)  \ , 
\end{align}
where $\binom{n}{m} = \frac{n!}{m!(n-m)!}$ is the binomial function.

We analyze the large $n$ limit of $I(\xi, n,r)$. The dominant contribution should behave like $\sim n!$, which may correspond to the integral regions near $\lambda = 0$ or $\lambda= \infty$ in Eq.~(\ref{eq:FTevenlog}), where $\ln^{n}\left(|\lambda| r e^{\gamma_E} \right)$ tends to $\infty$. Consider the integral near $\lambda= \infty$, there is a highly oscillatory factor $\exp \left(i \lambda (\xi-1)\right)$ for $\xi \neq 1$. As a distribution, $I(\xi, n,r)$ is not well defined at $\xi=1$ and we only consider a neighborhood excluding $\xi=1$. Thus the integral near $\lambda= \infty$ is suppressed by the oscillatory factor $\exp \left(i \lambda (\xi-1)\right)$. Thus the dominant contribution can only come from the integral region near $\lambda=0$. We restrict the integral within $[-\frac{\delta}{e^{\gamma_E} r}, \frac{\delta}{e^{\gamma_E} r}]$ where $\frac{\delta}{e^{\gamma_E} r} \ll 1$, 
\begin{align}\label{eq:Lnana}
&\int_{-\frac{\delta}{e^{\gamma_E} r}}^{+\frac{\delta}{e^{\gamma_E} r}} \frac{d \lambda}{2 \pi} \ln^{n}\left(|\lambda| e^{\gamma_E} r\right) \exp \left(i \lambda (\xi-1)\right) \approx \int_{-\frac{\delta}{e^{\gamma_E} r}}^{+\frac{\delta}{e^{\gamma_E} r}} \frac{d \lambda}{2 \pi} \ln^{n}\left(|\lambda| e^{\gamma_E} r\right) \nonumber\\
&= \frac{1}{e^{\gamma_E} r}\int_{-\delta}^{+\delta} \frac{d \lambda}{2 \pi} \ln^{n}\left(|\lambda|\right)
= \frac{1}{e^{\gamma_E} r}\frac{(-1)^{-n} \left(\Gamma (n+1,-\log (\delta ))+\left((-1)^{2 n}-1\right) \Gamma(n+1)\right)}{\pi} \nonumber\\
&\stackrel{n \rightarrow \infty}{=} \frac{1}{e^{\gamma_E} r}\frac{(-1)^{n}n!}{\pi}  \ ,
\end{align}
where the large $n$ expansion at the leading order is independent of $\delta$. The fact that the factorial growth comes from the short-distance region indicates its UV nature. The alternating factor $(-1)^n$ also confirms its UV nature. 

In the following, we calculate
these constants and study their large $n$ limits, based on which we explicitly calculate the large $n$ limit of $I(\xi, n,r)$.  
One can calculate the constant $C_p(n)$ as follows,
\begingroup
\allowdisplaybreaks
\begin{align}\label{eq:Cpexa}
&C_p(n) = \frac{d^n}{d\alpha^n} \left[ \sum_{m=0}^{+\infty} \frac{(-1)^m \sin\left[\frac{m \pi}{2}\right]}{m! \pi}\left(\alpha \frac{\pi}{2}\right)^m \exp\left(\sum_{k=2}^{+\infty} \frac{\zeta(k)}{k}(-\alpha)^{k}\right) \right] \Bigg|_{\alpha \rightarrow 0} \nonumber\\
&= \sum_{q=0}^{n} \sum_{m=0}^{+\infty} \frac{(-1)^m \sin\left[\frac{m \pi}{2}\right]}{m! \pi}\left(\frac{\pi}{2}\right)^m \binom{n}{q} \left[ \left(\frac{d^q\alpha^m}{d \alpha^q}\right) \left( \frac{d^{n-q} \exp\left(\sum_{k=2}^{+\infty} \frac{\zeta(k)}{k}(-\alpha)^{k}\right) }{d \alpha^{n-q}} \right)  \right] \Bigg|_{\alpha \rightarrow 0} \nonumber\\
&= \sum_{q=0}^{n} \sum_{m=0}^{+\infty} \frac{(-1)^m \sin\left[\frac{m \pi}{2}\right]}{m! \pi}\left(\frac{\pi}{2}\right)^m m! \delta_{m,q} \Bigg(\delta_{n-q,0} \Bigg.\nonumber\\
&\Bigg.\quad +(-1)^{n-q}B_{n-q}[0,\zeta(2),2!\zeta(3),...,(n-q-1)!\zeta(n-q)]\Bigg) \binom{n}{q} \nonumber\\
&= (-1)^{n} n! \sum_{{\rm odd} \, q=1}^{n} \frac{\sin\left[\frac{q \pi}{2}\right]}{q! \pi}\left(\frac{\pi}{2}\right)^q \Bigg(\delta_{n-q,0} \Bigg.\nonumber\\
&\Bigg.\quad+ \frac{1}{(n-q)!}B_{n-q}[0,\zeta(2),2!\zeta(3),...,(n-q-1)!\zeta(n-q)]\Bigg) \ ,
\end{align}
\endgroup
where $B_{n-q}[0,\zeta(2),2!\zeta(3),...,(n-q-1)!\zeta(n-q)]$ is the $(n-q)$th complete Bell polynomial, which is defined as follows
\begin{align}
B_{n}[x_1,x_2,x_3,...,x_n] = \left. \frac{\partial^n}{\partial t^n} \left[ \exp \left(\sum_{j=1}^n x_j \frac{t^j}{j !}\right) \right]\right|_{t=0} \ .
\end{align}
The large $n$ asymptotic form of $C_p(n)$ is 
\begin{align}\label{eq:CpLn}
C_p(n) \stackrel{n \rightarrow \infty}{=} (-1)^{n} n! \frac{e^{-\gamma_E}}{\pi} \ ,
\end{align}
which is checked numerically up to $n=30$. We only keep the dominant contribution to the $n!$ accuracy.
Following the same logic, one can obtain
\begin{align}\label{eq:Cdeltaexa}
&C_{\delta}(n) = (-1)^n n! \sum_{{\rm even} \,q=0}^{n} \frac{\cos \left[\frac{q \pi}{2}\right]}{(q+1)!} \left(-\frac{\pi}{2}\right)^q 
\Bigg( \delta_{n-q,0} \Bigg. \nonumber\\
&\Bigg. +\frac{1}{(n-q)!}B_{n-q}[0,\zeta(2),2!\zeta(3),...,(n-q-1)!\zeta(n-q)] \Bigg) \ ,
\end{align}
and the large $n$ asymptotic form
\begin{align}\label{eq:CdeltaLn}
C_{\delta}(n) \stackrel{n \rightarrow \infty}{=} (-1)^{n} n! \frac{2e^{-\gamma_E}}{\pi},
\end{align}
which is also checked numerically up to $n=30$.

Plugging Eq.~(\ref{eq:CpLn}) into Eq.~(\ref{eq:Ip&Idelta}), we obtain the large $n$ asymptotic form of $I_{p}(\xi,n,r)$
\begin{align}
I_{p}(\xi, n, r) &\stackrel{n \rightarrow \infty}{=} (-1)^n n! \frac{e^{-\gamma_E}}{\pi} \sum_{m=0}^{n} \frac{1}{m!} (-1)^{-m} \ln^{m}\left(\frac{r}{|\xi-1|}\right),
\end{align}
where the dominant contribution to $\sim n!$ comes from $m \ll n$ where $C_p(n-m) \approx (-1)^{n-m} (n-m)! \frac{e^{-\gamma_E}}{\pi}$ is a good approximation. For $m \sim n$, though $C_p(n-m) \approx (-1)^{n-m} (n-m)! \frac{e^{-\gamma_E}}{\pi}$ is not a good approximation, the contribution there is much smaller than $\sim n!$. Thus the $n!$ accuracy is preserved in the above approximation. Then one can obtain
\begin{align}
I_{p}(\xi, n,r) &\stackrel{n \rightarrow \infty}{=} (-1)^n n! \frac{e^{-\gamma_E}}{\pi} \sum_{m=0}^{+\infty} \frac{1}{m!} (-1)^{-m} \ln^{m}\left(\frac{r}{|\xi-1|}\right) = (-1)^n n! \frac{e^{-\gamma_E}}{\pi} \frac{|\xi-1|}{r},
\end{align}
where we include an extra part from $m=n+1$ to $\infty$ during the sum to obtain a simple result. This extra part is highly suppressed in the large $n$ limit so it does not influence the $n!$ accuracy.
Plugging Eq.~(\ref{eq:CdeltaLn}) into Eq.~(\ref{eq:Ip&Idelta}), we obtain the large $n$ asymptotic form of $I_{\delta}(n,r)$
\begin{align}
I_{\delta}(n,r) &\stackrel{n \rightarrow \infty}{=} (-1)^n n! \frac{2e^{-\gamma_E}}{\pi} \sum_{m=0}^{+\infty} \frac{1}{m!} (-1)^{-m} \ln^{m}\left(r\right) = (-1)^n n! \frac{2e^{-\gamma_E}}{\pi} \frac{1}{r} \ .
\end{align}
Plugging the above equations into Eq.~(\ref{eq:Inr}), we obtain the large $n$ asymptotic form of $I(\xi, n,r)$ to the $n!$ accuracy
\begin{align}\label{eq:FTinlargen}
&I(\xi, n,r) \stackrel{n \rightarrow \infty}{=} (-1)^n n! \frac{e^{-\gamma_E}}{\pi} \frac{1}{r}\left[{\cal P}(1) + 2\delta(\xi-1)\right] = (-1)^n n! \frac{e^{-\gamma_E}}{\pi} \frac{1}{r} \ ,
\end{align}
which is independent of $\xi$. The result is consistent with the analysis and short distance integral in Eq.~(\ref{eq:Lnana}).

Following the similar logic above, we study the Fourier transformation of 
logarithmic correlations in coordinate space $\lambda$ of type $i \, s_{\lambda} \ln^{n}\left(|\lambda|r e^{\gamma_E} \right)$,
\begin{align}\label{eq:FTlogodd}
&\tilde I(\xi, n,r) \equiv \int \frac{d \lambda}{2 \pi} \, i \, s_{\lambda} \ln^{n}\left(|\lambda| e^{\gamma_E} r\right) \exp \left(i \lambda (\xi-1)\right) \nonumber\\
&= \frac{d^n}{d\alpha^n} \left[ \int \frac{d \lambda}{2 \pi} \, i \, s_{\lambda} \left(|\lambda| e^{\gamma_E} r\right)^{\alpha} \exp \left(i \lambda (\xi-1)\right) \right] \bigg|_{\alpha \rightarrow 0} \nonumber\\
&= \frac{d^n}{d\alpha^n} \left[ \frac{-\cos\left(\frac{\pi \alpha}{2}\right)}{\pi} \Gamma (\alpha +1) \frac{{\rm sign}(\xi-1)}{|\xi-1|^{\alpha+1}} e^{\gamma_E  \alpha } r^{\alpha } \right] \Bigg|_{\alpha \rightarrow 0} \  ,
\end{align}
where $s_{\lambda} = {\rm sign}(\lambda) $ gives -1,0, or 1 depending on whether $\lambda$ is negative, zero, or positive. It appears in the calculation of perturbative matching kernel in the threshold limit in Eqs.~(\ref{eq:Hmom}),~(\ref{eq:ChybridTL}),~(\ref{eq:LCpdfq1loopfmtr}),~(\ref{eq:qpdfq1loopfmtr}) and~(\ref{eq:TRmatchUV}). 

The results for $n=0,1,...,3$ are
\begin{align}\label{eq:FTodd}
&\tilde I(\xi, 0,r) = \frac{1}{\pi}\frac{1}{1 -\xi} \nonumber\\
&\tilde I(\xi, 1,r) = \frac{1}{2\pi} \frac{\ln \left(\frac{r^2}{(\xi -1)^2}\right)}{1 - \xi } \nonumber\\
&\tilde I(\xi, 2,r) = \frac{1}{4\pi}\frac{ \ln ^2\left(\frac{r^2}{(\xi -1)^2}\right) - \frac{\pi^2}{3} }{1-\xi}  \nonumber\\
&\tilde I(\xi, 3,r) =  \frac{1}{8 \pi}\frac{\ln ^3\left(\frac{r^2}{(\xi -1)^2}\right)-\pi ^2 \ln \left(\frac{r^2}{(\xi
   -1)^2}\right)+8 \psi ^{(2)}(1)}{1-\xi} \ ,
\end{align}
where $\psi^{(2)}$ is the second derivative of the digamma function. 

We derive the general form of $\tilde I(\xi, n,r)$ as follows. Based on the definition Eq.~(\ref{eq:FTlogodd}), $\tilde I(\xi, n,r)$ satisfies the following recursive relation,
\begin{align}
    \frac{d \tilde I(\xi, n,r)}{d \ln r} = n \tilde I(\xi, n-1,r) \ .
\end{align}
Following the similar logic as Eq.~(\ref{eq:Ip&Idelta}), the solution to the recursive relation is
\begin{align}\label{eq:tIp}
\tilde I(\xi, n,r) = \sum_{m=0}^{n} \binom{n}{m} \tilde C_{p}(n-m) \frac{ \ln^m \frac{r}{|\xi-1|} }{1-\xi},
\end{align}
where the constant is
\begin{align}
\tilde C_{p}(n) = \frac{d^n}{d\alpha^n} \left[ \frac{\cos\left(\frac{\pi \alpha}{2}\right)}{\pi} \Gamma (\alpha +1) e^{\gamma_E  \alpha } \right] \Bigg|_{\alpha \rightarrow 0} \ ,  
\end{align}
which is obtained by setting $r=|\xi-1|$ in Eq.~(\ref{eq:FTlogodd}). 
Similar to Eq.~(\ref{eq:Cpexa}), $\tilde C_{p}(n)$ can be written in terms of the complete Bell functions,
\begin{align}\label{eq:tCpexa}
&\tilde C_p(n) = (-1)^{n} n! \sum_{{\rm even} \, q=0}^{n} \frac{\cos\left[\frac{q \pi}{2}\right]}{q! \pi}\left(\frac{\pi}{2}\right)^q 
\Bigg( \delta_{n-q,0} \Bigg. \nonumber\\
&\Bigg. \quad \quad \quad \quad +\frac{1}{(n-q)!}B_{n-q}[0,\zeta(2),2!\zeta(3),...,(n-q-1)!\zeta(n-q)] \Bigg) \ ,
\end{align}
where only even integers $q$ are needed in the sum. 

\bibliographystyle{apsrev4-1}
\bibliography{bibliography}

\begin{thebibliography}{210}%
\makeatletter
\providecommand \@ifxundefined [1]{%
 \@ifx{#1\undefined}
}%
\providecommand \@ifnum [1]{%
 \ifnum #1\expandafter \@firstoftwo
 \else \expandafter \@secondoftwo
 \fi
}%
\providecommand \@ifx [1]{%
 \ifx #1\expandafter \@firstoftwo
 \else \expandafter \@secondoftwo
 \fi
}%
\providecommand \natexlab [1]{#1}%
\providecommand \enquote  [1]{``#1''}%
\providecommand \bibnamefont  [1]{#1}%
\providecommand \bibfnamefont [1]{#1}%
\providecommand \citenamefont [1]{#1}%
\providecommand \href@noop [0]{\@secondoftwo}%
\providecommand \href [0]{\begingroup \@sanitize@url \@href}%
\providecommand \@href[1]{\@@startlink{#1}\@@href}%
\providecommand \@@href[1]{\endgroup#1\@@endlink}%
\providecommand \@sanitize@url [0]{\catcode `\\12\catcode `\$12\catcode
  `\&12\catcode `\#12\catcode `\^12\catcode `\_12\catcode `\%12\relax}%
\providecommand \@@startlink[1]{}%
\providecommand \@@endlink[0]{}%
\providecommand \url  [0]{\begingroup\@sanitize@url \@url }%
\providecommand \@url [1]{\endgroup\@href {#1}{\urlprefix }}%
\providecommand \urlprefix  [0]{URL }%
\providecommand \Eprint [0]{\href }%
\providecommand \doibase [0]{http://dx.doi.org/}%
\providecommand \selectlanguage [0]{\@gobble}%
\providecommand \bibinfo  [0]{\@secondoftwo}%
\providecommand \bibfield  [0]{\@secondoftwo}%
\providecommand \translation [1]{[#1]}%
\providecommand \BibitemOpen [0]{}%
\providecommand \bibitemStop [0]{}%
\providecommand \bibitemNoStop [0]{.\EOS\space}%
\providecommand \EOS [0]{\spacefactor3000\relax}%
\providecommand \BibitemShut  [1]{\csname bibitem#1\endcsname}%
\let\auto@bib@innerbib\@empty
\bibitem [{\citenamefont {Aad}\ \emph {et~al.}(2022)\citenamefont {Aad} \emph
  {et~al.}}]{ATLAS:2021vod}%
  \BibitemOpen
  \bibfield  {author} {\bibinfo {author} {\bibfnamefont {G.}~\bibnamefont
  {Aad}} \emph {et~al.} (\bibinfo {collaboration} {ATLAS}),\ }\href {\doibase
  10.1140/epjc/s10052-022-10217-z} {\bibfield  {journal} {\bibinfo  {journal}
  {Eur. Phys. J. C}\ }\textbf {\bibinfo {volume} {82}},\ \bibinfo {pages} {438}
  (\bibinfo {year} {2022})},\ \Eprint {http://arxiv.org/abs/2112.11266}
  {arXiv:2112.11266 [hep-ex]} \BibitemShut {NoStop}%
\bibitem [{\citenamefont {Ball}\ \emph {et~al.}(2022)\citenamefont {Ball} \emph
  {et~al.}}]{NNPDF:2021njg}%
  \BibitemOpen
  \bibfield  {author} {\bibinfo {author} {\bibfnamefont {R.~D.}\ \bibnamefont
  {Ball}} \emph {et~al.} (\bibinfo {collaboration} {NNPDF}),\ }\href {\doibase
  10.1140/epjc/s10052-022-10328-7} {\bibfield  {journal} {\bibinfo  {journal}
  {Eur. Phys. J. C}\ }\textbf {\bibinfo {volume} {82}},\ \bibinfo {pages} {428}
  (\bibinfo {year} {2022})},\ \Eprint {http://arxiv.org/abs/2109.02653}
  {arXiv:2109.02653 [hep-ph]} \BibitemShut {NoStop}%
\bibitem [{\citenamefont {Bailey}\ \emph {et~al.}(2021)\citenamefont {Bailey},
  \citenamefont {Cridge}, \citenamefont {Harland-Lang}, \citenamefont
  {Martin},\ and\ \citenamefont {Thorne}}]{Bailey:2020ooq}%
  \BibitemOpen
  \bibfield  {author} {\bibinfo {author} {\bibfnamefont {S.}~\bibnamefont
  {Bailey}}, \bibinfo {author} {\bibfnamefont {T.}~\bibnamefont {Cridge}},
  \bibinfo {author} {\bibfnamefont {L.~A.}\ \bibnamefont {Harland-Lang}},
  \bibinfo {author} {\bibfnamefont {A.~D.}\ \bibnamefont {Martin}}, \ and\
  \bibinfo {author} {\bibfnamefont {R.~S.}\ \bibnamefont {Thorne}},\ }\href
  {\doibase 10.1140/epjc/s10052-021-09057-0} {\bibfield  {journal} {\bibinfo
  {journal} {Eur. Phys. J. C}\ }\textbf {\bibinfo {volume} {81}},\ \bibinfo
  {pages} {341} (\bibinfo {year} {2021})},\ \Eprint
  {http://arxiv.org/abs/2012.04684} {arXiv:2012.04684 [hep-ph]} \BibitemShut
  {NoStop}%
\bibitem [{\citenamefont {Hou}\ \emph {et~al.}(2021)\citenamefont {Hou} \emph
  {et~al.}}]{Hou:2019efy}%
  \BibitemOpen
  \bibfield  {author} {\bibinfo {author} {\bibfnamefont {T.-J.}\ \bibnamefont
  {Hou}} \emph {et~al.},\ }\href {\doibase 10.1103/PhysRevD.103.014013}
  {\bibfield  {journal} {\bibinfo  {journal} {Phys. Rev. D}\ }\textbf {\bibinfo
  {volume} {103}},\ \bibinfo {pages} {014013} (\bibinfo {year} {2021})},\
  \Eprint {http://arxiv.org/abs/1912.10053} {arXiv:1912.10053 [hep-ph]}
  \BibitemShut {NoStop}%
\bibitem [{\citenamefont {Alekhin}\ \emph {et~al.}(2017)\citenamefont
  {Alekhin}, \citenamefont {Bl\"umlein}, \citenamefont {Moch},\ and\
  \citenamefont {Placakyte}}]{Alekhin:2017kpj}%
  \BibitemOpen
  \bibfield  {author} {\bibinfo {author} {\bibfnamefont {S.}~\bibnamefont
  {Alekhin}}, \bibinfo {author} {\bibfnamefont {J.}~\bibnamefont {Bl\"umlein}},
  \bibinfo {author} {\bibfnamefont {S.}~\bibnamefont {Moch}}, \ and\ \bibinfo
  {author} {\bibfnamefont {R.}~\bibnamefont {Placakyte}},\ }\href {\doibase
  10.1103/PhysRevD.96.014011} {\bibfield  {journal} {\bibinfo  {journal} {Phys.
  Rev. D}\ }\textbf {\bibinfo {volume} {96}},\ \bibinfo {pages} {014011}
  (\bibinfo {year} {2017})},\ \Eprint {http://arxiv.org/abs/1701.05838}
  {arXiv:1701.05838 [hep-ph]} \BibitemShut {NoStop}%
\bibitem [{\citenamefont {Abramowicz}\ \emph {et~al.}(2015)\citenamefont
  {Abramowicz} \emph {et~al.}}]{H1:2015ubc}%
  \BibitemOpen
  \bibfield  {author} {\bibinfo {author} {\bibfnamefont {H.}~\bibnamefont
  {Abramowicz}} \emph {et~al.} (\bibinfo {collaboration} {H1, ZEUS}),\ }\href
  {\doibase 10.1140/epjc/s10052-015-3710-4} {\bibfield  {journal} {\bibinfo
  {journal} {Eur. Phys. J. C}\ }\textbf {\bibinfo {volume} {75}},\ \bibinfo
  {pages} {580} (\bibinfo {year} {2015})},\ \Eprint
  {http://arxiv.org/abs/1506.06042} {arXiv:1506.06042 [hep-ex]} \BibitemShut
  {NoStop}%
\bibitem [{\citenamefont {Jimenez-Delgado}\ and\ \citenamefont
  {Reya}(2014)}]{Jimenez-Delgado:2014twa}%
  \BibitemOpen
  \bibfield  {author} {\bibinfo {author} {\bibfnamefont {P.}~\bibnamefont
  {Jimenez-Delgado}}\ and\ \bibinfo {author} {\bibfnamefont {E.}~\bibnamefont
  {Reya}},\ }\href {\doibase 10.1103/PhysRevD.89.074049} {\bibfield  {journal}
  {\bibinfo  {journal} {Phys. Rev. D}\ }\textbf {\bibinfo {volume} {89}},\
  \bibinfo {pages} {074049} (\bibinfo {year} {2014})},\ \Eprint
  {http://arxiv.org/abs/1403.1852} {arXiv:1403.1852 [hep-ph]} \BibitemShut
  {NoStop}%
\bibitem [{\citenamefont {Farrar}\ and\ \citenamefont
  {Jackson}(1975)}]{Farrar:1975yb}%
  \BibitemOpen
  \bibfield  {author} {\bibinfo {author} {\bibfnamefont {G.~R.}\ \bibnamefont
  {Farrar}}\ and\ \bibinfo {author} {\bibfnamefont {D.~R.}\ \bibnamefont
  {Jackson}},\ }\href {\doibase 10.1103/PhysRevLett.35.1416} {\bibfield
  {journal} {\bibinfo  {journal} {Phys. Rev. Lett.}\ }\textbf {\bibinfo
  {volume} {35}},\ \bibinfo {pages} {1416} (\bibinfo {year}
  {1975})}\BibitemShut {NoStop}%
\bibitem [{\citenamefont {Brodsky}\ \emph {et~al.}(1995)\citenamefont
  {Brodsky}, \citenamefont {Burkardt},\ and\ \citenamefont
  {Schmidt}}]{Brodsky:1994kg}%
  \BibitemOpen
  \bibfield  {author} {\bibinfo {author} {\bibfnamefont {S.~J.}\ \bibnamefont
  {Brodsky}}, \bibinfo {author} {\bibfnamefont {M.}~\bibnamefont {Burkardt}}, \
  and\ \bibinfo {author} {\bibfnamefont {I.}~\bibnamefont {Schmidt}},\ }\href
  {\doibase 10.1016/0550-3213(95)00009-H} {\bibfield  {journal} {\bibinfo
  {journal} {Nucl. Phys. B}\ }\textbf {\bibinfo {volume} {441}},\ \bibinfo
  {pages} {197} (\bibinfo {year} {1995})},\ \Eprint
  {http://arxiv.org/abs/hep-ph/9401328} {arXiv:hep-ph/9401328} \BibitemShut
  {NoStop}%
\bibitem [{\citenamefont {Isgur}(1999)}]{Isgur:1998yb}%
  \BibitemOpen
  \bibfield  {author} {\bibinfo {author} {\bibfnamefont {N.}~\bibnamefont
  {Isgur}},\ }\href {\doibase 10.1103/PhysRevD.59.034013} {\bibfield  {journal}
  {\bibinfo  {journal} {Phys. Rev. D}\ }\textbf {\bibinfo {volume} {59}},\
  \bibinfo {pages} {034013} (\bibinfo {year} {1999})},\ \Eprint
  {http://arxiv.org/abs/hep-ph/9809255} {arXiv:hep-ph/9809255} \BibitemShut
  {NoStop}%
\bibitem [{\citenamefont {Afnan}\ \emph {et~al.}(2003)\citenamefont {Afnan},
  \citenamefont {Bissey}, \citenamefont {Gomez}, \citenamefont {Katramatou},
  \citenamefont {Liuti}, \citenamefont {Melnitchouk}, \citenamefont
  {Petratos},\ and\ \citenamefont {Thomas}}]{Afnan:2003vh}%
  \BibitemOpen
  \bibfield  {author} {\bibinfo {author} {\bibfnamefont {I.~R.}\ \bibnamefont
  {Afnan}}, \bibinfo {author} {\bibfnamefont {F.~R.~P.}\ \bibnamefont
  {Bissey}}, \bibinfo {author} {\bibfnamefont {J.}~\bibnamefont {Gomez}},
  \bibinfo {author} {\bibfnamefont {A.~T.}\ \bibnamefont {Katramatou}},
  \bibinfo {author} {\bibfnamefont {S.}~\bibnamefont {Liuti}}, \bibinfo
  {author} {\bibfnamefont {W.}~\bibnamefont {Melnitchouk}}, \bibinfo {author}
  {\bibfnamefont {G.~G.}\ \bibnamefont {Petratos}}, \ and\ \bibinfo {author}
  {\bibfnamefont {A.~W.}\ \bibnamefont {Thomas}},\ }\href {\doibase
  10.1103/PhysRevC.68.035201} {\bibfield  {journal} {\bibinfo  {journal} {Phys.
  Rev. C}\ }\textbf {\bibinfo {volume} {68}},\ \bibinfo {pages} {035201}
  (\bibinfo {year} {2003})},\ \Eprint {http://arxiv.org/abs/nucl-th/0306054}
  {arXiv:nucl-th/0306054} \BibitemShut {NoStop}%
\bibitem [{\citenamefont {Tropiano}\ \emph {et~al.}(2019)\citenamefont
  {Tropiano}, \citenamefont {Ethier}, \citenamefont {Melnitchouk},\ and\
  \citenamefont {Sato}}]{Tropiano:2018quk}%
  \BibitemOpen
  \bibfield  {author} {\bibinfo {author} {\bibfnamefont {A.~J.}\ \bibnamefont
  {Tropiano}}, \bibinfo {author} {\bibfnamefont {J.~J.}\ \bibnamefont
  {Ethier}}, \bibinfo {author} {\bibfnamefont {W.}~\bibnamefont {Melnitchouk}},
  \ and\ \bibinfo {author} {\bibfnamefont {N.}~\bibnamefont {Sato}},\ }\href
  {\doibase 10.1103/PhysRevC.99.035201} {\bibfield  {journal} {\bibinfo
  {journal} {Phys. Rev. C}\ }\textbf {\bibinfo {volume} {99}},\ \bibinfo
  {pages} {035201} (\bibinfo {year} {2019})},\ \Eprint
  {http://arxiv.org/abs/1811.07668} {arXiv:1811.07668 [nucl-th]} \BibitemShut
  {NoStop}%
\bibitem [{\citenamefont {Abrams}\ \emph {et~al.}(2022)\citenamefont {Abrams}
  \emph {et~al.}}]{JeffersonLabHallATritium:2021usd}%
  \BibitemOpen
  \bibfield  {author} {\bibinfo {author} {\bibfnamefont {D.}~\bibnamefont
  {Abrams}} \emph {et~al.} (\bibinfo {collaboration} {Jefferson Lab Hall A
  Tritium}),\ }\href {\doibase 10.1103/PhysRevLett.128.132003} {\bibfield
  {journal} {\bibinfo  {journal} {Phys. Rev. Lett.}\ }\textbf {\bibinfo
  {volume} {128}},\ \bibinfo {pages} {132003} (\bibinfo {year} {2022})},\
  \Eprint {http://arxiv.org/abs/2104.05850} {arXiv:2104.05850 [hep-ex]}
  \BibitemShut {NoStop}%
\bibitem [{\citenamefont {Aubert}\ \emph {et~al.}(1983)\citenamefont {Aubert}
  \emph {et~al.}}]{EuropeanMuon:1983wih}%
  \BibitemOpen
  \bibfield  {author} {\bibinfo {author} {\bibfnamefont {J.~J.}\ \bibnamefont
  {Aubert}} \emph {et~al.} (\bibinfo {collaboration} {European Muon}),\ }\href
  {\doibase 10.1016/0370-2693(83)90437-9} {\bibfield  {journal} {\bibinfo
  {journal} {Phys. Lett. B}\ }\textbf {\bibinfo {volume} {123}},\ \bibinfo
  {pages} {275} (\bibinfo {year} {1983})}\BibitemShut {NoStop}%
\bibitem [{\citenamefont {Ke}\ \emph {et~al.}(2023)\citenamefont {Ke},
  \citenamefont {Zhang}, \citenamefont {Xing},\ and\ \citenamefont
  {Wang}}]{Ke:2023xeo}%
  \BibitemOpen
  \bibfield  {author} {\bibinfo {author} {\bibfnamefont {W.}~\bibnamefont
  {Ke}}, \bibinfo {author} {\bibfnamefont {Y.-Y.}\ \bibnamefont {Zhang}},
  \bibinfo {author} {\bibfnamefont {H.}~\bibnamefont {Xing}}, \ and\ \bibinfo
  {author} {\bibfnamefont {X.-N.}\ \bibnamefont {Wang}},\ }\href@noop {} {\
  (\bibinfo {year} {2023})},\ \Eprint {http://arxiv.org/abs/2304.10779}
  {arXiv:2304.10779 [hep-ph]} \BibitemShut {NoStop}%
\bibitem [{\citenamefont {Holt}\ and\ \citenamefont
  {Roberts}(2010)}]{Holt:2010vj}%
  \BibitemOpen
  \bibfield  {author} {\bibinfo {author} {\bibfnamefont {R.~J.}\ \bibnamefont
  {Holt}}\ and\ \bibinfo {author} {\bibfnamefont {C.~D.}\ \bibnamefont
  {Roberts}},\ }\href {\doibase 10.1103/RevModPhys.82.2991} {\bibfield
  {journal} {\bibinfo  {journal} {Rev. Mod. Phys.}\ }\textbf {\bibinfo {volume}
  {82}},\ \bibinfo {pages} {2991} (\bibinfo {year} {2010})},\ \Eprint
  {http://arxiv.org/abs/1002.4666} {arXiv:1002.4666 [nucl-th]} \BibitemShut
  {NoStop}%
\bibitem [{\citenamefont {Flay}\ \emph {et~al.}(2016)\citenamefont {Flay} \emph
  {et~al.}}]{JeffersonLabHallA:2016neg}%
  \BibitemOpen
  \bibfield  {author} {\bibinfo {author} {\bibfnamefont {D.}~\bibnamefont
  {Flay}} \emph {et~al.} (\bibinfo {collaboration} {Jefferson Lab Hall A}),\
  }\href {\doibase 10.1103/PhysRevD.94.052003} {\bibfield  {journal} {\bibinfo
  {journal} {Phys. Rev. D}\ }\textbf {\bibinfo {volume} {94}},\ \bibinfo
  {pages} {052003} (\bibinfo {year} {2016})},\ \Eprint
  {http://arxiv.org/abs/1603.03612} {arXiv:1603.03612 [nucl-ex]} \BibitemShut
  {NoStop}%
\bibitem [{\citenamefont {Adam}\ \emph {et~al.}(2019)\citenamefont {Adam} \emph
  {et~al.}}]{STAR:2019yqm}%
  \BibitemOpen
  \bibfield  {author} {\bibinfo {author} {\bibfnamefont {J.}~\bibnamefont
  {Adam}} \emph {et~al.} (\bibinfo {collaboration} {STAR}),\ }\href {\doibase
  10.1103/PhysRevD.100.052005} {\bibfield  {journal} {\bibinfo  {journal}
  {Phys. Rev. D}\ }\textbf {\bibinfo {volume} {100}},\ \bibinfo {pages}
  {052005} (\bibinfo {year} {2019})},\ \Eprint
  {http://arxiv.org/abs/1906.02740} {arXiv:1906.02740 [hep-ex]} \BibitemShut
  {NoStop}%
\bibitem [{\citenamefont {Friscic}\ \emph {et~al.}(2021)\citenamefont {Friscic}
  \emph {et~al.}}]{Friscic:2021oti}%
  \BibitemOpen
  \bibfield  {author} {\bibinfo {author} {\bibfnamefont {I.}~\bibnamefont
  {Friscic}} \emph {et~al.},\ }\href {\doibase 10.1016/j.physletb.2021.136726}
  {\bibfield  {journal} {\bibinfo  {journal} {Phys. Lett. B}\ }\textbf
  {\bibinfo {volume} {823}},\ \bibinfo {pages} {136726} (\bibinfo {year}
  {2021})},\ \Eprint {http://arxiv.org/abs/2106.08805} {arXiv:2106.08805
  [nucl-ex]} \BibitemShut {NoStop}%
\bibitem [{\citenamefont {Lagerquist}\ \emph {et~al.}(2023)\citenamefont
  {Lagerquist}, \citenamefont {Kuhn},\ and\ \citenamefont
  {Sato}}]{Lagerquist:2022tml}%
  \BibitemOpen
  \bibfield  {author} {\bibinfo {author} {\bibfnamefont {V.}~\bibnamefont
  {Lagerquist}}, \bibinfo {author} {\bibfnamefont {S.~E.}\ \bibnamefont
  {Kuhn}}, \ and\ \bibinfo {author} {\bibfnamefont {N.}~\bibnamefont {Sato}},\
  }\href {\doibase 10.1103/PhysRevC.107.045201} {\bibfield  {journal} {\bibinfo
   {journal} {Phys. Rev. C}\ }\textbf {\bibinfo {volume} {107}},\ \bibinfo
  {pages} {045201} (\bibinfo {year} {2023})},\ \Eprint
  {http://arxiv.org/abs/2205.01218} {arXiv:2205.01218 [nucl-ex]} \BibitemShut
  {NoStop}%
\bibitem [{\citenamefont {Kuhlmann}\ \emph {et~al.}(2000)\citenamefont
  {Kuhlmann}, \citenamefont {Huston}, \citenamefont {Morfin}, \citenamefont
  {Olness}, \citenamefont {Pumplin}, \citenamefont {Owens}, \citenamefont
  {Tung},\ and\ \citenamefont {Whitmore}}]{Kuhlmann:1999sf}%
  \BibitemOpen
  \bibfield  {author} {\bibinfo {author} {\bibfnamefont {S.}~\bibnamefont
  {Kuhlmann}}, \bibinfo {author} {\bibfnamefont {J.}~\bibnamefont {Huston}},
  \bibinfo {author} {\bibfnamefont {J.}~\bibnamefont {Morfin}}, \bibinfo
  {author} {\bibfnamefont {F.~I.}\ \bibnamefont {Olness}}, \bibinfo {author}
  {\bibfnamefont {J.}~\bibnamefont {Pumplin}}, \bibinfo {author} {\bibfnamefont
  {J.~F.}\ \bibnamefont {Owens}}, \bibinfo {author} {\bibfnamefont {W.~K.}\
  \bibnamefont {Tung}}, \ and\ \bibinfo {author} {\bibfnamefont {J.~J.}\
  \bibnamefont {Whitmore}},\ }\href {\doibase 10.1016/S0370-2693(00)00164-7}
  {\bibfield  {journal} {\bibinfo  {journal} {Phys. Lett. B}\ }\textbf
  {\bibinfo {volume} {476}},\ \bibinfo {pages} {291} (\bibinfo {year}
  {2000})},\ \Eprint {http://arxiv.org/abs/hep-ph/9912283}
  {arXiv:hep-ph/9912283} \BibitemShut {NoStop}%
\bibitem [{\citenamefont {Chatrchyan}\ \emph {et~al.}(2013)\citenamefont
  {Chatrchyan} \emph {et~al.}}]{CMS:2012ftr}%
  \BibitemOpen
  \bibfield  {author} {\bibinfo {author} {\bibfnamefont {S.}~\bibnamefont
  {Chatrchyan}} \emph {et~al.} (\bibinfo {collaboration} {CMS}),\ }\href
  {\doibase 10.1103/PhysRevD.87.112002} {\bibfield  {journal} {\bibinfo
  {journal} {Phys. Rev. D}\ }\textbf {\bibinfo {volume} {87}},\ \bibinfo
  {pages} {112002} (\bibinfo {year} {2013})},\ \bibinfo {note} {[Erratum:
  Phys.Rev.D 87, 119902 (2013)]},\ \Eprint {http://arxiv.org/abs/1212.6660}
  {arXiv:1212.6660 [hep-ex]} \BibitemShut {NoStop}%
\bibitem [{\citenamefont {Aad}\ \emph {et~al.}(2012)\citenamefont {Aad} \emph
  {et~al.}}]{ATLAS:2011juz}%
  \BibitemOpen
  \bibfield  {author} {\bibinfo {author} {\bibfnamefont {G.}~\bibnamefont
  {Aad}} \emph {et~al.} (\bibinfo {collaboration} {ATLAS}),\ }\href {\doibase
  10.1103/PhysRevD.86.014022} {\bibfield  {journal} {\bibinfo  {journal} {Phys.
  Rev. D}\ }\textbf {\bibinfo {volume} {86}},\ \bibinfo {pages} {014022}
  (\bibinfo {year} {2012})},\ \Eprint {http://arxiv.org/abs/1112.6297}
  {arXiv:1112.6297 [hep-ex]} \BibitemShut {NoStop}%
\bibitem [{\citenamefont {Brady}\ \emph {et~al.}(2012)\citenamefont {Brady},
  \citenamefont {Accardi}, \citenamefont {Melnitchouk},\ and\ \citenamefont
  {Owens}}]{Brady:2011hb}%
  \BibitemOpen
  \bibfield  {author} {\bibinfo {author} {\bibfnamefont {L.~T.}\ \bibnamefont
  {Brady}}, \bibinfo {author} {\bibfnamefont {A.}~\bibnamefont {Accardi}},
  \bibinfo {author} {\bibfnamefont {W.}~\bibnamefont {Melnitchouk}}, \ and\
  \bibinfo {author} {\bibfnamefont {J.~F.}\ \bibnamefont {Owens}},\ }\href
  {\doibase 10.1007/JHEP06(2012)019} {\bibfield  {journal} {\bibinfo  {journal}
  {JHEP}\ }\textbf {\bibinfo {volume} {06}},\ \bibinfo {pages} {019} (\bibinfo
  {year} {2012})},\ \Eprint {http://arxiv.org/abs/1110.5398} {arXiv:1110.5398
  [hep-ph]} \BibitemShut {NoStop}%
\bibitem [{\citenamefont {Abe}\ \emph {et~al.}(1996)\citenamefont {Abe} \emph
  {et~al.}}]{CDF:1996yow}%
  \BibitemOpen
  \bibfield  {author} {\bibinfo {author} {\bibfnamefont {F.}~\bibnamefont
  {Abe}} \emph {et~al.} (\bibinfo {collaboration} {CDF}),\ }\href {\doibase
  10.1103/PhysRevLett.77.438} {\bibfield  {journal} {\bibinfo  {journal} {Phys.
  Rev. Lett.}\ }\textbf {\bibinfo {volume} {77}},\ \bibinfo {pages} {438}
  (\bibinfo {year} {1996})},\ \Eprint {http://arxiv.org/abs/hep-ex/9601008}
  {arXiv:hep-ex/9601008} \BibitemShut {NoStop}%
\bibitem [{\citenamefont {Huston}\ \emph {et~al.}(1996)\citenamefont {Huston},
  \citenamefont {Kovacs}, \citenamefont {Kuhlmann}, \citenamefont {Lai},
  \citenamefont {Owens}, \citenamefont {Soper},\ and\ \citenamefont
  {Tung}}]{Huston:1995tw}%
  \BibitemOpen
  \bibfield  {author} {\bibinfo {author} {\bibfnamefont {J.}~\bibnamefont
  {Huston}}, \bibinfo {author} {\bibfnamefont {E.}~\bibnamefont {Kovacs}},
  \bibinfo {author} {\bibfnamefont {S.}~\bibnamefont {Kuhlmann}}, \bibinfo
  {author} {\bibfnamefont {H.~L.}\ \bibnamefont {Lai}}, \bibinfo {author}
  {\bibfnamefont {J.~F.}\ \bibnamefont {Owens}}, \bibinfo {author}
  {\bibfnamefont {D.~E.}\ \bibnamefont {Soper}}, \ and\ \bibinfo {author}
  {\bibfnamefont {W.~K.}\ \bibnamefont {Tung}},\ }\href {\doibase
  10.1103/PhysRevLett.77.444} {\bibfield  {journal} {\bibinfo  {journal} {Phys.
  Rev. Lett.}\ }\textbf {\bibinfo {volume} {77}},\ \bibinfo {pages} {444}
  (\bibinfo {year} {1996})},\ \Eprint {http://arxiv.org/abs/hep-ph/9511386}
  {arXiv:hep-ph/9511386} \BibitemShut {NoStop}%
\bibitem [{\citenamefont {Sterman}(1987)}]{Sterman:1986aj}%
  \BibitemOpen
  \bibfield  {author} {\bibinfo {author} {\bibfnamefont {G.~F.}\ \bibnamefont
  {Sterman}},\ }\href {\doibase 10.1016/0550-3213(87)90258-6} {\bibfield
  {journal} {\bibinfo  {journal} {Nucl. Phys. B}\ }\textbf {\bibinfo {volume}
  {281}},\ \bibinfo {pages} {310} (\bibinfo {year} {1987})}\BibitemShut
  {NoStop}%
\bibitem [{\citenamefont {Catani}\ and\ \citenamefont
  {Trentadue}(1989)}]{Catani:1989ne}%
  \BibitemOpen
  \bibfield  {author} {\bibinfo {author} {\bibfnamefont {S.}~\bibnamefont
  {Catani}}\ and\ \bibinfo {author} {\bibfnamefont {L.}~\bibnamefont
  {Trentadue}},\ }\href {\doibase 10.1016/0550-3213(89)90273-3} {\bibfield
  {journal} {\bibinfo  {journal} {Nucl. Phys. B}\ }\textbf {\bibinfo {volume}
  {327}},\ \bibinfo {pages} {323} (\bibinfo {year} {1989})}\BibitemShut
  {NoStop}%
\bibitem [{\citenamefont {Bauer}\ \emph
  {et~al.}(2002{\natexlab{a}})\citenamefont {Bauer}, \citenamefont {Fleming},
  \citenamefont {Pirjol}, \citenamefont {Rothstein},\ and\ \citenamefont
  {Stewart}}]{Bauer:2002nz}%
  \BibitemOpen
  \bibfield  {author} {\bibinfo {author} {\bibfnamefont {C.~W.}\ \bibnamefont
  {Bauer}}, \bibinfo {author} {\bibfnamefont {S.}~\bibnamefont {Fleming}},
  \bibinfo {author} {\bibfnamefont {D.}~\bibnamefont {Pirjol}}, \bibinfo
  {author} {\bibfnamefont {I.~Z.}\ \bibnamefont {Rothstein}}, \ and\ \bibinfo
  {author} {\bibfnamefont {I.~W.}\ \bibnamefont {Stewart}},\ }\href {\doibase
  10.1103/PhysRevD.66.014017} {\bibfield  {journal} {\bibinfo  {journal} {Phys.
  Rev. D}\ }\textbf {\bibinfo {volume} {66}},\ \bibinfo {pages} {014017}
  (\bibinfo {year} {2002}{\natexlab{a}})},\ \Eprint
  {http://arxiv.org/abs/hep-ph/0202088} {arXiv:hep-ph/0202088} \BibitemShut
  {NoStop}%
\bibitem [{\citenamefont {Manohar}(2003)}]{Manohar:2003vb}%
  \BibitemOpen
  \bibfield  {author} {\bibinfo {author} {\bibfnamefont {A.~V.}\ \bibnamefont
  {Manohar}},\ }\href {\doibase 10.1103/PhysRevD.68.114019} {\bibfield
  {journal} {\bibinfo  {journal} {Phys. Rev. D}\ }\textbf {\bibinfo {volume}
  {68}},\ \bibinfo {pages} {114019} (\bibinfo {year} {2003})},\ \Eprint
  {http://arxiv.org/abs/hep-ph/0309176} {arXiv:hep-ph/0309176} \BibitemShut
  {NoStop}%
\bibitem [{\citenamefont {Pecjak}(2005)}]{Pecjak:2005uh}%
  \BibitemOpen
  \bibfield  {author} {\bibinfo {author} {\bibfnamefont {B.~D.}\ \bibnamefont
  {Pecjak}},\ }\href {\doibase 10.1088/1126-6708/2005/10/040} {\bibfield
  {journal} {\bibinfo  {journal} {JHEP}\ }\textbf {\bibinfo {volume} {10}},\
  \bibinfo {pages} {040} (\bibinfo {year} {2005})},\ \Eprint
  {http://arxiv.org/abs/hep-ph/0506269} {arXiv:hep-ph/0506269} \BibitemShut
  {NoStop}%
\bibitem [{\citenamefont {Chay}\ and\ \citenamefont {Kim}(2007)}]{Chay:2005rz}%
  \BibitemOpen
  \bibfield  {author} {\bibinfo {author} {\bibfnamefont {J.}~\bibnamefont
  {Chay}}\ and\ \bibinfo {author} {\bibfnamefont {C.}~\bibnamefont {Kim}},\
  }\href {\doibase 10.1103/PhysRevD.75.016003} {\bibfield  {journal} {\bibinfo
  {journal} {Phys. Rev. D}\ }\textbf {\bibinfo {volume} {75}},\ \bibinfo
  {pages} {016003} (\bibinfo {year} {2007})},\ \Eprint
  {http://arxiv.org/abs/hep-ph/0511066} {arXiv:hep-ph/0511066} \BibitemShut
  {NoStop}%
\bibitem [{\citenamefont {Idilbi}\ \emph {et~al.}(2006)\citenamefont {Idilbi},
  \citenamefont {Ji},\ and\ \citenamefont {Yuan}}]{Idilbi:2006dg}%
  \BibitemOpen
  \bibfield  {author} {\bibinfo {author} {\bibfnamefont {A.}~\bibnamefont
  {Idilbi}}, \bibinfo {author} {\bibfnamefont {X.-d.}\ \bibnamefont {Ji}}, \
  and\ \bibinfo {author} {\bibfnamefont {F.}~\bibnamefont {Yuan}},\ }\href
  {\doibase 10.1016/j.nuclphysb.2006.07.002} {\bibfield  {journal} {\bibinfo
  {journal} {Nucl. Phys. B}\ }\textbf {\bibinfo {volume} {753}},\ \bibinfo
  {pages} {42} (\bibinfo {year} {2006})},\ \Eprint
  {http://arxiv.org/abs/hep-ph/0605068} {arXiv:hep-ph/0605068} \BibitemShut
  {NoStop}%
\bibitem [{\citenamefont {Chen}\ \emph {et~al.}(2007)\citenamefont {Chen},
  \citenamefont {Idilbi},\ and\ \citenamefont {Ji}}]{Chen:2006vd}%
  \BibitemOpen
  \bibfield  {author} {\bibinfo {author} {\bibfnamefont {P.-y.}\ \bibnamefont
  {Chen}}, \bibinfo {author} {\bibfnamefont {A.}~\bibnamefont {Idilbi}}, \ and\
  \bibinfo {author} {\bibfnamefont {X.-d.}\ \bibnamefont {Ji}},\ }\href
  {\doibase 10.1016/j.nuclphysb.2006.11.020} {\bibfield  {journal} {\bibinfo
  {journal} {Nucl. Phys. B}\ }\textbf {\bibinfo {volume} {763}},\ \bibinfo
  {pages} {183} (\bibinfo {year} {2007})},\ \Eprint
  {http://arxiv.org/abs/hep-ph/0607003} {arXiv:hep-ph/0607003} \BibitemShut
  {NoStop}%
\bibitem [{\citenamefont {Becher}\ \emph {et~al.}(2007)\citenamefont {Becher},
  \citenamefont {Neubert},\ and\ \citenamefont {Pecjak}}]{Becher:2006mr}%
  \BibitemOpen
  \bibfield  {author} {\bibinfo {author} {\bibfnamefont {T.}~\bibnamefont
  {Becher}}, \bibinfo {author} {\bibfnamefont {M.}~\bibnamefont {Neubert}}, \
  and\ \bibinfo {author} {\bibfnamefont {B.~D.}\ \bibnamefont {Pecjak}},\
  }\href {\doibase 10.1088/1126-6708/2007/01/076} {\bibfield  {journal}
  {\bibinfo  {journal} {JHEP}\ }\textbf {\bibinfo {volume} {01}},\ \bibinfo
  {pages} {076} (\bibinfo {year} {2007})},\ \Eprint
  {http://arxiv.org/abs/hep-ph/0607228} {arXiv:hep-ph/0607228} \BibitemShut
  {NoStop}%
\bibitem [{\citenamefont {Bonvini}\ \emph {et~al.}(2012)\citenamefont
  {Bonvini}, \citenamefont {Forte}, \citenamefont {Ghezzi},\ and\ \citenamefont
  {Ridolfi}}]{Bonvini:2012az}%
  \BibitemOpen
  \bibfield  {author} {\bibinfo {author} {\bibfnamefont {M.}~\bibnamefont
  {Bonvini}}, \bibinfo {author} {\bibfnamefont {S.}~\bibnamefont {Forte}},
  \bibinfo {author} {\bibfnamefont {M.}~\bibnamefont {Ghezzi}}, \ and\ \bibinfo
  {author} {\bibfnamefont {G.}~\bibnamefont {Ridolfi}},\ }\href {\doibase
  10.1016/j.nuclphysb.2012.04.010} {\bibfield  {journal} {\bibinfo  {journal}
  {Nucl. Phys. B}\ }\textbf {\bibinfo {volume} {861}},\ \bibinfo {pages} {337}
  (\bibinfo {year} {2012})},\ \Eprint {http://arxiv.org/abs/1201.6364}
  {arXiv:1201.6364 [hep-ph]} \BibitemShut {NoStop}%
\bibitem [{\citenamefont {Bauer}\ \emph {et~al.}(2001)\citenamefont {Bauer},
  \citenamefont {Fleming}, \citenamefont {Pirjol},\ and\ \citenamefont
  {Stewart}}]{Bauer:2000yr}%
  \BibitemOpen
  \bibfield  {author} {\bibinfo {author} {\bibfnamefont {C.~W.}\ \bibnamefont
  {Bauer}}, \bibinfo {author} {\bibfnamefont {S.}~\bibnamefont {Fleming}},
  \bibinfo {author} {\bibfnamefont {D.}~\bibnamefont {Pirjol}}, \ and\ \bibinfo
  {author} {\bibfnamefont {I.~W.}\ \bibnamefont {Stewart}},\ }\href {\doibase
  10.1103/PhysRevD.63.114020} {\bibfield  {journal} {\bibinfo  {journal} {Phys.
  Rev. D}\ }\textbf {\bibinfo {volume} {63}},\ \bibinfo {pages} {114020}
  (\bibinfo {year} {2001})},\ \Eprint {http://arxiv.org/abs/hep-ph/0011336}
  {arXiv:hep-ph/0011336} \BibitemShut {NoStop}%
\bibitem [{\citenamefont {Bauer}\ \emph
  {et~al.}(2002{\natexlab{b}})\citenamefont {Bauer}, \citenamefont {Pirjol},\
  and\ \citenamefont {Stewart}}]{Bauer:2001yt}%
  \BibitemOpen
  \bibfield  {author} {\bibinfo {author} {\bibfnamefont {C.~W.}\ \bibnamefont
  {Bauer}}, \bibinfo {author} {\bibfnamefont {D.}~\bibnamefont {Pirjol}}, \
  and\ \bibinfo {author} {\bibfnamefont {I.~W.}\ \bibnamefont {Stewart}},\
  }\href {\doibase 10.1103/PhysRevD.65.054022} {\bibfield  {journal} {\bibinfo
  {journal} {Phys. Rev. D}\ }\textbf {\bibinfo {volume} {65}},\ \bibinfo
  {pages} {054022} (\bibinfo {year} {2002}{\natexlab{b}})},\ \Eprint
  {http://arxiv.org/abs/hep-ph/0109045} {arXiv:hep-ph/0109045} \BibitemShut
  {NoStop}%
\bibitem [{\citenamefont {Becher}\ \emph {et~al.}(2015)\citenamefont {Becher},
  \citenamefont {Broggio},\ and\ \citenamefont {Ferroglia}}]{Becher:2014oda}%
  \BibitemOpen
  \bibfield  {author} {\bibinfo {author} {\bibfnamefont {T.}~\bibnamefont
  {Becher}}, \bibinfo {author} {\bibfnamefont {A.}~\bibnamefont {Broggio}}, \
  and\ \bibinfo {author} {\bibfnamefont {A.}~\bibnamefont {Ferroglia}},\ }\href
  {\doibase 10.1007/978-3-319-14848-9} {\emph {\bibinfo {title} {{Introduction
  to Soft-Collinear Effective Theory}}}},\ Vol.\ \bibinfo {volume} {896}\
  (\bibinfo  {publisher} {Springer},\ \bibinfo {year} {2015})\ \Eprint
  {http://arxiv.org/abs/1410.1892} {arXiv:1410.1892 [hep-ph]} \BibitemShut
  {NoStop}%
\bibitem [{\citenamefont {Becher}\ \emph {et~al.}(2008)\citenamefont {Becher},
  \citenamefont {Neubert},\ and\ \citenamefont {Xu}}]{Becher:2007ty}%
  \BibitemOpen
  \bibfield  {author} {\bibinfo {author} {\bibfnamefont {T.}~\bibnamefont
  {Becher}}, \bibinfo {author} {\bibfnamefont {M.}~\bibnamefont {Neubert}}, \
  and\ \bibinfo {author} {\bibfnamefont {G.}~\bibnamefont {Xu}},\ }\href
  {\doibase 10.1088/1126-6708/2008/07/030} {\bibfield  {journal} {\bibinfo
  {journal} {JHEP}\ }\textbf {\bibinfo {volume} {07}},\ \bibinfo {pages} {030}
  (\bibinfo {year} {2008})},\ \Eprint {http://arxiv.org/abs/0710.0680}
  {arXiv:0710.0680 [hep-ph]} \BibitemShut {NoStop}%
\bibitem [{\citenamefont {Das}\ \emph {et~al.}(2020{\natexlab{a}})\citenamefont
  {Das}, \citenamefont {Moch},\ and\ \citenamefont {Vogt}}]{Das:2019btv}%
  \BibitemOpen
  \bibfield  {author} {\bibinfo {author} {\bibfnamefont {G.}~\bibnamefont
  {Das}}, \bibinfo {author} {\bibfnamefont {S.-O.}\ \bibnamefont {Moch}}, \
  and\ \bibinfo {author} {\bibfnamefont {A.}~\bibnamefont {Vogt}},\ }\href
  {\doibase 10.1007/JHEP03(2020)116} {\bibfield  {journal} {\bibinfo  {journal}
  {JHEP}\ }\textbf {\bibinfo {volume} {03}},\ \bibinfo {pages} {116} (\bibinfo
  {year} {2020}{\natexlab{a}})},\ \Eprint {http://arxiv.org/abs/1912.12920}
  {arXiv:1912.12920 [hep-ph]} \BibitemShut {NoStop}%
\bibitem [{\citenamefont {Das}\ \emph {et~al.}(2020{\natexlab{b}})\citenamefont
  {Das}, \citenamefont {Moch},\ and\ \citenamefont {Vogt}}]{Das:2020adl}%
  \BibitemOpen
  \bibfield  {author} {\bibinfo {author} {\bibfnamefont {G.}~\bibnamefont
  {Das}}, \bibinfo {author} {\bibfnamefont {S.}~\bibnamefont {Moch}}, \ and\
  \bibinfo {author} {\bibfnamefont {A.}~\bibnamefont {Vogt}},\ }\href {\doibase
  10.1016/j.physletb.2020.135546} {\bibfield  {journal} {\bibinfo  {journal}
  {Phys. Lett. B}\ }\textbf {\bibinfo {volume} {807}},\ \bibinfo {pages}
  {135546} (\bibinfo {year} {2020}{\natexlab{b}})},\ \Eprint
  {http://arxiv.org/abs/2004.00563} {arXiv:2004.00563 [hep-ph]} \BibitemShut
  {NoStop}%
\bibitem [{\citenamefont {A~H}\ \emph {et~al.}(2020)\citenamefont {A~H},
  \citenamefont {Das}, \citenamefont {Kumar}, \citenamefont {Mukherjee},
  \citenamefont {Ravindran},\ and\ \citenamefont {Samanta}}]{AH:2020cok}%
  \BibitemOpen
  \bibfield  {author} {\bibinfo {author} {\bibfnamefont {A.}~\bibnamefont
  {A~H}}, \bibinfo {author} {\bibfnamefont {G.}~\bibnamefont {Das}}, \bibinfo
  {author} {\bibfnamefont {M.~C.}\ \bibnamefont {Kumar}}, \bibinfo {author}
  {\bibfnamefont {P.}~\bibnamefont {Mukherjee}}, \bibinfo {author}
  {\bibfnamefont {V.}~\bibnamefont {Ravindran}}, \ and\ \bibinfo {author}
  {\bibfnamefont {K.}~\bibnamefont {Samanta}},\ }\href {\doibase
  10.1007/JHEP10(2020)153} {\bibfield  {journal} {\bibinfo  {journal} {JHEP}\
  }\textbf {\bibinfo {volume} {10}},\ \bibinfo {pages} {153} (\bibinfo {year}
  {2020})},\ \Eprint {http://arxiv.org/abs/2001.11377} {arXiv:2001.11377
  [hep-ph]} \BibitemShut {NoStop}%
\bibitem [{\citenamefont {Das}(2023)}]{Das:2023bfi}%
  \BibitemOpen
  \bibfield  {author} {\bibinfo {author} {\bibfnamefont {G.}~\bibnamefont
  {Das}},\ }\href@noop {} {\  (\bibinfo {year} {2023})},\ \Eprint
  {http://arxiv.org/abs/2303.16578} {arXiv:2303.16578 [hep-ph]} \BibitemShut
  {NoStop}%
\bibitem [{\citenamefont {Banerjee}\ \emph
  {et~al.}(2018{\natexlab{a}})\citenamefont {Banerjee}, \citenamefont {Das},
  \citenamefont {Dhani},\ and\ \citenamefont {Ravindran}}]{Banerjee:2017cfc}%
  \BibitemOpen
  \bibfield  {author} {\bibinfo {author} {\bibfnamefont {P.}~\bibnamefont
  {Banerjee}}, \bibinfo {author} {\bibfnamefont {G.}~\bibnamefont {Das}},
  \bibinfo {author} {\bibfnamefont {P.~K.}\ \bibnamefont {Dhani}}, \ and\
  \bibinfo {author} {\bibfnamefont {V.}~\bibnamefont {Ravindran}},\ }\href
  {\doibase 10.1103/PhysRevD.97.054024} {\bibfield  {journal} {\bibinfo
  {journal} {Phys. Rev. D}\ }\textbf {\bibinfo {volume} {97}},\ \bibinfo
  {pages} {054024} (\bibinfo {year} {2018}{\natexlab{a}})},\ \Eprint
  {http://arxiv.org/abs/1708.05706} {arXiv:1708.05706 [hep-ph]} \BibitemShut
  {NoStop}%
\bibitem [{\citenamefont {Banerjee}\ \emph
  {et~al.}(2018{\natexlab{b}})\citenamefont {Banerjee}, \citenamefont {Das},
  \citenamefont {Dhani},\ and\ \citenamefont {Ravindran}}]{Banerjee:2018vvb}%
  \BibitemOpen
  \bibfield  {author} {\bibinfo {author} {\bibfnamefont {P.}~\bibnamefont
  {Banerjee}}, \bibinfo {author} {\bibfnamefont {G.}~\bibnamefont {Das}},
  \bibinfo {author} {\bibfnamefont {P.~K.}\ \bibnamefont {Dhani}}, \ and\
  \bibinfo {author} {\bibfnamefont {V.}~\bibnamefont {Ravindran}},\ }\href
  {\doibase 10.1103/PhysRevD.98.054018} {\bibfield  {journal} {\bibinfo
  {journal} {Phys. Rev. D}\ }\textbf {\bibinfo {volume} {98}},\ \bibinfo
  {pages} {054018} (\bibinfo {year} {2018}{\natexlab{b}})},\ \Eprint
  {http://arxiv.org/abs/1805.01186} {arXiv:1805.01186 [hep-ph]} \BibitemShut
  {NoStop}%
\bibitem [{\citenamefont {Moch}\ \emph
  {et~al.}(2005{\natexlab{a}})\citenamefont {Moch}, \citenamefont
  {Vermaseren},\ and\ \citenamefont {Vogt}}]{Moch:2005ba}%
  \BibitemOpen
  \bibfield  {author} {\bibinfo {author} {\bibfnamefont {S.}~\bibnamefont
  {Moch}}, \bibinfo {author} {\bibfnamefont {J.~A.~M.}\ \bibnamefont
  {Vermaseren}}, \ and\ \bibinfo {author} {\bibfnamefont {A.}~\bibnamefont
  {Vogt}},\ }\href {\doibase 10.1016/j.nuclphysb.2005.08.005} {\bibfield
  {journal} {\bibinfo  {journal} {Nucl. Phys. B}\ }\textbf {\bibinfo {volume}
  {726}},\ \bibinfo {pages} {317} (\bibinfo {year} {2005}{\natexlab{a}})},\
  \Eprint {http://arxiv.org/abs/hep-ph/0506288} {arXiv:hep-ph/0506288}
  \BibitemShut {NoStop}%
\bibitem [{\citenamefont {Corcella}\ and\ \citenamefont
  {Magnea}(2005)}]{Corcella:2005us}%
  \BibitemOpen
  \bibfield  {author} {\bibinfo {author} {\bibfnamefont {G.}~\bibnamefont
  {Corcella}}\ and\ \bibinfo {author} {\bibfnamefont {L.}~\bibnamefont
  {Magnea}},\ }\href {\doibase 10.1103/PhysRevD.72.074017} {\bibfield
  {journal} {\bibinfo  {journal} {Phys. Rev. D}\ }\textbf {\bibinfo {volume}
  {72}},\ \bibinfo {pages} {074017} (\bibinfo {year} {2005})},\ \Eprint
  {http://arxiv.org/abs/hep-ph/0506278} {arXiv:hep-ph/0506278} \BibitemShut
  {NoStop}%
\bibitem [{\citenamefont {Aicher}\ \emph {et~al.}(2010)\citenamefont {Aicher},
  \citenamefont {Schafer},\ and\ \citenamefont {Vogelsang}}]{Aicher:2010cb}%
  \BibitemOpen
  \bibfield  {author} {\bibinfo {author} {\bibfnamefont {M.}~\bibnamefont
  {Aicher}}, \bibinfo {author} {\bibfnamefont {A.}~\bibnamefont {Schafer}}, \
  and\ \bibinfo {author} {\bibfnamefont {W.}~\bibnamefont {Vogelsang}},\ }\href
  {\doibase 10.1103/PhysRevLett.105.252003} {\bibfield  {journal} {\bibinfo
  {journal} {Phys. Rev. Lett.}\ }\textbf {\bibinfo {volume} {105}},\ \bibinfo
  {pages} {252003} (\bibinfo {year} {2010})},\ \Eprint
  {http://arxiv.org/abs/1009.2481} {arXiv:1009.2481 [hep-ph]} \BibitemShut
  {NoStop}%
\bibitem [{\citenamefont {Bonvini}\ \emph {et~al.}(2015)\citenamefont
  {Bonvini}, \citenamefont {Marzani}, \citenamefont {Rojo}, \citenamefont
  {Rottoli}, \citenamefont {Ubiali}, \citenamefont {Ball}, \citenamefont
  {Bertone}, \citenamefont {Carrazza},\ and\ \citenamefont
  {Hartland}}]{Bonvini:2015ira}%
  \BibitemOpen
  \bibfield  {author} {\bibinfo {author} {\bibfnamefont {M.}~\bibnamefont
  {Bonvini}}, \bibinfo {author} {\bibfnamefont {S.}~\bibnamefont {Marzani}},
  \bibinfo {author} {\bibfnamefont {J.}~\bibnamefont {Rojo}}, \bibinfo {author}
  {\bibfnamefont {L.}~\bibnamefont {Rottoli}}, \bibinfo {author} {\bibfnamefont
  {M.}~\bibnamefont {Ubiali}}, \bibinfo {author} {\bibfnamefont {R.~D.}\
  \bibnamefont {Ball}}, \bibinfo {author} {\bibfnamefont {V.}~\bibnamefont
  {Bertone}}, \bibinfo {author} {\bibfnamefont {S.}~\bibnamefont {Carrazza}}, \
  and\ \bibinfo {author} {\bibfnamefont {N.~P.}\ \bibnamefont {Hartland}},\
  }\href {\doibase 10.1007/JHEP09(2015)191} {\bibfield  {journal} {\bibinfo
  {journal} {JHEP}\ }\textbf {\bibinfo {volume} {09}},\ \bibinfo {pages} {191}
  (\bibinfo {year} {2015})},\ \Eprint {http://arxiv.org/abs/1507.01006}
  {arXiv:1507.01006 [hep-ph]} \BibitemShut {NoStop}%
\bibitem [{\citenamefont {Westmark}\ and\ \citenamefont
  {Owens}(2017)}]{Westmark:2017uig}%
  \BibitemOpen
  \bibfield  {author} {\bibinfo {author} {\bibfnamefont {D.}~\bibnamefont
  {Westmark}}\ and\ \bibinfo {author} {\bibfnamefont {J.~F.}\ \bibnamefont
  {Owens}},\ }\href {\doibase 10.1103/PhysRevD.95.056024} {\bibfield  {journal}
  {\bibinfo  {journal} {Phys. Rev. D}\ }\textbf {\bibinfo {volume} {95}},\
  \bibinfo {pages} {056024} (\bibinfo {year} {2017})},\ \Eprint
  {http://arxiv.org/abs/1701.06716} {arXiv:1701.06716 [hep-ph]} \BibitemShut
  {NoStop}%
\bibitem [{\citenamefont {Barry}\ \emph {et~al.}(2021)\citenamefont {Barry},
  \citenamefont {Ji}, \citenamefont {Sato},\ and\ \citenamefont
  {Melnitchouk}}]{Barry:2021osv}%
  \BibitemOpen
  \bibfield  {author} {\bibinfo {author} {\bibfnamefont {P.~C.}\ \bibnamefont
  {Barry}}, \bibinfo {author} {\bibfnamefont {C.-R.}\ \bibnamefont {Ji}},
  \bibinfo {author} {\bibfnamefont {N.}~\bibnamefont {Sato}}, \ and\ \bibinfo
  {author} {\bibfnamefont {W.}~\bibnamefont {Melnitchouk}} (\bibinfo
  {collaboration} {Jefferson Lab Angular Momentum (JAM)}),\ }\href {\doibase
  10.1103/PhysRevLett.127.232001} {\bibfield  {journal} {\bibinfo  {journal}
  {Phys. Rev. Lett.}\ }\textbf {\bibinfo {volume} {127}},\ \bibinfo {pages}
  {232001} (\bibinfo {year} {2021})},\ \Eprint
  {http://arxiv.org/abs/2108.05822} {arXiv:2108.05822 [hep-ph]} \BibitemShut
  {NoStop}%
\bibitem [{\citenamefont {Ji}(2013)}]{Ji:2013dva}%
  \BibitemOpen
  \bibfield  {author} {\bibinfo {author} {\bibfnamefont {X.}~\bibnamefont
  {Ji}},\ }\href {\doibase 10.1103/PhysRevLett.110.262002} {\bibfield
  {journal} {\bibinfo  {journal} {Phys. Rev. Lett.}\ }\textbf {\bibinfo
  {volume} {110}},\ \bibinfo {pages} {262002} (\bibinfo {year} {2013})},\
  \Eprint {http://arxiv.org/abs/1305.1539} {arXiv:1305.1539 [hep-ph]}
  \BibitemShut {NoStop}%
\bibitem [{\citenamefont {Ji}(2014)}]{Ji:2014gla}%
  \BibitemOpen
  \bibfield  {author} {\bibinfo {author} {\bibfnamefont {X.}~\bibnamefont
  {Ji}},\ }\href {\doibase 10.1007/s11433-014-5492-3} {\bibfield  {journal}
  {\bibinfo  {journal} {Sci. China Phys. Mech. Astron.}\ }\textbf {\bibinfo
  {volume} {57}},\ \bibinfo {pages} {1407} (\bibinfo {year} {2014})},\ \Eprint
  {http://arxiv.org/abs/1404.6680} {arXiv:1404.6680 [hep-ph]} \BibitemShut
  {NoStop}%
\bibitem [{\citenamefont {Ji}(2024)}]{Ji:2024oka}%
  \BibitemOpen
  \bibfield  {author} {\bibinfo {author} {\bibfnamefont {X.}~\bibnamefont
  {Ji}},\ }\href@noop {} {\  (\bibinfo {year} {2024})},\ \Eprint
  {http://arxiv.org/abs/2408.03378} {arXiv:2408.03378 [hep-ph]} \BibitemShut
  {NoStop}%
\bibitem [{\citenamefont {Xiong}\ \emph {et~al.}(2014)\citenamefont {Xiong},
  \citenamefont {Ji}, \citenamefont {Zhang},\ and\ \citenamefont
  {Zhao}}]{Xiong:2013bka}%
  \BibitemOpen
  \bibfield  {author} {\bibinfo {author} {\bibfnamefont {X.}~\bibnamefont
  {Xiong}}, \bibinfo {author} {\bibfnamefont {X.}~\bibnamefont {Ji}}, \bibinfo
  {author} {\bibfnamefont {J.-H.}\ \bibnamefont {Zhang}}, \ and\ \bibinfo
  {author} {\bibfnamefont {Y.}~\bibnamefont {Zhao}},\ }\href {\doibase
  10.1103/PhysRevD.90.014051} {\bibfield  {journal} {\bibinfo  {journal} {Phys.
  Rev. D}\ }\textbf {\bibinfo {volume} {90}},\ \bibinfo {pages} {014051}
  (\bibinfo {year} {2014})},\ \Eprint {http://arxiv.org/abs/1310.7471}
  {arXiv:1310.7471 [hep-ph]} \BibitemShut {NoStop}%
\bibitem [{\citenamefont {Lin}\ \emph {et~al.}(2015)\citenamefont {Lin},
  \citenamefont {Chen}, \citenamefont {Cohen},\ and\ \citenamefont
  {Ji}}]{Lin:2014zya}%
  \BibitemOpen
  \bibfield  {author} {\bibinfo {author} {\bibfnamefont {H.-W.}\ \bibnamefont
  {Lin}}, \bibinfo {author} {\bibfnamefont {J.-W.}\ \bibnamefont {Chen}},
  \bibinfo {author} {\bibfnamefont {S.~D.}\ \bibnamefont {Cohen}}, \ and\
  \bibinfo {author} {\bibfnamefont {X.}~\bibnamefont {Ji}},\ }\href {\doibase
  10.1103/PhysRevD.91.054510} {\bibfield  {journal} {\bibinfo  {journal} {Phys.
  Rev. D}\ }\textbf {\bibinfo {volume} {91}},\ \bibinfo {pages} {054510}
  (\bibinfo {year} {2015})},\ \Eprint {http://arxiv.org/abs/1402.1462}
  {arXiv:1402.1462 [hep-ph]} \BibitemShut {NoStop}%
\bibitem [{\citenamefont {Alexandrou}\ \emph {et~al.}(2015)\citenamefont
  {Alexandrou}, \citenamefont {Cichy}, \citenamefont {Drach}, \citenamefont
  {Garcia-Ramos}, \citenamefont {Hadjiyiannakou}, \citenamefont {Jansen},
  \citenamefont {Steffens},\ and\ \citenamefont {Wiese}}]{Alexandrou:2015rja}%
  \BibitemOpen
  \bibfield  {author} {\bibinfo {author} {\bibfnamefont {C.}~\bibnamefont
  {Alexandrou}}, \bibinfo {author} {\bibfnamefont {K.}~\bibnamefont {Cichy}},
  \bibinfo {author} {\bibfnamefont {V.}~\bibnamefont {Drach}}, \bibinfo
  {author} {\bibfnamefont {E.}~\bibnamefont {Garcia-Ramos}}, \bibinfo {author}
  {\bibfnamefont {K.}~\bibnamefont {Hadjiyiannakou}}, \bibinfo {author}
  {\bibfnamefont {K.}~\bibnamefont {Jansen}}, \bibinfo {author} {\bibfnamefont
  {F.}~\bibnamefont {Steffens}}, \ and\ \bibinfo {author} {\bibfnamefont
  {C.}~\bibnamefont {Wiese}},\ }\href {\doibase 10.1103/PhysRevD.92.014502}
  {\bibfield  {journal} {\bibinfo  {journal} {Phys. Rev. D}\ }\textbf {\bibinfo
  {volume} {92}},\ \bibinfo {pages} {014502} (\bibinfo {year} {2015})},\
  \Eprint {http://arxiv.org/abs/1504.07455} {arXiv:1504.07455 [hep-lat]}
  \BibitemShut {NoStop}%
\bibitem [{\citenamefont {Chen}\ \emph {et~al.}(2016)\citenamefont {Chen},
  \citenamefont {Cohen}, \citenamefont {Ji}, \citenamefont {Lin},\ and\
  \citenamefont {Zhang}}]{Chen:2016utp}%
  \BibitemOpen
  \bibfield  {author} {\bibinfo {author} {\bibfnamefont {J.-W.}\ \bibnamefont
  {Chen}}, \bibinfo {author} {\bibfnamefont {S.~D.}\ \bibnamefont {Cohen}},
  \bibinfo {author} {\bibfnamefont {X.}~\bibnamefont {Ji}}, \bibinfo {author}
  {\bibfnamefont {H.-W.}\ \bibnamefont {Lin}}, \ and\ \bibinfo {author}
  {\bibfnamefont {J.-H.}\ \bibnamefont {Zhang}},\ }\href {\doibase
  10.1016/j.nuclphysb.2016.07.033} {\bibfield  {journal} {\bibinfo  {journal}
  {Nucl. Phys. B}\ }\textbf {\bibinfo {volume} {911}},\ \bibinfo {pages} {246}
  (\bibinfo {year} {2016})},\ \Eprint {http://arxiv.org/abs/1603.06664}
  {arXiv:1603.06664 [hep-ph]} \BibitemShut {NoStop}%
\bibitem [{\citenamefont {Alexandrou}\ \emph {et~al.}(2017)\citenamefont
  {Alexandrou}, \citenamefont {Cichy}, \citenamefont {Constantinou},
  \citenamefont {Hadjiyiannakou}, \citenamefont {Jansen}, \citenamefont
  {Steffens},\ and\ \citenamefont {Wiese}}]{Alexandrou:2016jqi}%
  \BibitemOpen
  \bibfield  {author} {\bibinfo {author} {\bibfnamefont {C.}~\bibnamefont
  {Alexandrou}}, \bibinfo {author} {\bibfnamefont {K.}~\bibnamefont {Cichy}},
  \bibinfo {author} {\bibfnamefont {M.}~\bibnamefont {Constantinou}}, \bibinfo
  {author} {\bibfnamefont {K.}~\bibnamefont {Hadjiyiannakou}}, \bibinfo
  {author} {\bibfnamefont {K.}~\bibnamefont {Jansen}}, \bibinfo {author}
  {\bibfnamefont {F.}~\bibnamefont {Steffens}}, \ and\ \bibinfo {author}
  {\bibfnamefont {C.}~\bibnamefont {Wiese}},\ }\href {\doibase
  10.1103/PhysRevD.96.014513} {\bibfield  {journal} {\bibinfo  {journal} {Phys.
  Rev. D}\ }\textbf {\bibinfo {volume} {96}},\ \bibinfo {pages} {014513}
  (\bibinfo {year} {2017})},\ \Eprint {http://arxiv.org/abs/1610.03689}
  {arXiv:1610.03689 [hep-lat]} \BibitemShut {NoStop}%
\bibitem [{\citenamefont {Alexandrou}\ \emph
  {et~al.}(2018{\natexlab{a}})\citenamefont {Alexandrou}, \citenamefont
  {Cichy}, \citenamefont {Constantinou}, \citenamefont {Jansen}, \citenamefont
  {Scapellato},\ and\ \citenamefont {Steffens}}]{Alexandrou:2018pbm}%
  \BibitemOpen
  \bibfield  {author} {\bibinfo {author} {\bibfnamefont {C.}~\bibnamefont
  {Alexandrou}}, \bibinfo {author} {\bibfnamefont {K.}~\bibnamefont {Cichy}},
  \bibinfo {author} {\bibfnamefont {M.}~\bibnamefont {Constantinou}}, \bibinfo
  {author} {\bibfnamefont {K.}~\bibnamefont {Jansen}}, \bibinfo {author}
  {\bibfnamefont {A.}~\bibnamefont {Scapellato}}, \ and\ \bibinfo {author}
  {\bibfnamefont {F.}~\bibnamefont {Steffens}},\ }\href {\doibase
  10.1103/PhysRevLett.121.112001} {\bibfield  {journal} {\bibinfo  {journal}
  {Phys. Rev. Lett.}\ }\textbf {\bibinfo {volume} {121}},\ \bibinfo {pages}
  {112001} (\bibinfo {year} {2018}{\natexlab{a}})},\ \Eprint
  {http://arxiv.org/abs/1803.02685} {arXiv:1803.02685 [hep-lat]} \BibitemShut
  {NoStop}%
\bibitem [{\citenamefont {Chen}\ \emph {et~al.}(2018)\citenamefont {Chen},
  \citenamefont {Jin}, \citenamefont {Lin}, \citenamefont {Liu}, \citenamefont
  {Yang}, \citenamefont {Zhang},\ and\ \citenamefont {Zhao}}]{Chen:2018xof}%
  \BibitemOpen
  \bibfield  {author} {\bibinfo {author} {\bibfnamefont {J.-W.}\ \bibnamefont
  {Chen}}, \bibinfo {author} {\bibfnamefont {L.}~\bibnamefont {Jin}}, \bibinfo
  {author} {\bibfnamefont {H.-W.}\ \bibnamefont {Lin}}, \bibinfo {author}
  {\bibfnamefont {Y.-S.}\ \bibnamefont {Liu}}, \bibinfo {author} {\bibfnamefont
  {Y.-B.}\ \bibnamefont {Yang}}, \bibinfo {author} {\bibfnamefont {J.-H.}\
  \bibnamefont {Zhang}}, \ and\ \bibinfo {author} {\bibfnamefont
  {Y.}~\bibnamefont {Zhao}},\ }\href@noop {} {\  (\bibinfo {year} {2018})},\
  \Eprint {http://arxiv.org/abs/1803.04393} {arXiv:1803.04393 [hep-lat]}
  \BibitemShut {NoStop}%
\bibitem [{\citenamefont {Lin}\ \emph {et~al.}(2018)\citenamefont {Lin},
  \citenamefont {Chen}, \citenamefont {Ji}, \citenamefont {Jin}, \citenamefont
  {Li}, \citenamefont {Liu}, \citenamefont {Yang}, \citenamefont {Zhang},\ and\
  \citenamefont {Zhao}}]{Lin:2018pvv}%
  \BibitemOpen
  \bibfield  {author} {\bibinfo {author} {\bibfnamefont {H.-W.}\ \bibnamefont
  {Lin}}, \bibinfo {author} {\bibfnamefont {J.-W.}\ \bibnamefont {Chen}},
  \bibinfo {author} {\bibfnamefont {X.}~\bibnamefont {Ji}}, \bibinfo {author}
  {\bibfnamefont {L.}~\bibnamefont {Jin}}, \bibinfo {author} {\bibfnamefont
  {R.}~\bibnamefont {Li}}, \bibinfo {author} {\bibfnamefont {Y.-S.}\
  \bibnamefont {Liu}}, \bibinfo {author} {\bibfnamefont {Y.-B.}\ \bibnamefont
  {Yang}}, \bibinfo {author} {\bibfnamefont {J.-H.}\ \bibnamefont {Zhang}}, \
  and\ \bibinfo {author} {\bibfnamefont {Y.}~\bibnamefont {Zhao}},\ }\href
  {\doibase 10.1103/PhysRevLett.121.242003} {\bibfield  {journal} {\bibinfo
  {journal} {Phys. Rev. Lett.}\ }\textbf {\bibinfo {volume} {121}},\ \bibinfo
  {pages} {242003} (\bibinfo {year} {2018})},\ \Eprint
  {http://arxiv.org/abs/1807.07431} {arXiv:1807.07431 [hep-lat]} \BibitemShut
  {NoStop}%
\bibitem [{\citenamefont {Liu}\ \emph {et~al.}(2020)\citenamefont {Liu} \emph
  {et~al.}}]{LatticeParton:2018gjr}%
  \BibitemOpen
  \bibfield  {author} {\bibinfo {author} {\bibfnamefont {Y.-S.}\ \bibnamefont
  {Liu}} \emph {et~al.} (\bibinfo {collaboration} {Lattice Parton}),\ }\href
  {\doibase 10.1103/PhysRevD.101.034020} {\bibfield  {journal} {\bibinfo
  {journal} {Phys. Rev. D}\ }\textbf {\bibinfo {volume} {101}},\ \bibinfo
  {pages} {034020} (\bibinfo {year} {2020})},\ \Eprint
  {http://arxiv.org/abs/1807.06566} {arXiv:1807.06566 [hep-lat]} \BibitemShut
  {NoStop}%
\bibitem [{\citenamefont {Alexandrou}\ \emph
  {et~al.}(2018{\natexlab{b}})\citenamefont {Alexandrou}, \citenamefont
  {Cichy}, \citenamefont {Constantinou}, \citenamefont {Jansen}, \citenamefont
  {Scapellato},\ and\ \citenamefont {Steffens}}]{Alexandrou:2018eet}%
  \BibitemOpen
  \bibfield  {author} {\bibinfo {author} {\bibfnamefont {C.}~\bibnamefont
  {Alexandrou}}, \bibinfo {author} {\bibfnamefont {K.}~\bibnamefont {Cichy}},
  \bibinfo {author} {\bibfnamefont {M.}~\bibnamefont {Constantinou}}, \bibinfo
  {author} {\bibfnamefont {K.}~\bibnamefont {Jansen}}, \bibinfo {author}
  {\bibfnamefont {A.}~\bibnamefont {Scapellato}}, \ and\ \bibinfo {author}
  {\bibfnamefont {F.}~\bibnamefont {Steffens}},\ }\href {\doibase
  10.1103/PhysRevD.98.091503} {\bibfield  {journal} {\bibinfo  {journal} {Phys.
  Rev. D}\ }\textbf {\bibinfo {volume} {98}},\ \bibinfo {pages} {091503}
  (\bibinfo {year} {2018}{\natexlab{b}})},\ \Eprint
  {http://arxiv.org/abs/1807.00232} {arXiv:1807.00232 [hep-lat]} \BibitemShut
  {NoStop}%
\bibitem [{\citenamefont {Liu}\ \emph {et~al.}(2018)\citenamefont {Liu},
  \citenamefont {Chen}, \citenamefont {Jin}, \citenamefont {Li}, \citenamefont
  {Lin}, \citenamefont {Yang}, \citenamefont {Zhang},\ and\ \citenamefont
  {Zhao}}]{Liu:2018hxv}%
  \BibitemOpen
  \bibfield  {author} {\bibinfo {author} {\bibfnamefont {Y.-S.}\ \bibnamefont
  {Liu}}, \bibinfo {author} {\bibfnamefont {J.-W.}\ \bibnamefont {Chen}},
  \bibinfo {author} {\bibfnamefont {L.}~\bibnamefont {Jin}}, \bibinfo {author}
  {\bibfnamefont {R.}~\bibnamefont {Li}}, \bibinfo {author} {\bibfnamefont
  {H.-W.}\ \bibnamefont {Lin}}, \bibinfo {author} {\bibfnamefont {Y.-B.}\
  \bibnamefont {Yang}}, \bibinfo {author} {\bibfnamefont {J.-H.}\ \bibnamefont
  {Zhang}}, \ and\ \bibinfo {author} {\bibfnamefont {Y.}~\bibnamefont {Zhao}},\
  }\href@noop {} {\  (\bibinfo {year} {2018})},\ \Eprint
  {http://arxiv.org/abs/1810.05043} {arXiv:1810.05043 [hep-lat]} \BibitemShut
  {NoStop}%
\bibitem [{\citenamefont {Zhang}\ \emph
  {et~al.}(2019{\natexlab{a}})\citenamefont {Zhang}, \citenamefont {Chen},
  \citenamefont {Jin}, \citenamefont {Lin}, \citenamefont {Sch\"afer},\ and\
  \citenamefont {Zhao}}]{Chen:2018fwa}%
  \BibitemOpen
  \bibfield  {author} {\bibinfo {author} {\bibfnamefont {J.-H.}\ \bibnamefont
  {Zhang}}, \bibinfo {author} {\bibfnamefont {J.-W.}\ \bibnamefont {Chen}},
  \bibinfo {author} {\bibfnamefont {L.}~\bibnamefont {Jin}}, \bibinfo {author}
  {\bibfnamefont {H.-W.}\ \bibnamefont {Lin}}, \bibinfo {author} {\bibfnamefont
  {A.}~\bibnamefont {Sch\"afer}}, \ and\ \bibinfo {author} {\bibfnamefont
  {Y.}~\bibnamefont {Zhao}},\ }\href {\doibase 10.1103/PhysRevD.100.034505}
  {\bibfield  {journal} {\bibinfo  {journal} {Phys. Rev. D}\ }\textbf {\bibinfo
  {volume} {100}},\ \bibinfo {pages} {034505} (\bibinfo {year}
  {2019}{\natexlab{a}})},\ \Eprint {http://arxiv.org/abs/1804.01483}
  {arXiv:1804.01483 [hep-lat]} \BibitemShut {NoStop}%
\bibitem [{\citenamefont {Izubuchi}\ \emph {et~al.}(2018)\citenamefont
  {Izubuchi}, \citenamefont {Ji}, \citenamefont {Jin}, \citenamefont
  {Stewart},\ and\ \citenamefont {Zhao}}]{Izubuchi:2018srq}%
  \BibitemOpen
  \bibfield  {author} {\bibinfo {author} {\bibfnamefont {T.}~\bibnamefont
  {Izubuchi}}, \bibinfo {author} {\bibfnamefont {X.}~\bibnamefont {Ji}},
  \bibinfo {author} {\bibfnamefont {L.}~\bibnamefont {Jin}}, \bibinfo {author}
  {\bibfnamefont {I.~W.}\ \bibnamefont {Stewart}}, \ and\ \bibinfo {author}
  {\bibfnamefont {Y.}~\bibnamefont {Zhao}},\ }\href {\doibase
  10.1103/PhysRevD.98.056004} {\bibfield  {journal} {\bibinfo  {journal} {Phys.
  Rev. D}\ }\textbf {\bibinfo {volume} {98}},\ \bibinfo {pages} {056004}
  (\bibinfo {year} {2018})},\ \Eprint {http://arxiv.org/abs/1801.03917}
  {arXiv:1801.03917 [hep-ph]} \BibitemShut {NoStop}%
\bibitem [{\citenamefont {Izubuchi}\ \emph {et~al.}(2019)\citenamefont
  {Izubuchi}, \citenamefont {Jin}, \citenamefont {Kallidonis}, \citenamefont
  {Karthik}, \citenamefont {Mukherjee}, \citenamefont {Petreczky},
  \citenamefont {Shugert},\ and\ \citenamefont {Syritsyn}}]{Izubuchi:2019lyk}%
  \BibitemOpen
  \bibfield  {author} {\bibinfo {author} {\bibfnamefont {T.}~\bibnamefont
  {Izubuchi}}, \bibinfo {author} {\bibfnamefont {L.}~\bibnamefont {Jin}},
  \bibinfo {author} {\bibfnamefont {C.}~\bibnamefont {Kallidonis}}, \bibinfo
  {author} {\bibfnamefont {N.}~\bibnamefont {Karthik}}, \bibinfo {author}
  {\bibfnamefont {S.}~\bibnamefont {Mukherjee}}, \bibinfo {author}
  {\bibfnamefont {P.}~\bibnamefont {Petreczky}}, \bibinfo {author}
  {\bibfnamefont {C.}~\bibnamefont {Shugert}}, \ and\ \bibinfo {author}
  {\bibfnamefont {S.}~\bibnamefont {Syritsyn}},\ }\href {\doibase
  10.1103/PhysRevD.100.034516} {\bibfield  {journal} {\bibinfo  {journal}
  {Phys. Rev. D}\ }\textbf {\bibinfo {volume} {100}},\ \bibinfo {pages}
  {034516} (\bibinfo {year} {2019})},\ \Eprint
  {http://arxiv.org/abs/1905.06349} {arXiv:1905.06349 [hep-lat]} \BibitemShut
  {NoStop}%
\bibitem [{\citenamefont {Shugert}\ \emph {et~al.}(2020)\citenamefont
  {Shugert}, \citenamefont {Gao}, \citenamefont {Izubichi}, \citenamefont
  {Jin}, \citenamefont {Kallidonis}, \citenamefont {Karthik}, \citenamefont
  {Mukherjee}, \citenamefont {Petreczky}, \citenamefont {Syritsyn},\ and\
  \citenamefont {Zhao}}]{Shugert:2020tgq}%
  \BibitemOpen
  \bibfield  {author} {\bibinfo {author} {\bibfnamefont {C.}~\bibnamefont
  {Shugert}}, \bibinfo {author} {\bibfnamefont {X.}~\bibnamefont {Gao}},
  \bibinfo {author} {\bibfnamefont {T.}~\bibnamefont {Izubichi}}, \bibinfo
  {author} {\bibfnamefont {L.}~\bibnamefont {Jin}}, \bibinfo {author}
  {\bibfnamefont {C.}~\bibnamefont {Kallidonis}}, \bibinfo {author}
  {\bibfnamefont {N.}~\bibnamefont {Karthik}}, \bibinfo {author} {\bibfnamefont
  {S.}~\bibnamefont {Mukherjee}}, \bibinfo {author} {\bibfnamefont
  {P.}~\bibnamefont {Petreczky}}, \bibinfo {author} {\bibfnamefont
  {S.}~\bibnamefont {Syritsyn}}, \ and\ \bibinfo {author} {\bibfnamefont
  {Y.}~\bibnamefont {Zhao}},\ }in\ \href@noop {} {\emph {\bibinfo {booktitle}
  {{37th International Symposium on Lattice Field Theory}}}}\ (\bibinfo {year}
  {2020})\ \Eprint {http://arxiv.org/abs/2001.11650} {arXiv:2001.11650
  [hep-lat]} \BibitemShut {NoStop}%
\bibitem [{\citenamefont {Chai}\ \emph {et~al.}(2020)\citenamefont {Chai} \emph
  {et~al.}}]{Chai:2020nxw}%
  \BibitemOpen
  \bibfield  {author} {\bibinfo {author} {\bibfnamefont {Y.}~\bibnamefont
  {Chai}} \emph {et~al.},\ }\href {\doibase 10.1103/PhysRevD.102.014508}
  {\bibfield  {journal} {\bibinfo  {journal} {Phys. Rev. D}\ }\textbf {\bibinfo
  {volume} {102}},\ \bibinfo {pages} {014508} (\bibinfo {year} {2020})},\
  \Eprint {http://arxiv.org/abs/2002.12044} {arXiv:2002.12044 [hep-lat]}
  \BibitemShut {NoStop}%
\bibitem [{\citenamefont {Lin}\ \emph {et~al.}(2021)\citenamefont {Lin},
  \citenamefont {Chen}, \citenamefont {Fan}, \citenamefont {Zhang},\ and\
  \citenamefont {Zhang}}]{Lin:2020ssv}%
  \BibitemOpen
  \bibfield  {author} {\bibinfo {author} {\bibfnamefont {H.-W.}\ \bibnamefont
  {Lin}}, \bibinfo {author} {\bibfnamefont {J.-W.}\ \bibnamefont {Chen}},
  \bibinfo {author} {\bibfnamefont {Z.}~\bibnamefont {Fan}}, \bibinfo {author}
  {\bibfnamefont {J.-H.}\ \bibnamefont {Zhang}}, \ and\ \bibinfo {author}
  {\bibfnamefont {R.}~\bibnamefont {Zhang}},\ }\href {\doibase
  10.1103/PhysRevD.103.014516} {\bibfield  {journal} {\bibinfo  {journal}
  {Phys. Rev. D}\ }\textbf {\bibinfo {volume} {103}},\ \bibinfo {pages}
  {014516} (\bibinfo {year} {2021})},\ \Eprint
  {http://arxiv.org/abs/2003.14128} {arXiv:2003.14128 [hep-lat]} \BibitemShut
  {NoStop}%
\bibitem [{\citenamefont {Fan}\ \emph {et~al.}(2020)\citenamefont {Fan},
  \citenamefont {Gao}, \citenamefont {Li}, \citenamefont {Lin}, \citenamefont
  {Karthik}, \citenamefont {Mukherjee}, \citenamefont {Petreczky},
  \citenamefont {Syritsyn}, \citenamefont {Yang},\ and\ \citenamefont
  {Zhang}}]{Fan:2020nzz}%
  \BibitemOpen
  \bibfield  {author} {\bibinfo {author} {\bibfnamefont {Z.}~\bibnamefont
  {Fan}}, \bibinfo {author} {\bibfnamefont {X.}~\bibnamefont {Gao}}, \bibinfo
  {author} {\bibfnamefont {R.}~\bibnamefont {Li}}, \bibinfo {author}
  {\bibfnamefont {H.-W.}\ \bibnamefont {Lin}}, \bibinfo {author} {\bibfnamefont
  {N.}~\bibnamefont {Karthik}}, \bibinfo {author} {\bibfnamefont
  {S.}~\bibnamefont {Mukherjee}}, \bibinfo {author} {\bibfnamefont
  {P.}~\bibnamefont {Petreczky}}, \bibinfo {author} {\bibfnamefont
  {S.}~\bibnamefont {Syritsyn}}, \bibinfo {author} {\bibfnamefont {Y.-B.}\
  \bibnamefont {Yang}}, \ and\ \bibinfo {author} {\bibfnamefont
  {R.}~\bibnamefont {Zhang}},\ }\href {\doibase 10.1103/PhysRevD.102.074504}
  {\bibfield  {journal} {\bibinfo  {journal} {Phys. Rev. D}\ }\textbf {\bibinfo
  {volume} {102}},\ \bibinfo {pages} {074504} (\bibinfo {year} {2020})},\
  \Eprint {http://arxiv.org/abs/2005.12015} {arXiv:2005.12015 [hep-lat]}
  \BibitemShut {NoStop}%
\bibitem [{\citenamefont {Gao}\ \emph {et~al.}(2021)\citenamefont {Gao},
  \citenamefont {Lee}, \citenamefont {Mukherjee}, \citenamefont {Shugert},\
  and\ \citenamefont {Zhao}}]{Gao:2021hxl}%
  \BibitemOpen
  \bibfield  {author} {\bibinfo {author} {\bibfnamefont {X.}~\bibnamefont
  {Gao}}, \bibinfo {author} {\bibfnamefont {K.}~\bibnamefont {Lee}}, \bibinfo
  {author} {\bibfnamefont {S.}~\bibnamefont {Mukherjee}}, \bibinfo {author}
  {\bibfnamefont {C.}~\bibnamefont {Shugert}}, \ and\ \bibinfo {author}
  {\bibfnamefont {Y.}~\bibnamefont {Zhao}},\ }\href {\doibase
  10.1103/PhysRevD.103.094504} {\bibfield  {journal} {\bibinfo  {journal}
  {Phys. Rev. D}\ }\textbf {\bibinfo {volume} {103}},\ \bibinfo {pages}
  {094504} (\bibinfo {year} {2021})},\ \Eprint
  {http://arxiv.org/abs/2102.01101} {arXiv:2102.01101 [hep-ph]} \BibitemShut
  {NoStop}%
\bibitem [{\citenamefont {Gao}\ \emph {et~al.}(2022{\natexlab{a}})\citenamefont
  {Gao}, \citenamefont {Hanlon}, \citenamefont {Mukherjee}, \citenamefont
  {Petreczky}, \citenamefont {Scior}, \citenamefont {Syritsyn},\ and\
  \citenamefont {Zhao}}]{Gao:2021dbh}%
  \BibitemOpen
  \bibfield  {author} {\bibinfo {author} {\bibfnamefont {X.}~\bibnamefont
  {Gao}}, \bibinfo {author} {\bibfnamefont {A.~D.}\ \bibnamefont {Hanlon}},
  \bibinfo {author} {\bibfnamefont {S.}~\bibnamefont {Mukherjee}}, \bibinfo
  {author} {\bibfnamefont {P.}~\bibnamefont {Petreczky}}, \bibinfo {author}
  {\bibfnamefont {P.}~\bibnamefont {Scior}}, \bibinfo {author} {\bibfnamefont
  {S.}~\bibnamefont {Syritsyn}}, \ and\ \bibinfo {author} {\bibfnamefont
  {Y.}~\bibnamefont {Zhao}},\ }\href {\doibase 10.1103/PhysRevLett.128.142003}
  {\bibfield  {journal} {\bibinfo  {journal} {Phys. Rev. Lett.}\ }\textbf
  {\bibinfo {volume} {128}},\ \bibinfo {pages} {142003} (\bibinfo {year}
  {2022}{\natexlab{a}})},\ \Eprint {http://arxiv.org/abs/2112.02208}
  {arXiv:2112.02208 [hep-lat]} \BibitemShut {NoStop}%
\bibitem [{\citenamefont {Gao}\ \emph {et~al.}(2022{\natexlab{b}})\citenamefont
  {Gao}, \citenamefont {Hanlon}, \citenamefont {Karthik}, \citenamefont
  {Mukherjee}, \citenamefont {Petreczky}, \citenamefont {Scior}, \citenamefont
  {Shi}, \citenamefont {Syritsyn}, \citenamefont {Zhao},\ and\ \citenamefont
  {Zhou}}]{Gao:2022iex}%
  \BibitemOpen
  \bibfield  {author} {\bibinfo {author} {\bibfnamefont {X.}~\bibnamefont
  {Gao}}, \bibinfo {author} {\bibfnamefont {A.~D.}\ \bibnamefont {Hanlon}},
  \bibinfo {author} {\bibfnamefont {N.}~\bibnamefont {Karthik}}, \bibinfo
  {author} {\bibfnamefont {S.}~\bibnamefont {Mukherjee}}, \bibinfo {author}
  {\bibfnamefont {P.}~\bibnamefont {Petreczky}}, \bibinfo {author}
  {\bibfnamefont {P.}~\bibnamefont {Scior}}, \bibinfo {author} {\bibfnamefont
  {S.}~\bibnamefont {Shi}}, \bibinfo {author} {\bibfnamefont {S.}~\bibnamefont
  {Syritsyn}}, \bibinfo {author} {\bibfnamefont {Y.}~\bibnamefont {Zhao}}, \
  and\ \bibinfo {author} {\bibfnamefont {K.}~\bibnamefont {Zhou}},\ }\href
  {\doibase 10.1103/PhysRevD.106.114510} {\bibfield  {journal} {\bibinfo
  {journal} {Phys. Rev. D}\ }\textbf {\bibinfo {volume} {106}},\ \bibinfo
  {pages} {114510} (\bibinfo {year} {2022}{\natexlab{b}})},\ \Eprint
  {http://arxiv.org/abs/2208.02297} {arXiv:2208.02297 [hep-lat]} \BibitemShut
  {NoStop}%
\bibitem [{\citenamefont {Su}\ \emph {et~al.}(2023)\citenamefont {Su},
  \citenamefont {Holligan}, \citenamefont {Ji}, \citenamefont {Yao},
  \citenamefont {Zhang},\ and\ \citenamefont {Zhang}}]{Su:2022fiu}%
  \BibitemOpen
  \bibfield  {author} {\bibinfo {author} {\bibfnamefont {Y.}~\bibnamefont
  {Su}}, \bibinfo {author} {\bibfnamefont {J.}~\bibnamefont {Holligan}},
  \bibinfo {author} {\bibfnamefont {X.}~\bibnamefont {Ji}}, \bibinfo {author}
  {\bibfnamefont {F.}~\bibnamefont {Yao}}, \bibinfo {author} {\bibfnamefont
  {J.-H.}\ \bibnamefont {Zhang}}, \ and\ \bibinfo {author} {\bibfnamefont
  {R.}~\bibnamefont {Zhang}},\ }\href {\doibase
  10.1016/j.nuclphysb.2023.116201} {\bibfield  {journal} {\bibinfo  {journal}
  {Nucl. Phys. B}\ }\textbf {\bibinfo {volume} {991}},\ \bibinfo {pages}
  {116201} (\bibinfo {year} {2023})},\ \Eprint
  {http://arxiv.org/abs/2209.01236} {arXiv:2209.01236 [hep-ph]} \BibitemShut
  {NoStop}%
\bibitem [{\citenamefont {Yao}\ \emph {et~al.}(2022)\citenamefont {Yao} \emph
  {et~al.}}]{LatticeParton:2022xsd}%
  \BibitemOpen
  \bibfield  {author} {\bibinfo {author} {\bibfnamefont {F.}~\bibnamefont
  {Yao}} \emph {et~al.} (\bibinfo {collaboration} {Lattice Parton}),\
  }\href@noop {} {\  (\bibinfo {year} {2022})},\ \Eprint
  {http://arxiv.org/abs/2208.08008} {arXiv:2208.08008 [hep-lat]} \BibitemShut
  {NoStop}%
\bibitem [{\citenamefont {Gao}\ \emph {et~al.}(2023{\natexlab{a}})\citenamefont
  {Gao}, \citenamefont {Hanlon}, \citenamefont {Holligan}, \citenamefont
  {Karthik}, \citenamefont {Mukherjee}, \citenamefont {Petreczky},
  \citenamefont {Syritsyn},\ and\ \citenamefont {Zhao}}]{Gao:2022uhg}%
  \BibitemOpen
  \bibfield  {author} {\bibinfo {author} {\bibfnamefont {X.}~\bibnamefont
  {Gao}}, \bibinfo {author} {\bibfnamefont {A.~D.}\ \bibnamefont {Hanlon}},
  \bibinfo {author} {\bibfnamefont {J.}~\bibnamefont {Holligan}}, \bibinfo
  {author} {\bibfnamefont {N.}~\bibnamefont {Karthik}}, \bibinfo {author}
  {\bibfnamefont {S.}~\bibnamefont {Mukherjee}}, \bibinfo {author}
  {\bibfnamefont {P.}~\bibnamefont {Petreczky}}, \bibinfo {author}
  {\bibfnamefont {S.}~\bibnamefont {Syritsyn}}, \ and\ \bibinfo {author}
  {\bibfnamefont {Y.}~\bibnamefont {Zhao}},\ }\href {\doibase
  10.1103/PhysRevD.107.074509} {\bibfield  {journal} {\bibinfo  {journal}
  {Phys. Rev. D}\ }\textbf {\bibinfo {volume} {107}},\ \bibinfo {pages}
  {074509} (\bibinfo {year} {2023}{\natexlab{a}})},\ \Eprint
  {http://arxiv.org/abs/2212.12569} {arXiv:2212.12569 [hep-lat]} \BibitemShut
  {NoStop}%
\bibitem [{\citenamefont {Chou}\ and\ \citenamefont
  {Chen}(2022)}]{Chou:2022drv}%
  \BibitemOpen
  \bibfield  {author} {\bibinfo {author} {\bibfnamefont {C.-Y.}\ \bibnamefont
  {Chou}}\ and\ \bibinfo {author} {\bibfnamefont {J.-W.}\ \bibnamefont
  {Chen}},\ }\href {\doibase 10.1103/PhysRevD.106.014507} {\bibfield  {journal}
  {\bibinfo  {journal} {Phys. Rev. D}\ }\textbf {\bibinfo {volume} {106}},\
  \bibinfo {pages} {014507} (\bibinfo {year} {2022})},\ \Eprint
  {http://arxiv.org/abs/2204.08343} {arXiv:2204.08343 [hep-lat]} \BibitemShut
  {NoStop}%
\bibitem [{\citenamefont {Gao}\ \emph {et~al.}(2023{\natexlab{b}})\citenamefont
  {Gao}, \citenamefont {Liu},\ and\ \citenamefont {Zhao}}]{Gao:2023lny}%
  \BibitemOpen
  \bibfield  {author} {\bibinfo {author} {\bibfnamefont {X.}~\bibnamefont
  {Gao}}, \bibinfo {author} {\bibfnamefont {W.-Y.}\ \bibnamefont {Liu}}, \ and\
  \bibinfo {author} {\bibfnamefont {Y.}~\bibnamefont {Zhao}},\ }\href@noop {}
  {\  (\bibinfo {year} {2023}{\natexlab{b}})},\ \Eprint
  {http://arxiv.org/abs/2306.14960} {arXiv:2306.14960 [hep-ph]} \BibitemShut
  {NoStop}%
\bibitem [{\citenamefont {Gao}\ \emph {et~al.}(2023{\natexlab{c}})\citenamefont
  {Gao}, \citenamefont {Hanlon}, \citenamefont {Mukherjee}, \citenamefont
  {Petreczky}, \citenamefont {Shi}, \citenamefont {Syritsyn},\ and\
  \citenamefont {Zhao}}]{Gao:2023ktu}%
  \BibitemOpen
  \bibfield  {author} {\bibinfo {author} {\bibfnamefont {X.}~\bibnamefont
  {Gao}}, \bibinfo {author} {\bibfnamefont {A.~D.}\ \bibnamefont {Hanlon}},
  \bibinfo {author} {\bibfnamefont {S.}~\bibnamefont {Mukherjee}}, \bibinfo
  {author} {\bibfnamefont {P.}~\bibnamefont {Petreczky}}, \bibinfo {author}
  {\bibfnamefont {Q.}~\bibnamefont {Shi}}, \bibinfo {author} {\bibfnamefont
  {S.}~\bibnamefont {Syritsyn}}, \ and\ \bibinfo {author} {\bibfnamefont
  {Y.}~\bibnamefont {Zhao}},\ }\href@noop {} {\  (\bibinfo {year}
  {2023}{\natexlab{c}})},\ \Eprint {http://arxiv.org/abs/2310.19047}
  {arXiv:2310.19047 [hep-lat]} \BibitemShut {NoStop}%
\bibitem [{\citenamefont {Chen}\ \emph {et~al.}(2024)\citenamefont {Chen},
  \citenamefont {Liu}, \citenamefont {Sun}, \citenamefont {Yang}, \citenamefont
  {Geng}, \citenamefont {Yao}, \citenamefont {Zhang},\ and\ \citenamefont
  {Zhang}}]{Chen:2024rgi}%
  \BibitemOpen
  \bibfield  {author} {\bibinfo {author} {\bibfnamefont {C.}~\bibnamefont
  {Chen}}, \bibinfo {author} {\bibfnamefont {L.}~\bibnamefont {Liu}}, \bibinfo
  {author} {\bibfnamefont {P.}~\bibnamefont {Sun}}, \bibinfo {author}
  {\bibfnamefont {Y.-B.}\ \bibnamefont {Yang}}, \bibinfo {author}
  {\bibfnamefont {Y.}~\bibnamefont {Geng}}, \bibinfo {author} {\bibfnamefont
  {F.}~\bibnamefont {Yao}}, \bibinfo {author} {\bibfnamefont {J.-H.}\
  \bibnamefont {Zhang}}, \ and\ \bibinfo {author} {\bibfnamefont
  {K.}~\bibnamefont {Zhang}},\ }\href@noop {} {\  (\bibinfo {year} {2024})},\
  \Eprint {http://arxiv.org/abs/2408.12819} {arXiv:2408.12819 [hep-lat]}
  \BibitemShut {NoStop}%
\bibitem [{\citenamefont {Holligan}\ and\ \citenamefont
  {Lin}(2024{\natexlab{a}})}]{Holligan:2024umc}%
  \BibitemOpen
  \bibfield  {author} {\bibinfo {author} {\bibfnamefont {J.}~\bibnamefont
  {Holligan}}\ and\ \bibinfo {author} {\bibfnamefont {H.-W.}\ \bibnamefont
  {Lin}},\ }\href {\doibase 10.1088/1361-6471/ad3162} {\bibfield  {journal}
  {\bibinfo  {journal} {J. Phys. G}\ }\textbf {\bibinfo {volume} {51}},\
  \bibinfo {pages} {065101} (\bibinfo {year} {2024}{\natexlab{a}})},\ \Eprint
  {http://arxiv.org/abs/2404.14525} {arXiv:2404.14525 [hep-lat]} \BibitemShut
  {NoStop}%
\bibitem [{\citenamefont {Holligan}\ and\ \citenamefont
  {Lin}(2024{\natexlab{b}})}]{Holligan:2024wpv}%
  \BibitemOpen
  \bibfield  {author} {\bibinfo {author} {\bibfnamefont {J.}~\bibnamefont
  {Holligan}}\ and\ \bibinfo {author} {\bibfnamefont {H.-W.}\ \bibnamefont
  {Lin}},\ }\href {\doibase 10.1016/j.physletb.2024.138731} {\bibfield
  {journal} {\bibinfo  {journal} {Phys. Lett. B}\ }\textbf {\bibinfo {volume}
  {854}},\ \bibinfo {pages} {138731} (\bibinfo {year} {2024}{\natexlab{b}})},\
  \Eprint {http://arxiv.org/abs/2405.18238} {arXiv:2405.18238 [hep-lat]}
  \BibitemShut {NoStop}%
\bibitem [{\citenamefont {Fan}\ \emph {et~al.}(2018)\citenamefont {Fan},
  \citenamefont {Yang}, \citenamefont {Anthony}, \citenamefont {Lin},\ and\
  \citenamefont {Liu}}]{Fan:2018dxu}%
  \BibitemOpen
  \bibfield  {author} {\bibinfo {author} {\bibfnamefont {Z.-Y.}\ \bibnamefont
  {Fan}}, \bibinfo {author} {\bibfnamefont {Y.-B.}\ \bibnamefont {Yang}},
  \bibinfo {author} {\bibfnamefont {A.}~\bibnamefont {Anthony}}, \bibinfo
  {author} {\bibfnamefont {H.-W.}\ \bibnamefont {Lin}}, \ and\ \bibinfo
  {author} {\bibfnamefont {K.-F.}\ \bibnamefont {Liu}},\ }\href {\doibase
  10.1103/PhysRevLett.121.242001} {\bibfield  {journal} {\bibinfo  {journal}
  {Phys. Rev. Lett.}\ }\textbf {\bibinfo {volume} {121}},\ \bibinfo {pages}
  {242001} (\bibinfo {year} {2018})},\ \Eprint
  {http://arxiv.org/abs/1808.02077} {arXiv:1808.02077 [hep-lat]} \BibitemShut
  {NoStop}%
\bibitem [{\citenamefont {Good}\ \emph {et~al.}(2024)\citenamefont {Good},
  \citenamefont {Hasan},\ and\ \citenamefont {Lin}}]{Good:2024iur}%
  \BibitemOpen
  \bibfield  {author} {\bibinfo {author} {\bibfnamefont {W.}~\bibnamefont
  {Good}}, \bibinfo {author} {\bibfnamefont {K.}~\bibnamefont {Hasan}}, \ and\
  \bibinfo {author} {\bibfnamefont {H.-W.}\ \bibnamefont {Lin}},\ }\href@noop
  {} {\  (\bibinfo {year} {2024})},\ \Eprint {http://arxiv.org/abs/2409.02750}
  {arXiv:2409.02750 [hep-lat]} \BibitemShut {NoStop}%
\bibitem [{\citenamefont {Chen}\ \emph {et~al.}(2020)\citenamefont {Chen},
  \citenamefont {Lin},\ and\ \citenamefont {Zhang}}]{Chen:2019lcm}%
  \BibitemOpen
  \bibfield  {author} {\bibinfo {author} {\bibfnamefont {J.-W.}\ \bibnamefont
  {Chen}}, \bibinfo {author} {\bibfnamefont {H.-W.}\ \bibnamefont {Lin}}, \
  and\ \bibinfo {author} {\bibfnamefont {J.-H.}\ \bibnamefont {Zhang}},\ }\href
  {\doibase 10.1016/j.nuclphysb.2020.114940} {\bibfield  {journal} {\bibinfo
  {journal} {Nucl. Phys. B}\ }\textbf {\bibinfo {volume} {952}},\ \bibinfo
  {pages} {114940} (\bibinfo {year} {2020})},\ \Eprint
  {http://arxiv.org/abs/1904.12376} {arXiv:1904.12376 [hep-lat]} \BibitemShut
  {NoStop}%
\bibitem [{\citenamefont {Alexandrou}\ \emph {et~al.}(2019)\citenamefont
  {Alexandrou}, \citenamefont {Cichy}, \citenamefont {Constantinou},
  \citenamefont {Hadjiyiannakou}, \citenamefont {Jansen}, \citenamefont
  {Scapellato},\ and\ \citenamefont {Steffens}}]{Alexandrou:2019dax}%
  \BibitemOpen
  \bibfield  {author} {\bibinfo {author} {\bibfnamefont {C.}~\bibnamefont
  {Alexandrou}}, \bibinfo {author} {\bibfnamefont {K.}~\bibnamefont {Cichy}},
  \bibinfo {author} {\bibfnamefont {M.}~\bibnamefont {Constantinou}}, \bibinfo
  {author} {\bibfnamefont {K.}~\bibnamefont {Hadjiyiannakou}}, \bibinfo
  {author} {\bibfnamefont {K.}~\bibnamefont {Jansen}}, \bibinfo {author}
  {\bibfnamefont {A.}~\bibnamefont {Scapellato}}, \ and\ \bibinfo {author}
  {\bibfnamefont {F.}~\bibnamefont {Steffens}},\ }\href {\doibase
  10.22323/1.363.0036} {\bibfield  {journal} {\bibinfo  {journal} {PoS}\
  }\textbf {\bibinfo {volume} {LATTICE2019}},\ \bibinfo {pages} {036} (\bibinfo
  {year} {2019})},\ \Eprint {http://arxiv.org/abs/1910.13229} {arXiv:1910.13229
  [hep-lat]} \BibitemShut {NoStop}%
\bibitem [{\citenamefont {Lin}(2021)}]{Lin:2020rxa}%
  \BibitemOpen
  \bibfield  {author} {\bibinfo {author} {\bibfnamefont {H.-W.}\ \bibnamefont
  {Lin}},\ }\href {\doibase 10.1103/PhysRevLett.127.182001} {\bibfield
  {journal} {\bibinfo  {journal} {Phys. Rev. Lett.}\ }\textbf {\bibinfo
  {volume} {127}},\ \bibinfo {pages} {182001} (\bibinfo {year} {2021})},\
  \Eprint {http://arxiv.org/abs/2008.12474} {arXiv:2008.12474 [hep-ph]}
  \BibitemShut {NoStop}%
\bibitem [{\citenamefont {Alexandrou}\ \emph {et~al.}(2020)\citenamefont
  {Alexandrou}, \citenamefont {Cichy}, \citenamefont {Constantinou},
  \citenamefont {Hadjiyiannakou}, \citenamefont {Jansen}, \citenamefont
  {Scapellato},\ and\ \citenamefont {Steffens}}]{Alexandrou:2020zbe}%
  \BibitemOpen
  \bibfield  {author} {\bibinfo {author} {\bibfnamefont {C.}~\bibnamefont
  {Alexandrou}}, \bibinfo {author} {\bibfnamefont {K.}~\bibnamefont {Cichy}},
  \bibinfo {author} {\bibfnamefont {M.}~\bibnamefont {Constantinou}}, \bibinfo
  {author} {\bibfnamefont {K.}~\bibnamefont {Hadjiyiannakou}}, \bibinfo
  {author} {\bibfnamefont {K.}~\bibnamefont {Jansen}}, \bibinfo {author}
  {\bibfnamefont {A.}~\bibnamefont {Scapellato}}, \ and\ \bibinfo {author}
  {\bibfnamefont {F.}~\bibnamefont {Steffens}},\ }\href {\doibase
  10.1103/PhysRevLett.125.262001} {\bibfield  {journal} {\bibinfo  {journal}
  {Phys. Rev. Lett.}\ }\textbf {\bibinfo {volume} {125}},\ \bibinfo {pages}
  {262001} (\bibinfo {year} {2020})},\ \Eprint
  {http://arxiv.org/abs/2008.10573} {arXiv:2008.10573 [hep-lat]} \BibitemShut
  {NoStop}%
\bibitem [{\citenamefont {Lin}(2022)}]{Lin:2021brq}%
  \BibitemOpen
  \bibfield  {author} {\bibinfo {author} {\bibfnamefont {H.-W.}\ \bibnamefont
  {Lin}},\ }\href {\doibase 10.1016/j.physletb.2021.136821} {\bibfield
  {journal} {\bibinfo  {journal} {Phys. Lett. B}\ }\textbf {\bibinfo {volume}
  {824}},\ \bibinfo {pages} {136821} (\bibinfo {year} {2022})},\ \Eprint
  {http://arxiv.org/abs/2112.07519} {arXiv:2112.07519 [hep-lat]} \BibitemShut
  {NoStop}%
\bibitem [{\citenamefont {Scapellato}\ \emph {et~al.}(2022)\citenamefont
  {Scapellato}, \citenamefont {Alexandrou}, \citenamefont {Cichy},
  \citenamefont {Constantinou}, \citenamefont {Hadjiyiannakou}, \citenamefont
  {Jansen},\ and\ \citenamefont {Steffens}}]{Scapellato:2022mai}%
  \BibitemOpen
  \bibfield  {author} {\bibinfo {author} {\bibfnamefont {A.}~\bibnamefont
  {Scapellato}}, \bibinfo {author} {\bibfnamefont {C.}~\bibnamefont
  {Alexandrou}}, \bibinfo {author} {\bibfnamefont {K.}~\bibnamefont {Cichy}},
  \bibinfo {author} {\bibfnamefont {M.}~\bibnamefont {Constantinou}}, \bibinfo
  {author} {\bibfnamefont {K.}~\bibnamefont {Hadjiyiannakou}}, \bibinfo
  {author} {\bibfnamefont {K.}~\bibnamefont {Jansen}}, \ and\ \bibinfo {author}
  {\bibfnamefont {F.}~\bibnamefont {Steffens}},\ }\href {\doibase
  10.31349/SuplRevMexFis.3.0308104} {\bibfield  {journal} {\bibinfo  {journal}
  {Rev. Mex. Fis. Suppl.}\ }\textbf {\bibinfo {volume} {3}},\ \bibinfo {pages}
  {0308104} (\bibinfo {year} {2022})},\ \Eprint
  {http://arxiv.org/abs/2201.06519} {arXiv:2201.06519 [hep-lat]} \BibitemShut
  {NoStop}%
\bibitem [{\citenamefont {Bhattacharya}\ \emph {et~al.}(2022)\citenamefont
  {Bhattacharya}, \citenamefont {Cichy}, \citenamefont {Constantinou},
  \citenamefont {Dodson}, \citenamefont {Gao}, \citenamefont {Metz},
  \citenamefont {Mukherjee}, \citenamefont {Scapellato}, \citenamefont
  {Steffens},\ and\ \citenamefont {Zhao}}]{Bhattacharya:2022aob}%
  \BibitemOpen
  \bibfield  {author} {\bibinfo {author} {\bibfnamefont {S.}~\bibnamefont
  {Bhattacharya}}, \bibinfo {author} {\bibfnamefont {K.}~\bibnamefont {Cichy}},
  \bibinfo {author} {\bibfnamefont {M.}~\bibnamefont {Constantinou}}, \bibinfo
  {author} {\bibfnamefont {J.}~\bibnamefont {Dodson}}, \bibinfo {author}
  {\bibfnamefont {X.}~\bibnamefont {Gao}}, \bibinfo {author} {\bibfnamefont
  {A.}~\bibnamefont {Metz}}, \bibinfo {author} {\bibfnamefont {S.}~\bibnamefont
  {Mukherjee}}, \bibinfo {author} {\bibfnamefont {A.}~\bibnamefont
  {Scapellato}}, \bibinfo {author} {\bibfnamefont {F.}~\bibnamefont
  {Steffens}}, \ and\ \bibinfo {author} {\bibfnamefont {Y.}~\bibnamefont
  {Zhao}},\ }\href {\doibase 10.1103/PhysRevD.106.114512} {\bibfield  {journal}
  {\bibinfo  {journal} {Phys. Rev. D}\ }\textbf {\bibinfo {volume} {106}},\
  \bibinfo {pages} {114512} (\bibinfo {year} {2022})},\ \Eprint
  {http://arxiv.org/abs/2209.05373} {arXiv:2209.05373 [hep-lat]} \BibitemShut
  {NoStop}%
\bibitem [{\citenamefont {Bhattacharya}\ \emph {et~al.}(2023)\citenamefont
  {Bhattacharya}, \citenamefont {Cichy}, \citenamefont {Constantinou},
  \citenamefont {Dodson}, \citenamefont {Metz}, \citenamefont {Scapellato},\
  and\ \citenamefont {Steffens}}]{Bhattacharya:2023nmv}%
  \BibitemOpen
  \bibfield  {author} {\bibinfo {author} {\bibfnamefont {S.}~\bibnamefont
  {Bhattacharya}}, \bibinfo {author} {\bibfnamefont {K.}~\bibnamefont {Cichy}},
  \bibinfo {author} {\bibfnamefont {M.}~\bibnamefont {Constantinou}}, \bibinfo
  {author} {\bibfnamefont {J.}~\bibnamefont {Dodson}}, \bibinfo {author}
  {\bibfnamefont {A.}~\bibnamefont {Metz}}, \bibinfo {author} {\bibfnamefont
  {A.}~\bibnamefont {Scapellato}}, \ and\ \bibinfo {author} {\bibfnamefont
  {F.}~\bibnamefont {Steffens}},\ }\href {\doibase 10.1103/PhysRevD.108.054501}
  {\bibfield  {journal} {\bibinfo  {journal} {Phys. Rev. D}\ }\textbf {\bibinfo
  {volume} {108}},\ \bibinfo {pages} {054501} (\bibinfo {year} {2023})},\
  \Eprint {http://arxiv.org/abs/2306.05533} {arXiv:2306.05533 [hep-lat]}
  \BibitemShut {NoStop}%
\bibitem [{\citenamefont {Bhattacharya}\ \emph {et~al.}(2024)\citenamefont
  {Bhattacharya} \emph {et~al.}}]{Bhattacharya:2023jsc}%
  \BibitemOpen
  \bibfield  {author} {\bibinfo {author} {\bibfnamefont {S.}~\bibnamefont
  {Bhattacharya}} \emph {et~al.},\ }\href {\doibase
  10.1103/PhysRevD.109.034508} {\bibfield  {journal} {\bibinfo  {journal}
  {Phys. Rev. D}\ }\textbf {\bibinfo {volume} {109}},\ \bibinfo {pages}
  {034508} (\bibinfo {year} {2024})},\ \Eprint
  {http://arxiv.org/abs/2310.13114} {arXiv:2310.13114 [hep-lat]} \BibitemShut
  {NoStop}%
\bibitem [{\citenamefont {Lin}(2023)}]{Lin:2023gxz}%
  \BibitemOpen
  \bibfield  {author} {\bibinfo {author} {\bibfnamefont {H.-W.}\ \bibnamefont
  {Lin}},\ }\href {\doibase 10.1016/j.physletb.2023.138181} {\bibfield
  {journal} {\bibinfo  {journal} {Phys. Lett. B}\ }\textbf {\bibinfo {volume}
  {846}},\ \bibinfo {pages} {138181} (\bibinfo {year} {2023})},\ \Eprint
  {http://arxiv.org/abs/2310.10579} {arXiv:2310.10579 [hep-lat]} \BibitemShut
  {NoStop}%
\bibitem [{\citenamefont {Holligan}\ and\ \citenamefont
  {Lin}(2024{\natexlab{c}})}]{Holligan:2023jqh}%
  \BibitemOpen
  \bibfield  {author} {\bibinfo {author} {\bibfnamefont {J.}~\bibnamefont
  {Holligan}}\ and\ \bibinfo {author} {\bibfnamefont {H.-W.}\ \bibnamefont
  {Lin}},\ }\href {\doibase 10.1103/PhysRevD.110.034503} {\bibfield  {journal}
  {\bibinfo  {journal} {Phys. Rev. D}\ }\textbf {\bibinfo {volume} {110}},\
  \bibinfo {pages} {034503} (\bibinfo {year} {2024}{\natexlab{c}})},\ \Eprint
  {http://arxiv.org/abs/2312.10829} {arXiv:2312.10829 [hep-lat]} \BibitemShut
  {NoStop}%
\bibitem [{\citenamefont {Ding}\ \emph {et~al.}(2024)\citenamefont {Ding},
  \citenamefont {Gao}, \citenamefont {Mukherjee}, \citenamefont {Petreczky},
  \citenamefont {Shi}, \citenamefont {Syritsyn},\ and\ \citenamefont
  {Zhao}}]{Ding:2024hkz}%
  \BibitemOpen
  \bibfield  {author} {\bibinfo {author} {\bibfnamefont {H.-T.}\ \bibnamefont
  {Ding}}, \bibinfo {author} {\bibfnamefont {X.}~\bibnamefont {Gao}}, \bibinfo
  {author} {\bibfnamefont {S.}~\bibnamefont {Mukherjee}}, \bibinfo {author}
  {\bibfnamefont {P.}~\bibnamefont {Petreczky}}, \bibinfo {author}
  {\bibfnamefont {Q.}~\bibnamefont {Shi}}, \bibinfo {author} {\bibfnamefont
  {S.}~\bibnamefont {Syritsyn}}, \ and\ \bibinfo {author} {\bibfnamefont
  {Y.}~\bibnamefont {Zhao}},\ }\href@noop {} {\  (\bibinfo {year} {2024})},\
  \Eprint {http://arxiv.org/abs/2407.03516} {arXiv:2407.03516 [hep-lat]}
  \BibitemShut {NoStop}%
\bibitem [{\citenamefont {Zhang}\ \emph {et~al.}(2017)\citenamefont {Zhang},
  \citenamefont {Chen}, \citenamefont {Ji}, \citenamefont {Jin},\ and\
  \citenamefont {Lin}}]{Zhang:2017bzy}%
  \BibitemOpen
  \bibfield  {author} {\bibinfo {author} {\bibfnamefont {J.-H.}\ \bibnamefont
  {Zhang}}, \bibinfo {author} {\bibfnamefont {J.-W.}\ \bibnamefont {Chen}},
  \bibinfo {author} {\bibfnamefont {X.}~\bibnamefont {Ji}}, \bibinfo {author}
  {\bibfnamefont {L.}~\bibnamefont {Jin}}, \ and\ \bibinfo {author}
  {\bibfnamefont {H.-W.}\ \bibnamefont {Lin}},\ }\href {\doibase
  10.1103/PhysRevD.95.094514} {\bibfield  {journal} {\bibinfo  {journal} {Phys.
  Rev. D}\ }\textbf {\bibinfo {volume} {95}},\ \bibinfo {pages} {094514}
  (\bibinfo {year} {2017})},\ \Eprint {http://arxiv.org/abs/1702.00008}
  {arXiv:1702.00008 [hep-lat]} \BibitemShut {NoStop}%
\bibitem [{\citenamefont {Zhang}\ \emph
  {et~al.}(2019{\natexlab{b}})\citenamefont {Zhang}, \citenamefont {Jin},
  \citenamefont {Lin}, \citenamefont {Sch\"afer}, \citenamefont {Sun},
  \citenamefont {Yang}, \citenamefont {Zhang}, \citenamefont {Zhao},\ and\
  \citenamefont {Chen}}]{Chen:2017gck}%
  \BibitemOpen
  \bibfield  {author} {\bibinfo {author} {\bibfnamefont {J.-H.}\ \bibnamefont
  {Zhang}}, \bibinfo {author} {\bibfnamefont {L.}~\bibnamefont {Jin}}, \bibinfo
  {author} {\bibfnamefont {H.-W.}\ \bibnamefont {Lin}}, \bibinfo {author}
  {\bibfnamefont {A.}~\bibnamefont {Sch\"afer}}, \bibinfo {author}
  {\bibfnamefont {P.}~\bibnamefont {Sun}}, \bibinfo {author} {\bibfnamefont
  {Y.-B.}\ \bibnamefont {Yang}}, \bibinfo {author} {\bibfnamefont
  {R.}~\bibnamefont {Zhang}}, \bibinfo {author} {\bibfnamefont
  {Y.}~\bibnamefont {Zhao}}, \ and\ \bibinfo {author} {\bibfnamefont {J.-W.}\
  \bibnamefont {Chen}} (\bibinfo {collaboration} {LP3}),\ }\href {\doibase
  10.1016/j.nuclphysb.2018.12.020} {\bibfield  {journal} {\bibinfo  {journal}
  {Nucl. Phys. B}\ }\textbf {\bibinfo {volume} {939}},\ \bibinfo {pages} {429}
  (\bibinfo {year} {2019}{\natexlab{b}})},\ \Eprint
  {http://arxiv.org/abs/1712.10025} {arXiv:1712.10025 [hep-ph]} \BibitemShut
  {NoStop}%
\bibitem [{\citenamefont {Zhang}\ \emph
  {et~al.}(2020{\natexlab{a}})\citenamefont {Zhang}, \citenamefont {Honkala},
  \citenamefont {Lin},\ and\ \citenamefont {Chen}}]{Zhang:2020gaj}%
  \BibitemOpen
  \bibfield  {author} {\bibinfo {author} {\bibfnamefont {R.}~\bibnamefont
  {Zhang}}, \bibinfo {author} {\bibfnamefont {C.}~\bibnamefont {Honkala}},
  \bibinfo {author} {\bibfnamefont {H.-W.}\ \bibnamefont {Lin}}, \ and\
  \bibinfo {author} {\bibfnamefont {J.-W.}\ \bibnamefont {Chen}},\ }\href
  {\doibase 10.1103/PhysRevD.102.094519} {\bibfield  {journal} {\bibinfo
  {journal} {Phys. Rev. D}\ }\textbf {\bibinfo {volume} {102}},\ \bibinfo
  {pages} {094519} (\bibinfo {year} {2020}{\natexlab{a}})},\ \Eprint
  {http://arxiv.org/abs/2005.13955} {arXiv:2005.13955 [hep-lat]} \BibitemShut
  {NoStop}%
\bibitem [{\citenamefont {Hua}\ \emph {et~al.}(2021)\citenamefont {Hua},
  \citenamefont {Chu}, \citenamefont {Sun}, \citenamefont {Wang}, \citenamefont
  {Xu}, \citenamefont {Yang}, \citenamefont {Zhang},\ and\ \citenamefont
  {Zhang}}]{Hua:2020gnw}%
  \BibitemOpen
  \bibfield  {author} {\bibinfo {author} {\bibfnamefont {J.}~\bibnamefont
  {Hua}}, \bibinfo {author} {\bibfnamefont {M.-H.}\ \bibnamefont {Chu}},
  \bibinfo {author} {\bibfnamefont {P.}~\bibnamefont {Sun}}, \bibinfo {author}
  {\bibfnamefont {W.}~\bibnamefont {Wang}}, \bibinfo {author} {\bibfnamefont
  {J.}~\bibnamefont {Xu}}, \bibinfo {author} {\bibfnamefont {Y.-B.}\
  \bibnamefont {Yang}}, \bibinfo {author} {\bibfnamefont {J.-H.}\ \bibnamefont
  {Zhang}}, \ and\ \bibinfo {author} {\bibfnamefont {Q.-A.}\ \bibnamefont
  {Zhang}} (\bibinfo {collaboration} {Lattice Parton}),\ }\href {\doibase
  10.1103/PhysRevLett.127.062002} {\bibfield  {journal} {\bibinfo  {journal}
  {Phys. Rev. Lett.}\ }\textbf {\bibinfo {volume} {127}},\ \bibinfo {pages}
  {062002} (\bibinfo {year} {2021})},\ \Eprint
  {http://arxiv.org/abs/2011.09788} {arXiv:2011.09788 [hep-lat]} \BibitemShut
  {NoStop}%
\bibitem [{\citenamefont {Hua}\ \emph {et~al.}(2022)\citenamefont {Hua} \emph
  {et~al.}}]{LatticeParton:2022zqc}%
  \BibitemOpen
  \bibfield  {author} {\bibinfo {author} {\bibfnamefont {J.}~\bibnamefont
  {Hua}} \emph {et~al.} (\bibinfo {collaboration} {Lattice Parton}),\ }\href
  {\doibase 10.1103/PhysRevLett.129.132001} {\bibfield  {journal} {\bibinfo
  {journal} {Phys. Rev. Lett.}\ }\textbf {\bibinfo {volume} {129}},\ \bibinfo
  {pages} {132001} (\bibinfo {year} {2022})},\ \Eprint
  {http://arxiv.org/abs/2201.09173} {arXiv:2201.09173 [hep-lat]} \BibitemShut
  {NoStop}%
\bibitem [{\citenamefont {Gao}\ \emph {et~al.}(2022{\natexlab{c}})\citenamefont
  {Gao}, \citenamefont {Hanlon}, \citenamefont {Karthik}, \citenamefont
  {Mukherjee}, \citenamefont {Petreczky}, \citenamefont {Scior}, \citenamefont
  {Syritsyn},\ and\ \citenamefont {Zhao}}]{Gao:2022vyh}%
  \BibitemOpen
  \bibfield  {author} {\bibinfo {author} {\bibfnamefont {X.}~\bibnamefont
  {Gao}}, \bibinfo {author} {\bibfnamefont {A.~D.}\ \bibnamefont {Hanlon}},
  \bibinfo {author} {\bibfnamefont {N.}~\bibnamefont {Karthik}}, \bibinfo
  {author} {\bibfnamefont {S.}~\bibnamefont {Mukherjee}}, \bibinfo {author}
  {\bibfnamefont {P.}~\bibnamefont {Petreczky}}, \bibinfo {author}
  {\bibfnamefont {P.}~\bibnamefont {Scior}}, \bibinfo {author} {\bibfnamefont
  {S.}~\bibnamefont {Syritsyn}}, \ and\ \bibinfo {author} {\bibfnamefont
  {Y.}~\bibnamefont {Zhao}},\ }\href {\doibase 10.1103/PhysRevD.106.074505}
  {\bibfield  {journal} {\bibinfo  {journal} {Phys. Rev. D}\ }\textbf {\bibinfo
  {volume} {106}},\ \bibinfo {pages} {074505} (\bibinfo {year}
  {2022}{\natexlab{c}})},\ \Eprint {http://arxiv.org/abs/2206.04084}
  {arXiv:2206.04084 [hep-lat]} \BibitemShut {NoStop}%
\bibitem [{\citenamefont {Xu}\ and\ \citenamefont {Zhang}(2022)}]{Xu:2022guw}%
  \BibitemOpen
  \bibfield  {author} {\bibinfo {author} {\bibfnamefont {J.}~\bibnamefont
  {Xu}}\ and\ \bibinfo {author} {\bibfnamefont {X.-R.}\ \bibnamefont {Zhang}},\
  }\href {\doibase 10.1103/PhysRevD.106.114019} {\bibfield  {journal} {\bibinfo
   {journal} {Phys. Rev. D}\ }\textbf {\bibinfo {volume} {106}},\ \bibinfo
  {pages} {114019} (\bibinfo {year} {2022})},\ \Eprint
  {http://arxiv.org/abs/2209.10719} {arXiv:2209.10719 [hep-ph]} \BibitemShut
  {NoStop}%
\bibitem [{\citenamefont {Holligan}\ \emph {et~al.}(2023)\citenamefont
  {Holligan}, \citenamefont {Ji}, \citenamefont {Lin}, \citenamefont {Su},\
  and\ \citenamefont {Zhang}}]{Holligan:2023rex}%
  \BibitemOpen
  \bibfield  {author} {\bibinfo {author} {\bibfnamefont {J.}~\bibnamefont
  {Holligan}}, \bibinfo {author} {\bibfnamefont {X.}~\bibnamefont {Ji}},
  \bibinfo {author} {\bibfnamefont {H.-W.}\ \bibnamefont {Lin}}, \bibinfo
  {author} {\bibfnamefont {Y.}~\bibnamefont {Su}}, \ and\ \bibinfo {author}
  {\bibfnamefont {R.}~\bibnamefont {Zhang}},\ }\href {\doibase
  10.1016/j.nuclphysb.2023.116282} {\bibfield  {journal} {\bibinfo  {journal}
  {Nucl. Phys. B}\ }\textbf {\bibinfo {volume} {993}},\ \bibinfo {pages}
  {116282} (\bibinfo {year} {2023})},\ \Eprint
  {http://arxiv.org/abs/2301.10372} {arXiv:2301.10372 [hep-lat]} \BibitemShut
  {NoStop}%
\bibitem [{\citenamefont {Deng}\ \emph {et~al.}(2023)\citenamefont {Deng},
  \citenamefont {Han}, \citenamefont {Wang}, \citenamefont {Zeng},\ and\
  \citenamefont {Zhang}}]{Deng:2023csv}%
  \BibitemOpen
  \bibfield  {author} {\bibinfo {author} {\bibfnamefont {Z.-F.}\ \bibnamefont
  {Deng}}, \bibinfo {author} {\bibfnamefont {C.}~\bibnamefont {Han}}, \bibinfo
  {author} {\bibfnamefont {W.}~\bibnamefont {Wang}}, \bibinfo {author}
  {\bibfnamefont {J.}~\bibnamefont {Zeng}}, \ and\ \bibinfo {author}
  {\bibfnamefont {J.-L.}\ \bibnamefont {Zhang}},\ }\href {\doibase
  10.1007/JHEP07(2023)191} {\bibfield  {journal} {\bibinfo  {journal} {JHEP}\
  }\textbf {\bibinfo {volume} {07}},\ \bibinfo {pages} {191} (\bibinfo {year}
  {2023})},\ \Eprint {http://arxiv.org/abs/2304.09004} {arXiv:2304.09004
  [hep-ph]} \BibitemShut {NoStop}%
\bibitem [{\citenamefont {Han}\ \emph {et~al.}(2023)\citenamefont {Han},
  \citenamefont {Su}, \citenamefont {Wang},\ and\ \citenamefont
  {Zhang}}]{Han:2023xbl}%
  \BibitemOpen
  \bibfield  {author} {\bibinfo {author} {\bibfnamefont {C.}~\bibnamefont
  {Han}}, \bibinfo {author} {\bibfnamefont {Y.}~\bibnamefont {Su}}, \bibinfo
  {author} {\bibfnamefont {W.}~\bibnamefont {Wang}}, \ and\ \bibinfo {author}
  {\bibfnamefont {J.-L.}\ \bibnamefont {Zhang}},\ }\href {\doibase
  10.1007/JHEP12(2023)044} {\bibfield  {journal} {\bibinfo  {journal} {JHEP}\
  }\textbf {\bibinfo {volume} {12}},\ \bibinfo {pages} {044} (\bibinfo {year}
  {2023})},\ \Eprint {http://arxiv.org/abs/2308.16793} {arXiv:2308.16793
  [hep-ph]} \BibitemShut {NoStop}%
\bibitem [{\citenamefont {Han}\ and\ \citenamefont
  {Zhang}(2023)}]{Han:2023hgy}%
  \BibitemOpen
  \bibfield  {author} {\bibinfo {author} {\bibfnamefont {C.}~\bibnamefont
  {Han}}\ and\ \bibinfo {author} {\bibfnamefont {J.}~\bibnamefont {Zhang}},\
  }\href@noop {} {\  (\bibinfo {year} {2023})},\ \Eprint
  {http://arxiv.org/abs/2311.02669} {arXiv:2311.02669 [hep-ph]} \BibitemShut
  {NoStop}%
\bibitem [{\citenamefont {Han}\ \emph {et~al.}(2024{\natexlab{a}})\citenamefont
  {Han}, \citenamefont {Hua}, \citenamefont {Ji}, \citenamefont {L\"u},
  \citenamefont {Wang}, \citenamefont {Xu}, \citenamefont {Zhang},\ and\
  \citenamefont {Zhao}}]{Han:2024min}%
  \BibitemOpen
  \bibfield  {author} {\bibinfo {author} {\bibfnamefont {X.-Y.}\ \bibnamefont
  {Han}}, \bibinfo {author} {\bibfnamefont {J.}~\bibnamefont {Hua}}, \bibinfo
  {author} {\bibfnamefont {X.}~\bibnamefont {Ji}}, \bibinfo {author}
  {\bibfnamefont {C.-D.}\ \bibnamefont {L\"u}}, \bibinfo {author}
  {\bibfnamefont {W.}~\bibnamefont {Wang}}, \bibinfo {author} {\bibfnamefont
  {J.}~\bibnamefont {Xu}}, \bibinfo {author} {\bibfnamefont {Q.-A.}\
  \bibnamefont {Zhang}}, \ and\ \bibinfo {author} {\bibfnamefont
  {S.}~\bibnamefont {Zhao}},\ }\href@noop {} {\  (\bibinfo {year}
  {2024}{\natexlab{a}})},\ \Eprint {http://arxiv.org/abs/2403.17492}
  {arXiv:2403.17492 [hep-ph]} \BibitemShut {NoStop}%
\bibitem [{\citenamefont {Han}\ \emph {et~al.}(2024{\natexlab{b}})\citenamefont
  {Han}, \citenamefont {Wang}, \citenamefont {Zeng},\ and\ \citenamefont
  {Zhang}}]{Han:2024ucv}%
  \BibitemOpen
  \bibfield  {author} {\bibinfo {author} {\bibfnamefont {C.}~\bibnamefont
  {Han}}, \bibinfo {author} {\bibfnamefont {W.}~\bibnamefont {Wang}}, \bibinfo
  {author} {\bibfnamefont {J.}~\bibnamefont {Zeng}}, \ and\ \bibinfo {author}
  {\bibfnamefont {J.-L.}\ \bibnamefont {Zhang}},\ }\href {\doibase
  10.1007/JHEP07(2024)019} {\bibfield  {journal} {\bibinfo  {journal} {JHEP}\
  }\textbf {\bibinfo {volume} {07}},\ \bibinfo {pages} {019} (\bibinfo {year}
  {2024}{\natexlab{b}})},\ \Eprint {http://arxiv.org/abs/2404.04855}
  {arXiv:2404.04855 [hep-ph]} \BibitemShut {NoStop}%
\bibitem [{\citenamefont {Baker}\ \emph {et~al.}(2024)\citenamefont {Baker},
  \citenamefont {Bollweg}, \citenamefont {Boyle}, \citenamefont {Clo\"et},
  \citenamefont {Gao}, \citenamefont {Mukherjee}, \citenamefont {Petreczky},
  \citenamefont {Zhang},\ and\ \citenamefont {Zhao}}]{Baker:2024zcd}%
  \BibitemOpen
  \bibfield  {author} {\bibinfo {author} {\bibfnamefont {E.}~\bibnamefont
  {Baker}}, \bibinfo {author} {\bibfnamefont {D.}~\bibnamefont {Bollweg}},
  \bibinfo {author} {\bibfnamefont {P.}~\bibnamefont {Boyle}}, \bibinfo
  {author} {\bibfnamefont {I.}~\bibnamefont {Clo\"et}}, \bibinfo {author}
  {\bibfnamefont {X.}~\bibnamefont {Gao}}, \bibinfo {author} {\bibfnamefont
  {S.}~\bibnamefont {Mukherjee}}, \bibinfo {author} {\bibfnamefont
  {P.}~\bibnamefont {Petreczky}}, \bibinfo {author} {\bibfnamefont
  {R.}~\bibnamefont {Zhang}}, \ and\ \bibinfo {author} {\bibfnamefont
  {Y.}~\bibnamefont {Zhao}},\ }\href {\doibase 10.1007/JHEP07(2024)211}
  {\bibfield  {journal} {\bibinfo  {journal} {JHEP}\ }\textbf {\bibinfo
  {volume} {07}},\ \bibinfo {pages} {211} (\bibinfo {year} {2024})},\ \Eprint
  {http://arxiv.org/abs/2405.20120} {arXiv:2405.20120 [hep-lat]} \BibitemShut
  {NoStop}%
\bibitem [{\citenamefont {Cloet}\ \emph {et~al.}(2024)\citenamefont {Cloet},
  \citenamefont {Gao}, \citenamefont {Mukherjee}, \citenamefont {Syritsyn},
  \citenamefont {Karthik}, \citenamefont {Petreczky}, \citenamefont {Zhang},\
  and\ \citenamefont {Zhao}}]{Cloet:2024vbv}%
  \BibitemOpen
  \bibfield  {author} {\bibinfo {author} {\bibfnamefont {I.}~\bibnamefont
  {Cloet}}, \bibinfo {author} {\bibfnamefont {X.}~\bibnamefont {Gao}}, \bibinfo
  {author} {\bibfnamefont {S.}~\bibnamefont {Mukherjee}}, \bibinfo {author}
  {\bibfnamefont {S.}~\bibnamefont {Syritsyn}}, \bibinfo {author}
  {\bibfnamefont {N.}~\bibnamefont {Karthik}}, \bibinfo {author} {\bibfnamefont
  {P.}~\bibnamefont {Petreczky}}, \bibinfo {author} {\bibfnamefont
  {R.}~\bibnamefont {Zhang}}, \ and\ \bibinfo {author} {\bibfnamefont
  {Y.}~\bibnamefont {Zhao}},\ }\href@noop {} {\  (\bibinfo {year} {2024})},\
  \Eprint {http://arxiv.org/abs/2407.00206} {arXiv:2407.00206 [hep-lat]}
  \BibitemShut {NoStop}%
\bibitem [{\citenamefont {Han}\ \emph {et~al.}(2024{\natexlab{c}})\citenamefont
  {Han}, \citenamefont {Wang}, \citenamefont {Zhang},\ and\ \citenamefont
  {Zhang}}]{Han:2024cht}%
  \BibitemOpen
  \bibfield  {author} {\bibinfo {author} {\bibfnamefont {C.}~\bibnamefont
  {Han}}, \bibinfo {author} {\bibfnamefont {W.}~\bibnamefont {Wang}}, \bibinfo
  {author} {\bibfnamefont {J.-L.}\ \bibnamefont {Zhang}}, \ and\ \bibinfo
  {author} {\bibfnamefont {J.-H.}\ \bibnamefont {Zhang}},\ }\href@noop {} {\
  (\bibinfo {year} {2024}{\natexlab{c}})},\ \Eprint
  {http://arxiv.org/abs/2408.13486} {arXiv:2408.13486 [hep-ph]} \BibitemShut
  {NoStop}%
\bibitem [{\citenamefont {Deng}\ \emph {et~al.}(2024)\citenamefont {Deng},
  \citenamefont {Wang}, \citenamefont {Wei},\ and\ \citenamefont
  {Zeng}}]{Deng:2024dkd}%
  \BibitemOpen
  \bibfield  {author} {\bibinfo {author} {\bibfnamefont {Z.-F.}\ \bibnamefont
  {Deng}}, \bibinfo {author} {\bibfnamefont {W.}~\bibnamefont {Wang}}, \bibinfo
  {author} {\bibfnamefont {Y.-B.}\ \bibnamefont {Wei}}, \ and\ \bibinfo
  {author} {\bibfnamefont {J.}~\bibnamefont {Zeng}},\ }\href@noop {} {\
  (\bibinfo {year} {2024})},\ \Eprint {http://arxiv.org/abs/2409.00632}
  {arXiv:2409.00632 [hep-ph]} \BibitemShut {NoStop}%
\bibitem [{\citenamefont {Ji}\ \emph {et~al.}(2015)\citenamefont {Ji},
  \citenamefont {Sun}, \citenamefont {Xiong},\ and\ \citenamefont
  {Yuan}}]{Ji:2014hxa}%
  \BibitemOpen
  \bibfield  {author} {\bibinfo {author} {\bibfnamefont {X.}~\bibnamefont
  {Ji}}, \bibinfo {author} {\bibfnamefont {P.}~\bibnamefont {Sun}}, \bibinfo
  {author} {\bibfnamefont {X.}~\bibnamefont {Xiong}}, \ and\ \bibinfo {author}
  {\bibfnamefont {F.}~\bibnamefont {Yuan}},\ }\href {\doibase
  10.1103/PhysRevD.91.074009} {\bibfield  {journal} {\bibinfo  {journal} {Phys.
  Rev. D}\ }\textbf {\bibinfo {volume} {91}},\ \bibinfo {pages} {074009}
  (\bibinfo {year} {2015})},\ \Eprint {http://arxiv.org/abs/1405.7640}
  {arXiv:1405.7640 [hep-ph]} \BibitemShut {NoStop}%
\bibitem [{\citenamefont {Shanahan}\ \emph
  {et~al.}(2020{\natexlab{a}})\citenamefont {Shanahan}, \citenamefont
  {Wagman},\ and\ \citenamefont {Zhao}}]{Shanahan:2019zcq}%
  \BibitemOpen
  \bibfield  {author} {\bibinfo {author} {\bibfnamefont {P.}~\bibnamefont
  {Shanahan}}, \bibinfo {author} {\bibfnamefont {M.~L.}\ \bibnamefont
  {Wagman}}, \ and\ \bibinfo {author} {\bibfnamefont {Y.}~\bibnamefont
  {Zhao}},\ }\href {\doibase 10.1103/PhysRevD.101.074505} {\bibfield  {journal}
  {\bibinfo  {journal} {Phys. Rev. D}\ }\textbf {\bibinfo {volume} {101}},\
  \bibinfo {pages} {074505} (\bibinfo {year} {2020}{\natexlab{a}})},\ \Eprint
  {http://arxiv.org/abs/1911.00800} {arXiv:1911.00800 [hep-lat]} \BibitemShut
  {NoStop}%
\bibitem [{\citenamefont {Shanahan}\ \emph
  {et~al.}(2020{\natexlab{b}})\citenamefont {Shanahan}, \citenamefont
  {Wagman},\ and\ \citenamefont {Zhao}}]{Shanahan:2020zxr}%
  \BibitemOpen
  \bibfield  {author} {\bibinfo {author} {\bibfnamefont {P.}~\bibnamefont
  {Shanahan}}, \bibinfo {author} {\bibfnamefont {M.}~\bibnamefont {Wagman}}, \
  and\ \bibinfo {author} {\bibfnamefont {Y.}~\bibnamefont {Zhao}},\ }\href
  {\doibase 10.1103/PhysRevD.102.014511} {\bibfield  {journal} {\bibinfo
  {journal} {Phys. Rev. D}\ }\textbf {\bibinfo {volume} {102}},\ \bibinfo
  {pages} {014511} (\bibinfo {year} {2020}{\natexlab{b}})},\ \Eprint
  {http://arxiv.org/abs/2003.06063} {arXiv:2003.06063 [hep-lat]} \BibitemShut
  {NoStop}%
\bibitem [{\citenamefont {Zhang}\ \emph
  {et~al.}(2020{\natexlab{b}})\citenamefont {Zhang} \emph
  {et~al.}}]{Zhang:2020dbb}%
  \BibitemOpen
  \bibfield  {author} {\bibinfo {author} {\bibfnamefont {Q.-A.}\ \bibnamefont
  {Zhang}} \emph {et~al.} (\bibinfo {collaboration} {Lattice Parton}),\ }\href
  {\doibase 10.22323/1.396.0477} {\bibfield  {journal} {\bibinfo  {journal}
  {Phys. Rev. Lett.}\ }\textbf {\bibinfo {volume} {125}},\ \bibinfo {pages}
  {192001} (\bibinfo {year} {2020}{\natexlab{b}})},\ \Eprint
  {http://arxiv.org/abs/2005.14572} {arXiv:2005.14572 [hep-lat]} \BibitemShut
  {NoStop}%
\bibitem [{\citenamefont {Ji}\ and\ \citenamefont {Liu}(2022)}]{Ji:2021znw}%
  \BibitemOpen
  \bibfield  {author} {\bibinfo {author} {\bibfnamefont {X.}~\bibnamefont
  {Ji}}\ and\ \bibinfo {author} {\bibfnamefont {Y.}~\bibnamefont {Liu}},\
  }\href {\doibase 10.1103/PhysRevD.105.076014} {\bibfield  {journal} {\bibinfo
   {journal} {Phys. Rev. D}\ }\textbf {\bibinfo {volume} {105}},\ \bibinfo
  {pages} {076014} (\bibinfo {year} {2022})},\ \Eprint
  {http://arxiv.org/abs/2106.05310} {arXiv:2106.05310 [hep-ph]} \BibitemShut
  {NoStop}%
\bibitem [{\citenamefont {Chu}\ \emph {et~al.}(2022)\citenamefont {Chu} \emph
  {et~al.}}]{LatticePartonLPC:2022eev}%
  \BibitemOpen
  \bibfield  {author} {\bibinfo {author} {\bibfnamefont {M.-H.}\ \bibnamefont
  {Chu}} \emph {et~al.} (\bibinfo {collaboration} {Lattice Parton (LPC)}),\
  }\href {\doibase 10.1103/PhysRevD.106.034509} {\bibfield  {journal} {\bibinfo
   {journal} {Phys. Rev. D}\ }\textbf {\bibinfo {volume} {106}},\ \bibinfo
  {pages} {034509} (\bibinfo {year} {2022})},\ \Eprint
  {http://arxiv.org/abs/2204.00200} {arXiv:2204.00200 [hep-lat]} \BibitemShut
  {NoStop}%
\bibitem [{\citenamefont {Liu}(2022)}]{Liu:2022nnk}%
  \BibitemOpen
  \bibfield  {author} {\bibinfo {author} {\bibfnamefont {Y.}~\bibnamefont
  {Liu}},\ }\href {\doibase 10.5506/APhysPolB.53.4-A2} {\bibfield  {journal}
  {\bibinfo  {journal} {Acta Phys. Polon. B}\ }\textbf {\bibinfo {volume}
  {53}},\ \bibinfo {pages} {4} (\bibinfo {year} {2022})}\BibitemShut {NoStop}%
\bibitem [{\citenamefont {Zhang}\ \emph {et~al.}(2022)\citenamefont {Zhang},
  \citenamefont {Ji}, \citenamefont {Yang}, \citenamefont {Yao},\ and\
  \citenamefont {Zhang}}]{Zhang:2022xuw}%
  \BibitemOpen
  \bibfield  {author} {\bibinfo {author} {\bibfnamefont {K.}~\bibnamefont
  {Zhang}}, \bibinfo {author} {\bibfnamefont {X.}~\bibnamefont {Ji}}, \bibinfo
  {author} {\bibfnamefont {Y.-B.}\ \bibnamefont {Yang}}, \bibinfo {author}
  {\bibfnamefont {F.}~\bibnamefont {Yao}}, \ and\ \bibinfo {author}
  {\bibfnamefont {J.-H.}\ \bibnamefont {Zhang}} (\bibinfo {collaboration}
  {[Lattice Parton Collaboration (LPC)]}),\ }\href {\doibase
  10.1103/PhysRevLett.129.082002} {\bibfield  {journal} {\bibinfo  {journal}
  {Phys. Rev. Lett.}\ }\textbf {\bibinfo {volume} {129}},\ \bibinfo {pages}
  {082002} (\bibinfo {year} {2022})},\ \Eprint
  {http://arxiv.org/abs/2205.13402} {arXiv:2205.13402 [hep-lat]} \BibitemShut
  {NoStop}%
\bibitem [{\citenamefont {Deng}\ \emph {et~al.}(2022)\citenamefont {Deng},
  \citenamefont {Wang},\ and\ \citenamefont {Zeng}}]{Deng:2022gzi}%
  \BibitemOpen
  \bibfield  {author} {\bibinfo {author} {\bibfnamefont {Z.-F.}\ \bibnamefont
  {Deng}}, \bibinfo {author} {\bibfnamefont {W.}~\bibnamefont {Wang}}, \ and\
  \bibinfo {author} {\bibfnamefont {J.}~\bibnamefont {Zeng}},\ }\href {\doibase
  10.1007/JHEP09(2022)046} {\bibfield  {journal} {\bibinfo  {journal} {JHEP}\
  }\textbf {\bibinfo {volume} {09}},\ \bibinfo {pages} {046} (\bibinfo {year}
  {2022})},\ \Eprint {http://arxiv.org/abs/2207.07280} {arXiv:2207.07280
  [hep-th]} \BibitemShut {NoStop}%
\bibitem [{\citenamefont {Zhu}\ \emph {et~al.}(2023)\citenamefont {Zhu},
  \citenamefont {Ji}, \citenamefont {Zhang},\ and\ \citenamefont
  {Zhao}}]{Zhu:2022bja}%
  \BibitemOpen
  \bibfield  {author} {\bibinfo {author} {\bibfnamefont {R.}~\bibnamefont
  {Zhu}}, \bibinfo {author} {\bibfnamefont {Y.}~\bibnamefont {Ji}}, \bibinfo
  {author} {\bibfnamefont {J.-H.}\ \bibnamefont {Zhang}}, \ and\ \bibinfo
  {author} {\bibfnamefont {S.}~\bibnamefont {Zhao}},\ }\href {\doibase
  10.1007/JHEP02(2023)114} {\bibfield  {journal} {\bibinfo  {journal} {JHEP}\
  }\textbf {\bibinfo {volume} {02}},\ \bibinfo {pages} {114} (\bibinfo {year}
  {2023})},\ \Eprint {http://arxiv.org/abs/2209.05443} {arXiv:2209.05443
  [hep-ph]} \BibitemShut {NoStop}%
\bibitem [{\citenamefont {He}\ \emph {et~al.}(2024)\citenamefont {He},
  \citenamefont {Chu}, \citenamefont {Hua}, \citenamefont {Ji}, \citenamefont
  {Sch\"afer}, \citenamefont {Su}, \citenamefont {Wang}, \citenamefont {Yang},
  \citenamefont {Zhang},\ and\ \citenamefont
  {Zhang}}]{LatticePartonCollaborationLPC:2022myp}%
  \BibitemOpen
  \bibfield  {author} {\bibinfo {author} {\bibfnamefont {J.-C.}\ \bibnamefont
  {He}}, \bibinfo {author} {\bibfnamefont {M.-H.}\ \bibnamefont {Chu}},
  \bibinfo {author} {\bibfnamefont {J.}~\bibnamefont {Hua}}, \bibinfo {author}
  {\bibfnamefont {X.}~\bibnamefont {Ji}}, \bibinfo {author} {\bibfnamefont
  {A.}~\bibnamefont {Sch\"afer}}, \bibinfo {author} {\bibfnamefont
  {Y.}~\bibnamefont {Su}}, \bibinfo {author} {\bibfnamefont {W.}~\bibnamefont
  {Wang}}, \bibinfo {author} {\bibfnamefont {Y.-B.}\ \bibnamefont {Yang}},
  \bibinfo {author} {\bibfnamefont {J.-H.}\ \bibnamefont {Zhang}}, \ and\
  \bibinfo {author} {\bibfnamefont {Q.-A.}\ \bibnamefont {Zhang}} (\bibinfo
  {collaboration} {Lattice Parton Collaboration (LPC)}),\ }\href {\doibase
  10.1103/PhysRevD.109.114513} {\bibfield  {journal} {\bibinfo  {journal}
  {Phys. Rev. D}\ }\textbf {\bibinfo {volume} {109}},\ \bibinfo {pages}
  {114513} (\bibinfo {year} {2024})},\ \Eprint
  {http://arxiv.org/abs/2211.02340} {arXiv:2211.02340 [hep-lat]} \BibitemShut
  {NoStop}%
\bibitem [{\citenamefont {Rodini}\ and\ \citenamefont
  {Vladimirov}(2023)}]{Rodini:2022wic}%
  \BibitemOpen
  \bibfield  {author} {\bibinfo {author} {\bibfnamefont {S.}~\bibnamefont
  {Rodini}}\ and\ \bibinfo {author} {\bibfnamefont {A.}~\bibnamefont
  {Vladimirov}},\ }\href {\doibase 10.1007/JHEP09(2023)117} {\bibfield
  {journal} {\bibinfo  {journal} {JHEP}\ }\textbf {\bibinfo {volume} {09}},\
  \bibinfo {pages} {117} (\bibinfo {year} {2023})},\ \Eprint
  {http://arxiv.org/abs/2211.04494} {arXiv:2211.04494 [hep-ph]} \BibitemShut
  {NoStop}%
\bibitem [{\citenamefont {Shu}\ \emph {et~al.}(2023)\citenamefont {Shu},
  \citenamefont {Schlemmer}, \citenamefont {Sizmann}, \citenamefont
  {Vladimirov}, \citenamefont {Walter}, \citenamefont {Engelhardt},
  \citenamefont {Sch\"afer},\ and\ \citenamefont {Yang}}]{Shu:2023cot}%
  \BibitemOpen
  \bibfield  {author} {\bibinfo {author} {\bibfnamefont {H.-T.}\ \bibnamefont
  {Shu}}, \bibinfo {author} {\bibfnamefont {M.}~\bibnamefont {Schlemmer}},
  \bibinfo {author} {\bibfnamefont {T.}~\bibnamefont {Sizmann}}, \bibinfo
  {author} {\bibfnamefont {A.}~\bibnamefont {Vladimirov}}, \bibinfo {author}
  {\bibfnamefont {L.}~\bibnamefont {Walter}}, \bibinfo {author} {\bibfnamefont
  {M.}~\bibnamefont {Engelhardt}}, \bibinfo {author} {\bibfnamefont
  {A.}~\bibnamefont {Sch\"afer}}, \ and\ \bibinfo {author} {\bibfnamefont
  {Y.-B.}\ \bibnamefont {Yang}},\ }\href {\doibase 10.1103/PhysRevD.108.074519}
  {\bibfield  {journal} {\bibinfo  {journal} {Phys. Rev. D}\ }\textbf {\bibinfo
  {volume} {108}},\ \bibinfo {pages} {074519} (\bibinfo {year} {2023})},\
  \Eprint {http://arxiv.org/abs/2302.06502} {arXiv:2302.06502 [hep-lat]}
  \BibitemShut {NoStop}%
\bibitem [{\citenamefont {Chu}\ \emph {et~al.}(2024)\citenamefont {Chu} \emph
  {et~al.}}]{LatticeParton:2023xdl}%
  \BibitemOpen
  \bibfield  {author} {\bibinfo {author} {\bibfnamefont {M.-H.}\ \bibnamefont
  {Chu}} \emph {et~al.} (\bibinfo {collaboration} {Lattice Parton}),\ }\href
  {\doibase 10.1103/PhysRevD.109.L091503} {\bibfield  {journal} {\bibinfo
  {journal} {Phys. Rev. D}\ }\textbf {\bibinfo {volume} {109}},\ \bibinfo
  {pages} {L091503} (\bibinfo {year} {2024})},\ \Eprint
  {http://arxiv.org/abs/2302.09961} {arXiv:2302.09961 [hep-lat]} \BibitemShut
  {NoStop}%
\bibitem [{\citenamefont {del R\'\i{}o}\ and\ \citenamefont
  {Vladimirov}(2023)}]{delRio:2023pse}%
  \BibitemOpen
  \bibfield  {author} {\bibinfo {author} {\bibfnamefont {O.}~\bibnamefont {del
  R\'\i{}o}}\ and\ \bibinfo {author} {\bibfnamefont {A.}~\bibnamefont
  {Vladimirov}},\ }\href@noop {} {\  (\bibinfo {year} {2023})},\ \Eprint
  {http://arxiv.org/abs/2304.14440} {arXiv:2304.14440 [hep-ph]} \BibitemShut
  {NoStop}%
\bibitem [{\citenamefont {Chu}\ \emph {et~al.}(2023)\citenamefont {Chu} \emph
  {et~al.}}]{LatticePartonLPC:2023pdv}%
  \BibitemOpen
  \bibfield  {author} {\bibinfo {author} {\bibfnamefont {M.-H.}\ \bibnamefont
  {Chu}} \emph {et~al.} (\bibinfo {collaboration} {Lattice Parton (LPC)}),\
  }\href {\doibase 10.1007/JHEP08(2023)172} {\bibfield  {journal} {\bibinfo
  {journal} {JHEP}\ }\textbf {\bibinfo {volume} {08}},\ \bibinfo {pages} {172}
  (\bibinfo {year} {2023})},\ \Eprint {http://arxiv.org/abs/2306.06488}
  {arXiv:2306.06488 [hep-lat]} \BibitemShut {NoStop}%
\bibitem [{\citenamefont {Alexandrou}\ \emph {et~al.}(2023)\citenamefont
  {Alexandrou} \emph {et~al.}}]{Alexandrou:2023ucc}%
  \BibitemOpen
  \bibfield  {author} {\bibinfo {author} {\bibfnamefont {C.}~\bibnamefont
  {Alexandrou}} \emph {et~al.},\ }\href {\doibase 10.1103/PhysRevD.108.114503}
  {\bibfield  {journal} {\bibinfo  {journal} {Phys. Rev. D}\ }\textbf {\bibinfo
  {volume} {108}},\ \bibinfo {pages} {114503} (\bibinfo {year} {2023})},\
  \Eprint {http://arxiv.org/abs/2305.11824} {arXiv:2305.11824 [hep-lat]}
  \BibitemShut {NoStop}%
\bibitem [{\citenamefont {Avkhadiev}\ \emph {et~al.}(2023)\citenamefont
  {Avkhadiev}, \citenamefont {Shanahan}, \citenamefont {Wagman},\ and\
  \citenamefont {Zhao}}]{Avkhadiev:2023poz}%
  \BibitemOpen
  \bibfield  {author} {\bibinfo {author} {\bibfnamefont {A.}~\bibnamefont
  {Avkhadiev}}, \bibinfo {author} {\bibfnamefont {P.}~\bibnamefont {Shanahan}},
  \bibinfo {author} {\bibfnamefont {M.}~\bibnamefont {Wagman}}, \ and\ \bibinfo
  {author} {\bibfnamefont {Y.}~\bibnamefont {Zhao}},\ }\href@noop {} {\
  (\bibinfo {year} {2023})},\ \Eprint {http://arxiv.org/abs/2307.12359}
  {arXiv:2307.12359 [hep-lat]} \BibitemShut {NoStop}%
\bibitem [{\citenamefont {Zhao}(2023)}]{Zhao:2023ptv}%
  \BibitemOpen
  \bibfield  {author} {\bibinfo {author} {\bibfnamefont {Y.}~\bibnamefont
  {Zhao}},\ }\href@noop {} {\  (\bibinfo {year} {2023})},\ \Eprint
  {http://arxiv.org/abs/2311.01391} {arXiv:2311.01391 [hep-ph]} \BibitemShut
  {NoStop}%
\bibitem [{\citenamefont {Avkhadiev}\ \emph {et~al.}(2024)\citenamefont
  {Avkhadiev}, \citenamefont {Shanahan}, \citenamefont {Wagman},\ and\
  \citenamefont {Zhao}}]{Avkhadiev:2024mgd}%
  \BibitemOpen
  \bibfield  {author} {\bibinfo {author} {\bibfnamefont {A.}~\bibnamefont
  {Avkhadiev}}, \bibinfo {author} {\bibfnamefont {P.~E.}\ \bibnamefont
  {Shanahan}}, \bibinfo {author} {\bibfnamefont {M.~L.}\ \bibnamefont
  {Wagman}}, \ and\ \bibinfo {author} {\bibfnamefont {Y.}~\bibnamefont
  {Zhao}},\ }\href {\doibase 10.1103/PhysRevLett.132.231901} {\bibfield
  {journal} {\bibinfo  {journal} {Phys. Rev. Lett.}\ }\textbf {\bibinfo
  {volume} {132}},\ \bibinfo {pages} {231901} (\bibinfo {year} {2024})},\
  \Eprint {http://arxiv.org/abs/2402.06725} {arXiv:2402.06725 [hep-lat]}
  \BibitemShut {NoStop}%
\bibitem [{\citenamefont {Bollweg}\ \emph {et~al.}(2024)\citenamefont
  {Bollweg}, \citenamefont {Gao}, \citenamefont {Mukherjee},\ and\
  \citenamefont {Zhao}}]{Bollweg:2024zet}%
  \BibitemOpen
  \bibfield  {author} {\bibinfo {author} {\bibfnamefont {D.}~\bibnamefont
  {Bollweg}}, \bibinfo {author} {\bibfnamefont {X.}~\bibnamefont {Gao}},
  \bibinfo {author} {\bibfnamefont {S.}~\bibnamefont {Mukherjee}}, \ and\
  \bibinfo {author} {\bibfnamefont {Y.}~\bibnamefont {Zhao}},\ }\href {\doibase
  10.1016/j.physletb.2024.138617} {\bibfield  {journal} {\bibinfo  {journal}
  {Phys. Lett. B}\ }\textbf {\bibinfo {volume} {852}},\ \bibinfo {pages}
  {138617} (\bibinfo {year} {2024})},\ \Eprint
  {http://arxiv.org/abs/2403.00664} {arXiv:2403.00664 [hep-lat]} \BibitemShut
  {NoStop}%
\bibitem [{\citenamefont {Spanoudes}\ \emph {et~al.}(2024)\citenamefont
  {Spanoudes}, \citenamefont {Constantinou},\ and\ \citenamefont
  {Panagopoulos}}]{Spanoudes:2024kpb}%
  \BibitemOpen
  \bibfield  {author} {\bibinfo {author} {\bibfnamefont {G.}~\bibnamefont
  {Spanoudes}}, \bibinfo {author} {\bibfnamefont {M.}~\bibnamefont
  {Constantinou}}, \ and\ \bibinfo {author} {\bibfnamefont {H.}~\bibnamefont
  {Panagopoulos}},\ }\href {\doibase 10.1103/PhysRevD.109.114501} {\bibfield
  {journal} {\bibinfo  {journal} {Phys. Rev. D}\ }\textbf {\bibinfo {volume}
  {109}},\ \bibinfo {pages} {114501} (\bibinfo {year} {2024})},\ \Eprint
  {http://arxiv.org/abs/2401.01182} {arXiv:2401.01182 [hep-lat]} \BibitemShut
  {NoStop}%
\bibitem [{\citenamefont {Zhang}(2023)}]{Zhang:2023wea}%
  \BibitemOpen
  \bibfield  {author} {\bibinfo {author} {\bibfnamefont {J.-H.}\ \bibnamefont
  {Zhang}},\ }\href@noop {} {\  (\bibinfo {year} {2023})},\ \Eprint
  {http://arxiv.org/abs/2304.12481} {arXiv:2304.12481 [hep-ph]} \BibitemShut
  {NoStop}%
\bibitem [{\citenamefont {Jaarsma}\ \emph {et~al.}(2023)\citenamefont
  {Jaarsma}, \citenamefont {Rahn},\ and\ \citenamefont
  {Waalewijn}}]{Jaarsma:2023woo}%
  \BibitemOpen
  \bibfield  {author} {\bibinfo {author} {\bibfnamefont {M.}~\bibnamefont
  {Jaarsma}}, \bibinfo {author} {\bibfnamefont {R.}~\bibnamefont {Rahn}}, \
  and\ \bibinfo {author} {\bibfnamefont {W.~J.}\ \bibnamefont {Waalewijn}},\
  }\href {\doibase 10.1007/JHEP12(2023)014} {\bibfield  {journal} {\bibinfo
  {journal} {JHEP}\ }\textbf {\bibinfo {volume} {12}},\ \bibinfo {pages} {014}
  (\bibinfo {year} {2023})},\ \Eprint {http://arxiv.org/abs/2305.09716}
  {arXiv:2305.09716 [hep-ph]} \BibitemShut {NoStop}%
\bibitem [{\citenamefont {Cichy}\ and\ \citenamefont
  {Constantinou}(2019)}]{Cichy:2018mum}%
  \BibitemOpen
  \bibfield  {author} {\bibinfo {author} {\bibfnamefont {K.}~\bibnamefont
  {Cichy}}\ and\ \bibinfo {author} {\bibfnamefont {M.}~\bibnamefont
  {Constantinou}},\ }\href {\doibase 10.1155/2019/3036904} {\bibfield
  {journal} {\bibinfo  {journal} {Adv. High Energy Phys.}\ }\textbf {\bibinfo
  {volume} {2019}},\ \bibinfo {pages} {3036904} (\bibinfo {year} {2019})},\
  \Eprint {http://arxiv.org/abs/1811.07248} {arXiv:1811.07248 [hep-lat]}
  \BibitemShut {NoStop}%
\bibitem [{\citenamefont {Ji}\ \emph {et~al.}(2021{\natexlab{a}})\citenamefont
  {Ji}, \citenamefont {Liu}, \citenamefont {Liu}, \citenamefont {Zhang},\ and\
  \citenamefont {Zhao}}]{Ji:2020ect}%
  \BibitemOpen
  \bibfield  {author} {\bibinfo {author} {\bibfnamefont {X.}~\bibnamefont
  {Ji}}, \bibinfo {author} {\bibfnamefont {Y.-S.}\ \bibnamefont {Liu}},
  \bibinfo {author} {\bibfnamefont {Y.}~\bibnamefont {Liu}}, \bibinfo {author}
  {\bibfnamefont {J.-H.}\ \bibnamefont {Zhang}}, \ and\ \bibinfo {author}
  {\bibfnamefont {Y.}~\bibnamefont {Zhao}},\ }\href {\doibase
  10.1103/RevModPhys.93.035005} {\bibfield  {journal} {\bibinfo  {journal}
  {Rev. Mod. Phys.}\ }\textbf {\bibinfo {volume} {93}},\ \bibinfo {pages}
  {035005} (\bibinfo {year} {2021}{\natexlab{a}})},\ \Eprint
  {http://arxiv.org/abs/2004.03543} {arXiv:2004.03543 [hep-ph]} \BibitemShut
  {NoStop}%
\bibitem [{\citenamefont {Ji}\ \emph {et~al.}(2023)\citenamefont {Ji},
  \citenamefont {Liu},\ and\ \citenamefont {Su}}]{Ji:2023pba}%
  \BibitemOpen
  \bibfield  {author} {\bibinfo {author} {\bibfnamefont {X.}~\bibnamefont
  {Ji}}, \bibinfo {author} {\bibfnamefont {Y.}~\bibnamefont {Liu}}, \ and\
  \bibinfo {author} {\bibfnamefont {Y.}~\bibnamefont {Su}},\ }\href {\doibase
  10.1007/JHEP08(2023)037} {\bibfield  {journal} {\bibinfo  {journal} {JHEP}\
  }\textbf {\bibinfo {volume} {08}},\ \bibinfo {pages} {037} (\bibinfo {year}
  {2023})},\ \Eprint {http://arxiv.org/abs/2305.04416} {arXiv:2305.04416
  [hep-ph]} \BibitemShut {NoStop}%
\bibitem [{\citenamefont {Ji}\ \emph {et~al.}(2020)\citenamefont {Ji},
  \citenamefont {Liu},\ and\ \citenamefont {Liu}}]{Ji:2019ewn}%
  \BibitemOpen
  \bibfield  {author} {\bibinfo {author} {\bibfnamefont {X.}~\bibnamefont
  {Ji}}, \bibinfo {author} {\bibfnamefont {Y.}~\bibnamefont {Liu}}, \ and\
  \bibinfo {author} {\bibfnamefont {Y.-S.}\ \bibnamefont {Liu}},\ }\href
  {\doibase 10.1016/j.physletb.2020.135946} {\bibfield  {journal} {\bibinfo
  {journal} {Phys. Lett. B}\ }\textbf {\bibinfo {volume} {811}},\ \bibinfo
  {pages} {135946} (\bibinfo {year} {2020})},\ \Eprint
  {http://arxiv.org/abs/1911.03840} {arXiv:1911.03840 [hep-ph]} \BibitemShut
  {NoStop}%
\bibitem [{\citenamefont {Liu}\ and\ \citenamefont {Su}(2024)}]{Liu:2023onm}%
  \BibitemOpen
  \bibfield  {author} {\bibinfo {author} {\bibfnamefont {Y.}~\bibnamefont
  {Liu}}\ and\ \bibinfo {author} {\bibfnamefont {Y.}~\bibnamefont {Su}},\
  }\href {\doibase 10.1007/JHEP02(2024)204} {\bibfield  {journal} {\bibinfo
  {journal} {JHEP}\ }\textbf {\bibinfo {volume} {2024}},\ \bibinfo {pages}
  {204} (\bibinfo {year} {2024})},\ \Eprint {http://arxiv.org/abs/2311.06907}
  {arXiv:2311.06907 [hep-ph]} \BibitemShut {NoStop}%
\bibitem [{\citenamefont {Zhang}\ \emph {et~al.}(2023)\citenamefont {Zhang},
  \citenamefont {Holligan}, \citenamefont {Ji},\ and\ \citenamefont
  {Su}}]{Zhang:2023bxs}%
  \BibitemOpen
  \bibfield  {author} {\bibinfo {author} {\bibfnamefont {R.}~\bibnamefont
  {Zhang}}, \bibinfo {author} {\bibfnamefont {J.}~\bibnamefont {Holligan}},
  \bibinfo {author} {\bibfnamefont {X.}~\bibnamefont {Ji}}, \ and\ \bibinfo
  {author} {\bibfnamefont {Y.}~\bibnamefont {Su}},\ }\href@noop {} {\
  (\bibinfo {year} {2023})},\ \Eprint {http://arxiv.org/abs/2305.05212}
  {arXiv:2305.05212 [hep-lat]} \BibitemShut {NoStop}%
\bibitem [{\citenamefont {Moch}\ \emph {et~al.}(2004)\citenamefont {Moch},
  \citenamefont {Vermaseren},\ and\ \citenamefont {Vogt}}]{Moch:2004pa}%
  \BibitemOpen
  \bibfield  {author} {\bibinfo {author} {\bibfnamefont {S.}~\bibnamefont
  {Moch}}, \bibinfo {author} {\bibfnamefont {J.~A.~M.}\ \bibnamefont
  {Vermaseren}}, \ and\ \bibinfo {author} {\bibfnamefont {A.}~\bibnamefont
  {Vogt}},\ }\href {\doibase 10.1016/j.nuclphysb.2004.03.030} {\bibfield
  {journal} {\bibinfo  {journal} {Nucl. Phys. B}\ }\textbf {\bibinfo {volume}
  {688}},\ \bibinfo {pages} {101} (\bibinfo {year} {2004})},\ \Eprint
  {http://arxiv.org/abs/hep-ph/0403192} {arXiv:hep-ph/0403192} \BibitemShut
  {NoStop}%
\bibitem [{\citenamefont {Vogt}\ \emph {et~al.}(2004)\citenamefont {Vogt},
  \citenamefont {Moch},\ and\ \citenamefont {Vermaseren}}]{Vogt:2004mw}%
  \BibitemOpen
  \bibfield  {author} {\bibinfo {author} {\bibfnamefont {A.}~\bibnamefont
  {Vogt}}, \bibinfo {author} {\bibfnamefont {S.}~\bibnamefont {Moch}}, \ and\
  \bibinfo {author} {\bibfnamefont {J.~A.~M.}\ \bibnamefont {Vermaseren}},\
  }\href {\doibase 10.1016/j.nuclphysb.2004.04.024} {\bibfield  {journal}
  {\bibinfo  {journal} {Nucl. Phys. B}\ }\textbf {\bibinfo {volume} {691}},\
  \bibinfo {pages} {129} (\bibinfo {year} {2004})},\ \Eprint
  {http://arxiv.org/abs/hep-ph/0404111} {arXiv:hep-ph/0404111} \BibitemShut
  {NoStop}%
\bibitem [{\citenamefont {Bl\"umlein}\ \emph {et~al.}(2021)\citenamefont
  {Bl\"umlein}, \citenamefont {Marquard}, \citenamefont {Schneider},\ and\
  \citenamefont {Sch\"onwald}}]{Blumlein:2021enk}%
  \BibitemOpen
  \bibfield  {author} {\bibinfo {author} {\bibfnamefont {J.}~\bibnamefont
  {Bl\"umlein}}, \bibinfo {author} {\bibfnamefont {P.}~\bibnamefont
  {Marquard}}, \bibinfo {author} {\bibfnamefont {C.}~\bibnamefont {Schneider}},
  \ and\ \bibinfo {author} {\bibfnamefont {K.}~\bibnamefont {Sch\"onwald}},\
  }\href {\doibase 10.1016/j.nuclphysb.2021.115542} {\bibfield  {journal}
  {\bibinfo  {journal} {Nucl. Phys. B}\ }\textbf {\bibinfo {volume} {971}},\
  \bibinfo {pages} {115542} (\bibinfo {year} {2021})},\ \Eprint
  {http://arxiv.org/abs/2107.06267} {arXiv:2107.06267 [hep-ph]} \BibitemShut
  {NoStop}%
\bibitem [{\citenamefont {Moch}\ \emph {et~al.}(2017)\citenamefont {Moch},
  \citenamefont {Ruijl}, \citenamefont {Ueda}, \citenamefont {Vermaseren},\
  and\ \citenamefont {Vogt}}]{Moch:2017uml}%
  \BibitemOpen
  \bibfield  {author} {\bibinfo {author} {\bibfnamefont {S.}~\bibnamefont
  {Moch}}, \bibinfo {author} {\bibfnamefont {B.}~\bibnamefont {Ruijl}},
  \bibinfo {author} {\bibfnamefont {T.}~\bibnamefont {Ueda}}, \bibinfo {author}
  {\bibfnamefont {J.~A.~M.}\ \bibnamefont {Vermaseren}}, \ and\ \bibinfo
  {author} {\bibfnamefont {A.}~\bibnamefont {Vogt}},\ }\href {\doibase
  10.1007/JHEP10(2017)041} {\bibfield  {journal} {\bibinfo  {journal} {JHEP}\
  }\textbf {\bibinfo {volume} {10}},\ \bibinfo {pages} {041} (\bibinfo {year}
  {2017})},\ \Eprint {http://arxiv.org/abs/1707.08315} {arXiv:1707.08315
  [hep-ph]} \BibitemShut {NoStop}%
\bibitem [{\citenamefont {Falcioni}\ \emph
  {et~al.}(2023{\natexlab{a}})\citenamefont {Falcioni}, \citenamefont {Herzog},
  \citenamefont {Moch},\ and\ \citenamefont {Vogt}}]{Falcioni:2023vqq}%
  \BibitemOpen
  \bibfield  {author} {\bibinfo {author} {\bibfnamefont {G.}~\bibnamefont
  {Falcioni}}, \bibinfo {author} {\bibfnamefont {F.}~\bibnamefont {Herzog}},
  \bibinfo {author} {\bibfnamefont {S.}~\bibnamefont {Moch}}, \ and\ \bibinfo
  {author} {\bibfnamefont {A.}~\bibnamefont {Vogt}},\ }\href {\doibase
  10.1016/j.physletb.2023.138215} {\bibfield  {journal} {\bibinfo  {journal}
  {Phys. Lett. B}\ }\textbf {\bibinfo {volume} {846}},\ \bibinfo {pages}
  {138215} (\bibinfo {year} {2023}{\natexlab{a}})},\ \Eprint
  {http://arxiv.org/abs/2307.04158} {arXiv:2307.04158 [hep-ph]} \BibitemShut
  {NoStop}%
\bibitem [{\citenamefont {Falcioni}\ \emph
  {et~al.}(2023{\natexlab{b}})\citenamefont {Falcioni}, \citenamefont {Herzog},
  \citenamefont {Moch},\ and\ \citenamefont {Vogt}}]{Falcioni:2023luc}%
  \BibitemOpen
  \bibfield  {author} {\bibinfo {author} {\bibfnamefont {G.}~\bibnamefont
  {Falcioni}}, \bibinfo {author} {\bibfnamefont {F.}~\bibnamefont {Herzog}},
  \bibinfo {author} {\bibfnamefont {S.}~\bibnamefont {Moch}}, \ and\ \bibinfo
  {author} {\bibfnamefont {A.}~\bibnamefont {Vogt}},\ }\href {\doibase
  10.1016/j.physletb.2023.137944} {\bibfield  {journal} {\bibinfo  {journal}
  {Phys. Lett. B}\ }\textbf {\bibinfo {volume} {842}},\ \bibinfo {pages}
  {137944} (\bibinfo {year} {2023}{\natexlab{b}})},\ \Eprint
  {http://arxiv.org/abs/2302.07593} {arXiv:2302.07593 [hep-ph]} \BibitemShut
  {NoStop}%
\bibitem [{\citenamefont {Moch}\ \emph {et~al.}(2024)\citenamefont {Moch},
  \citenamefont {Ruijl}, \citenamefont {Ueda}, \citenamefont {Vermaseren},\
  and\ \citenamefont {Vogt}}]{Moch:2023tdj}%
  \BibitemOpen
  \bibfield  {author} {\bibinfo {author} {\bibfnamefont {S.}~\bibnamefont
  {Moch}}, \bibinfo {author} {\bibfnamefont {B.}~\bibnamefont {Ruijl}},
  \bibinfo {author} {\bibfnamefont {T.}~\bibnamefont {Ueda}}, \bibinfo {author}
  {\bibfnamefont {J.}~\bibnamefont {Vermaseren}}, \ and\ \bibinfo {author}
  {\bibfnamefont {A.}~\bibnamefont {Vogt}},\ }\href {\doibase
  10.1016/j.physletb.2024.138468} {\bibfield  {journal} {\bibinfo  {journal}
  {Phys. Lett. B}\ }\textbf {\bibinfo {volume} {849}},\ \bibinfo {pages}
  {138468} (\bibinfo {year} {2024})},\ \Eprint
  {http://arxiv.org/abs/2310.05744} {arXiv:2310.05744 [hep-ph]} \BibitemShut
  {NoStop}%
\bibitem [{\citenamefont {Moch}\ \emph {et~al.}(2022)\citenamefont {Moch},
  \citenamefont {Ruijl}, \citenamefont {Ueda}, \citenamefont {Vermaseren},\
  and\ \citenamefont {Vogt}}]{Moch:2021qrk}%
  \BibitemOpen
  \bibfield  {author} {\bibinfo {author} {\bibfnamefont {S.}~\bibnamefont
  {Moch}}, \bibinfo {author} {\bibfnamefont {B.}~\bibnamefont {Ruijl}},
  \bibinfo {author} {\bibfnamefont {T.}~\bibnamefont {Ueda}}, \bibinfo {author}
  {\bibfnamefont {J.~A.~M.}\ \bibnamefont {Vermaseren}}, \ and\ \bibinfo
  {author} {\bibfnamefont {A.}~\bibnamefont {Vogt}},\ }\href {\doibase
  10.1016/j.physletb.2021.136853} {\bibfield  {journal} {\bibinfo  {journal}
  {Phys. Lett. B}\ }\textbf {\bibinfo {volume} {825}},\ \bibinfo {pages}
  {136853} (\bibinfo {year} {2022})},\ \Eprint
  {http://arxiv.org/abs/2111.15561} {arXiv:2111.15561 [hep-ph]} \BibitemShut
  {NoStop}%
\bibitem [{\citenamefont {Falcioni}\ \emph
  {et~al.}(2024{\natexlab{a}})\citenamefont {Falcioni}, \citenamefont {Herzog},
  \citenamefont {Moch}, \citenamefont {Vermaseren},\ and\ \citenamefont
  {Vogt}}]{Falcioni:2023tzp}%
  \BibitemOpen
  \bibfield  {author} {\bibinfo {author} {\bibfnamefont {G.}~\bibnamefont
  {Falcioni}}, \bibinfo {author} {\bibfnamefont {F.}~\bibnamefont {Herzog}},
  \bibinfo {author} {\bibfnamefont {S.}~\bibnamefont {Moch}}, \bibinfo {author}
  {\bibfnamefont {J.}~\bibnamefont {Vermaseren}}, \ and\ \bibinfo {author}
  {\bibfnamefont {A.}~\bibnamefont {Vogt}},\ }\href {\doibase
  10.1016/j.physletb.2023.138351} {\bibfield  {journal} {\bibinfo  {journal}
  {Phys. Lett. B}\ }\textbf {\bibinfo {volume} {848}},\ \bibinfo {pages}
  {138351} (\bibinfo {year} {2024}{\natexlab{a}})},\ \Eprint
  {http://arxiv.org/abs/2310.01245} {arXiv:2310.01245 [hep-ph]} \BibitemShut
  {NoStop}%
\bibitem [{\citenamefont {Gehrmann}\ \emph
  {et~al.}(2024{\natexlab{a}})\citenamefont {Gehrmann}, \citenamefont {von
  Manteuffel}, \citenamefont {Sotnikov},\ and\ \citenamefont
  {Yang}}]{Gehrmann:2023cqm}%
  \BibitemOpen
  \bibfield  {author} {\bibinfo {author} {\bibfnamefont {T.}~\bibnamefont
  {Gehrmann}}, \bibinfo {author} {\bibfnamefont {A.}~\bibnamefont {von
  Manteuffel}}, \bibinfo {author} {\bibfnamefont {V.}~\bibnamefont {Sotnikov}},
  \ and\ \bibinfo {author} {\bibfnamefont {T.-Z.}\ \bibnamefont {Yang}},\
  }\href {\doibase 10.1007/JHEP01(2024)029} {\bibfield  {journal} {\bibinfo
  {journal} {JHEP}\ }\textbf {\bibinfo {volume} {01}},\ \bibinfo {pages} {029}
  (\bibinfo {year} {2024}{\natexlab{a}})},\ \Eprint
  {http://arxiv.org/abs/2308.07958} {arXiv:2308.07958 [hep-ph]} \BibitemShut
  {NoStop}%
\bibitem [{\citenamefont {Gehrmann}\ \emph
  {et~al.}(2024{\natexlab{b}})\citenamefont {Gehrmann}, \citenamefont {von
  Manteuffel}, \citenamefont {Sotnikov},\ and\ \citenamefont
  {Yang}}]{Gehrmann:2023iah}%
  \BibitemOpen
  \bibfield  {author} {\bibinfo {author} {\bibfnamefont {T.}~\bibnamefont
  {Gehrmann}}, \bibinfo {author} {\bibfnamefont {A.}~\bibnamefont {von
  Manteuffel}}, \bibinfo {author} {\bibfnamefont {V.}~\bibnamefont {Sotnikov}},
  \ and\ \bibinfo {author} {\bibfnamefont {T.-Z.}\ \bibnamefont {Yang}},\
  }\href {\doibase 10.1016/j.physletb.2023.138427} {\bibfield  {journal}
  {\bibinfo  {journal} {Phys. Lett. B}\ }\textbf {\bibinfo {volume} {849}},\
  \bibinfo {pages} {138427} (\bibinfo {year} {2024}{\natexlab{b}})},\ \Eprint
  {http://arxiv.org/abs/2310.12240} {arXiv:2310.12240 [hep-ph]} \BibitemShut
  {NoStop}%
\bibitem [{\citenamefont {Falcioni}\ \emph
  {et~al.}(2024{\natexlab{b}})\citenamefont {Falcioni}, \citenamefont {Herzog},
  \citenamefont {Moch}, \citenamefont {Pelloni},\ and\ \citenamefont
  {Vogt}}]{Falcioni:2024qpd}%
  \BibitemOpen
  \bibfield  {author} {\bibinfo {author} {\bibfnamefont {G.}~\bibnamefont
  {Falcioni}}, \bibinfo {author} {\bibfnamefont {F.}~\bibnamefont {Herzog}},
  \bibinfo {author} {\bibfnamefont {S.}~\bibnamefont {Moch}}, \bibinfo {author}
  {\bibfnamefont {A.}~\bibnamefont {Pelloni}}, \ and\ \bibinfo {author}
  {\bibfnamefont {A.}~\bibnamefont {Vogt}},\ }\href@noop {} {\  (\bibinfo
  {year} {2024}{\natexlab{b}})},\ \Eprint {http://arxiv.org/abs/2410.08089}
  {arXiv:2410.08089 [hep-ph]} \BibitemShut {NoStop}%
\bibitem [{\citenamefont {Ma}\ and\ \citenamefont {Qiu}(2018)}]{Ma:2014jla}%
  \BibitemOpen
  \bibfield  {author} {\bibinfo {author} {\bibfnamefont {Y.-Q.}\ \bibnamefont
  {Ma}}\ and\ \bibinfo {author} {\bibfnamefont {J.-W.}\ \bibnamefont {Qiu}},\
  }\href {\doibase 10.1103/PhysRevD.98.074021} {\bibfield  {journal} {\bibinfo
  {journal} {Phys. Rev. D}\ }\textbf {\bibinfo {volume} {98}},\ \bibinfo
  {pages} {074021} (\bibinfo {year} {2018})},\ \Eprint
  {http://arxiv.org/abs/1404.6860} {arXiv:1404.6860 [hep-ph]} \BibitemShut
  {NoStop}%
\bibitem [{\citenamefont {Li}\ \emph {et~al.}(2021)\citenamefont {Li},
  \citenamefont {Ma},\ and\ \citenamefont {Qiu}}]{Li:2020xml}%
  \BibitemOpen
  \bibfield  {author} {\bibinfo {author} {\bibfnamefont {Z.-Y.}\ \bibnamefont
  {Li}}, \bibinfo {author} {\bibfnamefont {Y.-Q.}\ \bibnamefont {Ma}}, \ and\
  \bibinfo {author} {\bibfnamefont {J.-W.}\ \bibnamefont {Qiu}},\ }\href
  {\doibase 10.1103/PhysRevLett.126.072001} {\bibfield  {journal} {\bibinfo
  {journal} {Phys. Rev. Lett.}\ }\textbf {\bibinfo {volume} {126}},\ \bibinfo
  {pages} {072001} (\bibinfo {year} {2021})},\ \Eprint
  {http://arxiv.org/abs/2006.12370} {arXiv:2006.12370 [hep-ph]} \BibitemShut
  {NoStop}%
\bibitem [{\citenamefont {Chen}\ \emph {et~al.}(2021)\citenamefont {Chen},
  \citenamefont {Wang},\ and\ \citenamefont {Zhu}}]{Chen:2020ody}%
  \BibitemOpen
  \bibfield  {author} {\bibinfo {author} {\bibfnamefont {L.-B.}\ \bibnamefont
  {Chen}}, \bibinfo {author} {\bibfnamefont {W.}~\bibnamefont {Wang}}, \ and\
  \bibinfo {author} {\bibfnamefont {R.}~\bibnamefont {Zhu}},\ }\href {\doibase
  10.1103/PhysRevLett.126.072002} {\bibfield  {journal} {\bibinfo  {journal}
  {Phys. Rev. Lett.}\ }\textbf {\bibinfo {volume} {126}},\ \bibinfo {pages}
  {072002} (\bibinfo {year} {2021})},\ \Eprint
  {http://arxiv.org/abs/2006.14825} {arXiv:2006.14825 [hep-ph]} \BibitemShut
  {NoStop}%
\bibitem [{\citenamefont {Cheng}\ \emph {et~al.}(2024)\citenamefont {Cheng},
  \citenamefont {Huang}, \citenamefont {Li},\ and\ \citenamefont
  {Ma}}]{Cheng:2024wyu}%
  \BibitemOpen
  \bibfield  {author} {\bibinfo {author} {\bibfnamefont {C.}~\bibnamefont
  {Cheng}}, \bibinfo {author} {\bibfnamefont {L.-H.}\ \bibnamefont {Huang}},
  \bibinfo {author} {\bibfnamefont {X.}~\bibnamefont {Li}}, \ and\ \bibinfo
  {author} {\bibfnamefont {Y.-Q.}\ \bibnamefont {Ma}},\ }\href@noop {} {\
  (\bibinfo {year} {2024})},\ \Eprint {http://arxiv.org/abs/2410.05141}
  {arXiv:2410.05141 [hep-ph]} \BibitemShut {NoStop}%
\bibitem [{\citenamefont {Ji}\ and\ \citenamefont {Musolf}(1991)}]{Ji:1991pr}%
  \BibitemOpen
  \bibfield  {author} {\bibinfo {author} {\bibfnamefont {X.-D.}\ \bibnamefont
  {Ji}}\ and\ \bibinfo {author} {\bibfnamefont {M.~J.}\ \bibnamefont
  {Musolf}},\ }\href {\doibase 10.1016/0370-2693(91)91916-J} {\bibfield
  {journal} {\bibinfo  {journal} {Phys. Lett. B}\ }\textbf {\bibinfo {volume}
  {257}},\ \bibinfo {pages} {409} (\bibinfo {year} {1991})}\BibitemShut
  {NoStop}%
\bibitem [{\citenamefont {Chetyrkin}\ and\ \citenamefont
  {Grozin}(2003)}]{Chetyrkin:2003vi}%
  \BibitemOpen
  \bibfield  {author} {\bibinfo {author} {\bibfnamefont {K.~G.}\ \bibnamefont
  {Chetyrkin}}\ and\ \bibinfo {author} {\bibfnamefont {A.~G.}\ \bibnamefont
  {Grozin}},\ }\href {\doibase 10.1016/S0550-3213(03)00490-5} {\bibfield
  {journal} {\bibinfo  {journal} {Nucl. Phys. B}\ }\textbf {\bibinfo {volume}
  {666}},\ \bibinfo {pages} {289} (\bibinfo {year} {2003})},\ \Eprint
  {http://arxiv.org/abs/hep-ph/0303113} {arXiv:hep-ph/0303113} \BibitemShut
  {NoStop}%
\bibitem [{\citenamefont {Braun}\ \emph {et~al.}(2020)\citenamefont {Braun},
  \citenamefont {Chetyrkin},\ and\ \citenamefont {Kniehl}}]{Braun:2020ymy}%
  \BibitemOpen
  \bibfield  {author} {\bibinfo {author} {\bibfnamefont {V.~M.}\ \bibnamefont
  {Braun}}, \bibinfo {author} {\bibfnamefont {K.~G.}\ \bibnamefont
  {Chetyrkin}}, \ and\ \bibinfo {author} {\bibfnamefont {B.~A.}\ \bibnamefont
  {Kniehl}},\ }\href {\doibase 10.1007/JHEP07(2020)161} {\bibfield  {journal}
  {\bibinfo  {journal} {JHEP}\ }\textbf {\bibinfo {volume} {07}},\ \bibinfo
  {pages} {161} (\bibinfo {year} {2020})},\ \Eprint
  {http://arxiv.org/abs/2004.01043} {arXiv:2004.01043 [hep-ph]} \BibitemShut
  {NoStop}%
\bibitem [{\citenamefont {Grozin}(2023{\natexlab{a}})}]{Grozin:2023dlk}%
  \BibitemOpen
  \bibfield  {author} {\bibinfo {author} {\bibfnamefont {A.}~\bibnamefont
  {Grozin}},\ }\href@noop {} {\  (\bibinfo {year} {2023}{\natexlab{a}})},\
  \Eprint {http://arxiv.org/abs/2311.09894} {arXiv:2311.09894 [hep-ph]}
  \BibitemShut {NoStop}%
\bibitem [{\citenamefont {Manohar}\ and\ \citenamefont
  {Stewart}(2007)}]{Manohar:2006nz}%
  \BibitemOpen
  \bibfield  {author} {\bibinfo {author} {\bibfnamefont {A.~V.}\ \bibnamefont
  {Manohar}}\ and\ \bibinfo {author} {\bibfnamefont {I.~W.}\ \bibnamefont
  {Stewart}},\ }\href {\doibase 10.1103/PhysRevD.76.074002} {\bibfield
  {journal} {\bibinfo  {journal} {Phys. Rev. D}\ }\textbf {\bibinfo {volume}
  {76}},\ \bibinfo {pages} {074002} (\bibinfo {year} {2007})},\ \Eprint
  {http://arxiv.org/abs/hep-ph/0605001} {arXiv:hep-ph/0605001} \BibitemShut
  {NoStop}%
\bibitem [{\citenamefont {Korchemsky}(1989)}]{Korchemsky:1988si}%
  \BibitemOpen
  \bibfield  {author} {\bibinfo {author} {\bibfnamefont {G.~P.}\ \bibnamefont
  {Korchemsky}},\ }\href {\doibase 10.1142/S0217732389001453} {\bibfield
  {journal} {\bibinfo  {journal} {Mod. Phys. Lett. A}\ }\textbf {\bibinfo
  {volume} {4}},\ \bibinfo {pages} {1257} (\bibinfo {year} {1989})}\BibitemShut
  {NoStop}%
\bibitem [{\citenamefont {Berger}(2002)}]{Berger:2002sv}%
  \BibitemOpen
  \bibfield  {author} {\bibinfo {author} {\bibfnamefont {C.~F.}\ \bibnamefont
  {Berger}},\ }\href {\doibase 10.1103/PhysRevD.66.116002} {\bibfield
  {journal} {\bibinfo  {journal} {Phys. Rev. D}\ }\textbf {\bibinfo {volume}
  {66}},\ \bibinfo {pages} {116002} (\bibinfo {year} {2002})},\ \Eprint
  {http://arxiv.org/abs/hep-ph/0209107} {arXiv:hep-ph/0209107} \BibitemShut
  {NoStop}%
\bibitem [{\citenamefont {Ji}\ \emph {et~al.}(2005)\citenamefont {Ji},
  \citenamefont {Ma},\ and\ \citenamefont {Yuan}}]{Ji:2004hz}%
  \BibitemOpen
  \bibfield  {author} {\bibinfo {author} {\bibfnamefont {X.-d.}\ \bibnamefont
  {Ji}}, \bibinfo {author} {\bibfnamefont {J.-P.}\ \bibnamefont {Ma}}, \ and\
  \bibinfo {author} {\bibfnamefont {F.}~\bibnamefont {Yuan}},\ }\href {\doibase
  10.1016/j.physletb.2005.02.019} {\bibfield  {journal} {\bibinfo  {journal}
  {Phys. Lett. B}\ }\textbf {\bibinfo {volume} {610}},\ \bibinfo {pages} {247}
  (\bibinfo {year} {2005})},\ \Eprint {http://arxiv.org/abs/hep-ph/0411382}
  {arXiv:hep-ph/0411382} \BibitemShut {NoStop}%
\bibitem [{\citenamefont {Catani}\ \emph {et~al.}(1996)\citenamefont {Catani},
  \citenamefont {Mangano}, \citenamefont {Nason},\ and\ \citenamefont
  {Trentadue}}]{Catani:1996yz}%
  \BibitemOpen
  \bibfield  {author} {\bibinfo {author} {\bibfnamefont {S.}~\bibnamefont
  {Catani}}, \bibinfo {author} {\bibfnamefont {M.~L.}\ \bibnamefont {Mangano}},
  \bibinfo {author} {\bibfnamefont {P.}~\bibnamefont {Nason}}, \ and\ \bibinfo
  {author} {\bibfnamefont {L.}~\bibnamefont {Trentadue}},\ }\href {\doibase
  10.1016/0550-3213(96)00399-9} {\bibfield  {journal} {\bibinfo  {journal}
  {Nucl. Phys. B}\ }\textbf {\bibinfo {volume} {478}},\ \bibinfo {pages} {273}
  (\bibinfo {year} {1996})},\ \Eprint {http://arxiv.org/abs/hep-ph/9604351}
  {arXiv:hep-ph/9604351} \BibitemShut {NoStop}%
\bibitem [{\citenamefont {Ji}\ \emph {et~al.}(2021{\natexlab{b}})\citenamefont
  {Ji}, \citenamefont {Liu}, \citenamefont {Sch\"afer}, \citenamefont {Wang},
  \citenamefont {Yang}, \citenamefont {Zhang},\ and\ \citenamefont
  {Zhao}}]{Ji:2020brr}%
  \BibitemOpen
  \bibfield  {author} {\bibinfo {author} {\bibfnamefont {X.}~\bibnamefont
  {Ji}}, \bibinfo {author} {\bibfnamefont {Y.}~\bibnamefont {Liu}}, \bibinfo
  {author} {\bibfnamefont {A.}~\bibnamefont {Sch\"afer}}, \bibinfo {author}
  {\bibfnamefont {W.}~\bibnamefont {Wang}}, \bibinfo {author} {\bibfnamefont
  {Y.-B.}\ \bibnamefont {Yang}}, \bibinfo {author} {\bibfnamefont {J.-H.}\
  \bibnamefont {Zhang}}, \ and\ \bibinfo {author} {\bibfnamefont
  {Y.}~\bibnamefont {Zhao}},\ }\href {\doibase 10.1016/j.nuclphysb.2021.115311}
  {\bibfield  {journal} {\bibinfo  {journal} {Nucl. Phys. B}\ }\textbf
  {\bibinfo {volume} {964}},\ \bibinfo {pages} {115311} (\bibinfo {year}
  {2021}{\natexlab{b}})},\ \Eprint {http://arxiv.org/abs/2008.03886}
  {arXiv:2008.03886 [hep-ph]} \BibitemShut {NoStop}%
\bibitem [{\citenamefont {Huo}\ \emph {et~al.}(2021)\citenamefont {Huo} \emph
  {et~al.}}]{LatticePartonCollaborationLPC:2021xdx}%
  \BibitemOpen
  \bibfield  {author} {\bibinfo {author} {\bibfnamefont {Y.-K.}\ \bibnamefont
  {Huo}} \emph {et~al.} (\bibinfo {collaboration} {Lattice Parton Collaboration
  (LPC)}),\ }\href {\doibase 10.1016/j.nuclphysb.2021.115443} {\bibfield
  {journal} {\bibinfo  {journal} {Nucl. Phys. B}\ }\textbf {\bibinfo {volume}
  {969}},\ \bibinfo {pages} {115443} (\bibinfo {year} {2021})},\ \Eprint
  {http://arxiv.org/abs/2103.02965} {arXiv:2103.02965 [hep-lat]} \BibitemShut
  {NoStop}%
\bibitem [{\citenamefont {Abbate}\ \emph {et~al.}(2011)\citenamefont {Abbate},
  \citenamefont {Fickinger}, \citenamefont {Hoang}, \citenamefont {Mateu},\
  and\ \citenamefont {Stewart}}]{Abbate:2010xh}%
  \BibitemOpen
  \bibfield  {author} {\bibinfo {author} {\bibfnamefont {R.}~\bibnamefont
  {Abbate}}, \bibinfo {author} {\bibfnamefont {M.}~\bibnamefont {Fickinger}},
  \bibinfo {author} {\bibfnamefont {A.~H.}\ \bibnamefont {Hoang}}, \bibinfo
  {author} {\bibfnamefont {V.}~\bibnamefont {Mateu}}, \ and\ \bibinfo {author}
  {\bibfnamefont {I.~W.}\ \bibnamefont {Stewart}},\ }\href {\doibase
  10.1103/PhysRevD.83.074021} {\bibfield  {journal} {\bibinfo  {journal} {Phys.
  Rev. D}\ }\textbf {\bibinfo {volume} {83}},\ \bibinfo {pages} {074021}
  (\bibinfo {year} {2011})},\ \Eprint {http://arxiv.org/abs/1006.3080}
  {arXiv:1006.3080 [hep-ph]} \BibitemShut {NoStop}%
\bibitem [{\citenamefont {Hoang}\ \emph {et~al.}(2015)\citenamefont {Hoang},
  \citenamefont {Kolodrubetz}, \citenamefont {Mateu},\ and\ \citenamefont
  {Stewart}}]{Hoang:2014wka}%
  \BibitemOpen
  \bibfield  {author} {\bibinfo {author} {\bibfnamefont {A.~H.}\ \bibnamefont
  {Hoang}}, \bibinfo {author} {\bibfnamefont {D.~W.}\ \bibnamefont
  {Kolodrubetz}}, \bibinfo {author} {\bibfnamefont {V.}~\bibnamefont {Mateu}},
  \ and\ \bibinfo {author} {\bibfnamefont {I.~W.}\ \bibnamefont {Stewart}},\
  }\href {\doibase 10.1103/PhysRevD.91.094017} {\bibfield  {journal} {\bibinfo
  {journal} {Phys. Rev. D}\ }\textbf {\bibinfo {volume} {91}},\ \bibinfo
  {pages} {094017} (\bibinfo {year} {2015})},\ \Eprint
  {http://arxiv.org/abs/1411.6633} {arXiv:1411.6633 [hep-ph]} \BibitemShut
  {NoStop}%
\bibitem [{\citenamefont {Gao}\ \emph {et~al.}(2020)\citenamefont {Gao},
  \citenamefont {Jin}, \citenamefont {Kallidonis}, \citenamefont {Karthik},
  \citenamefont {Mukherjee}, \citenamefont {Petreczky}, \citenamefont
  {Shugert}, \citenamefont {Syritsyn},\ and\ \citenamefont
  {Zhao}}]{Gao:2020ito}%
  \BibitemOpen
  \bibfield  {author} {\bibinfo {author} {\bibfnamefont {X.}~\bibnamefont
  {Gao}}, \bibinfo {author} {\bibfnamefont {L.}~\bibnamefont {Jin}}, \bibinfo
  {author} {\bibfnamefont {C.}~\bibnamefont {Kallidonis}}, \bibinfo {author}
  {\bibfnamefont {N.}~\bibnamefont {Karthik}}, \bibinfo {author} {\bibfnamefont
  {S.}~\bibnamefont {Mukherjee}}, \bibinfo {author} {\bibfnamefont
  {P.}~\bibnamefont {Petreczky}}, \bibinfo {author} {\bibfnamefont
  {C.}~\bibnamefont {Shugert}}, \bibinfo {author} {\bibfnamefont
  {S.}~\bibnamefont {Syritsyn}}, \ and\ \bibinfo {author} {\bibfnamefont
  {Y.}~\bibnamefont {Zhao}},\ }\href {\doibase 10.1103/PhysRevD.102.094513}
  {\bibfield  {journal} {\bibinfo  {journal} {Phys. Rev. D}\ }\textbf {\bibinfo
  {volume} {102}},\ \bibinfo {pages} {094513} (\bibinfo {year} {2020})},\
  \Eprint {http://arxiv.org/abs/2007.06590} {arXiv:2007.06590 [hep-lat]}
  \BibitemShut {NoStop}%
\bibitem [{\citenamefont {Gao}\ \emph {et~al.}(2023{\natexlab{d}})\citenamefont
  {Gao}, \citenamefont {Hanlon}, \citenamefont {Mukherjee}, \citenamefont
  {Petreczky}, \citenamefont {Scior}, \citenamefont {Syritsyn},\ and\
  \citenamefont {Zhao}}]{Gao:2022ytj}%
  \BibitemOpen
  \bibfield  {author} {\bibinfo {author} {\bibfnamefont {X.}~\bibnamefont
  {Gao}}, \bibinfo {author} {\bibfnamefont {A.~D.}\ \bibnamefont {Hanlon}},
  \bibinfo {author} {\bibfnamefont {S.}~\bibnamefont {Mukherjee}}, \bibinfo
  {author} {\bibfnamefont {P.}~\bibnamefont {Petreczky}}, \bibinfo {author}
  {\bibfnamefont {P.}~\bibnamefont {Scior}}, \bibinfo {author} {\bibfnamefont
  {S.}~\bibnamefont {Syritsyn}}, \ and\ \bibinfo {author} {\bibfnamefont
  {Y.}~\bibnamefont {Zhao}},\ }\href {\doibase 10.22323/1.430.0104} {\bibfield
  {journal} {\bibinfo  {journal} {PoS}\ }\textbf {\bibinfo {volume}
  {LATTICE2022}},\ \bibinfo {pages} {104} (\bibinfo {year}
  {2023}{\natexlab{d}})}\BibitemShut {NoStop}%
\bibitem [{\citenamefont {Beneke}\ and\ \citenamefont
  {Braun}(1994)}]{Beneke:1994sw}%
  \BibitemOpen
  \bibfield  {author} {\bibinfo {author} {\bibfnamefont {M.}~\bibnamefont
  {Beneke}}\ and\ \bibinfo {author} {\bibfnamefont {V.~M.}\ \bibnamefont
  {Braun}},\ }\href {\doibase 10.1016/0550-3213(94)90314-X} {\bibfield
  {journal} {\bibinfo  {journal} {Nucl. Phys. B}\ }\textbf {\bibinfo {volume}
  {426}},\ \bibinfo {pages} {301} (\bibinfo {year} {1994})},\ \Eprint
  {http://arxiv.org/abs/hep-ph/9402364} {arXiv:hep-ph/9402364} \BibitemShut
  {NoStop}%
\bibitem [{\citenamefont {Beneke}(1999)}]{Beneke:1998ui}%
  \BibitemOpen
  \bibfield  {author} {\bibinfo {author} {\bibfnamefont {M.}~\bibnamefont
  {Beneke}},\ }\href {\doibase 10.1016/S0370-1573(98)00130-6} {\bibfield
  {journal} {\bibinfo  {journal} {Phys. Rept.}\ }\textbf {\bibinfo {volume}
  {317}},\ \bibinfo {pages} {1} (\bibinfo {year} {1999})},\ \Eprint
  {http://arxiv.org/abs/hep-ph/9807443} {arXiv:hep-ph/9807443} \BibitemShut
  {NoStop}%
\bibitem [{\citenamefont {Beneke}\ and\ \citenamefont
  {Braun}(2000)}]{Beneke:2000kc}%
  \BibitemOpen
  \bibfield  {author} {\bibinfo {author} {\bibfnamefont {M.}~\bibnamefont
  {Beneke}}\ and\ \bibinfo {author} {\bibfnamefont {V.~M.}\ \bibnamefont
  {Braun}},\ }\href {\doibase 10.1142/9789812810458_0036} {\ ,\ \bibinfo
  {pages} {1719} (\bibinfo {year} {2000})},\ \Eprint
  {http://arxiv.org/abs/hep-ph/0010208} {arXiv:hep-ph/0010208} \BibitemShut
  {NoStop}%
\bibitem [{\citenamefont {Shifman}(2015)}]{Shifman:2013uka}%
  \BibitemOpen
  \bibfield  {author} {\bibinfo {author} {\bibfnamefont {M.}~\bibnamefont
  {Shifman}},\ }\href {\doibase 10.1142/S0217751X15430010} {\bibfield
  {journal} {\bibinfo  {journal} {Int. J. Mod. Phys. A}\ }\textbf {\bibinfo
  {volume} {30}},\ \bibinfo {pages} {1543001} (\bibinfo {year} {2015})},\
  \Eprint {http://arxiv.org/abs/1310.1966} {arXiv:1310.1966 [hep-th]}
  \BibitemShut {NoStop}%
\bibitem [{\citenamefont {Braun}\ \emph {et~al.}(2019)\citenamefont {Braun},
  \citenamefont {Vladimirov},\ and\ \citenamefont {Zhang}}]{Braun:2018brg}%
  \BibitemOpen
  \bibfield  {author} {\bibinfo {author} {\bibfnamefont {V.~M.}\ \bibnamefont
  {Braun}}, \bibinfo {author} {\bibfnamefont {A.}~\bibnamefont {Vladimirov}}, \
  and\ \bibinfo {author} {\bibfnamefont {J.-H.}\ \bibnamefont {Zhang}},\ }\href
  {\doibase 10.1103/PhysRevD.99.014013} {\bibfield  {journal} {\bibinfo
  {journal} {Phys. Rev. D}\ }\textbf {\bibinfo {volume} {99}},\ \bibinfo
  {pages} {014013} (\bibinfo {year} {2019})},\ \Eprint
  {http://arxiv.org/abs/1810.00048} {arXiv:1810.00048 [hep-ph]} \BibitemShut
  {NoStop}%
\bibitem [{\citenamefont {Pineda}(2001)}]{Pineda:2001zq}%
  \BibitemOpen
  \bibfield  {author} {\bibinfo {author} {\bibfnamefont {A.}~\bibnamefont
  {Pineda}},\ }\href {\doibase 10.1088/1126-6708/2001/06/022} {\bibfield
  {journal} {\bibinfo  {journal} {JHEP}\ }\textbf {\bibinfo {volume} {06}},\
  \bibinfo {pages} {022} (\bibinfo {year} {2001})},\ \Eprint
  {http://arxiv.org/abs/hep-ph/0105008} {arXiv:hep-ph/0105008} \BibitemShut
  {NoStop}%
\bibitem [{\citenamefont {Pineda}(2003)}]{Pineda:2002se}%
  \BibitemOpen
  \bibfield  {author} {\bibinfo {author} {\bibfnamefont {A.}~\bibnamefont
  {Pineda}},\ }\href {\doibase 10.1088/0954-3899/29/2/313} {\bibfield
  {journal} {\bibinfo  {journal} {J. Phys. G}\ }\textbf {\bibinfo {volume}
  {29}},\ \bibinfo {pages} {371} (\bibinfo {year} {2003})},\ \Eprint
  {http://arxiv.org/abs/hep-ph/0208031} {arXiv:hep-ph/0208031} \BibitemShut
  {NoStop}%
\bibitem [{\citenamefont {Beneke}\ \emph {et~al.}(1994)\citenamefont {Beneke},
  \citenamefont {Braun},\ and\ \citenamefont {Zakharov}}]{Beneke:1994bc}%
  \BibitemOpen
  \bibfield  {author} {\bibinfo {author} {\bibfnamefont {M.}~\bibnamefont
  {Beneke}}, \bibinfo {author} {\bibfnamefont {V.~M.}\ \bibnamefont {Braun}}, \
  and\ \bibinfo {author} {\bibfnamefont {V.~I.}\ \bibnamefont {Zakharov}},\
  }\href {\doibase 10.1103/PhysRevLett.73.3058} {\bibfield  {journal} {\bibinfo
   {journal} {Phys. Rev. Lett.}\ }\textbf {\bibinfo {volume} {73}},\ \bibinfo
  {pages} {3058} (\bibinfo {year} {1994})},\ \Eprint
  {http://arxiv.org/abs/hep-ph/9405304} {arXiv:hep-ph/9405304} \BibitemShut
  {NoStop}%
\bibitem [{\citenamefont {Bigi}\ \emph {et~al.}(1994)\citenamefont {Bigi},
  \citenamefont {Shifman}, \citenamefont {Uraltsev},\ and\ \citenamefont
  {Vainshtein}}]{Bigi:1994em}%
  \BibitemOpen
  \bibfield  {author} {\bibinfo {author} {\bibfnamefont {I.~I.~Y.}\
  \bibnamefont {Bigi}}, \bibinfo {author} {\bibfnamefont {M.~A.}\ \bibnamefont
  {Shifman}}, \bibinfo {author} {\bibfnamefont {N.~G.}\ \bibnamefont
  {Uraltsev}}, \ and\ \bibinfo {author} {\bibfnamefont {A.~I.}\ \bibnamefont
  {Vainshtein}},\ }\href {\doibase 10.1103/PhysRevD.50.2234} {\bibfield
  {journal} {\bibinfo  {journal} {Phys. Rev. D}\ }\textbf {\bibinfo {volume}
  {50}},\ \bibinfo {pages} {2234} (\bibinfo {year} {1994})},\ \Eprint
  {http://arxiv.org/abs/hep-ph/9402360} {arXiv:hep-ph/9402360} \BibitemShut
  {NoStop}%
\bibitem [{\citenamefont {Beneke}(1995)}]{Beneke:1994rs}%
  \BibitemOpen
  \bibfield  {author} {\bibinfo {author} {\bibfnamefont {M.}~\bibnamefont
  {Beneke}},\ }\href {\doibase 10.1016/0370-2693(94)01505-7} {\bibfield
  {journal} {\bibinfo  {journal} {Phys. Lett. B}\ }\textbf {\bibinfo {volume}
  {344}},\ \bibinfo {pages} {341} (\bibinfo {year} {1995})},\ \Eprint
  {http://arxiv.org/abs/hep-ph/9408380} {arXiv:hep-ph/9408380} \BibitemShut
  {NoStop}%
\bibitem [{\citenamefont {Bali}\ \emph {et~al.}(2013)\citenamefont {Bali},
  \citenamefont {Bauer}, \citenamefont {Pineda},\ and\ \citenamefont
  {Torrero}}]{Bali:2013pla}%
  \BibitemOpen
  \bibfield  {author} {\bibinfo {author} {\bibfnamefont {G.~S.}\ \bibnamefont
  {Bali}}, \bibinfo {author} {\bibfnamefont {C.}~\bibnamefont {Bauer}},
  \bibinfo {author} {\bibfnamefont {A.}~\bibnamefont {Pineda}}, \ and\ \bibinfo
  {author} {\bibfnamefont {C.}~\bibnamefont {Torrero}},\ }\href {\doibase
  10.1103/PhysRevD.87.094517} {\bibfield  {journal} {\bibinfo  {journal} {Phys.
  Rev. D}\ }\textbf {\bibinfo {volume} {87}},\ \bibinfo {pages} {094517}
  (\bibinfo {year} {2013})},\ \Eprint {http://arxiv.org/abs/1303.3279}
  {arXiv:1303.3279 [hep-lat]} \BibitemShut {NoStop}%
\bibitem [{\citenamefont {Ayala}\ \emph {et~al.}(2014)\citenamefont {Ayala},
  \citenamefont {Cveti\v{c}},\ and\ \citenamefont {Pineda}}]{Ayala:2014yxa}%
  \BibitemOpen
  \bibfield  {author} {\bibinfo {author} {\bibfnamefont {C.}~\bibnamefont
  {Ayala}}, \bibinfo {author} {\bibfnamefont {G.}~\bibnamefont {Cveti\v{c}}}, \
  and\ \bibinfo {author} {\bibfnamefont {A.}~\bibnamefont {Pineda}},\ }\href
  {\doibase 10.1007/JHEP09(2014)045} {\bibfield  {journal} {\bibinfo  {journal}
  {JHEP}\ }\textbf {\bibinfo {volume} {09}},\ \bibinfo {pages} {045} (\bibinfo
  {year} {2014})},\ \Eprint {http://arxiv.org/abs/1407.2128} {arXiv:1407.2128
  [hep-ph]} \BibitemShut {NoStop}%
\bibitem [{\citenamefont {Melnikov}\ and\ \citenamefont
  {Ritbergen}(2000)}]{Melnikov:2000qh}%
  \BibitemOpen
  \bibfield  {author} {\bibinfo {author} {\bibfnamefont {K.}~\bibnamefont
  {Melnikov}}\ and\ \bibinfo {author} {\bibfnamefont {T.~v.}\ \bibnamefont
  {Ritbergen}},\ }\href {\doibase 10.1016/S0370-2693(00)00507-4} {\bibfield
  {journal} {\bibinfo  {journal} {Phys. Lett. B}\ }\textbf {\bibinfo {volume}
  {482}},\ \bibinfo {pages} {99} (\bibinfo {year} {2000})},\ \Eprint
  {http://arxiv.org/abs/hep-ph/9912391} {arXiv:hep-ph/9912391} \BibitemShut
  {NoStop}%
\bibitem [{\citenamefont {Hoang}\ \emph {et~al.}(2018)\citenamefont {Hoang},
  \citenamefont {Jain}, \citenamefont {Lepenik}, \citenamefont {Mateu},
  \citenamefont {Preisser}, \citenamefont {Scimemi},\ and\ \citenamefont
  {Stewart}}]{Hoang:2017suc}%
  \BibitemOpen
  \bibfield  {author} {\bibinfo {author} {\bibfnamefont {A.~H.}\ \bibnamefont
  {Hoang}}, \bibinfo {author} {\bibfnamefont {A.}~\bibnamefont {Jain}},
  \bibinfo {author} {\bibfnamefont {C.}~\bibnamefont {Lepenik}}, \bibinfo
  {author} {\bibfnamefont {V.}~\bibnamefont {Mateu}}, \bibinfo {author}
  {\bibfnamefont {M.}~\bibnamefont {Preisser}}, \bibinfo {author}
  {\bibfnamefont {I.}~\bibnamefont {Scimemi}}, \ and\ \bibinfo {author}
  {\bibfnamefont {I.~W.}\ \bibnamefont {Stewart}},\ }\href {\doibase
  10.1007/JHEP04(2018)003} {\bibfield  {journal} {\bibinfo  {journal} {JHEP}\
  }\textbf {\bibinfo {volume} {04}},\ \bibinfo {pages} {003} (\bibinfo {year}
  {2018})},\ \Eprint {http://arxiv.org/abs/1704.01580} {arXiv:1704.01580
  [hep-ph]} \BibitemShut {NoStop}%
\bibitem [{\citenamefont {Chyla}(1992)}]{Chyla:1990na}%
  \BibitemOpen
  \bibfield  {author} {\bibinfo {author} {\bibfnamefont {J.}~\bibnamefont
  {Chyla}},\ }\href {\doibase 10.1007/BF01598424} {\bibfield  {journal}
  {\bibinfo  {journal} {Czech. J. Phys.}\ }\textbf {\bibinfo {volume} {42}},\
  \bibinfo {pages} {263} (\bibinfo {year} {1992})}\BibitemShut {NoStop}%
\bibitem [{\citenamefont {Cvetic}(2003)}]{Cvetic:2002qf}%
  \BibitemOpen
  \bibfield  {author} {\bibinfo {author} {\bibfnamefont {G.}~\bibnamefont
  {Cvetic}},\ }\href {\doibase 10.1103/PhysRevD.67.074022} {\bibfield
  {journal} {\bibinfo  {journal} {Phys. Rev. D}\ }\textbf {\bibinfo {volume}
  {67}},\ \bibinfo {pages} {074022} (\bibinfo {year} {2003})},\ \Eprint
  {http://arxiv.org/abs/hep-ph/0211226} {arXiv:hep-ph/0211226} \BibitemShut
  {NoStop}%
\bibitem [{\citenamefont {Ayala}\ \emph {et~al.}(2019)\citenamefont {Ayala},
  \citenamefont {Lobregat},\ and\ \citenamefont {Pineda}}]{Ayala:2019uaw}%
  \BibitemOpen
  \bibfield  {author} {\bibinfo {author} {\bibfnamefont {C.}~\bibnamefont
  {Ayala}}, \bibinfo {author} {\bibfnamefont {X.}~\bibnamefont {Lobregat}}, \
  and\ \bibinfo {author} {\bibfnamefont {A.}~\bibnamefont {Pineda}},\ }\href
  {\doibase 10.1103/PhysRevD.99.074019} {\bibfield  {journal} {\bibinfo
  {journal} {Phys. Rev. D}\ }\textbf {\bibinfo {volume} {99}},\ \bibinfo
  {pages} {074019} (\bibinfo {year} {2019})},\ \Eprint
  {http://arxiv.org/abs/1902.07736} {arXiv:1902.07736 [hep-th]} \BibitemShut
  {NoStop}%
\bibitem [{\citenamefont {Hoang}\ \emph {et~al.}(2008)\citenamefont {Hoang},
  \citenamefont {Jain}, \citenamefont {Scimemi},\ and\ \citenamefont
  {Stewart}}]{Hoang:2008yj}%
  \BibitemOpen
  \bibfield  {author} {\bibinfo {author} {\bibfnamefont {A.~H.}\ \bibnamefont
  {Hoang}}, \bibinfo {author} {\bibfnamefont {A.}~\bibnamefont {Jain}},
  \bibinfo {author} {\bibfnamefont {I.}~\bibnamefont {Scimemi}}, \ and\
  \bibinfo {author} {\bibfnamefont {I.~W.}\ \bibnamefont {Stewart}},\ }\href
  {\doibase 10.1103/PhysRevLett.101.151602} {\bibfield  {journal} {\bibinfo
  {journal} {Phys. Rev. Lett.}\ }\textbf {\bibinfo {volume} {101}},\ \bibinfo
  {pages} {151602} (\bibinfo {year} {2008})},\ \Eprint
  {http://arxiv.org/abs/0803.4214} {arXiv:0803.4214 [hep-ph]} \BibitemShut
  {NoStop}%
\bibitem [{\citenamefont {Hoang}\ \emph {et~al.}(2010)\citenamefont {Hoang},
  \citenamefont {Jain}, \citenamefont {Scimemi},\ and\ \citenamefont
  {Stewart}}]{Hoang:2009yr}%
  \BibitemOpen
  \bibfield  {author} {\bibinfo {author} {\bibfnamefont {A.~H.}\ \bibnamefont
  {Hoang}}, \bibinfo {author} {\bibfnamefont {A.}~\bibnamefont {Jain}},
  \bibinfo {author} {\bibfnamefont {I.}~\bibnamefont {Scimemi}}, \ and\
  \bibinfo {author} {\bibfnamefont {I.~W.}\ \bibnamefont {Stewart}},\ }\href
  {\doibase 10.1103/PhysRevD.82.011501} {\bibfield  {journal} {\bibinfo
  {journal} {Phys. Rev. D}\ }\textbf {\bibinfo {volume} {82}},\ \bibinfo
  {pages} {011501} (\bibinfo {year} {2010})},\ \Eprint
  {http://arxiv.org/abs/0908.3189} {arXiv:0908.3189 [hep-ph]} \BibitemShut
  {NoStop}%
\bibitem [{\citenamefont {Benitez-Rathgeb}\ \emph {et~al.}(2022)\citenamefont
  {Benitez-Rathgeb}, \citenamefont {Boito}, \citenamefont {Hoang},\ and\
  \citenamefont {Jamin}}]{Benitez-Rathgeb:2022yqb}%
  \BibitemOpen
  \bibfield  {author} {\bibinfo {author} {\bibfnamefont {M.~A.}\ \bibnamefont
  {Benitez-Rathgeb}}, \bibinfo {author} {\bibfnamefont {D.}~\bibnamefont
  {Boito}}, \bibinfo {author} {\bibfnamefont {A.~H.}\ \bibnamefont {Hoang}}, \
  and\ \bibinfo {author} {\bibfnamefont {M.}~\bibnamefont {Jamin}},\ }\href
  {\doibase 10.1007/JHEP07(2022)016} {\bibfield  {journal} {\bibinfo  {journal}
  {JHEP}\ }\textbf {\bibinfo {volume} {07}},\ \bibinfo {pages} {016} (\bibinfo
  {year} {2022})},\ \Eprint {http://arxiv.org/abs/2202.10957} {arXiv:2202.10957
  [hep-ph]} \BibitemShut {NoStop}%
\bibitem [{\citenamefont {Liu}(2024)}]{Liu:2024omb}%
  \BibitemOpen
  \bibfield  {author} {\bibinfo {author} {\bibfnamefont {Y.}~\bibnamefont
  {Liu}},\ }\href {\doibase 10.1007/JHEP09(2024)093} {\bibfield  {journal}
  {\bibinfo  {journal} {JHEP}\ }\textbf {\bibinfo {volume} {09}},\ \bibinfo
  {pages} {093} (\bibinfo {year} {2024})},\ \Eprint
  {http://arxiv.org/abs/2403.06787} {arXiv:2403.06787 [hep-th]} \BibitemShut
  {NoStop}%
\bibitem [{\citenamefont {Liu}\ and\ \citenamefont {Chen}(2021)}]{Liu:2020rqi}%
  \BibitemOpen
  \bibfield  {author} {\bibinfo {author} {\bibfnamefont {W.-Y.}\ \bibnamefont
  {Liu}}\ and\ \bibinfo {author} {\bibfnamefont {J.-W.}\ \bibnamefont {Chen}},\
  }\href {\doibase 10.1103/PhysRevD.104.094501} {\bibfield  {journal} {\bibinfo
   {journal} {Phys. Rev. D}\ }\textbf {\bibinfo {volume} {104}},\ \bibinfo
  {pages} {094501} (\bibinfo {year} {2021})},\ \Eprint
  {http://arxiv.org/abs/2010.06623} {arXiv:2010.06623 [hep-ph]} \BibitemShut
  {NoStop}%
\bibitem [{\citenamefont {Beneke}\ \emph {et~al.}(2023)\citenamefont {Beneke},
  \citenamefont {Finauri}, \citenamefont {Vos},\ and\ \citenamefont
  {Wei}}]{Beneke:2023nmj}%
  \BibitemOpen
  \bibfield  {author} {\bibinfo {author} {\bibfnamefont {M.}~\bibnamefont
  {Beneke}}, \bibinfo {author} {\bibfnamefont {G.}~\bibnamefont {Finauri}},
  \bibinfo {author} {\bibfnamefont {K.~K.}\ \bibnamefont {Vos}}, \ and\
  \bibinfo {author} {\bibfnamefont {Y.}~\bibnamefont {Wei}},\ }\href {\doibase
  10.1007/JHEP09(2023)066} {\bibfield  {journal} {\bibinfo  {journal} {JHEP}\
  }\textbf {\bibinfo {volume} {09}},\ \bibinfo {pages} {066} (\bibinfo {year}
  {2023})},\ \Eprint {http://arxiv.org/abs/2305.06401} {arXiv:2305.06401
  [hep-ph]} \BibitemShut {NoStop}%
\bibitem [{\citenamefont {Collins}(2013)}]{Collins:2011zzd}%
  \BibitemOpen
  \bibfield  {author} {\bibinfo {author} {\bibfnamefont {J.}~\bibnamefont
  {Collins}},\ }\href {\doibase 10.1017/9781009401845} {\emph {\bibinfo {title}
  {{Foundations of perturbative QCD}}}},\ Vol.~\bibinfo {volume} {32}\
  (\bibinfo  {publisher} {Cambridge University Press},\ \bibinfo {year}
  {2013})\BibitemShut {NoStop}%
\bibitem [{\citenamefont {Jain}\ \emph {et~al.}(2008)\citenamefont {Jain},
  \citenamefont {Scimemi},\ and\ \citenamefont {Stewart}}]{Jain:2008gb}%
  \BibitemOpen
  \bibfield  {author} {\bibinfo {author} {\bibfnamefont {A.}~\bibnamefont
  {Jain}}, \bibinfo {author} {\bibfnamefont {I.}~\bibnamefont {Scimemi}}, \
  and\ \bibinfo {author} {\bibfnamefont {I.~W.}\ \bibnamefont {Stewart}},\
  }\href {\doibase 10.1103/PhysRevD.77.094008} {\bibfield  {journal} {\bibinfo
  {journal} {Phys. Rev. D}\ }\textbf {\bibinfo {volume} {77}},\ \bibinfo
  {pages} {094008} (\bibinfo {year} {2008})},\ \Eprint
  {http://arxiv.org/abs/0801.0743} {arXiv:0801.0743 [hep-ph]} \BibitemShut
  {NoStop}%
\bibitem [{\citenamefont {Korchemskaya}\ and\ \citenamefont
  {Korchemsky}(1992)}]{Korchemskaya:1992je}%
  \BibitemOpen
  \bibfield  {author} {\bibinfo {author} {\bibfnamefont {I.~A.}\ \bibnamefont
  {Korchemskaya}}\ and\ \bibinfo {author} {\bibfnamefont {G.~P.}\ \bibnamefont
  {Korchemsky}},\ }\href {\doibase 10.1016/0370-2693(92)91895-G} {\bibfield
  {journal} {\bibinfo  {journal} {Phys. Lett. B}\ }\textbf {\bibinfo {volume}
  {287}},\ \bibinfo {pages} {169} (\bibinfo {year} {1992})}\BibitemShut
  {NoStop}%
\bibitem [{\citenamefont {Henn}\ \emph {et~al.}(2020)\citenamefont {Henn},
  \citenamefont {Korchemsky},\ and\ \citenamefont
  {Mistlberger}}]{Henn:2019swt}%
  \BibitemOpen
  \bibfield  {author} {\bibinfo {author} {\bibfnamefont {J.~M.}\ \bibnamefont
  {Henn}}, \bibinfo {author} {\bibfnamefont {G.~P.}\ \bibnamefont
  {Korchemsky}}, \ and\ \bibinfo {author} {\bibfnamefont {B.}~\bibnamefont
  {Mistlberger}},\ }\href {\doibase 10.1007/JHEP04(2020)018} {\bibfield
  {journal} {\bibinfo  {journal} {JHEP}\ }\textbf {\bibinfo {volume} {04}},\
  \bibinfo {pages} {018} (\bibinfo {year} {2020})},\ \Eprint
  {http://arxiv.org/abs/1911.10174} {arXiv:1911.10174 [hep-th]} \BibitemShut
  {NoStop}%
\bibitem [{\citenamefont {von Manteuffel}\ \emph {et~al.}(2020)\citenamefont
  {von Manteuffel}, \citenamefont {Panzer},\ and\ \citenamefont
  {Schabinger}}]{vonManteuffel:2020vjv}%
  \BibitemOpen
  \bibfield  {author} {\bibinfo {author} {\bibfnamefont {A.}~\bibnamefont {von
  Manteuffel}}, \bibinfo {author} {\bibfnamefont {E.}~\bibnamefont {Panzer}}, \
  and\ \bibinfo {author} {\bibfnamefont {R.~M.}\ \bibnamefont {Schabinger}},\
  }\href {\doibase 10.1103/PhysRevLett.124.162001} {\bibfield  {journal}
  {\bibinfo  {journal} {Phys. Rev. Lett.}\ }\textbf {\bibinfo {volume} {124}},\
  \bibinfo {pages} {162001} (\bibinfo {year} {2020})},\ \Eprint
  {http://arxiv.org/abs/2002.04617} {arXiv:2002.04617 [hep-ph]} \BibitemShut
  {NoStop}%
\bibitem [{\citenamefont {Grozin}(2023{\natexlab{b}})}]{Grozin:2022umo}%
  \BibitemOpen
  \bibfield  {author} {\bibinfo {author} {\bibfnamefont {A.}~\bibnamefont
  {Grozin}},\ }\href {\doibase 10.1142/S0217751X23300041} {\bibfield  {journal}
  {\bibinfo  {journal} {Int. J. Mod. Phys.}\ }\textbf {\bibinfo {volume} {38}}
  (\bibinfo {year} {2023}{\natexlab{b}}),\ 10.1142/S0217751X23300041},\ \Eprint
  {http://arxiv.org/abs/2212.05290} {arXiv:2212.05290 [hep-ph]} \BibitemShut
  {NoStop}%
\bibitem [{\citenamefont {Moch}\ \emph
  {et~al.}(2005{\natexlab{b}})\citenamefont {Moch}, \citenamefont
  {Vermaseren},\ and\ \citenamefont {Vogt}}]{Moch:2005id}%
  \BibitemOpen
  \bibfield  {author} {\bibinfo {author} {\bibfnamefont {S.}~\bibnamefont
  {Moch}}, \bibinfo {author} {\bibfnamefont {J.~A.~M.}\ \bibnamefont
  {Vermaseren}}, \ and\ \bibinfo {author} {\bibfnamefont {A.}~\bibnamefont
  {Vogt}},\ }\href {\doibase 10.1088/1126-6708/2005/08/049} {\bibfield
  {journal} {\bibinfo  {journal} {JHEP}\ }\textbf {\bibinfo {volume} {08}},\
  \bibinfo {pages} {049} (\bibinfo {year} {2005}{\natexlab{b}})},\ \Eprint
  {http://arxiv.org/abs/hep-ph/0507039} {arXiv:hep-ph/0507039} \BibitemShut
  {NoStop}%
\bibitem [{\citenamefont {Korchemsky}\ and\ \citenamefont
  {Marchesini}(1993)}]{Korchemsky:1992xv}%
  \BibitemOpen
  \bibfield  {author} {\bibinfo {author} {\bibfnamefont {G.~P.}\ \bibnamefont
  {Korchemsky}}\ and\ \bibinfo {author} {\bibfnamefont {G.}~\bibnamefont
  {Marchesini}},\ }\href {\doibase 10.1016/0550-3213(93)90167-N} {\bibfield
  {journal} {\bibinfo  {journal} {Nucl. Phys. B}\ }\textbf {\bibinfo {volume}
  {406}},\ \bibinfo {pages} {225} (\bibinfo {year} {1993})},\ \Eprint
  {http://arxiv.org/abs/hep-ph/9210281} {arXiv:hep-ph/9210281} \BibitemShut
  {NoStop}%
\bibitem [{\citenamefont {Br\"user}\ \emph {et~al.}(2020)\citenamefont
  {Br\"user}, \citenamefont {Liu},\ and\ \citenamefont
  {Stahlhofen}}]{Bruser:2019yjk}%
  \BibitemOpen
  \bibfield  {author} {\bibinfo {author} {\bibfnamefont {R.}~\bibnamefont
  {Br\"user}}, \bibinfo {author} {\bibfnamefont {Z.~L.}\ \bibnamefont {Liu}}, \
  and\ \bibinfo {author} {\bibfnamefont {M.}~\bibnamefont {Stahlhofen}},\
  }\href {\doibase 10.1007/JHEP03(2020)071} {\bibfield  {journal} {\bibinfo
  {journal} {JHEP}\ }\textbf {\bibinfo {volume} {03}},\ \bibinfo {pages} {071}
  (\bibinfo {year} {2020})},\ \Eprint {http://arxiv.org/abs/1911.04494}
  {arXiv:1911.04494 [hep-ph]} \BibitemShut {NoStop}%
\bibitem [{\citenamefont {Moult}\ \emph {et~al.}(2022)\citenamefont {Moult},
  \citenamefont {Zhu},\ and\ \citenamefont {Zhu}}]{Moult:2022xzt}%
  \BibitemOpen
  \bibfield  {author} {\bibinfo {author} {\bibfnamefont {I.}~\bibnamefont
  {Moult}}, \bibinfo {author} {\bibfnamefont {H.~X.}\ \bibnamefont {Zhu}}, \
  and\ \bibinfo {author} {\bibfnamefont {Y.~J.}\ \bibnamefont {Zhu}},\ }\href
  {\doibase 10.1007/JHEP08(2022)280} {\bibfield  {journal} {\bibinfo  {journal}
  {JHEP}\ }\textbf {\bibinfo {volume} {08}},\ \bibinfo {pages} {280} (\bibinfo
  {year} {2022})},\ \Eprint {http://arxiv.org/abs/2205.02249} {arXiv:2205.02249
  [hep-ph]} \BibitemShut {NoStop}%
\end{thebibliography}%

\end{document}